%% file: wrap.tex
\documentclass[paper=a4, fontsize=12pt]{book}

\pdfoutput=1

\usepackage[top=2.9cm, bottom=2.9cm, outer=2.9cm, inner=3.4cm]{geometry}

\usepackage{minitoc} 
\usepackage[round,authoryear,comma,numbers]{natbib} 
\usepackage[bottom]{footmisc}
\usepackage{enumerate}
\usepackage[english]{babel}  
\usepackage{caption}
\usepackage{subcaption}
\usepackage[utf8]{inputenc}  
\usepackage[T1]{fontenc}
\usepackage{graphicx,import}
\usepackage{url,tabularx,array}
\usepackage{sectsty} 
\usepackage[normalem]{ulem}
\usepackage{xcolor}
\usepackage{hyperref}
\hypersetup{
     colorlinks   = true,
     linkcolor    = red,
     urlcolor  = magenta,
     citecolor = blue,
     anchorcolor = green	
}
\usepackage{bbold}
\usepackage{transparent}
\usepackage{rotating}
\usepackage{wrapfig}
\usepackage{setspace}
\usepackage{floatrow}
\usepackage{appendix}
\usepackage{footnote}
\usepackage{tablefootnote}
\usepackage[flushleft]{threeparttable}

\usepackage{stmaryrd}

\newcommand\abstractname{Abstract}  


\newcommand{\colorulem}[1][black]{\bgroup
\ifdim\ULdepth=\maxdimen\settodepth\ULdepth{(j}\advance\ULdepth.4pt\fi
\markoverwith{\kern0em\vtop{\kern\ULdepth {\color{#1}\hrule width .4em}}\kern0em}\ULon}
\usepackage{color} 
\usepackage{cancel}
\definecolor{heading}{rgb}{0.5,1,0}
\usepackage{fourier} 
\usepackage{amsmath,amsfonts,amsthm,amssymb} 
\usepackage{bm}

\usepackage{fancyhdr} 
\usepackage{xspace}
\usepackage[absolute]{textpos} 

\newcommand{\sgx}{Sg\textsc{xb}\xspace}
\newcommand*{\eg}{e.g.\@\xspace}
\newcommand*{\ie}{i.e.\@\xspace}
\newcommand*{\aka}{a.k.a.\@\xspace}
\newcommand*{\rhs}{r.h.s.\@\xspace}
\newcommand*{\lhs}{l.h.s.\@\xspace}
\newcommand*{\rlof}{\textsc{rlof}\@\xspace}
\newcommand*{\lmxb}{\textsc{lmxb}\@\xspace}
\newcommand*{\hmxb}{\textsc{hmxb}\@\xspace}
\newcommand*{\bexb}{Be\textsc{xb}\@\xspace}
\newcommand*{\sfxt}{\textsc{sfxt}\@\xspace}
\newcommand*{\agb}{\textsc{agb}\@\xspace}
\newcommand*{\cemp}{\textsc{cemp}\@\xspace}
\newcommand*{\bh}{\textsc{bh}\@\xspace}
\newcommand*{\ns}{\textsc{ns}\@\xspace}
\newcommand*{\whd}{\textsc{wd}\@\xspace}
\newcommand*{\msun}{$M_{\odot}$\@\xspace}
\newcommand*{\rsun}{$R_{\odot}$\@\xspace}

\newcommand*{\bhl}{\textsc{bhl}\@\xspace}
\newcommand*{\gr}{\textsc{gr}\@\xspace}
\newcommand*{\sr}{\textsc{sr}\@\xspace}
\renewcommand{\d}{\text{d}}
\newcommand*{\cak}{\textsc{cak}\@\xspace}
\newcommand*{\wlr}{\textsc{wlr}\@\xspace}
\newcommand*{\cpu}{\textsc{cpu}\@\xspace}
\newcommand*{\cpus}{\textsc{cpu}s\@\xspace}
\newcommand*{\vac}{\texttt{VAC}\@\xspace}
\newcommand*{\cfl}{\textsc{cfl}\@\xspace}
\newcommand*{\amr}{\textsc{amr}\@\xspace}
\newcommand*{\ats}{\textsc{ats}\@\xspace}
\newcommand*{\tde}{\textsc{tde}\@\xspace}

\newcommand*{\eos}{e.o.s.\@\xspace}
\newcommand*{\hd}{\textsc{hd}\@\xspace}

\newcommand*{\kms}{km$\cdot$s$^{-1}$\@\xspace}

\newcommand*{\INTEGRAL}{\textsc{integral}\@\xspace}

\begin{document}

\newgeometry{top=0.9cm, bottom=0.9cm, outer=2cm, inner=2cm}
\begin{titlepage}

\begin{figure}
\begin{textblock*}{3.5cm}(2mm,3mm) 
\includegraphics[scale=0.9]{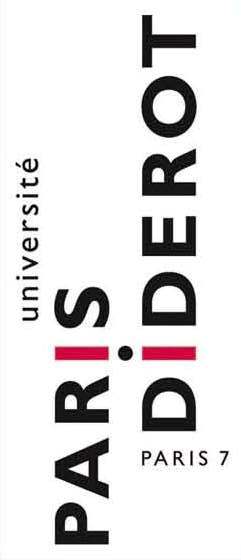} 
\end{textblock*} 
\begin{textblock*}{21cm}(2mm,3mm) 
\includegraphics[scale=0.5]{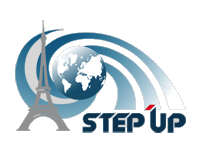} 
\end{textblock*} 
\begin{textblock*}{37cm}(2mm,3mm) 
\includegraphics[scale=0.11]{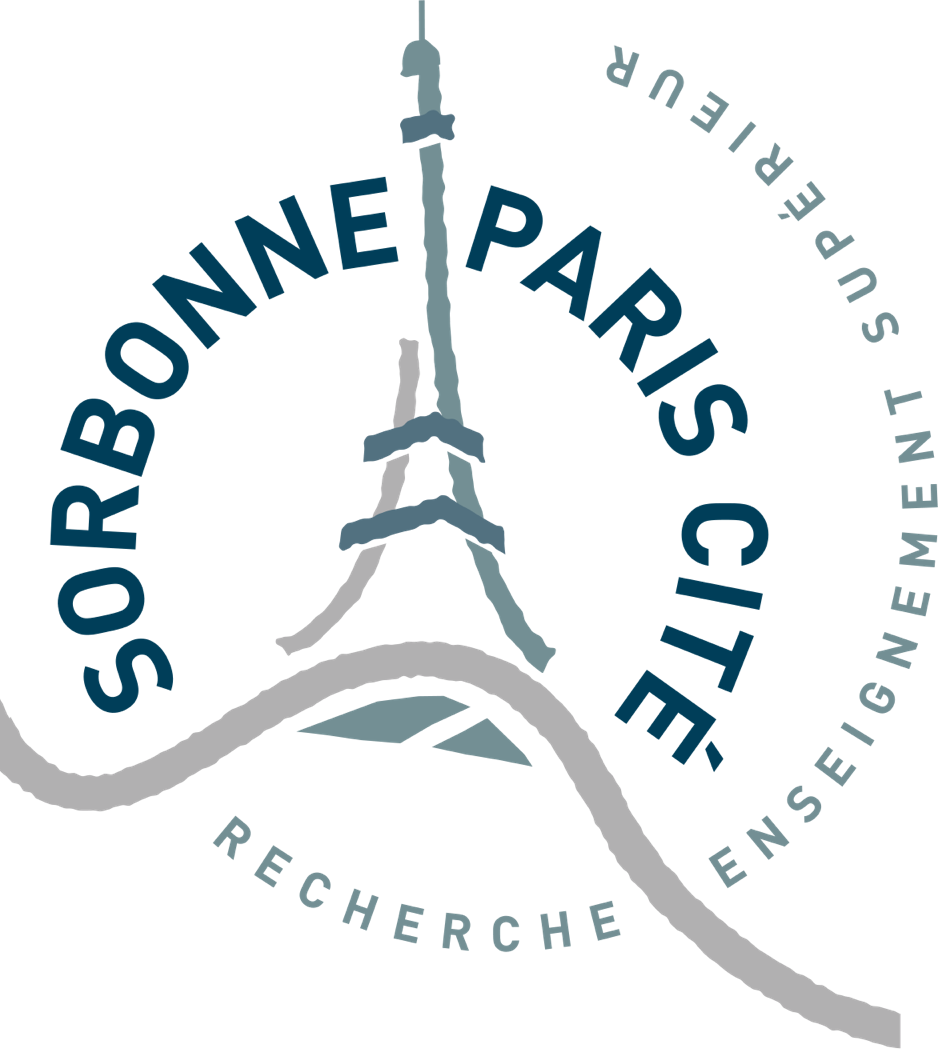} 
\end{textblock*} 
\end{figure}
\vspace*{2.4cm}
\centering
\textsc{\Large Universit\'e Paris 7 Diderot - Sorbonne Paris Cit\'e}\\
\vspace*{0.5cm}
\large{Ecole Doctorale 560 - STEP'UP\\``Sciences de la Terre et de l'Environnement\\ et Physique de l'Univers de Paris''}\\
\vspace{0.7cm}
\textsc{\Large \textsc{ Thèse de Doctorat}}\\
\large{Astrophysique}\\
\vspace{0.8cm}

\hrulefill

\vspace{1cm}
\textbf{\huge Wind accretion onto compact objects}\\
\vspace{0.5cm}

\hrulefill

\vspace{0.5cm}
par\\
\vspace{0.25cm}
{\Large \bfseries Ileyk} \textsc{\Large \bfseries El Mellah}\\
\vspace{0.75cm}
\large{pour l'obtention du titre de}\\
\vspace{0.25cm}
\textsc{\large Docteur de l'Universit\'e Paris 7 Diderot - Sorbonne Paris Cit\'e}\\
\vspace{1.125cm}
Thèse dirigée par Andrea \textsc{Goldwurm} \& Fabien \textsc{Casse}\\
au laboratoire AstroParticule et Cosmologie \\
\vspace{1.125cm}
pr\'esent\'ee et soutenue publiquement le 7 Septembre 2016 devant le jury compos\'e de :
\begin{center}
\noindent  
\begin{tabular}{llcl}
      \textit{Pr\'esident :}	& Pr St\'ephane \textsc{Corbel}		& - & AIM CEA Saclay\\
      \textit{Directeur :}	& Dr Andrea \textsc{Goldwurm}		& - & APC Paris 7\\
      \textit{Co-directeur :}	& Dr Fabien \textsc{Casse}		& - & APC Paris 7\\
      \textit{Rapporteurs :}	& Pr Maximilian \textsc{Ruffert}		& - & University of Edinburgh\\
				& Dr Guillaume \textsc{Dubus}		& - & IPAG\\
      \textit{Examinateurs :} & Dr Thierry \textsc{Foglizzo}   & - & AIM CEA Saclay\\
      		& Pr Rony \textsc{Keppens}	& - & KU Leuven\\
\end{tabular}
\end{center}

%
\end{titlepage}
\restoregeometry

\dominitoc

\pagestyle{plain}


\chapter*{Remerciements}
\vspace*{-0.8cm}
Je tiens \`a exprimer mes plus vifs remerciements \`a Fabien Casse et Andrea Goldwurm qui furent pour moi des directeurs de th\`ese attentifs et disponibles malgr\'e leurs nombreuses responsabilit\'es. Leur comp\'etence, leur rigueur scientifique et leur clairvoyance m’ont beaucoup appris. Ils ont \'et\'e et resteront des moteurs de mon travail de chercheur.\\

I thank Guillaume Dubus and Maximilian Ruffert for accepting to be the rapporteurs of this work and to review my manuscript. Their insightful comments played a major role in the improvements brought to this document. Merci \'egalement aux examinateurs pour leur participation \`a mon jury de th\`ese : Rony Keppens pour les enrichissants s\'ejours pass\'es \`a Leuven \`a explorer les arcanes du code, St\'ephane Corbel pour son mentorat tout au long de ma th\`ese et Thierry Foglizzo pour avoir raviv\'e ma curiosit\'e physique en m'ayant accord\'e d'inestimables s\'eances de r\'eflexion.\\

I would like to express my deepest gratitude to Saul Rappaport for having granted me a decisive first experience in Research. This year at \textsc{mit} has been the funding step of my career thanks to the incredibly stimulating environment I have found there. Thank you for having been such an inspiring figure Saul, the intelligence, culture and imagination which fuel your work are inexhaustible sources of motivation.\\

Mes remerciements vont aussi \`a Peggy Varni\`ere et Fabrice Dodu, pour leur soutien tant professionel qu'amical au quotidien. Le num\'erique est un art retors au large duquel je n'aurais pas manqu\'e de m'\'echouer sans leur aide bienvenue.\\

C'est m\^u d'une gratitude profonde que je me tourne maintenant vers celles et ceux qui m'ont offert leurs conseils et assistance durant ma th\`ese : J\'er\^ome Rodriguez, Zakaria Meliani, H\'eloïse M\'eheut, Allard Jan Van Marle, R\'emi Kazeroni, Fr\'ed\'eric Vincent, R\'egis Terrier, Alexis Coleiro, Sasha Tchekhovskoy, Geoffroy Lesur, Beno\^it Commer\c con et de nombreuses autres personnes que je m'excuse de ne pas citer ici mais \`a qui je souhaite dire combien je suis conscient de tout ce que je leur dois. Merci aussi aux relecteurs attentifs du pr\'esent manuscrit pour le temps pr\'ecieux qu'ils m'ont \'epargn\'e en traquant les coquilles et autres fautes de frappe.\\

Merci aux personnels techniques, administratifs et s\^uret\'e du bâtiment Condorcet pour leur pr\'esence, leur r\'eactivit\'e et leur amabilit\'e, \`a Martine Piochaud pour sa bonne humeur, \`a Brahim Mazzar pour les \'echanges conviviaux autour de caf\'es nocturnes, \`a Nabouana Ciss\'e pour ses ritournelles matinales, \`a Fran\c cois Carr\'e, Ribeiro Lima Reinaldo, Jo\"{e}lle Taieb, Ludovic Davila, Sabine Tesson, pour m'avoir tir\'e d'emb\^uches administratives inextricables.\\

Un grand merci \`a tous les formidables th\'esards et post-docs, pass\'es et pr\'esents, pour avoir fait de la vie sociale \`a l'\textsc{apc} un plaisir sans cesse renouvel\'e. Les moments de complicit\'e pass\'es ensemble n'ont pas \'et\'e pour peu dans ma capacit\'e \`a mener \`a bien ce travail de th\`ese.\\

Pour leur soutien ind\'efectible et leur compr\'ehension sans borne pour mes absences et mes \'ecarts, je remercie tout particuli\`erement mes amis et mes proches sans qui rien de tout cela n'aurait \'et\'e possible. Des regards bienveillants aux encouragements, je ne sais par quoi commencer ; je n'oublie rien de la tendresse et de la profondeur de nos relations sans lesquels ma vie n'aurait pas la m\^eme saveur.\\

Et \`a ma bonne \'etoile, pour m'avoir montr\'e le chemin, merci.\\

\hypersetup{linkcolor=black}

\tableofcontents

\hypersetup{linkcolor=red}


\pagestyle{fancy}
\chapter*{Avant-propos}
\addcontentsline{toc}{part}{Avant-propos}
\adjustmtc 

Que les derni\`eres lueurs du jour s'\'evanouissent sous l'horizon et voil\`a que surgissent de l'obscurit\'e des myriades d'\'eclats qui jettent sur le monde une clart\'e nouvelle. Si d’aventure quelques marginaux s'\'egarent hors des sentiers battus, l’immense majorit\'e de ces vacillantes lanternes se pla\^it \`a embarquer, chaque soir, pour un man\`ege c\'eleste \`a la m\'ecanique bien rod\'ee. L’apparente rotation solide de la voûte supra-planétaire a longtemps induit en erreur les penseurs qui s’imaginaient le ciel perc\'e de trous donnant sur une immensit\'e baign\'ee de lumi\`ere. Il fallut l’audace de quelques uns pour donner du relief \`a ce dôme d’arri\`ere-plan. Les mesures de parallaxe prenaient soudain tout leur sens et suspendaient chaque \'etoile dans un espace dont les dimensions frayaient avec l’infini. A la grande toile passive succ\'edait un vide qui insultait tant le finalisme que l’anthropocentrisme : si les \'etoiles \'etaient anim\'ees d’une existence propre, peut-\^etre n’\'etaient-elles pas si diff\'erentes de ce Soleil qui, d\'ej\`a, avait ravi \`a la Terre sa centralit\'e. L’id\'ee d’une pluralit\'e des mondes f\^it son chemin mais les distances en jeu moquaient nos difficult\'es \`a se repr\'esenter l’extension du microcosme solaire. Dans les ann\'ees 1830, Auguste Comte prend \`a t\'emoin les \'etoiles pour hi\'erarchiser l'\'epist\'emologie positiviste en une physique universelle et fondamentale d'un c\^ot\'e et une physique incarn\'ee de l'autre :

\begin{quote}
"Les ph\'enom\`enes astronomiques \'etant les plus g\'en\'eraux, les plus simples, les plus abstraits de tous, c'est \'evidemment par leur \'etude que doit commencer la philosophie naturelle, puisque les lois auxquelles ils sont assujettis influent sur celles de tous les autres ph\'enom\`enes, dont elles-m\^emes sont, au contraire, essentiellement ind\'ependantes. [...] lorsqu'on analyse le ph\'enom\`ene terrestre le plus simple, non seulement en prenant un ph\'enom\`ene chimique, mais en choisissant m\^eme un ph\'enom\`ene purement m\'ecanique, on le trouve constamment plus compos\'e que le ph\'enom\`ene c\'eleste le plus compliqu\'e. [...] Une telle consid\'eration montre clairement combien il est indispensable de s\'eparer nettement la physique c\'eleste et la physique terrestre [...]."
\end{quote}

L’on peut conna\^itre le c\oe{}ur des hommes\footnote{La physique sociale que Comte d\'eveloppe jette les bases d’une sociologie m\'ecaniste.} mais pas la nature intime des \'etoiles. Quelques d\'ecennies plus tard, l’essor de la spectroscopie r\'efutera cette malheureuse pr\'ediction. Les \'etoiles se voient par\'ees de traits propres ; \`a chacune son spectre, sa signature dans laquelle se cachent les affres d’une histoire singuli\`ere. Car oui, le si\`ecle qui vit na\^itre l’Histoire\footnote{L’on pense ici tant \`a la naissance de l’historicisme qu’\`a l’engouement que suscitent les vestiges expos\'es dans les mus\'ees qui ravissent la vedette aux cabinets de curiosit\'es des encyclop\'edistes du XVIII$^{\text{\`eme}}$.} interroge la permanence de ces repr\'esentants d\'echus d’un sacr\'e \`a l’abri de la corruption. Pour la premi\`ere fois, l’on se demande s\'erieusement de quel bois se nourrissent ces bûchers c\'elestes. Alors que la seconde R\'evolution industrielle souligne l’importance strat\'egique des \'energies fossiles, l’opulence des astres suscite des convoitises. Comment donc de si prodigieuses luminosit\'es peuvent-elles \^etres entretenues sur des \'echelles de temps dont la mesure \'echappe \`a nos calendriers? Avant m\^eme que la Physique nucl\'eaire n’offre une r\'eponse \`a cette question germe l’id\'ee sublime d’un contournement de la difficult\'e chronologique : si l’hypoth\`ese ergodique s’est montr\'ee si prolifique en r\'eduisant les moyennes temporelles \`a des moyennes d’ensemble, pourquoi ne pas voir derri\`ere la diversit\'e des \'etoiles la manifestation d’une temporalit\'e? Les diff\'erences observ\'ees seraient alors \`a mettre sur le compte d’âges diff\'erents et, mis bout \`a bout, les instantan\'es dessineraient l’histoire d’une \'etoile qui recouvrerait par l\`a-m\^eme son universalit\'e. Si le principe fait flor\`es, il r\'ev\`ele aussi une s\'erie d’\'ecarts syst\'ematiques, comme autant de trappes par lesquelles se d\'erobe de nouveau l’\'etoile pour laisser place \`a des familles d’\'etoiles\footnote{Via le diagramme d’Hertzprung-Russel et la classification en types spectraux.}. Une fois coupl\'ees aux d\'ecouvertes sur les particules et leurs int\'eractions avec les radiations se tisse la premi\`ere trame de l’histoire des astres en fonction de leur masse initiale. L’exploit scientifique acte la naissance de l’Astrophysique.

La r\'econciliation des physiques c\'eleste et terrestre inspire un foisonnement de travaux sur la multitude des corps astraux, d’autant qu’avec la M\'ecanique Quantique et la Relativit\'e G\'en\'erale s’ouvrent des portes vers des paysages insoupçonn\'es. Pendant que la Cosmologie prend son envol, les astrophysiciens s’interrogent sur le reliquat que laisse une \'etoile consomm\'ee. Une croisi\`ere et un coup de g\'enie plus tard, la naine blanche fait son entr\'ee au panth\'eon des corps compacts. Elle inaugure la famille des vestiges stellaires \`a laquelle viendront se joindre \'etoiles \`a neutrons et trous noirs, deux autres compagnons de route \`a la densit\'e d\'efiant tout concurrence. Si les naines blanches se signalent rapidement dans les observations, les \'etoiles \`a neutrons et trous noirs laissent longtemps la communaut\'e perplexe. Comment des objets \`a l’extension spatiale si r\'eduite voire raval\'ee pourraient-ils jamais \^etre observ\'es? C’\'etait sans compter l’impact du transfert de masse dans les syst\`emes binaires et la prodigieuse efficacit\'e du processus d’accr\'etion sur un objet compact pour convertir l’\'energie potentielle gravitationnelle du flot en rayons X\footnote{La d\'ecouverte concomitante des pulsars et des quasars apporte aussi de l’eau au moulin compact.}. Les progr\`es de l’astronomie haute \'energie r\'ev\`elent les premi\`eres sources extra-solaires dans les ann\'ees 1960 et une d\'ecennie plus tard sont jet\'ees les bases th\'eoriques de l’accr\'etion sur objet compact : l’existence des \'etoiles \`a neutrons est ent\'erin\'ee et celle des trous noirs est s\'erieusement envisag\'ee. \\

\begin{center}
\rule{0.5\textwidth}{.4pt}
\end{center}
\vspace*{0.5cm}

L’\'epop\'ee astrale ne s’arr\^ete pourtant pas l\`a. Bien que l’\'evolution stellaire suive les tendances esquiss\'ees par les mod\`eles, les observations contemporaines soulignent des anomalies syst\'ematiques \`a l’\'epreuve de la th\'eorie. En particulier, le rôle de la binarit\'e dans le cheminement d’une \'etoile, de l’ignition des r\'eactions thermonucl\'eaires en son c\oe{}ur \`a son existence compacte, reste un sujet d’\'etude. Les binaires X offrent un pr\'ecieux instantan\'e de cette \'evolution binôm\'ee. La valse m\'elancolique qui s’y danse entre une \'etoile et feu son compagnon r\'eduit \`a l’\'etat d’\'etoile \`a neutrons ou de trou noir, est la sc\`ene privil\'egi\'ee d’un festin gargantuesque. En lieu et place des transferts de masse classiques entre \'etoiles, l’on trouve un flot qui s’enfonce loin dans le champ de pesanteur de l’accr\'eteur compact avant d’y rencontrer qui d’une magn\'etosph\`ere, qui d’un horizon des \'ev\`enements. L’\'energie potentielle ainsi perdue sera convertie d’abord en \'energie cin\'etique puis possiblement en radiations responsables des spectres observ\'es. 

Seulement, l\`a encore, tous les transferts de masse ne se valent pas. La source de mati\`ere a longtemps \'et\'e incrimin\'ee pour expliquer la diversit\'e des profils d’accr\'etion, menant \`a une classification en binaires X de faible et forte masse, en r\'ef\'erence \`a la bimodalit\'e des mensurations stellaires observ\'ees dans ces syst\`emes : d’un côt\'e des \'etoiles de types spectraux F ou plus tardif et de l’autre, des \'etoiles de plusieurs masses solaires de type O ou B, la premi\`ere population \'etant associ\'ee \`a des p\'eriodes orbitales nettement inf\'erieures \`a la seconde. L’on peut montrer qu’une \'etoile est d’autant plus susceptible de d\'eposer directement sa mati\`ere dans le puits de potentiel de son compagnon compact que la p\'eriode orbitale est faible. En outre, si la situation venait \`a se pr\'esenter dans une configuration où le donneur stellaire est plus massif que l’accr\'eteur compact (g\'en\'eralement de masse inf\'erieure \`a 2\msun), le transfert serait g\'en\'eralement instable. Ces deux remarques tendent \`a immuniser les \'etoiles massives contre l’app\'etit vorace du vestige compact alors que dans les binaires X de faible masse, la mati\`ere stellaire s’\'ecoule volontiers le long d’un canal \'etroit et vient former un disque mince autour de l’accr\'eteur. Le gaz y spirale lentement en s’\'echauffant sous l’action d’une viscosit\'e turbulente et atteint des temp\'eratures dantesques associ\'ees \`a une \'emission X. 

Pourtant, l’observation m\^eme de binaires X où l’\'etoile compagnon est une \'etoile de forte masse trop profond\'ement enfouie dans son champ de pesanteur pour que l’objet compact ne sirote la mati\`ere \`a m\^eme la photosph\`ere montre qu’il existe d’autres modes de transfert de masse. Le sc\'enario privil\'egi\'e dans ce cas s’inspire de travaux des ann\'ees 1970 qui soulignent la capacit\'e des \'etoiles massives \`a entretenir un flot sortant transonique autrement plus important que la brise \'emise par des \'etoiles moins massives comme le Soleil. L’accr\'etion de masse par l’objet compact peut alors se faire de façon opportuniste compte tenu de l’\'energie cin\'etique importante qu’acquiert le vent durant son lancement : seule une fraction du vent verra son cours s\'erieusement d\'evi\'e par l’intervention fortuite du champ de pesanteur de l’objet compact. Il sera alors envisageable pour cette fraction de former un choc autour de l’accr\'eteur qui dissipera suffisamment d’\'energie pour que certaines particules de fluide n’en r\'echappent pas. Les taux de perte de masse des \'etoiles OB sont estim\'es \`a des niveaux si importants qu’en d\'epit de l’infime fraction de vent finalement captur\'ee, la quantit\'e absolue de masse accr\'et\'ee par unit\'e de temps peut atteindre des niveaux quasi comparables \`a ceux des binaires X de faible masse. Mais qu’en est-il de la g\'eom\'etrie du flot accr\'et\'e, modalit\'e qui conditionne le spectre \'emis? L’\'emission de rayons X est-elle associ\'ee aux chocs dans le vent stellaire, autour de l’objet compact, au niveau des pôles magn\'etiques de l’\'etoile \`a neutrons? A-t’elle lieu dans un environnement optiquement \'epais ou mince? Est-elle retrait\'ee par un milieu ambiant, une couronne? \\

\begin{center}
\rule{0.5\textwidth}{.4pt}
\end{center}
\vspace*{0.5cm}

Avec la question de la source d’\'emission se pose celle de notre capacit\'e \`a r\'esoudre les diff\'erentes structures qui ne manquent pas de se manifester par une \'emission de radiations dans les binaires X. Toute tentative de r\'esolution spatiale est \`a proscrire tant la taille angulaire de ces syst\`emes est faible face aux r\'esolutions angulaires limit\'ees auxquelles nos instruments ont acc\`es \`a hautes \'energies. Restent les r\'esolutions spectrales\footnote{Ici, un spectre renvoie \`a la distribution d’\'energie des photons, pas \`a la transform\'ee de Fourier d’un signal temporel.} et temporelles.

La premi\`ere se justifie d\`es lors que chaque structure est \`a une temp\'erature diff\'erente et que l’on est en mesure de diff\'erencier rayonnements thermique et non-thermique\footnote{Un rayonnement thermique a un spectre de corps noir, centr\'e sur une longueur d’onde, alors qu’un spectre non-thermique est g\'en\'eralement ajust\'e par une loi de puissance.}. Si l’\'etoile et les r\'egions internes du disque \'emettent effectivement dans des gammes de longueur d’onde bien distinctes dans les binaires X de faible masse, il est plus difficile de s’assurer de cette s\'eparation dans le cas des binaires X de forte masse où l’\'etoile est plus chaude et la g\'eom\'etrie du flot accr\'et\'e, plus incertaine. De surcro\^it, d’autres structures tel qu’un jet peuvent aussi contribuer au spectre dans des intervalles et proportions variables. Parce que le spectre observ\'e est int\'egr\'e sur une zone de l’espace où cohabitent une multitude de structures aux propri\'et\'es d’\'emission parfois semblables, il n’est pas toujours possible d’identifier chacune des structures pr\'esentes. Une approche empirique largement r\'epandue consiste \`a juxtaposer des zones d’\'emission pr\'e-d\'efinies (couronne, disque, point chaud, etc) dont on ajuste les param\`etres pour \'epouser le spectre observ\'e. Le grand nombre de variables en jeu et l’absence de justifications physiques au choix de ces structures rend l’exercice p\'erilleux ; difficile de s’assurer que la combinaison identifi\'ee soit unique ou que l’ensemble des possibilit\'es aff\'erentes au cas \'etudi\'e ait \'et\'e largement explor\'e sinon circonscris. L’explication est alors valable mais pas forc\'ement satisfaisante si un mod\`ele physique en amont ne motive pas la pr\'esence des \'emetteurs n\'ecessaires pour reproduire le spectre. Il arrive que la robustesse des ajustements ne soit pas tant gage de validit\'e que de d\'eg\'en\'erescence : \`a trop \'etoffer le proc\'ed\'e l’on d\'esengage la cause. L’on verra dans un instant le rôle privil\'egi\'e que peuvent jouer les simulations num\'eriques dans la r\'esolution de cette tension \'epist\'emologique. 

Le fait m\^eme que la question de la r\'esolution temporelle puisse se poser dans les binaires X refl\`ete la fascinante richesse des ph\'enom\`enes \`a l’\oe{}uvre dans ces syst\`emes. Au Cosmos immuable des Anciens se substituent avantageusement des objets dont la diversit\'e n’a d’\'egale que la variabilit\'e. Les binaires X pr\'esentent tout un spectre de variations photom\'etriques et spectroscopiques, des \'echelles s\'eculaires associ\'ees aux \'evolutions stellaires (\eg expansion de l’enveloppe ou changement de temp\'erature effective de la photosph\`ere) et orbitales (\eg circularisation ou changement de la s\'eparation orbitale) \`a des \'echelles dont la dur\'ee est suffisamment courte pour qu’il s’agisse vraisemblablement d’\'ev\`enements d’extension spatiale comparable aux dimensions de l’accr\'eteur compact\footnote{Dans la mesure où la vitesse de la lumi\`ere $c$ fixe une limite sup\'erieure aux vitesses de propagation d’une information susceptible de mener \`a une variation globale des propri\'et\'es d’\'emission, l’on peut \'evaluer la dimension maximale $\Delta x$ d’une zone pr\'esentant une variation temporelle de temps caract\'eristique $\Delta t$ via $\Delta x \sim c \Delta t$.}, possiblement dans son voisinage. La caract\'erisation de la variabilit\'e temporelle accessible aux missions d’observation est un enjeu qui peut nous permettre de diff\'erencier des structures de taille et d’inertie importantes et dont les propri\'et\'es sont r\'egul\'ees par les param\`etres g\'en\'eraux du syst\`eme et des structures plus modestes, peut-\^etre multiples et dues au d\'eclenchement d’instabilit\'es locales d’ampleur raisonnable. La physique non-lin\'eaire qui sous-tend l’\'evolution des instabilit\'es s’illustre par sa capacit\'e \`a coupler diverses \'echelles d’espace et de temps pour alimenter l’\'emergence voire l’entretien\footnote{Par instabilit\'es nous entendons aussi les cycles limites tel que l’oscillateur de Van der Pol, le cycle proie-pr\'edateur de Lotka \& Volterra ou, possiblement, le cycle Q dans les diagrammes intensit\'e - duret\'e de certaines binaires X.} d’un ph\'enom\`ene \`a m\^eme de modifier les modalit\'es d’expression du syst\`eme. La multiplicit\'e des couplages en jeu et la difficult\'e - voire l’impossibilit\'e - d’expliciter des solutions analytiques menacent l\`a aussi notre habilit\'e \`a produire des explications satisfaisantes \ie intelligibles, causales et reproductibles\footnote{Ces trois conditions guarantissent en principe la falsifiabilit\'e du propos d\'evelopp\'e : si le cheminement causal est d\'eterministe (\'eventuellement au sens statistique) et suffisamment d\'etaill\'e pour pouvoir \^etre suivi et critiqu\'e de proche en proche (intelligibilit\'e), alors il est possible d’identifier des cons\'equences secondaires du raisonnement principal. L’invalidation de ces derni\`eres entra\^inerait, par contrapos\'ee, l’invalidation des pr\'emisses.}. \\

\begin{center}
\rule{0.5\textwidth}{.4pt}
\end{center}
\vspace*{0.5cm}

Ce travail de th\`ese s'attache \`a l'\'etude des propri\'et\'es du processus d’accr\'etion d’un flot \`a grande vitesse\footnote{Appel\'e “wind accretion” dans la suite de ce manuscrit.} sur un objet compact ainsi qu’au couplage de ce ph\'enom\`ene avec les propri\'et\'es stellaires et orbitales dans une binaire X où le vent provient d’une superg\'eante OB. La superposition commode entre classification selon le mode de transfert de masse ou selon le type spectral du compagnon a montr\'e ses limites : des superg\'eantes OB pr\'esentent des flots d’accr\'etion focalis\'es sinon amplifi\'es en direction de l’accr\'eteur et des \'etoiles sur la s\'equence principale alimentent un compagnon compact via leurs vents. Dans ce manuscrit, l’on s’attache \`a construire une description homog\`ene des flots d’accr\'etion, ind\'ependante de la distinction entre accr\'etion par vent ou par remplissage de lobe de Roche. La question de la formation et des caract\'eristiques d’un disque d’accr\'etion peut alors se poser sans pr\'esumer du mode de transfert de masse. L’approche que nous adoptons est semi-analytique avec des mod\`eles dont la conception ob\'eit \`a la fois aux principes physiques et aux contraintes num\'eriques. Compte tenu des nombres de Reynolds \'elev\'es auxquels nous sommes confront\'es, l’usage de l’exp\'erimentation au sens usuel n’est pas imm\'ediat. Nous lui pr\'ef\`ererons l’exp\'erimentation num\'erique (ou simulation) qui repose sur les capacit\'es technologiques et algorithmiques du calcul haute performance pour produire des images fid\`eles de solutions physiques inaccessibles aux manipulations exp\'erimentales directes. Si le recours \`a l’univers symbolique des nombres nous lib\`ere des contraintes mat\'erielles de la paillasse, il nous astreint \`a suivre un code, un programme dont le langage impr\`egne non seulement le propos final mais aussi la pens\'ee en amont ; l’on ne construit pas un mod\`ele de la m\^eme mani\`ere selon qu’il a vocation \`a \^etre \'eprouv\'e et explor\'e dans un processeur ou un acc\'el\'erateur. Se laisser griser par le potentiel d\'emiurgique des simulations num\'eriques, c’est prendre le risque de perdre le lien explicatif avec le sujet physique en produisant des r\'esultats qui n’ont plus qu’eux-m\^emes pour objet. 

Dans un premier temps, nous pr\'esentons une repr\'esentation num\'erique de l’\'ecoulement de Bondi-Hoyle-Lyttleton sur un vaste \'eventail d’\'echelles spatiales, du sillage choqu\'e au voisinage de l’objet compact (part II). La coh\'erence physique des r\'esultats nous am\`ene ensuite \`a introduire des effets orbitaux que l’on trouve dans les binaires X et le lancement du vent stellaire par une superg\'eante OB (part III). L’\'elaboration d’un mod\`ele \'el\'ementaire nous permet d’explorer les configurations possibles et d’expliciter les param\`etres physiques essentiels d’un tel syst\`eme. Des tendances quant \`a la g\'eom\'etrie et \`a l’ampleur du flot d’accr\'etion apparaissent alors.

\part{Physical objects \& numerical tools}


\chapter*{Introduction}
\addcontentsline{toc}{chapter}{Introduction}

In this preliminary part, we introduce basic notions about the definition of a compact object and what makes its gravitational influence so special compared to other bodies of similar mass. Ultimate stages of the stellar evolution, compact objects harbor physical conditions on the fringe of current fundamental laws at our disposal. Those cheeky collapsed remnants in an expanding Universe set among the privileged stages to enlarge the realm of our imagination. They bridge the gap between the so-called "two infinities" with macroscopic objects displaying quantum behaviors such as neutron stars. For lack of an adapted theoretical framework and of accessible analogous objects to rely on, the way matter is organized in those objects is still subject to speculation. Even more challenging is the physical censorship which sets the close vicinity of a black hole out of straightforward observational reach : a major technological leap forward would not be enough to make the contents of an event horizon directly observable. \par
\indent This pessimistic statement can however be bypassed if one agrees to appeal to an intellectual work-around, similar to the one which formerly unveiled the physical origin of Thermodynamics. Unsatisfied with the answers of the XIX$^{\text{th}}$ century Physics, the atomists with Boltzmann ahead resorted to a discrete representation of fluids, in spite of the impossibility for direct observations to back up the atomic hypothesis. The reluctance of the community to corpuscular\footnote{Sealed by the accomplishments of Wave Optics and later on, Electromagnetism.} approaches was reinforced by the apparent renouncements it seemed to involve : the numbers of particles at stake in available systems were so large that kinetic theories would bar Newtonian Mechanics from being of any concrete help. Statistical Physics could offer distributions but no singular prediction. The indeterministic cat\footnote{In this case, it is an indeterminism introduced by the modeling : the dynamics of a given particle is fully determined by the equations and the initial conditions but our effective incapacity to follow it makes any singular prediction about this particle meaningless. A few decades later, another cat would engrave indeterminism in the physical laws themselves.} was among the positivist pigeons... The epistemological profitability of the atomic approach forgave this apostasy afterwhile. \par
\indent The observational inaccessibility of compact objects has fueled the inventiveness of modelers, theoreticians and observers for decades in an attempt to grasp their surroundings. Numerical physicists can now join the struggle by making a previously chimeric dream come true : designing essential ersatzes which encapsulate all the potentialities set by the physical laws encoded, and delegating to the computing facilities the responsibility to deploy a contingent seed into an effective system. Modern supercomputers enable us to push computing limits above levels unthinkable a decade ago. The scope of application of numerical simulations expands day after day and brings up fundamental questions about the very object of Science. The post-Newtonian theories of Physics plunged the observer into the formalism : by making the measurement a physical problem, Quantum Mechanics seized a notoriously informal domain, while General Relativity was shaking the convenient absoluteness of static space and time. Numerical simulations involve the physicist a step farther by making her summon virtual avatars of the world she intends to understand. The technical core of this PhD thesis thus deserved at least an overview to familiarize the reader with the everyday concerns of Numerical Astrophysics. \par
\indent In the first chapter, we describe the intrinsic properties of accreting compact objects and illustrate their capacity to convert gravitational potential energy into radiation. It will be the occasion to shed some light on the privileged status of high energy astronomy to observe those systems. We then proceed with a description of several environments in which those objects are found, with an emphasis on X-ray binaries, the astrophysical counterpart of the model developed in the last part of this manuscript. The evolutionary questions pertaining to compact objects are postponed to Chapter \ref{chap:roche} where we address the formation of X-ray binary systems. For the moment, we embrace a static approach to describe compact bodies, with an emphasis on their generic properties in nominal regimes whose characteristic drift time is large compared to the observational times of X-ray binaries. The numerical tools used to investigate those systems are briefly discussed in the last Chapter of this first part. 


\chapter{Accreting compact objects}
\label{chap:acc_comp_obj}
\chaptermark{Accreting compact objects}
\hypersetup{linkcolor=black}
\minitoc
\hypersetup{linkcolor=red}
\setlength{\parskip}{1ex} 

\section{Compact objects}
\label{sec:CO}


\subsection{Compactness}
\label{sec:compacity}

The theory of Relativity, glimpsed by Lorentz, Minkowski \& Poincar\'e at the turn of the century, sketched by Einstein in 1905 and firmly established in 1915 by Einstein \& Grossmann\footnote{Who brought a decisive Mathematical support to extend Special Relativity (\sr) into General Relativity (\gr) in the early 1910's.}, obeys a correspondence principle : it retrieves the results derived and experimentally not invalidated\footnote{Given the levels of precision which can be reached in a given context and in the meaning of Popper's falsifiability theory.} in the Newtonian framework, which justifies the use of the latter in the physical conditions qualified as "Newtonian"\footnote{The term "classical" is usually reserved to an opposition with a "quantum" framework. In this sense, \gr is a classical theory of Physics.}. However, because it takes the transformations of the electric and magnetic fields by change of frame at its word\footnote{See \cite{Semay2010} for a wonderfully elegant derivation of Lorentz transformations from symmetry principles and the need for causality only.} (\sr) and promotes space and time to the status of physical dynamical objects\footnote{To quote \href{http://www.canal-u.tv/video/cerimes/de_la_relativite_a_la_chronogeometrie.9118}{Jean-Marc L\'evy-Leblond}, \gr would be better called "Chronogeometry" \citep{Levy-Leblond}.} (\gr), Relativity enlarges the scope of our understanding of phenomena occurring in "relativistic" physical conditions namely :
\begin{enumerate}
\item as the relative velocities at stake approach the speed of light\footnote{As explained in \cite{Semay2010}, this maximum velocity is actually the one of a mass less particle. Given the very constraining limits on the mass of the photon \citep{Wu2016}, consistent with a zero mass, we can afford to call it the speed of light.} $c$ \citep[see \eg the jets of GRS1915+105][]{Mirabel1994}.
\item as the interaction potential energy between components approach the rest mass energy of those components (nuclear Physics and neutron stars for example).
\item as the thermal energy density approaches the rest mass energy density (\eg the \href{http://www-lmj.cea.fr/en/lmj/index.htm}{Laser M\'egajoule} with plasmas of temperature a few 100MK). 
\item as the gravitational potential differences (within the system considered or between an observer and the subject) approach $c^2$ (\eg precession of Mercury on long periods and the vicinity of compact objects).
\end{enumerate}
All those situations can be understood as energetic comparisons with respect to the rest mass energy of the physical system considered.

In the physical topics covered in this PhD, gravity will play the leading role which makes the latter interpretation the most insightful. If we set the potential at infinity to zero, the difference between the Newtonian gravitational potential\footnote{Whose expression is obtained by solving the differential form of Gauss' law for gravity outside a spherical distribution of mass density $\rho\left(\mathbf{r}\right)$, of total mass $M$ and of spatial extension $R$ : 
\begin{equation}
\begin{cases}
\Delta \Phi = 4\pi G \rho \\
\int \int \int \rho \left(\mathbf{r}\right) \d ^3 r = M \\
\rho \left(\mathbf{r}\right) = 0 \text{ for } r>R
\end{cases}
\end{equation}} at the surface $r=R$ of a spherical astrophysical body and of an observer at infinity is given by, when compared to $c^2$ :
\begin{equation}
\frac{-\Phi (r=R)}{c^2}=\frac{GM}{Rc^2}\hat{=}\Xi
\end{equation}
where $M$ and $R$ are the mass and the radius of the body respectively, $G$ the gravitational constant and $\Xi$ is a dimensionless number called the compactness parameter : in agreement with the aforementioned exclusive regimes of application of Relativity, the closer $\Xi$ gets to 1, the more pressing is the need for a relativistic framework. Note that $\Xi$ relies on a non relativistic expression of the gravitational potential but it can be shown that below moderately high values of the compactness parameter ($\sim$10\%), its value is essentially unaltered by the use or not of Relativity\footnote{Beware, this comment must be considered with precautions when one gets interested in a secular effect of \gr like the precession of Mercury : the instantaneous influence of \gr over a Newtonian approach is vanishingly small but once integrated over a large number of dynamical periods, it plays a decisive role.}. We can also interpret this parameter in terms of velocities by considering the Newtonian escape velocity at the surface of the body with respect to the speed of light :
\begin{equation}
\frac{v_{\text{esc}}}{c}=\sqrt{\frac{2GM}{Rc^2}}=\sqrt{2\Xi}
\end{equation}
which motivates the introduction of a characteristic radius\footnote{It is also the radius of the event horizon of a black hole (\bh). The radius of the \bh itself is set, by convention, to $GM/c^2$ such as $\Xi=1$ for a \bh.}, the Schwarzschild radius $R_{\text{Schw}}$, such as for $R<R_{\text{Schw}}$, no particle, massive\footnote{By "massive" we mean "which has a measured mass strictly positive".} or not, can escape the body since $v_{\text{esc}}>c$ :
\begin{equation}
\label{eq:RS}
R_{\text{Schw}}=\frac{2GM}{c^2}=2\Xi R
\end{equation}
In \ref{sec:KH_extraction}, on the occasion of a discussion of the accretion phenomenon, we provide an energetic interpretation of the compactness parameter.

\begin{table}
 \caption{Physical objects with representative radii and masses, along with their compactness parameter. The presence of the electron in this table serves to convince the doubtful reader about the non redundancy between the concepts of mass density and compactness. The radius given for the \bh is the one of its event horizon (for a Schwarzschild \bh - \ie non-rotating) and the compactness is set to 1 by convention.}
 \label{tab:xi}
 \begin{tabular}{lccc}
  \hline
  & & & \\
         Object & Radius (km) & Mass (\msun) & Compactness\\
           & & & \\	
  \hline
   \phantom{Stellar parameters...} & & &\\
   Stellar mass \bh & 50 & 17 & 1\\
      \phantom{Stellar parameters...} & & &\\
   Neutron star & 10 & 1.4 & 0.2\\
   \phantom{Stellar parameters...} & & &\\
   White dwarf & 10$^4$ & 1 & 10$^{-5}$\\
   \phantom{Stellar parameters...} & & &\\
   Sun & 7$\cdot$10$^5$ & 1 & 10$^{-6}$\\
   \phantom{Stellar parameters...} & & &\\
   Electron\tablefootnote{Where we considered the classical radius obtained by balancing the rest mass energy with the electrostatic binding energy} & 10$^{-18}$ & 4.6$\cdot$10$^{-61}$ & 10$^{-43}$\\  
      \phantom{Stellar parameters...} & & &\\
  \hline
 \end{tabular}
\end{table}
 
Concerning the typical numerical values of $\Xi$, we provide a set of evaluations in Table \ref{tab:xi} for different physical objects. From now on, any system with $\Xi>10^{-1}$ will qualify as "compact". This restrictive definition\footnote{Usually, we start to speak about compact bodies when the degenerancy pressure starts to play a role, for $\Xi>10^{-4}$, which makes the white dwarfs (in cataclysmic variables for instance) belong to this category.} confine the compact objects to neutron stars and black holes\footnote{More exotic theoretical objects such as boson stars \citep{Grandclement2014} can also be considered as compact objects.}, the two main kinds of accretors studied in this PhD thesis. If those objects themselves must be studied through the prism of \gr, their surroundings do not always require a relativistic treatment per se. Sometimes, a pseudo-Newtonian approach can be sufficient to grasp preliminary features. A handy way to implement this intermediate framework\footnote{Initially introduced to model a Schwarzschild black hole but extensions to a rotating black hole exist \citep[see references in the review by ][]{Abramowicz2013}.} is to write the gravitational potential produced by a point mass $M$ with the Paczy\'nski-Wiita potential \citep{Paczynsky1980} :
\begin{equation}
\Phi_{\textsc{pw}}\left(\mathbf{r}\right)=-\frac{GM}{\left|\mathbf{r}\right|- a} \quad \text{    for    } \left|\mathbf{r}\right|>a
\end{equation}
where $a$ is a degree of freedom which stands for a domain of extension of the relativistic regime, of the order of the Schwarzschild radius. The associated force is still centered and conservative.


\subsection{Formation}
\label{sec:formation}

\begin{figure}
\begin{center} %
\includegraphics[height=7cm, width=9.8cm]{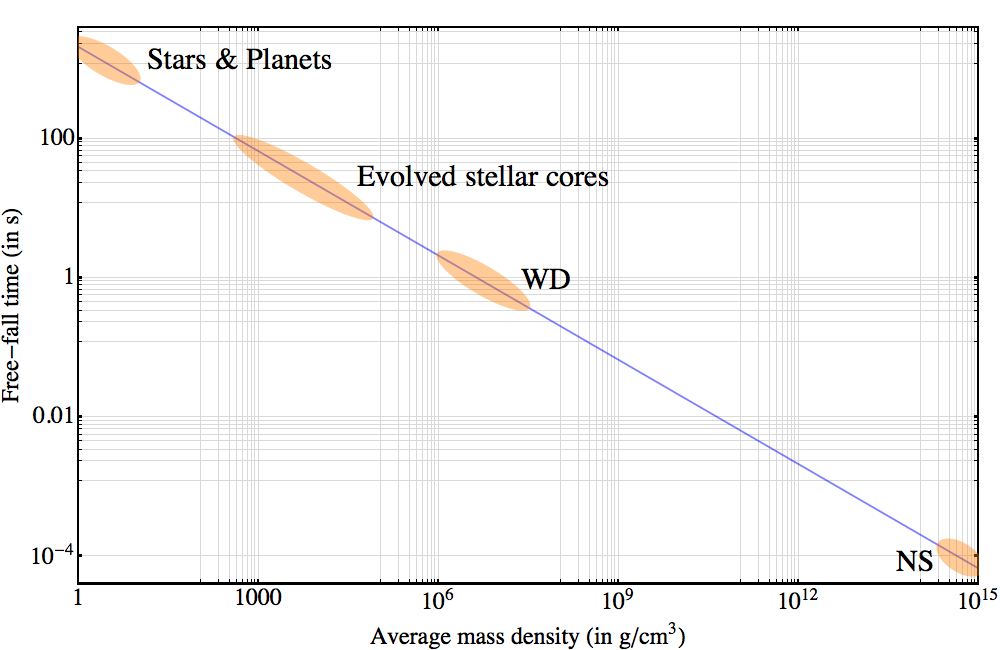}	
\caption{Free-fall (or dynamical) time as a function of the average density of the system. It corresponds to the time it takes to collapse if gravity is the only force on stage. It ranges from a few hours for a Sun-like star to a fraction of a millisecond for a pulsar.}
\label{fig:free-fall}
\end{center}
\end{figure}

All other forces neglected, any massive system is doomed to collapse over a dynamical timescale $\tau_d$ under the action of gravity. The shrinking of the radius at a constant mass makes the compactness parameter rise and it soon reaches relativistic values ($\lesssim$ 1). The characteristic duration of this collapse for a system of mass $M$ and initial radius and density $R$ and $\rho$ respectively can be derived using Kepler's third law applied to a test-mass on a degenerate orbit of eccentricity almost 1 and of semi-major axis $R/2$ :
\begin{equation}
\tau _d \sim \frac{1}{2} \times 2\pi \sqrt{\frac{\left(R/2\right)^3}{GM}} = \sqrt{\frac{3\pi}{32G\rho}}
\end{equation}
As visible in Figure\,\ref{fig:free-fall}, this dynamical time is always much smaller than the actual life expectancy of the system considered\footnote{Apart maybe for collapsing molecular clouds with densities of a few millions atoms per cubic centimeter, where the free-fall time is of the order of a few hundreds of thousands years.}. For the sake of the Newtonian world, the intercession of (effectively) repulsing forces guarantees the existence of a broad range of intermediate equilibria not doomed to enter the relativistic realm within a dynamical timescale. For instance, stars can maintain as long as the thermonuclear fusion in their core sustains thermal (or, for the most massive ones, radiative) pressure to counterbalance gravity. Each time it runs out of nuclear fuel, after a laps given by the nuclear timescale\footnote{Which can be computed by estimating the efficiency of, \eg, the proton fusion and the rest-mass of the combustible (typically 10\% of the total mass) with respect to the stellar luminosity. The former requires to follow Gamow's elegant computation of the tunnelling effect associated to the fusion between protons - see, for the German-speaking people \cite{Gamow1928} and for the French-speaking people, \cite{Daigne2015}.}, the core contracts (while the outer layers expand) and the higher temperature reached triggers the ignition of a larger atomic number element\footnote{See Aston's packing fraction curve to quantify the energy released (or needed beyond Iron) by the fusion of different nuclear elements.}, buying some additional time to the star\footnote{Stellar evolution, albeit a fundamental motivation of this work and a permanent interpretation background of our results, will not be addressed in detail in this manuscript. The interested reader can get more details in the comprehensive manual by \cite{Prialnik2009}.}. At some point though,the core reaches the electronic (for \whd) or neutron (for \ns) degeneracy limit when the distance between particles becomes of the order of the thermal De Broglie wavelength. If the electronic pressure of degeneracy can take over and stabilize the \whd, for \ns, it takes the strong interaction between baryons to compensate for the gravitational field. As the outer layers bounce back on the surface of the newly formed \ns, they are ejected at important speeds : a core-collapse supernova occurs. As specified in Table \ref{tab:xi}, the radii which result from these combined actions\footnote{Other forces can also come into play like the centrifugal force for rapidly rotating stars \citep{Maeder2009}.} are above the Schwarzschild radii of the objects (\ie $\Xi<1$) ; the stellar structure for instance can be appreciated up to a sophisticated level of understanding without a relativistic description. Nevertheless, if the mass of the collapsing core is too important, the strong interaction between baryons is not enough to stop gravity from bringing the compactness parameter to fully relativistic values : a black hole is formed, surrounded by an event horizon\footnote{The event horizon is an exclusive property of black holes (see the cosmic censorship hypothesis which bans the possibility of a naked singularity - \ie a singularity without event horizon). Visualizing it to confirm the \bh nature of the \bh candidates - by opposition to boson stars \eg - is one of the main goals of the contemporary Event Horizon Telescope \citep{Ricarte2014} and \textsc{gravity} mission \citep{Eisenhauer2011}.}.

To summarize, in a final stand (whose modalities strongly depend on the mass of the star but also on its metallicity, its angular momentum...), the exhausted star collapses into either :
\begin{enumerate}
\item a white dwarf for initial stellar masses below 10\msun
\item a neutron star for initial stellar masses between 10 and 15-40\msun \citep{Fryer1999,Kochanek2014,Daigne2015}
\item a black hole for initial stellar masses above 15-40\msun 
\end{enumerate}


\subsection{Population synthesis}
\label{sec:pop}

Computing the expected populations of \ns and \bh in an environment (globular clusters, galactic center, etc) given the prior of its statistical properties (\eg mean particle density, temperature, etc) requires preliminary insights on the following astrophysical questions.

\subsubsection{Formation models}

As described in the previous section, we have at our disposal theoretical and numerical representations of the ultimate stages of stellar evolutionary tracks. However, the levels of accuracy we can currently reach leaves room to large\footnote{With respect to the observational capacities.} uncertainties on the properties of the progenitor. The question "which star yields which compact object?" must be answered with more robustness and precision to extrapolate the fate of a given set of stars.

\subsubsection{Stellar evolution}

If the initial stellar mass is the key parameter to appreciate stellar evolution with a precision of an order-of-magnitude, observations highlighted systematic biases from a population to another\footnote{For instance between stars of the Milky Way and the peripheral galaxies \textsc{smc} (Small Magellanic Cloud) and \textsc{lmc} (Large Magellanic Clouds), the latter two displaying lower metallicities than our Galaxy \citep{Wood1998}.} \citep[see \eg][for low mass stars]{Demory2009}. It made the case for additional not-so-hidden parameters in the stellar models such as the chemical composition, the angular momentum, the magnetic field, the binarity, etc. In particular, for high mass stars (\ie above a few solar masses), the mass is believed to significantly decrease from the initial one over the stellar lifetime\footnote{In spite of the dramatically shorter life expectancy of massive stars.} which itself must be narrowly determined to know when the compact object forms. Indeed, while the Sun emits a thermally driven wind \citep[see Parker's model described in][]{Lamers1999} at a rate of $\sim 10^{-14}M_{\odot}\cdot$yr$^{-1}$, massive stars undergo radiatively driven winds with much larger mass loss rates (see Chapter \ref{chap:wind}). This very non conservativeness of the mass of stellar bodies among the most likely to give compact objects emphasizes the need to surpass the classical models ; describing the collapse of a star takes a reliable stage. 

\subsubsection{Initial Mass Function}

The mechanism which leads from a cold and dense molecular cloud to a proto-star conditions the Initial Mass Function (\textsc{imf}), the initial mass distribution of a stellar population just formed - see the recent review by \cite{Kroupa2011} and the widely used \textsc{imf} given by \cite{Chabrier2003}. The underlying fragmentation of the turbulent interstellar medium is an ongoing topic of research which makes the most of multi-scale and multi-Physics simulations \citep{Abel2000}. It is believed that this \textsc{imf}, which nowadays strongly favours low-mass stars in the Milky Way\footnote{$\sim$ 97\% of presently formed stars are below the threshold of $\sim$10\msun to ultimately form a \ns (and, a fortiori, a \bh).}, was different in the past with a bias towards much larger initial masses\footnote{With gravitational collapses of $\sim$100\msun stars being possible physical sources of some of the observed hypernovas.} \citep{Schneider2002}. 

\subsubsection{Stellar Formation Rate}

The Stellar Formation Rate (\textsc{sfr}) indicates how fast free gas in the interstellar medium is converted into stars. Similarly to the \textsc{imf}, it can be constrained through the combined investigation of large scale numerical simulations \citep{Vogelsberger2014} and galactic observations. Starburst galaxies suggest that this rate can punctually skyrocket, for instance due to galactic mergers \citep{Genzel2010}.


\subsection{Objects}
\label{sec:obj}

\subsubsection{Neutron stars}

\begin{wrapfigure}{r}{0.5\textwidth}
\begin{center}
\includegraphics[height=8.5cm, width=0.95\textwidth]{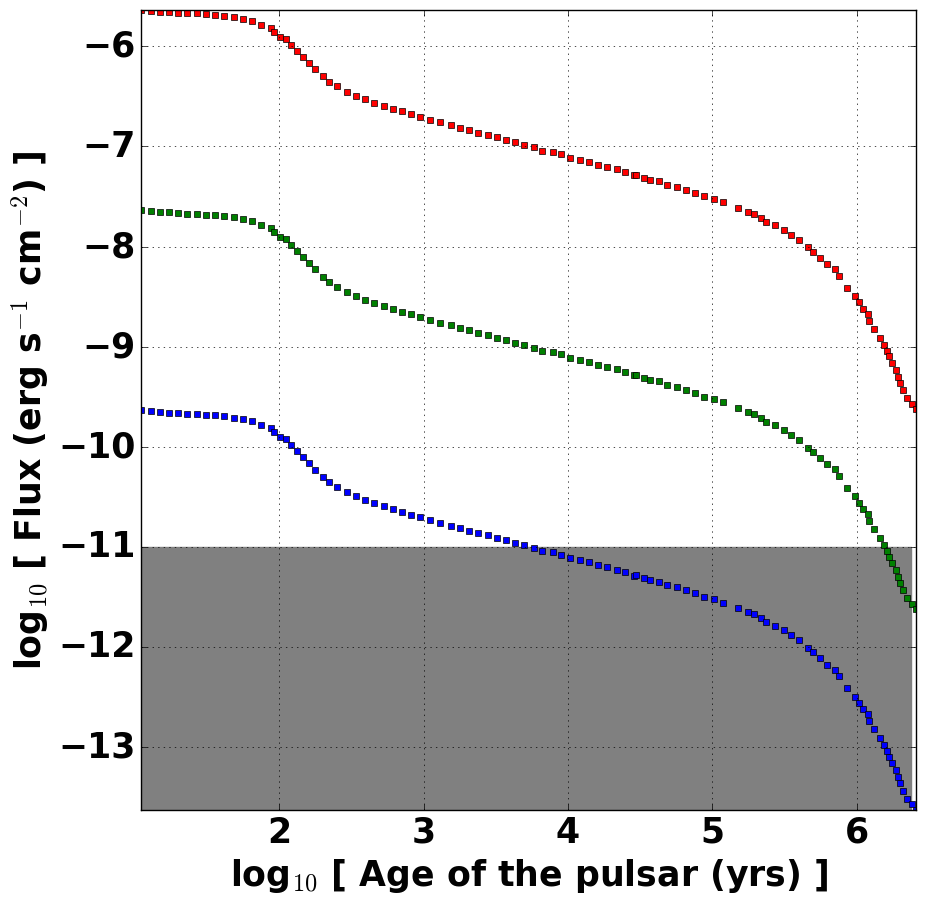}	
\caption{Fluxes of the black body surface thermal emission for a 10km neutron star at 10, 100 and 1,000pc from top to bottom. Based on the cooling curve in Figure 1 in \cite{Yakovlev2004} ; the associated data used to draw the latter cooling curve have been extracted using the webtool \href{http://arohatgi.info/WebPlotDigitizer/}{WebPlotDigitizer}. The shaded area at the bottom corresponds to an estimate of the current photometric sensitivity limit (see text for details).}
\label{fig:cooling_NS}
\end{center}
\end{wrapfigure}
\begin{figure}[!t]
\begin{center} %
\includegraphics[height=10.2cm, width=12.1cm]{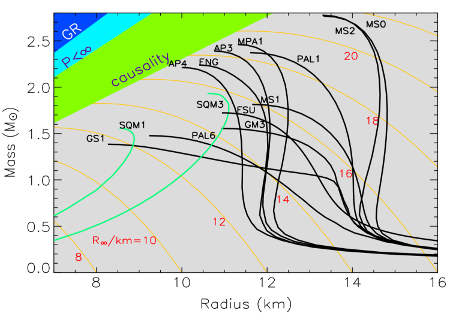}	
\caption{Mass-radius relationships (solid black and green) deduced from different equations-of-state of the matter in a \ns. They all display different maximum masses albeit in a narrow range compared to our capacity to accurately measure the mass of a \ns. From \cite{Lattimer2010}.}
\label{fig:mass-radius_NS}
\end{center}
\end{figure}

Neutron stars are compact remnants of the Iron core of massive stars whose collapse was stopped by the disruption of the nuclei into nucleons, the subsequent neutronization of the protons by electronic capture and the resulting repulsive force between baryons\footnote{Related to the Pauli exclusion principle since baryons belong to fermions \citep{Martin2007}.}. The average mass density of those objects exceeds the nucleus mass density (a few 10$^{14}$g$\cdot$cm$^{-3}$) and the compactness parameter reaches a couple of 10\%. Given the extreme conditions associated to this macroscopic object whose structure must be studied using Quantum Chromodynamics in a relativistic framework, the equation of state of the ultra-dense matter they contain (well approached by a barotropic one) is still an active field of research. A strong observational constrain on the equation of state is the maximum mass of the neutron star \citep{Chamel2013}. Indeed, like the Chandrasekhar limit for \whd, there is a mass above which stable \ns can not exist. The first computation of this upper limit was made by Tolman, Oppenheimer \& Volkoff in 1939 (the \textsc{tov} limit) and although it gave a value too low to validate their model, it paved the way to more elaborated models of the interactions at stake in this exotic state of matter. Each of them gave birth to a mass-radius relationship (some of them being represented in Figure\,\ref{fig:mass-radius_NS}) and more importantly, to a maximum mass which can be confronted to observations. The maximum masses derived all lie approximately within 1.6 and 3\msun ; any body heavier than the latter is thus considered to be a black hole candidate\footnote{By the denomination "candidate", we want to emphasize the fact that the exclusive property of \bh, the existence of an event horizon, has not been confirmed yet for any of the available candidates. However, the recent detection of two gravitational wave signals in spectacular agreement with the signals expected for a binary \bh coalescence provides strong if not definitive support in favor of the existence of black holes \citep{Abbott2016,Abbott2016a}.}, not a neutron star.

\begin{figure}
\floatbox[{\capbeside\thisfloatsetup{capbesideposition={left,top},capbesidewidth=8cm}}]{figure}[\FBwidth]
{\caption{Measured decay rates of the radio and millisecond pulsars spin period as a function of the spin periods. The iso-magnetic field, iso-luminosity and iso-time dashed lines are guidelines issued from theoretical models not described in the present manuscript \citep[see \eg][]{Daigne2015}. From \cite{Lorimer2012}.}
\label{fig:p-pdot}}
{\includegraphics[width=8cm]{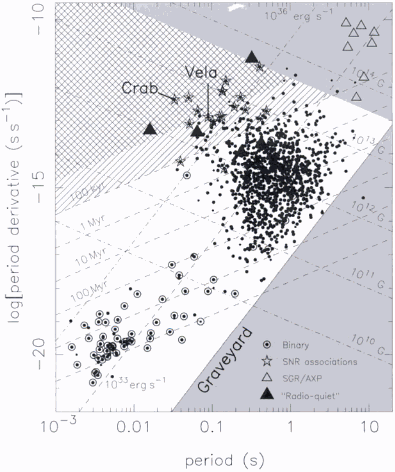}}
\end{figure}

If \ns have an initial surface temperature much higher than the stellar photospheres, they quickly cool down \cite[Figure\,\ref{fig:cooling_NS}, see also][for a review on thermal emission from isolated neutron stars]{Zavlin2007}. Due to their small radii ($\sim$ 10km), their glare does not carry very far : the \ns which are detected only from their surface thermal emission all lie within a few 100pc. Indeed, let us take, as an indicative guideline, the sensitivity of a state-of-the-art satellite as Kepler, although its nominal waveband is not the one adequate for the observation of a hot young neutron star. The Kepler satellite observes up to an apparent magnitude\footnote{With the flux of Vega of the order of 2 erg$\cdot$s$^{-1}\cdot$cm$ {-2}$ to scale it.} of 15, as a reference for the bolometric limit, it corresponds to a conservative\footnote{In practice, for low flux objects, the issue is not the detectability itself but the physical back and foreground luminosity levels (the "blends") : the lower the luminosity of the source, the larger the relative contribution of the neighbourhood along the line-of-sight.} flux detectability limit of approximately 10$^{-11}$ erg$\cdot$s$^{-1}\cdot$cm$^{-2}$. Given the Figure\,\ref{fig:cooling_NS}, it means that we can only detect very young (< 10,000 years) neutron stars at a kiloparsec, but we can go up to one million years at 100pc. We count approximately 10 objects detected only with their surface thermal emission (including the so-called "Magnificient Seven"). Most isolated \ns are actually observed thanks to their large magnetic field and rotation frequency, and are called radio pulsars\footnote{PSR B1919+21, \aka LMG-1 (for Little Green Man - 1) being the first pulsar recognized as such. The pulsar lying at the core of the Crab Nebula is probably one of the more studied of them.}. If the magnetic and rotation axis are misaligned, the beamed\footnote{Along the magnetic axis.} light emitted by the rotating dipole will sweep a specific circular region of the sky, analogous to the way a lighthouse sweeps its beam of light around in a circle. It taps the kinematic energy of rotation of the solid body and induces a magnetic braking which is measured and consistent with this scenario (see Figure\,\ref{fig:p-pdot}). Thus, isolated neutron stars\footnote{See \cite{Mereghetti2010} for a review on isolated neutron stars.} do not have a rich enough environment to interact with to make the measure of decisive parameters such as their mass robust.


Only a fraction of \ns are expected to be found orbiting a stellar companion. Indeed, when the neutron star forms, the departure from isotropicity of the explosion\footnote{See \cite{Scheck2004} and \cite{Scheck2006} for two-dimensional numerical simulations of neutrino-driven core-collapse supernovae and \cite{Wongwathanarat2010a} and \cite{Wongwathanarat2013} for an extension to three-dimensional simulations.} provides a non zero kick velocity to the compact object with respect to its companion, possibly large enough to disrupt the binary system \citep{Tauris1998,Janka2013,Grefenstette2014,Hirai:2014tn}. The spatial distribution of isolated pulsars in the Milky Way, with a larger dispersion around the Galactic plane than the gas and population I stars, supports this scenario. The couples which survive the supernova explosion are goldmines since the mass transfer and orbital modulation of the emission are additional phenomena where the neutron star properties play a major if not a leading role : it becomes possible, provided the influence of the plentiful parameters can be disentangled, to trace back, for instance, the mass of the neutron star.

Eventually, as mentioned earlier, neutron stars present huge magnetic fields and short spin periods. Indeed, conservation of angular momentum and the freezing-flux theorem \citep{Shu1992} guarantee, provided the fraction of the total mass ejected during the collapse remains modest :
\begin{equation}
\begin{cases}
R_*^2 / P_* = R_{\bullet}^2 / P_{\bullet}\\
B_* R_*^2 = B_{\bullet} R_{\bullet}^2
\end{cases}
\end{equation}
where $P$, $B$ and $R$ refer to the spin period (associated to a spin angular velocity $\Omega$), the magnetic field and the radius respectively of either the star (subscript $*$) or the compact remnant (subscript $\bullet$). If we take the solar orders-of-magnitude as guidelines with a spin period of 25 days and a polar magnetic field of 1G, applied to a 10\rsun progenitor, we get the following values for the newly formed neutron star :
\begin{equation}
\begin{cases}
P_{\bullet} \sim 4\cdot 10^{-6}\text{s}\\
B_{\bullet} \sim 5\cdot 10^{11}\text{G}
\end{cases}
\end{equation}
If the magnetic field is consistent with what is observed for young X-ray binaries pulsars\footnote{Slightly below according to Figure\,\ref{fig:p-pdot}, which would suggest, for example, larger magnetic fields for more massive stars.}, the spin period is well below the actual initial spin periods at birth of those objects (of the order of a few 10ms). Actually, with this spin period, the centrifugal force disrupts the body. Indeed, if we compare the gravitational binding energy of an object of mass $M$ and the rotational energy\footnote{Obtained using the moment of inertia.} $E_{\text{rot}}\sim MR^2\Omega ^2$, we get the following necessary condition to guarantee the integrity of the body at a fixed angular momentum $J$ :
\begin{equation}
\frac{E_{\text{rot}}}{E_{p,g}}\sim \frac{MR^2\Omega ^2}{GM^2/R}=\frac{J^2}{GM^3}\frac{1}{R}<1 \quad \text{ if and only if } \quad R>R_{\text{min}}=\frac{J^2}{GM^3}
\end{equation}
For the aforementioned numerical values applied to a 10\msun star, the minimum radius is of one thousand times larger than the size of the neutron star. A more realistic approach would have been to consider only the core or the central sphere containing $M_{\bullet}=1.5$\msun (the future mass of the \ns). If we refer to Table 1 of \cite{Heger2005} who estimate the pre-supernova angular momentum of this sphere to 10$^{49}$erg$\cdot$s, we have, if we assume a differential rotation (\ie $P\sim 1/R^2$) for the core :
\begin{equation}
\begin{aligned}
\frac{M_{\bullet}R_{\bullet}^2}{P_{\bullet}}&\sim 8\cdot 10^{49}\text{erg}\cdot\text{s}\\
P_{\bullet} &\sim 200\text{ms}
\end{aligned}
\end{equation}
We retrieve a more realistic value of the initial spin period of a \ns. A first comment to be made though is that the iron core is believed to follow a more solid than differential rotation ; more importantly, \cite{Blondin2007} have shown that angular momentum is redistributed during the collapse because of the Standing Accretion Shock Instability\footnote{See \cite{Foglizzo2011} for a spectacular analogue of this instability in the shallow water context and \cite{Kazeroni2015} for new insights on the spin-up of a \ns during core-collapse.} \citep[\textsc{sasi,}][]{Blondin2003}.


Concerning the structure of the magnetic field around the neutron star, different models exist \citep{Daigne2015}. We expect the flow structure to be shaped by the magnetic field once it reaches the magnetosphere of the body, whose radius can be evaluated in the following way \citep[see][section 6.3]{Frank2002}. Balancing ram pressure (since the inflow is highly supersonic) and magnetic pressure leads to a localisation of the magnetosphere at a radius $R_m$ (\aka the Alfven radius) such as, for a magnetic field of magnitude $B$ (and with a literal expression in \textsc{mksa} units) :
\begin{equation}
\label{eq:ns_magnetosphere}
R_m=\left[ \frac{\pi\sqrt{2G}}{\mu_0} B^2R^5\frac{\sqrt{M}}{L_{\text{acc}}} \right]^{2/7} \sim 1,560\text{km} \left( \frac{B}{10^{11}\text{G}} \right)^{\frac{4}{7}} \left( \frac{R}{10\text{km}} \right) ^{\frac{10}{7}} \left( \frac{M}{1.5M_{\odot}} \right) ^{\frac{1}{7}} \left( \frac{L_{\text{acc}}}{10^{36}\text{erg}\cdot\text{s}^{-1}} \right) ^{-\frac{2}{7}}
\end{equation}
where the accretion luminosity $L_{\text{acc}}$, representative of the X-ray luminosity of the system, is introduced in section \ref{sec:lum_spec}. Within this region, the dynamics of the flow is dominated by the coupling between the ionized gas and the magnetic field. The plasma is channelled along the field lines which breaks up an eventual geometrically thin and optically thick structure of the flow (see section \ref{sec:lum_spec}). The impact of the inflow on the polar caps of the neutron stars obeys column accretion models\footnote{For more details, see the seminal paper by \cite{Davidson1973} along with references in \href{http://pulsar.sternwarte.uni-erlangen.de/wilms/teach/xrb/handouts.html}{R.-S. Bamberg's course}.}.

\subsubsection{Black holes}

\begin{figure}[!b]
\begin{center} %
\includegraphics[height=7.2cm, width=9.1cm]{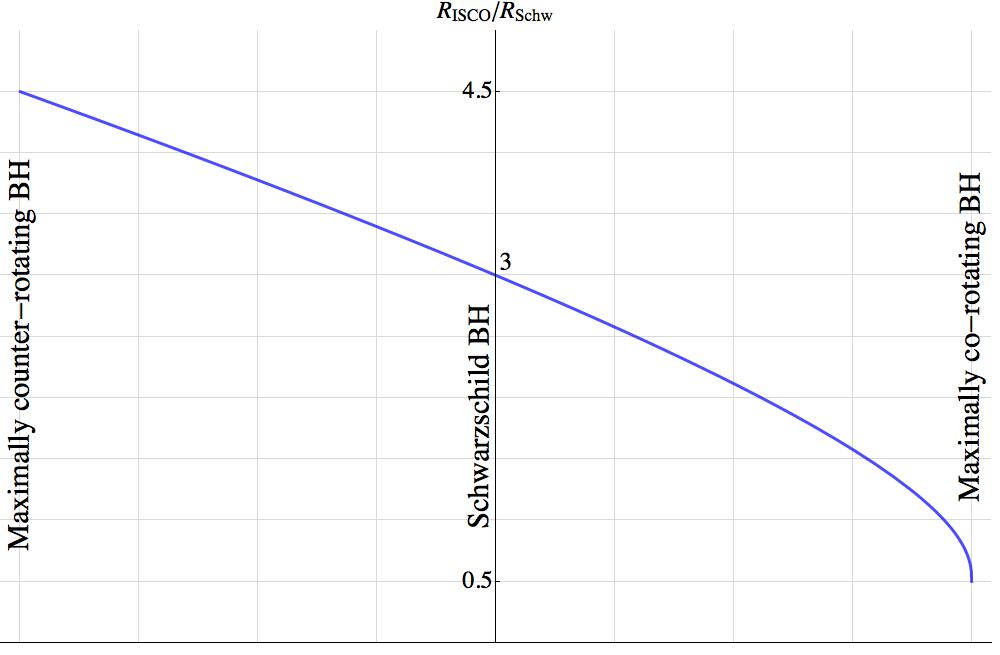}	
\caption{Position of the \textsc{isco} with respect to the Schwarzschild radius as a function of a parameter which quantifies the \bh spin. Along the central line, the \bh has no angular momentum (Schwarzschild metric) and the \textsc{isco} is located three times farther than the event horizon which has a size equal to the Schwarzschild radius. For a rotating \bh (Kerr \bh), a rotating flow which swings from co to counter-rotating can extend down to a different minimal radius due to the non invariance of the \textsc{isco} by a swing in \bh spin (\ie equivalently in the rotation direction of the flow).}
\label{fig:spin_BH}
\end{center}
\end{figure}
\begin{figure}
\begin{center} %
\includegraphics[height=20.2cm, width=7.1cm]{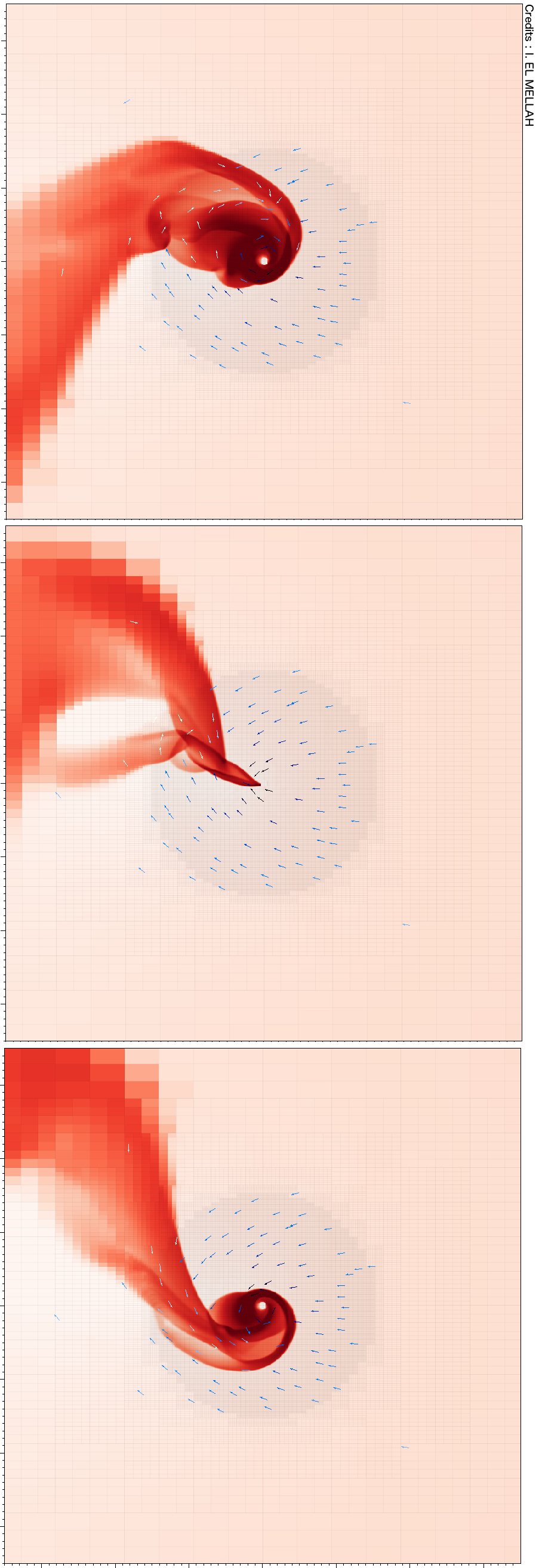}	
\caption{Zoomed in time series (from top to bottom) of a two-dimensional simulation of a supersonic planar flow on a point-mass in cylindrical geometry realized with \textsc{mpi-amrvac} (see Chapter \ref{chap:num_tools}). The colormap represents the gas density and the arrows stand for the velocity field (the bluer the faster). The mesh, with 6 levels of refinement, is overplotted. A disk structure forms around the accretor and after a sudden flushing phase, a similar disk with reverted angular momentum forms.}
\label{fig:flip-flop}
\end{center}
\end{figure}

This solution of \gr has long been considered as merely theoretical until accretion\footnote{By stellar mass \bh such as Cyg X-1 but also by super-massive \bh in active galactic nuclei \citep{Beckmann2012}.} and later on, the stellar orbits in the vicinity of the Galactic center Sgr A* \citep{Genzel2010a}, brought it back in a more meaningful spotlight. We now believe that \bh are the most likely option when a compact object above a few solar masses\footnote{Above the maximum mass of a \ns (see previous section).} is identified. Stellar mass \bh (< a few 10\msun) are the remnants of collapsed cores of massive stars while Super-Massive Black Holes (\textsc{smbh}, with masses above 10$^6$\msun) are believed to lie preferentially at the center of galaxies. The formation of \textsc{smbh}, by direct collapse of gas or by agglomeration of stellar mass \bh and runaway accretion, is still an active matter of debate and so is the puzzling scarcity of intermediate-mass \bh \citep[\textsc{imbh}, ][]{VanderMarel2003a} : although none has been definitely confirmed, super-Eddington\footnote{See section \ref{sec:eddington_lim} in Chapter \ref{chap:wind} for the Eddington luminosity.} accreting systems such as Ultra and Hyper Luminous X-ray sources might host the long awaited \textsc{imbh} \citep{Webb2014}. Their unity-value compactness makes a \bh a unique laboratory not only to contemplate \gr at stake\footnote{The deformation of the Iron 6.4keV emission line for instance \citep{Fabian2000} encapsulates several effects such as the relativistic Doppler boosting and the gravitational reddening but also the spectacular rotational frame dragging by the spinning \bh \citep{Ingram2012}, \aka the Lense-Thirring effect \citep{Gourgoulhon2014}.} but also to study strong-field gravity and lay the empirical foundations of the next physical theories. The no-hair theorem guarantees that a \bh is entirely defined by only three parameters :
\begin{enumerate}
\item its mass. It sets its Schwarzschild radius given by \eqref{eq:RS} and which corresponds to the radius of the event horizon for a non-rotating \bh (Schwarzschild metric).
\item its spin. When it is not zero (\ie when we deal with a Kerr \bh), an intermediate region outside of the event horizon exists called the ergosphere. This domain is believed to play a major role in the launching of the jets observed in quasars and microquasars \citep[see \eg the mechanism introduced by][]{Blandford1977}.
\item its electric charge, a parameter often considered as miscellaneous due to the absence of observation of processes responsible for piling up of electric charges in a \bh.
\end{enumerate}  
Spin measurements, albeit still discussed within the community, have started to blossom \citep{Reynolds:2013va} to unveil the second key parameter of \bh\footnote{Gravitational waves now offer a new way to measure the \bh spin \citep{Abbott2016,Abbott2016a}.}. They invite us to think that rapidly rotating\footnote{\ie those for which the angular momentum is close from a theoretical limit written $J_{\textsc{max}}$ in the following.} \bh are not rare. An important feature of rapidly rotating \bh which impacts the accretion  phenomenon and in particular wind accretion is the existence of an innermost stable circular orbit (\textsc{isco}) whose position differs for a co and a counter rotating disk\footnote{Which is equivalent, from the point of view of the accreted flow, to switch the sign of the \bh spin.} around a Kerr \bh (see Figure\,\ref{fig:spin_BH} and the applet \href{https://duetosymmetry.com/tool/kerr-isco-calculator/}{designed by Leo C. Stein}). Below this orbit, the disk structure can no longer subsist and the flow quickly spirals in. Since most of the light emitted by an accretion disk comes from its inner parts (see section \ref{sec:lum_spec}) and if the inner edge of the disk matches the \textsc{isco}\footnote{It is also possible, as suggested by \cite{Yamada2013} to account for the low/hard states of Cygnus X-1, that the inner disk is truncated well beyond the \textsc{isco} which would jeopardize the analysis sketched here.}, it means that switches between the two regimes of rotation imply observationally distinguishable spectra and luminosity. The main trigger of this swing could be a large scale hydrodynamical instability called the "flip-flop" instability\footnote{Although it is sometimes compared to the oscillations running along a flag or the von K\'arm\'an vortex street \citep{1991MNRAS.253..633L}, one must keep in mind that in the current case (i) the geometry is spherical and three-dimensional and (ii) the conditions at the surface of the accretor are not trivially reflecting ones (see section \ref{sec:num_impl} in Chapter \ref{chap:num_sim_BHL}).}, illustrated in Figure\,\ref{fig:flip-flop} \citep{Foglizzo2005}.


\section{Accretion}
\label{sec:accretion}


\subsection{The Kelvin-Helmholtz conversion process}
\label{sec:KH_extraction}

The ubiquity of gravitation in Astrophysics promotes accretion to the rank of the unavoidable mechanisms. In this manuscript, we use the word "accretion" to refer to the fall of a continuous medium (gas or plasma) in a gravitational potential produced by an external\footnote{Self-gravity of the accreted flow can be neglected in X-ray binaries. The skeptical reader could verify it by using the equation (5.49) of \cite{Frank2002} to evaluate the mass stored in a disk (for instance) and compare it to the mass of the accretor.} massive body. Before considering a proper fluid, we compute the change in the specific\footnote{Applied to an energy, this adjective means "per unit mass" in this manuscript.} gravitational potential energy of a test-mass which would go from infinity to the surface of the compact object of radius $R$ :
\begin{equation}
\Delta E_{p,g} = \int_{E_{p,g}\left(+\infty\right)}^{E_{p,g}\left(R\right)}\d E_{p,g}=E_{p,g}\left(R\right)-\underbrace{E_{p,g}\left(+\infty\right)}_{=0 \text{ by convention}}=-\frac{GM}{R}\sim-\Xi c^2
\end{equation} 
Since the mechanical specific energy of the test-mass, $E_m$, is conserved, this decrease in the absolute value of gravitational potential energy results in an increase in the specific kinetic energy of the particle of the order of, for a particle with a negligible velocity at infinity compared to the one at the surface of the object :
\begin{equation}
\Delta E_m = 0 = \Delta E_c + \Delta E_{p,g}
\end{equation}
Hence :
\begin{equation}
\Delta E_c = \int_{E_c\left(+\infty\right)}^{E_c\left(R\right)} \d E_c \sim E_c\left(R\right) \sim \Xi c^2
\end{equation}
This rise in specific kinetic energy is of the order of the specific rest mass of test-mass ($c^2$) for a compact object\footnote{Once again, this Newtonian argumentation is qualitatively unchanged in the relativistic framework.} : a few 10\% of the rest mass of the particle has been converted into kinetic energy, which makes accretion onto a compact object the most efficient energy conversion process\footnote{In comparison, nuclear fusion of Hydrogen into Helium peaks at 0.7\% (see the aforementioned Gamow's derivation).}. In a fluid, this sudden surge in velocity will go with a heating of the flow and thus, with a significant emission of light (see section \ref{sec:lum_spec}). We then conveniently define the accretion luminosity $L_{\text{acc}}$ as the maximum luminosity associated to an object of compactness $\Xi$ accreting with a mass accretion rate $\dot{M}$ :
\begin{equation}
\label{eq:Lacc}
L_{\text{acc}}=\Xi \dot{M} c^2 \sim 10^{36}\text{erg}\cdot \text{s}^{-1} \left( \frac{\dot{M}}{10^{-10}M_{\odot}\cdot\text{yr}^{-1}} \right)
\end{equation}
Part of the gravitational potential energy can also be reinjected in the jets (kinetic energy) or "swallowed" by the \bh since the energy carried by the flow, as it crosses the event horizon, takes an infinite amount of time to be released for the observer\footnote{Energy is effectively trapped including under its radiative form, because of the relativistic gravitational reddening.} \citep[see the Advection Dominated Accretion Flow model for instance with][]{Narayan1998}.

This energy conversion process was first designed well before the relativistic notion of compactness bore any physical relevance. In the XIX$^{\text{th}}$ century, the rising interest in Geology brought up the question of the age of the Earth and, by then, of the Sun whose formation is believed to be concomitant\footnote{See the Laplace nebular theory of planetary formation \citep{Woolfson1993}.}. Kelvin \& Helmholtz suggested that the luminosity $L$ of the latter was due to the gravitational contraction of the whole body which provides the following age estimate, for a constant luminosity through time :
\begin{equation}
\tau \sim -\frac{\int_{E_{p,g\text{,ini}}}^{E_{p,g\text{,now}}}\d E_{p,g}}{L} \sim \frac{E_{p,g\text{,ini}}-E_{p,g\text{,now}}}{L} \sim -\frac{E_{p,g\text{,now}}}{L} \quad \text{ with } \quad \left|E_{p,g\text{,ini}}\right|\ll \left|E_{p,g\text{,now}}\right|
\end{equation}
where $E_{p,g\text{,now}}$ is the gravitational binding energy of the Sun today. For a uniform sphere of gas density $\rho$ and radius $R$, the gravitational binding energy can be derived by considering the successive addition of layers of infinitesimal mass $\d m=4\pi r^2 \d r \cdot \rho$ :
\begin{equation}
E_{p,g\text{,now}}=-\int_{r=0}^R \frac{GM\left(<r\right)}{r}\d m = - \frac{3}{5}\frac{GM^2}{R} \quad \text{ where } \quad M\left(<r\right)=M\left(\frac{r}{R}\right)^3 \quad \text{ is the mass already added }
\end{equation}
Using the virial theorem \citep{Daigne2015}, it can be shown that up to half of this energy can be radiated as the body contracts\footnote{The other half is added to the internal energy of the body.}, which gives an estimate of the age of the Earth :
\begin{equation}
\tau \sim 20 \text{Myr}
\end{equation}
This estimate was already incompatible with the geological and Darwinian timescales required to explain the observations available at that time which were advocating in favour of larger (and eventually, more realistic) ages. Nuclear power turned out, a few decades later, to be a more realistic process to account for stellar luminosities, at least on the main sequence. Indeed, the Kelvin-Helmholtz mechanism is believed to be the main culprit to explain the luminosity of proto-stars\footnote{Before the core reaches temperatures high enough to trigger on thermonuclear fusion ($\sim$10$^6$K for Hydrogen), the cloud keeps collapsing under the action of gravity.}. 

\subsection{Light emission}
\label{sec:lum_spec}

\subsubsection{Luminosity}

We have seen that the luminosity of accreting compact objects could reach $10^{36}$erg$\cdot$s$^{-1}$ ($\sim 260$L$_{\odot}$), provided they were fed significant amounts of gas ($\dot{M}\sim10^{-10}M_{\odot}\cdot$yr$^{-1}$). We leave the question of the source of matter for Chapter \ref{chap:x-ray_sources} and \ref{chap:roche} and portray the case of the steady $\alpha$-disk\footnote{$\alpha$ refers to a degree of freedom of the problem which stands for an effective viscosity due to the turbulence triggered by the magneto-rotational instability for instance \citep{Balbus1991}. It is a necessary feature to guarantee the transport of angular momentum outwards and, by then, the possibility for matter to spiral in. Indeed, molecular viscosity can be shown to not be large enough to account for outwards transport of angular momentum \citep[see section 4.7 of][]{Frank2002}.} model introduced by \cite{Shakura1973b} : the angular momentum of the accreted flow is high enough to make the inflow form a geometrically thin and optically thick disk around the accretor. This landscape which serves as a reference for accretion theory will not be described in detail in this manuscript but is precisely documented in \cite{Frank2002}. The computation we carry on in this section also owes much to J. N. Winn's course 8.901 given at \textsc{mit} during the second term of year 2011-2012.

\begin{figure}
\begin{center} %
\includegraphics[height=7.2cm, width=11.1cm]{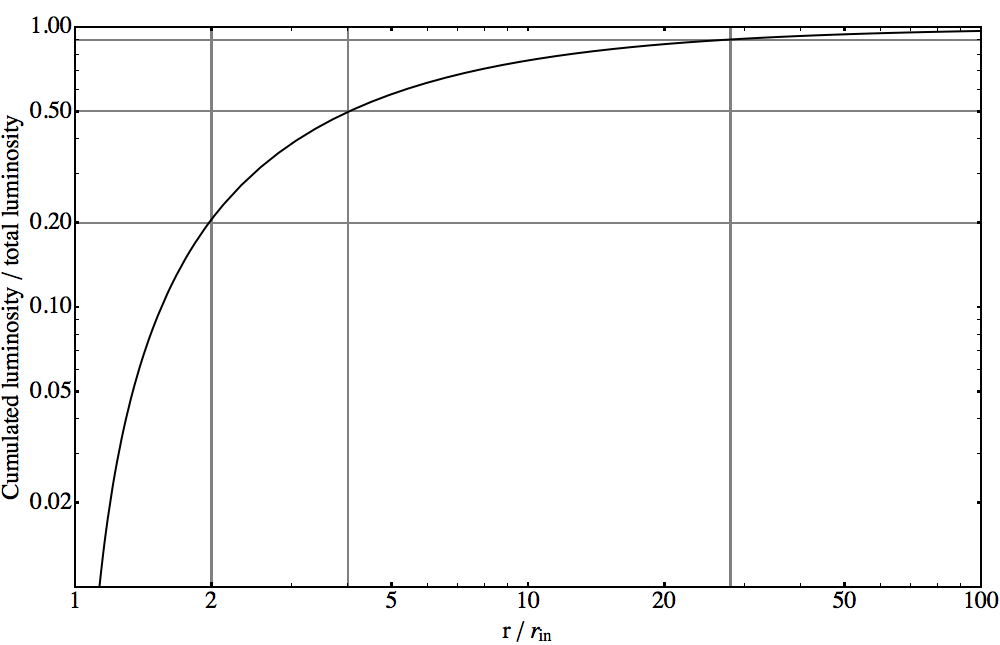}	
\caption{Fractional luminosity of a steady $\alpha$-disk below a radius $r$ in units of the inner radius of the disk, according to \eqref{eq:cum_lum}. Most of the light is emitted by the inner parts of the disk, hotter, due to the strong dependence of the black body flux on the temperature ($\propto T^4$).}
\label{fig:cumulated_disk_luminosity}
\end{center}
\end{figure}

Luckily enough, we will see in a moment that most of the light emitted by this structure turns out to be, when the accretor is a compact object, in the X-ray domain - well above the energy range of the stellar spectrum of the companion in an X-ray binary for instance. The disk itself is not the only structure expected around a compact accretor\footnote{The system can display, among others, jets \citep{Markoff2003}, internal shocks \citep{Cohen2014a}, a stellar and a disk wind, a corona or a hot spot in stream-dominated mass transfers.}, in particular for a low angular momentum flow, nor the only one to radiate high energy photons. However, the $\alpha$-disk model is a convenient framework which proved useful to explain a substantial number of observational configurations and more specifically, the multi-color black body observed in many cases (see next section for synthetic spectra). In Chapter 5, \cite{Frank2002} shows that the effective black body temperature of the disk at a distance $r$ of the accretor of mass $M$ accreting at a rate $\dot{M}$ is given by :
\begin{equation}
\label{eq:temp_eff}
T\left(r\right)=\left[ \frac{3GM\dot{M}}{8\pi \sigma r^3} \left( 1-\sqrt{\frac{r_{\text{in}}}{r}} \right) \right]^{1/4}
\end{equation} 
where $\sigma$ is the Stefan-Boltzmann constant and $r_{\text{in}}$ is the position of the inner edge of the disk. That being said, we can start to quantify more rigorously\footnote{In the previous section \ref{sec:KH_extraction}, we used a test-mass but did not address the flow as a continuous medium.} the accretion luminosity associated to this configuration. If the outer edge of the disk is much farther away from the accretor than the inner edge, we have :
\begin{equation}
\label{eq:Ltot}
L_{\text{tot}}\sim\int_{r=r_{\text{in}}}^{r_{\text{out}}=+\infty} \sigma T^4\left(r\right)\times 2 \times 2\pi r \d r = \frac{1}{2} \frac{GM\dot{M}}{r_{\text{in}}}
\end{equation}
where we summed over all the black body rings of infinitesimal surface $2\pi r \d r \times 2$ (since upper and lower faces) and where $r_{\text{out}}$ refers to the outer border of the disk. Without surprise, if the disk extends down to the surface of the accretor (\ie $r_{\text{in}}=R$), we retrieve the standard expression of the accretion luminosity given by equation \eqref{eq:Lacc}, the factor $1/2$ being due to the fact that we now work on a system which has an internal structure (and energy) \ie where thermodynamics and the virial theorem preclude that more than half of the variation in gravitational potential energy goes in the bulk motion of the flow. A decisive feature of those discs, at least from an observational point of view, is that most of the luminosity comes from the inner parts of the disk. Indeed, if we compute the ratio $L\left(<r\right)/L_{\text{tot}}$ of the luminosity being emitted by the disk below a certain radius $r$ - \ie the cumulative luminosity - with respect to the total luminosity, we get :
\begin{equation}
\label{eq:cum_lum}
\begin{aligned}
\frac{L\left(<r\right)}{L_{\text{tot}}}&=	\frac{2r_{\text{in}}}{GM\dot{M}}\cdot \int_{r_{\text{in}}}^{r} \sigma T^4\left(r'\right)\cdot 4\pi r' \d r' \\[10pt]
&=3 \int_1^x \frac{1-\sqrt{1/x'}}{x'^2}\d x' \quad \text{ with } \quad x=r/r_{\text{in}}\\[10pt]
&=1-3\frac{r_{\text{in}}}{r}\left( 1-\frac{2}{3}\sqrt{\frac{r_{\text{in}}}{r}} \right)
\end{aligned}
\end{equation}
in Figure\,\ref{fig:cumulated_disk_luminosity}, we notice that wherever the inner edge of the disk is, a fifth of the total light is emitted within 2$r_{\text{in}}$, half within 4$r_{\text{in}}$ and 90\% within 30$r_{\text{in}}$.

\subsubsection{Spectrum}

A classical preliminary remark about the spectral range of the emitted light due to the accretion process is to distinguish the optically thick case, treated in more details below, and the optically thin case. In the latter configuration, where the disk is secondary if not absent, the emission is simply associated to the thermal energy of all the particles (of mean weight per particle $\mu$ of the order of the mass of a proton $m_p$) related to the change in gravitational potential energy via the virial theorem :
\begin{equation}
\frac{3}{2}k_B T_{\text{thin}} \sim \frac{1}{2}\frac{GM\mu}{R}
\end{equation} 
\begin{equation}
k_BT_{\text{thin}}\sim \frac{GMm_p}{3R} \sim 69\text{MeV}
\end{equation} 
where $k_B$ is Boltzmann's constant. As mentioned in section \ref{sec:obj}, the diversion of the rotating flow within the magnetosphere of a neutron star can lead to the accretion of an optically thin flow susceptible to account for very high energy photons. This gamma-ray emission is indeed observed in several systems but is not always believed to be the main source of light which often finds its origin in the process we now describe. 

Let us compute the spectrum itself of the source\footnote{An advanced way to do what we just sketched here is provided by the code \href{https://heasarc.gsfc.nasa.gov/xanadu/xspec/}{\textsc{xspec}} \citep{Arnaud1996}.} assuming that the emitted spectrum at the surface of the disk is not altered by any other structure\footnote{In practice, the emitted photons are reprocessed in the hot corona surrounding the disk. It can be used, once the reflection on the disk is studied, to estimate the inner radius of the disk and, by then, the spin of the \bh \citep{Reynolds:2013va}.}. The radiative theory notions used here are introduced and developed in much details in \cite{Rybicki1981}. The specific intensity $I_{\nu}$ (in erg$\cdot$s$^{-1}\cdot$Hz$^{-1}\cdot$sr$^{-1}$) of the disk surface follows Planck's black body photon frequency distribution :
\begin{equation}
I_{\nu}\left( r \right) = \frac{2h\nu ^3}{c^2} \frac{1}{e^{\frac{h\nu}{kT\left(r\right)}}-1}
\end{equation}
where $T\left(r\right)$ is given by equation \eqref{eq:temp_eff}.

\begin{figure*}
\hspace*{-1cm}
\begin{subfigure}{0.49\textwidth}
\begin{center}
\includegraphics[width=7.8cm, height=7cm]{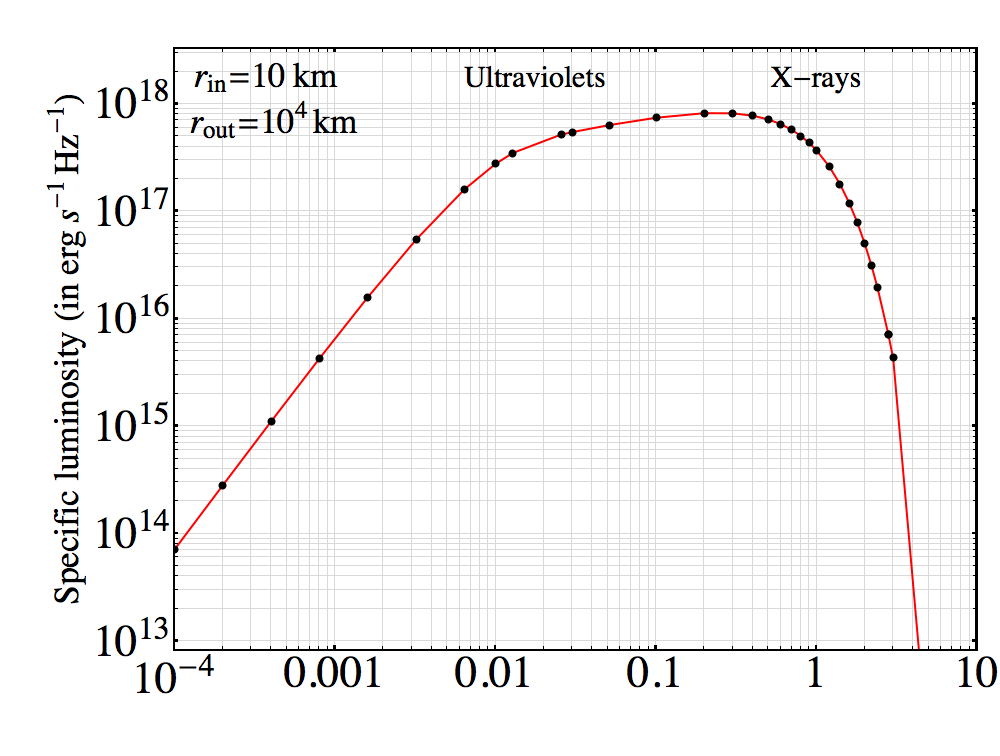} 
\label{fig:subim1}
\end{center}
\vspace*{-1cm}
\end{subfigure}
\hspace*{0.5cm}
\begin{subfigure}{0.49\textwidth}
\begin{center}
\includegraphics[width=7.5cm, height=7cm]{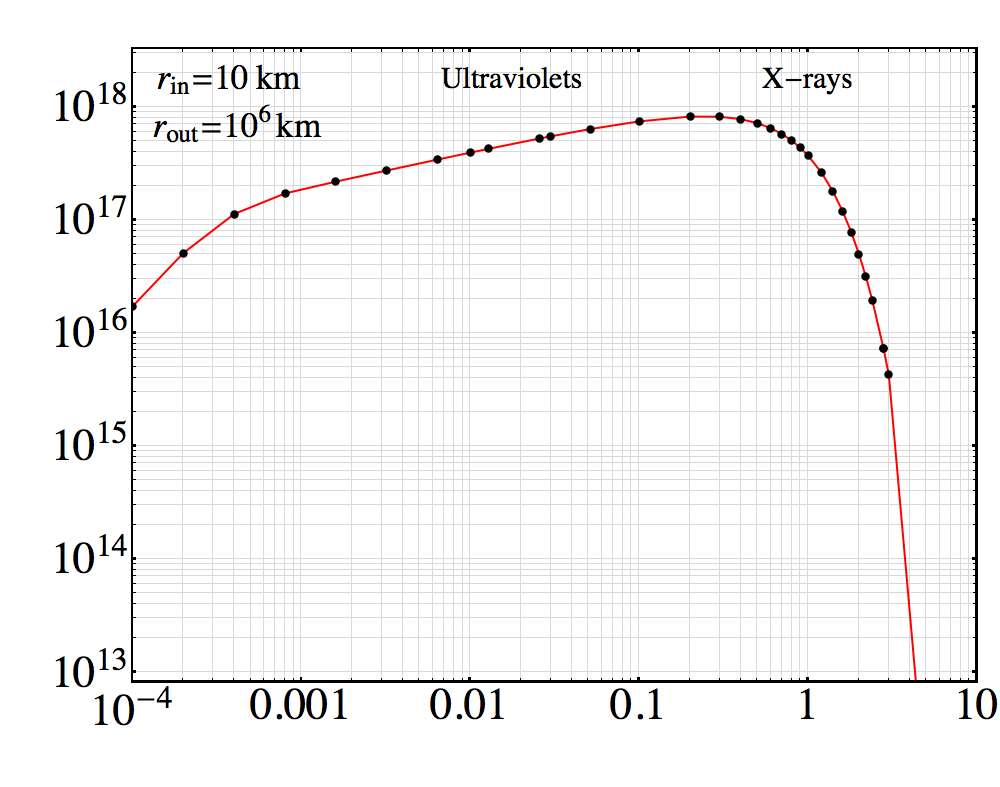}
\label{fig:subim2}
\end{center}
\vspace*{-1cm}
\end{subfigure}
\hspace*{-1cm}
\begin{subfigure}{0.49\textwidth}
\begin{center}
\includegraphics[width=7.8cm, height=7cm]{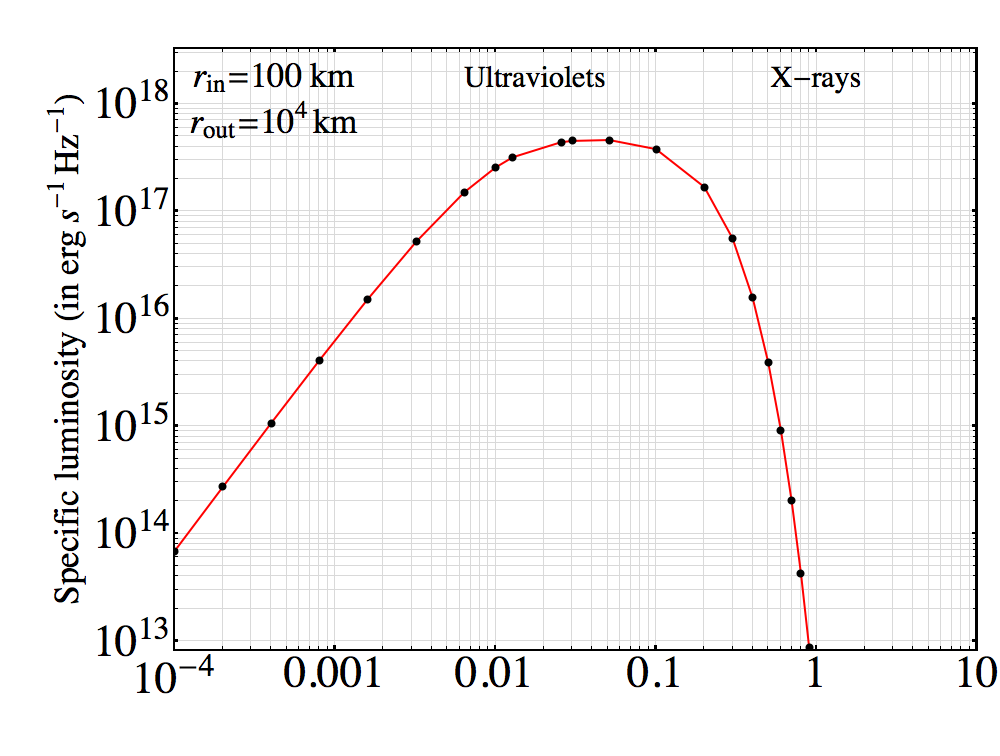}
\label{fig:subim2}
\end{center}
\vspace*{-1cm}
\end{subfigure}
\hspace*{0.5cm}
\begin{subfigure}{0.49\textwidth}
\begin{center}
\includegraphics[width=7.5cm, height=7cm]{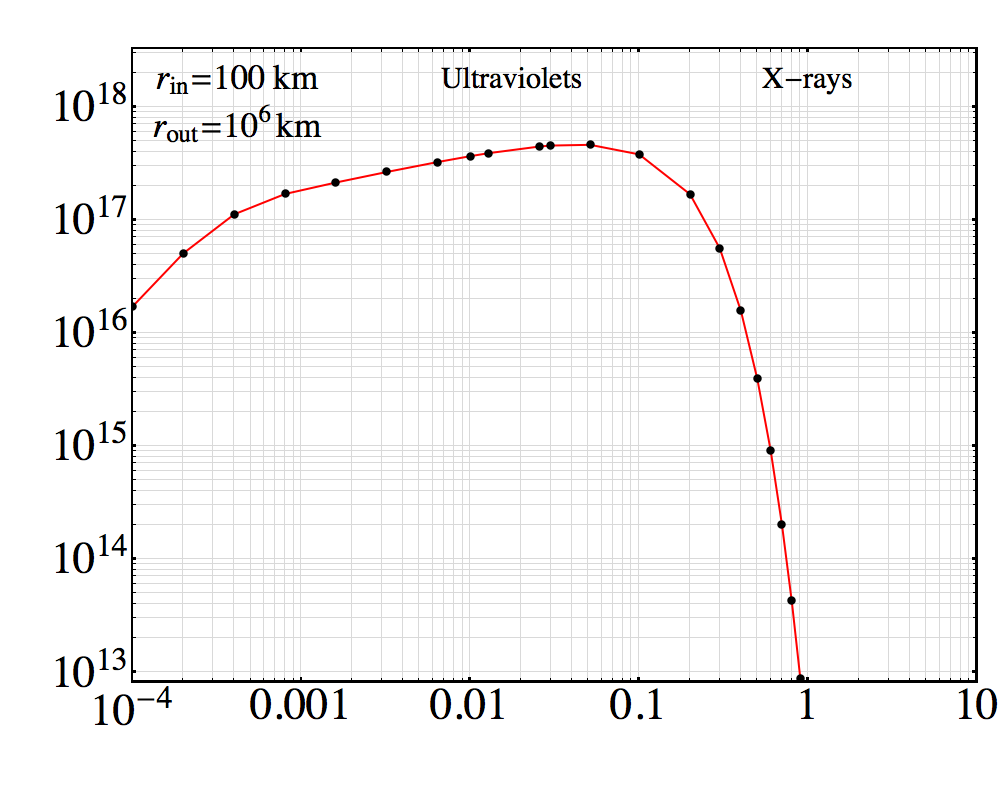} 
\label{fig:subim1}
\end{center}
\vspace*{-1cm}
\end{subfigure}
\hspace*{-1cm}
\begin{subfigure}{0.49\textwidth}
\begin{center}
\includegraphics[width=7.8cm, height=7cm]{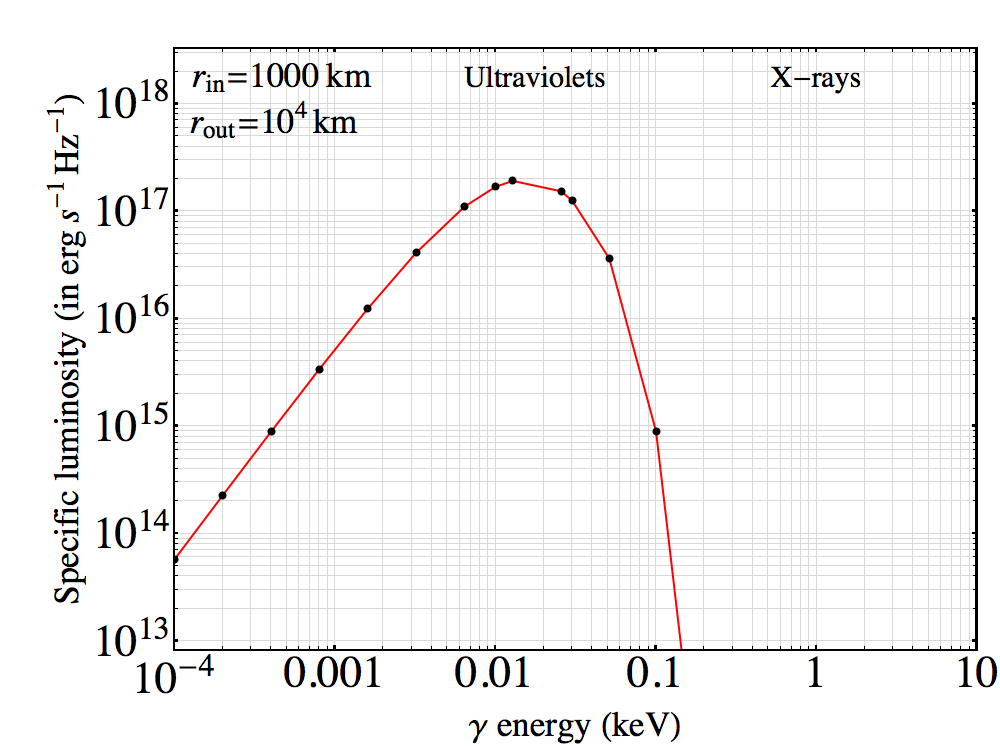}
\label{fig:subim2}
\end{center}
\end{subfigure}
\hspace*{0.5cm}
\begin{subfigure}{0.49\textwidth}
\begin{center}
\includegraphics[width=7.5cm, height=7cm]{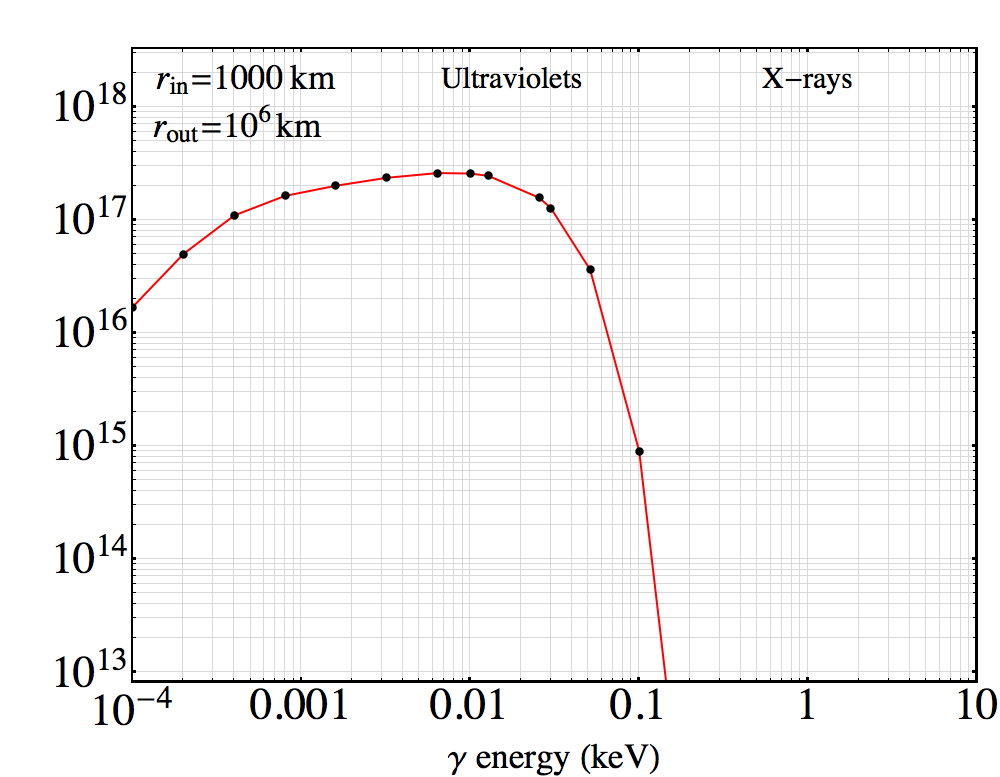}
\label{fig:subim2}
\end{center}
\end{subfigure}
\caption{Synthetic spectra of a steady $\alpha$-disk \citep{Shakura1973b} for an inner edge at 10 (up), 100 (middle) and 1,000km (bottom) of the center of the compact accretor. The right column corresponds to discs with a larger spatial extension (extending up to one million kilometers) than the left column (10,000km). The red line is a linear interpolation between the black dots which have been numerically evaluated using equation \eqref{eq:Lnu_spectrum}. The vertical and horizontal scales are the same in all 6 configurations and the domains of ultraviolets and X-rays have been indicated.}
\label{fig:spectra}
\end{figure*}

The associated specific luminosity $L_{\nu}$ (\ie luminosity per frequency unit, in erg$\cdot$s$^{-1}\cdot$Hz$^{-1}$) of an annulus at radius $r$ (for both sides) is :
\begin{equation}
\label{eq:Lnu_spectrum}
L_{\nu}=\int_{r_{\text{in}}}^{r_{\text{out}}}I_{\nu}\left( r \right) \cdot 4\pi r \d r = \frac{8\pi}{h^2 c^2}r_{\text{in}}^2 E^3 \int_{1}^{r_{\text{out}}/r_{\text{in}}} \frac{x\d x}{e^{\beta / f\left( x \right)}-1}
\end{equation}
where $E$ is the photon energy $h\nu$, $x$ is still $r/r_{in}$ and :
\begin{equation}
\begin{cases}
\beta=\frac{E}{k_B}\left[ \frac{8\pi\sigma r^3_{\text{in}}}{3GM\dot{M}} \right]^{1/4}\sim 1.6 \left( \frac{E}{1\text{keV}} \right) \left( \frac{r_{\text{in}}}{\text{10km}} \right)^{\frac{3}{4}} \left( \frac{M}{1.5M_{\odot}} \right)^{-\frac{1}{4}} \left( \frac{\dot{M}}{10^{-10}M_{\odot}\cdot\text{yr}^{-1}} \right)^{-\frac{1}{4}} \\
f\left( x \right) = \frac{1}{x^{3/4}}\left( 1-\sqrt{\frac{1}{x}} \right)^{1/4}
\end{cases}
\end{equation}
Spectra corresponding to the typical numerical values for a disk around a \ns (of mass 1.5\msun) orbiting a stellar companion (and accreting at a rate of $\dot{M}=10^{-10}M_{\odot}\cdot$yr$^{-1}$) have been sampled and represented in Figure\,\ref{fig:spectra}. The left column corresponds to the case of a disk of smaller radial extension (\ie lower angular momentum) than the right column where the outer edge is of the order of the radius of the Roche lobe (see Chapter \ref{chap:roche}). Three distinct regime can be seen : a first fast rise in $\nu^2$ (corresponding to the low-energy tail of the outermost black body emission at a temperature $T\left( r_{\text{out}}\right)$), a flat multi-color black body component (made of the piling up of black bodies with temperatures between $T\left( r_{\text{out}}\right)$ and $T\left( r_{\text{in}}\right)$) and an exponential cut-off beyond the energy corresponding to the innermost temperature $T\left( r_{\text{in}}\right)$. Notice that given its low luminosity, the position of the outer edge has little influence on the spectrum (since it does not modify the high energy part) : most of the time, it is not observationally constrained directly with the spectrum \citep[but rather with a host spot or with the theoretical upper limit of 70\% of the Roche lobe radius, \aka the disk tidal truncation radius given by][]{Paczynski1977}. Indeed, if the spectrum is at the same level of specific intensity, the actual luminosity\footnote{\ie after having integrated over frequencies, much narrower for the radiation emitted by the outer edge. It is a reason why observers sometimes prefer to plot the product of the frequency by the specific flux.} is several orders smaller than the portion associated to the high energy part of the spectrum\footnote{Hence a total luminosity in equation \eqref{eq:Ltot} which essentially depends on the position of the inner edge of the disk, not of the outer edge.}. A more visible feature is the cutoff at the inner edge of the disk which moves from 10km ($\sim$ surface of a \ns) on the upper row to 1,000km ($\sim$ surface of a white dwarf or the \ns magnetosphere - see section \ref{sec:obj}) on the lower row, where the emission in X-rays is vanishingly small ; since the stellar companion in an X-ray binary emits partly in ultraviolets (especially for a hot companion in a High-Mass X-ray Binary, see section \ref{sec:HMLMXB} in the next chapter), it challenges our very capacity to distinguish the contribution of the disk, in addition of being on the whole less luminous according to equation \eqref{eq:Ltot} for the total luminosity.

A final comment before presenting accreting compact objects in their environment : as suggested by Figure\,\ref{fig:spin_BH}, if the position of the inner edge of the disk follows the one of the \textsc{isco}, a sudden swing in the angular momentum of the flow which feeds the disk would have dramatic effects around a maximally rotating \bh. Indeed, the inner border would then be almost ten times farther away from the accretor when the disk is counter-rotating (with respect to the \bh spin) than when it is co-rotating. Photometrically, it would correspond to a luminosity almost ten times smaller\footnote{Like for an eclipsing system though, dilution effects play a role if there are other X-ray sources in the system such as a corona around the disk (which is usually fitted by a power-law) ; as usual, the integrated luminosity is merely a hint of what the full light distribution (\ie the spectrum) looks like.} while spectroscopically, the cutoff drops by a factor of a few for an inner edge rising from 10 to 100kms. 
	

\newpage



\setlength{\parskip}{0ex} 


\chapter{X-ray sources}
\label{chap:x-ray_sources}
\chaptermark{X-ray sources}
\hypersetup{linkcolor=black}
\minitoc
\hypersetup{linkcolor=red}
\setlength{\parskip}{1ex} 

Historically, the first extrasolar X-ray source discovered by \cite{Giacconi1962}, Sco X-1, turned out to be an accreting compact object in a binary system. Isolated, compact objects generally display signals too quiet to be detected with the current technology but in interaction with their environment, a rich phenomenology blossoms. In a first section, we introduce the physical stage which is the central one in this manuscript, the X-ray binaries. The vast family of the X-ray binaries has long ramified into observational and physical branches that we will briefly summarize. The accent will be put on successive dichotomies which include the archetype of systems where wind accretion occurs, Supergiant X-ray Binaries. Concerning the X-ray sources not related to binary systems, the second half of this chapter is devoted to a few astronomical objects susceptible to be well represented by some of the numerical setups we designed and where the present numerical work might be relevant.


\section{X-ray binaries}
\label{sec:x-ray_binaries}


\subsection{The accretor}
\label{sec:the_accretor}

As explained in section \ref{sec:lum_spec}, the accretor has to be a compact object to account for such an emission of high and very high energy photons (\ie X-rays and beyond). The accreting body plays a role through its gravitational influence and is hardly responsible directly for the radiations we observe : what we witness is the emission of the flow being accreted. It is why it has been suggested that the environment of the compact object could serve as a probe to investigate its Dantean neighborhood and maybe, unveil unexpected behaviour in strong gravity regime. However, such an aim can only be reached if we manage to delimit the variability inherent to the inflow and dig for systematics behind the scenes.

The vast majority of compact accretors are neutron stars\footnote{The two first suspected candidates were the accretors in the Roche lobe overflow high and intermediate mass X-ray binaries, respectively Cen X-3 and Her X-1. The few accreting black hole candidates will be mentioned in \ref{sec:BH_RLOF_SgXB}.}, a diagnosis which can rely on several arguments. First, many of those neutron stars manifest themselves as accreting pulsars (in particular in high mass X-ray binaries - see section below), not to be confused with radio pulsars which tap their emission energy from the \ns rotation. The magnetic field is large enough (typically $\gtrsim $10$^{11}$G) to funnel the accreted gas to the magnetic poles where it produces tiny regions of bright and high energy emission \citep[see the accretion columns in][]{Pringle1972,Davidson1973}. As they spin with the neutron star, since the magnetic field is not aligned with the spin axis, the amplitude of the emission is modulated and pulses at the neutron star spin period appear in the light signal. The magnetic field can even be evaluated, validating the above scenario, by using the cyclotron resonant scattering absorption lines observed in those systems\footnote{They are caused by the scattering of hard X-ray photons on electrons whose energy is quantified by the magnetic field according to the Landau levels - see \eg \cite{Wasserman1983} for the theoretical framework and \cite{Walter15} for an overview of observations of this feature in Galactic \hmxb.}. Second, the dynamically deduced masses are consistent with the conclusions drawn by core-collapse models and the maximum masses set by the available equations-of-state for the ultra-condensed matter of \ns. Finally,  spectroscopic arguments can be used to differentiate an accreting \ns from an accreting \bh. The former is characterized by a power-law of photon index 0.3 to 2 and a high energy exponential cutoff with cyclotron resonance scattering features at higher energies \citep{Walter15}.


\subsection{\hmxb \& \lmxb}
\label{sec:HMLMXB}

The two main families of X-ray binaries depend on the mass of the companion star :
\begin{enumerate}
\item \underline{Low mass X-ray binaries (\lmxb) :} the companion star is of low mass (at most a couple of solar masses but more generally below one solar mass). The corresponding spectral type is A or later.
\item \underline{High mass X-ray binaries (\hmxb) :} the companion star is of high mass (above several solar masses, usually above 10\msun) The corresponding spectral type is B or earlier.
\end{enumerate} 
The \lmxb then host stars older than \hmxb due to the slower evolution of lower mass stars. While the age of the star is typically of less than 10 millions years in \hmxb, \lmxb host stars older than a few billions years\footnote{See section \ref{sec:massTransfer} for a more detailed discussion on the secular evolution of a stellar binary system into an X-ray binary. In particular, the remarkable bimodality of the stellar mass distribution of those systems (with few intermediate mass X-ray binaries) is discussed there.}. Since the \hmxb are younger, the magnetic field of the \ns they host (if it is not a \bh) is still large. The accretion columns are brighter and more extended and a pulsar is more likely to be detected than in \lmxb where the magnetic field is lower.

\begin{figure}
\begin{center}
\includegraphics[height=9cm, width=12cm]{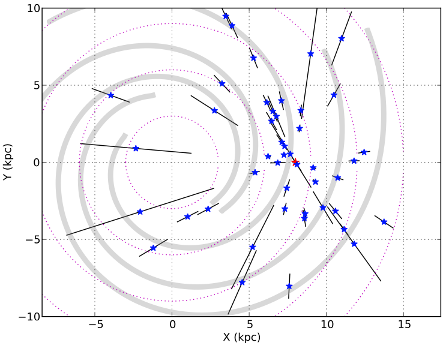}	
\caption{Zoom in around the Sun (red star set at 8.5kpc of the Galactic center) with the spiral arms in light gray. The blue stars with the black lines represent the positions of the \hmxb in \cite{Coleiro2013} with the uncertainty on the distance. From \cite{Coleiro2013a}.}
\label{fig:milky_way_hmxb}
\end{center}
\end{figure}

Concerning their spatial distribution in the Milky Way, \hmxb appear very close to the Galactic plane while \lmxb have a broader Galactic latitude distribution. If the more venerable age of \lmxb partly explains this discrepancy, it is not a sufficient reason as emphasized by \cite{Podsiadlowski2010}. Indeed, the main element is the kick provided to the system by the non spherical collapse of the naked Helium core into a neutron star. For systems which remain bounded after the associated Supernova explosion, it results in larger kicks for lower mass systems \ie the \lmxb. In addition, because of the longer lifetime of low mass stars, a \lmxb will be able to travel further away than a \hmxb. \cite{Brandt1995} showed that X-ray binaries born in the Galactic plane with realistic mass distributions reproduce well the observations when the same \ns kick velocity distribution is adopted for all binaries. For the spatial distribution within the Galactic plane, \cite{Coleiro2013} used a spectral energy distribution fitting procedure in optical and near infrared to derive the distance and absorption of a statistical sample of \hmxb. As visible in Figure\,\ref{fig:milky_way_hmxb}, there is a correlation between the position of the \hmxb and the stellar formation zones (\eg the spiral arms).

Another feature concerns the mass transfer from the star to the compact companion responsible for the X-ray bright emission. Most of the time, the spectrum is softer in \lmxb (with photon energies $\lesssim$ 10keV) than in \hmxb ($\gtrsim$ 15keV) which invite us, according to the arguments developed in section \ref{sec:lum_spec}, to support the existence of a robust and large optically thick geometrically thin disk-like structure around the compact object in \lmxb, not necessarily in \hmxb. The presence of this disk in \lmxb is supported by the mechanism of Roche lobe overflow (\rlof) described in Chapter \ref{chap:roche}, more likely to be stable in \lmxb than in \hmxb. In the latter, wind accretion (see Chapter \ref{chap:SgXB}) is believed to be the main culprit for mass transfer, with much lower subsequent angular momentum for the inflow.

Eventually, jets have been observed mostly in \lmxb but also in \hmxb\footnote{Cygnus X-1, Cygnus X-3 and SS433.}. The compact accretor can be either a \bh (\eg GRS1015+105) or a \ns (\eg Sco X-1). Different mechanisms exist to explain those ejections of matter highly beamed and at velocities sometimes of the order of the speed of light : an electromagnetic equivalent of the Penrose extraction process to tap the spin energy of Kerr \bh in the ergosphere \citep{Blandford1977} or a wind from a magnetized accretion-ejection structure \citep{Blandford1982,Casse2002}. Those systems are called microquasars which refers to a subfamily of active galactic nuclei, the quasars.


\subsection{\sgx \& Be\textsc{xb}}
\label{sec:SgxBeXB}

\begin{figure}
\begin{center}
\includegraphics[height=9.5cm, width=9cm]{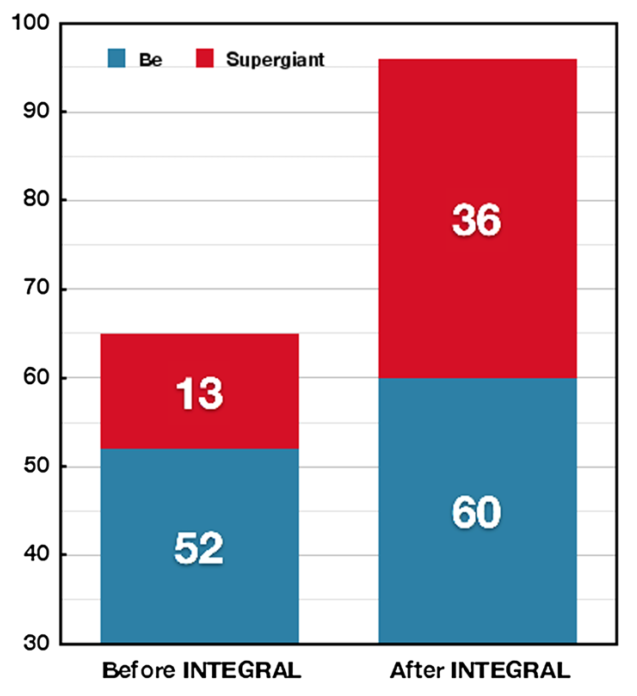}	
\caption{Estimates of the numbers of observed \hmxb in the Milky Way classified as \bexb or \sgx, before and after the \INTEGRAL mission. From \cite{Walter15}.}
\label{fig:prop_Be_VS_Sg}
\end{center}
\end{figure}

\begin{figure}
\begin{center}
\includegraphics[height=8cm, width=10cm]{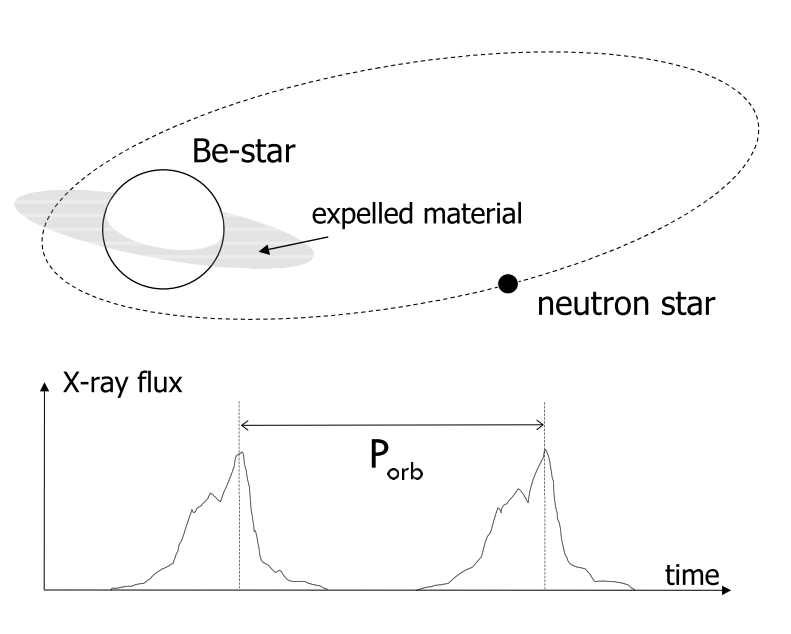}	
\caption{Representation of a \bexb with a neutron star on an eccentric orbit around a non-\rlof Be star. The misalignement between the stellar and orbital spins implies intermittent crossings of the decretion disk by the neutron star, which produces an orbital time modulation of the X-ray flux (bottom part). From \cite{Tauris2003}.}
\label{fig:BeXB}
\end{center}
\end{figure}

\begin{figure}
\begin{center}
\includegraphics[height=14cm, width=15cm]{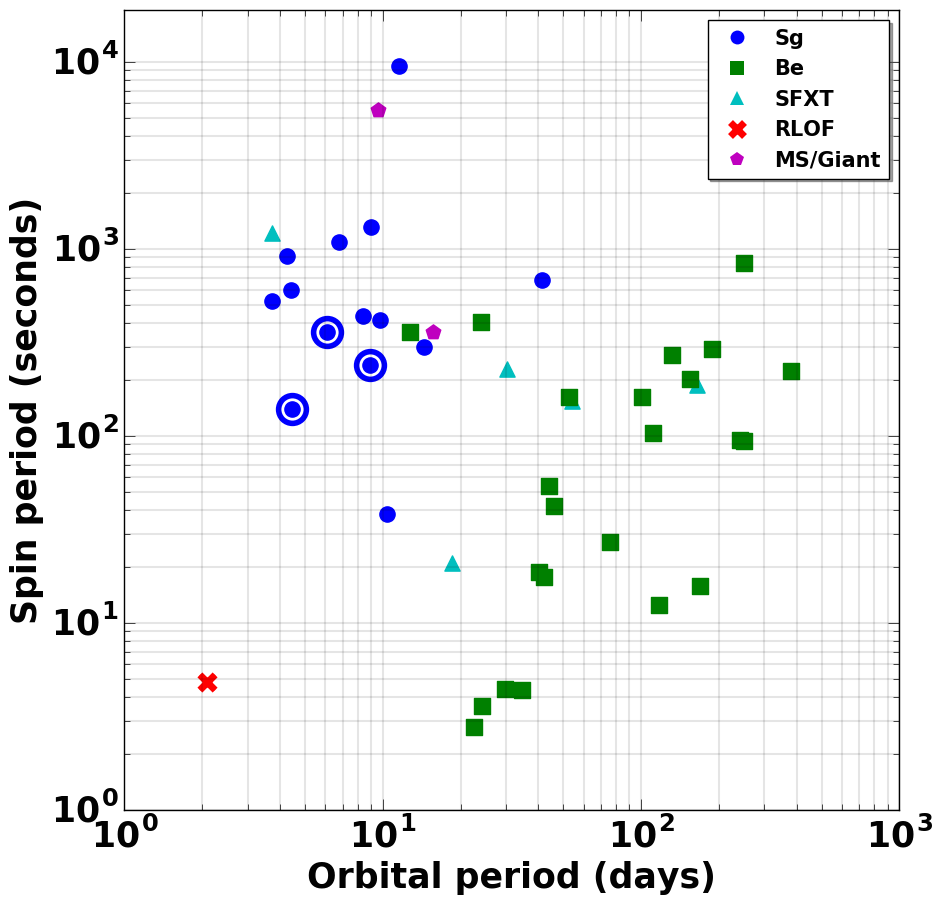}	
\caption{Corbet diagram of Galactic \hmxb hosting a \ns. The \ns spin is represented as a function of the orbital period of the binary, for systems where both are measured. For \bexb (green squares), a correlation exists between the two, not in the case of \sgx (blue dots). In the bottom left part of the Figure is the only \rlof system among this sample (red cross), Cen X-3. See \citep{VanderMeer2007} for a joint study of two other \rlof \hmxb hosting a neutron star, SMC X-1 and LMC X-4 (both extragalactic). \sfxt and systems with an atypical donor star have also been represented. The three circled \sgx are the sources investigated in more details in Chapter \ref{chap:SgXB}.}
\label{fig:corbet_diag}
\end{center}
\end{figure}

Among \hmxb, an additional distinction is made between two main categories depending on the nature of the high mass star \citep[see also the review by][]{Chaty2011a}. Indeed, until the \INTEGRAL satellite\footnote{For \textsc{inte}rnational \textsc{g}amma-\textsc{r}ay \textsc{a}strophysics \textsc{l}aboratory, launched in 2002.}, most \hmxb observed hosted a specific kind of B star\footnote{More precisely, the star is generally between O9 and B3.} characterized by high spin rates, many strong Balmer lines in emission and an infrared excess \citep{Coe1999,Porter2003,Reig2011a}. They are called Be stars\footnote{Where the \textit{e} stands for \textit{emission lines}.} and the \hmxb where they are found, \bexb. The star spins fast enough to get close from its breaking limit and a decretion disk, responsible for the emission lines and the infrared excess, forms in the equatorial plane (see the skecth in Figure\,\ref{fig:BeXB}). All \bexb host a \ns as an accretor, on an eccentric orbit ($>$0.3), usually inclined with respect to the stellar equatorial plane. \bexb are generally transient sources (with a quiescent level at 10$^{34}$erg$\cdot$s$^{-1}$). They display periodic X-ray type I bursts (up to 10$^{36}$erg$\cdot$s$^{-1}$) at the orbital period of the orbiting neutron star (bottom panel in Figure\,\ref{fig:BeXB}) which can be explained by the passage of the \ns through the decretion disk where the mass density is higher. The X-ray luminosity levels reached are consistent with a wind accretion process\footnote{Through this manuscript, by wind accretion we mean not only the accretion of a stellar wind but also any mode of accretion where the kinetic energy of the flow with respect to the accretor at large distance of it is important compared to the magnitude of the gravitational potential energy of the compact object at the Bondi radius (\ie the flow is supersonic at infinity) and compared to the angular rotation speed of the system if it is in a binary. It covers the framework established by Bondi, Hoyle and Lyttleton for instance - see Chapter \ref{chap:acc_pt-mass}.}. However, some of those sources also display non periodic larger bursts (type II), up to an X-ray luminosity of 10$^{37}$erg$\cdot$s$^{-1}$, which are believed to feature transient massive accretion of the stellar extended atmosphere in a way reminiscent of the Roche lobe overflow in \lmxb\footnote{Indeed, with an eccentric orbit, a star can fill or not its Roche lobe depending on the phase of the orbit.}. These events are typically accompanied by large spin-up rates that suggest the formation of a transient accretion disk around the \ns (see the correlation in Figure\,\ref{fig:corbet_diag}).

The second family, generally more persistent, contains the archetypes of the wind accreting X-ray sources. The companion star is an evolved OB Supergiant (hence the name \sgx) and the accretor, on a low eccentricity orbit ($<$0.2), is either a \ns (most of the time) or a \bh. The orbital periods are shorter than in \bexb ($<$10 days). In \sgx hosting a \ns, a handy tool to differentiate \sgx and \bexb is the Corbet diagram \citep{Corbet1984} which represents the spin period of the neutron star as a function of the orbital period for observed values. Figure\,\ref{fig:corbet_diag} is an updated version of this diagram for Galactic \hmxb, mostly based on the measured gathered in \cite{Walter15}. The main results highlighted by this Figure are the following :
\begin{enumerate}
\item \bexb present a correlation between the orbital period and the spin of the \ns which suggests that the angular momentum transfer is efficient and the mass transfer, conservative. It was suggested by \cite{Waters1989} that it was a manifestation of the propeller effect introduced by \cite{Illarionov1975} : a negative feedback locks the \ns spin period such as the corotation velocity at the magnetospheric radius matches the Keplerian velocity. The orbital periods span from a couple of 10 days up to a few hundred days.
\item the few \hmxb undergoing Roche lobe overflow (only one in the Milky Way, Cen X-3) with short orbital periods (see section \ref{sec:BH_RLOF_SgXB}) host fast rotating pulsars with very short spin periods. An accretion disk is believed to be a favoured transitional structure to guarantee an optimal transfer of orbital angular momentum to the spin of the \ns, significantly accelerating it.
\item the \sgx systems show no trace of correlation between orbital and \ns spin periods, suggesting an inefficient mass transfer mechanism, consistent with wind accretion (highly non conservative). They also have shorter orbital periods, consistent with the secular evolutionary tracks. The \ns spin periods are rather long (above 100s).
\end{enumerate} 
The position in this diagram is, by itself, an argument to classify a source as a \sgx or a \bexb. When we differentiate between the two based on the eccentricity of the \ns orbit, one has to keep in mind that the circularization of the orbit is an on/off mechanism with a rather sharp threshold (see section \ref{sec:tidal_interactions} and Figure\,\ref{fig:obs_circularization} in particular). Indeed, the capacity of an orbit to circularize within the stellar lifetime (see equation \eqref{eq:tides_circ}) is very dependent on the ratio of the stellar radius by the orbital separation and the stars in \bexb are both smaller and associated to longer periods, which makes circularization over the nuclear timescale essentially ineffective, contrary to \sgx which turn out to be below the threshold for circularization. However, notice that non eccentric \bexb would lead to truncated decretion disks with lower outer radii and thus, much lower peak X-ray luminosities as the \ns intersects a lower density environment \citep{Negueruela2000} ; even if they existed, their detection would be observationally challenging.

\subsection{Classical \sgx, obscured \sgx \& \textsc{sfxt}}
\label{sec:class_sfxt}

Among the \sgx are found several classes depending on the time variability of the X-ray emission and the absorption properties. Due to the reduced number of \sgx, it is the last available classification up to now, with 3 main families : in Figure\,\ref{fig:prop_Be_VS_Sg}, among the 36 confirmed \sgx in 2015, 11 to 18 are persistent \sgx (the classical ones), 6 to 9 are highly obscured \sgx and the rest are mostly SuperFast X-ray Transients \citep[\sfxt, ][]{Walter15}. 

The latter family presents fast outbursts with rise times of the order of a few ten minutes and with typical durations of a few hours \citep{Negueruela2006}. Their quiescent level is much lower than the other two families, at an X-ray luminosity of 10$^{32}$ to 10$^{34}$erg$\cdot$s$^{-1}$ \citep{Sidoli2008}, of the order of the observational limit of sensitivity. However, the dynamic range of the flares is also much higher (between 10$^2$ and 10$^4$), which makes them transiently detectable at X-ray luminosity levels up to 10$^{37}$erg$\cdot$s$^{-1}$ : since they are in a non detectable quiescent state most of the time, it took a field of view as large as the one provided by the \INTEGRAL satellite to identify those low duty cycle systems, the flares being separated by weeks. The outburst spectra are consistent with accreting \ns, even if the possibility that some systems host a \bh can not be ruled out when no pulsation has been detected. Their positions in the Corbet diagram (see Figure\,\ref{fig:corbet_diag}) does not follow neither the classical \sgx nor the \bexb ones, which brings up the question of their evolutionary bonds with those two families. It has been proposed that those flares can be caused by serendipitous accretion of clumps in the wind of the Supergiant stellar companion \citep{Walter2007,Ducci2009}, albeit the overdensities alone are not expected to match the high dynamic range factors.

A physical distinction between \sfxt and classical \sgx has been proposed by \cite{Negueruela2008a} relying on the relative size of the accretion radius of the compact object (see Chapter \ref{chap:acc_pt-mass}) relatively to the characteristic size of the clumps in the wind of the Supergiant stars. Because the latter varies with the distance to the star, the orbital separation would make the \ns lie either in a homogeneous or in a heterogeneous environment (relatively to its propensity to accrete matter).

The obscured \sgx are sources with Hydrogen column densities tens times larger than the values associated to other \sgx, up to several 10$^{24}$cm$^{-2}$ \citep{Chaty2011a}. \cite{Filliatre2004} showed that the absorption of X-rays was due to material in the vicinity of the compact object while the absorption in infrared and optical was linked to a cocoon of dust enshrouding the whole system \citep{Chaty2012}.  


\subsection{Outliers and \bh in \sgx}
\label{sec:BH_RLOF_SgXB}

Given the large number of parameters able to influence the behaviour of a wind accreting binary, exceptions are the rule. Four main kinds of exceptions deserve to be mentioned \citep{Walter15} :
\begin{enumerate}
\item The systems where the stellar parameters do not fit the frame previously sketched for \sgx. GX301-2 is one of them, hosting a very evolved hypergiant star, possibly on its way to a Wolf-Rayet stage. Due to the enormous mass loss rate of the star, the system can afford a much larger orbital separation than the other \sgx (see Figure\,\ref{fig:wind_accretion}) and still be detectable. On the other extremity of the range are found more modest main sequence and giant donor stars. 
\item The systems where the stellar wind present atypical properties. For instance, OAO 1657-415 hosts a star with a wind anomalously low at 250\kms \citep{VanLoon2001}.
\item The systems where the accretor is not a \ns but a \bh candidate, which is the likely scenario for a few \sgx : Cyg X-1 \citep{Orosz2011}, Cyg X-3 and SS433\footnote{In the latter two systems, the nature of the stellar companion is not yet clearly identified ; those systems might be intermediate mass X-ray binaries.}. If the gravitational influence of a \ns and a \bh are extremely similar beyond a few Schwarzschild radii of the object\footnote{Due to their similar compactness parameters - see section \ref{sec:compacity}.}, the objects themselves present structural differences\footnote{Especially in terms of self-emission and magnetic properties.} which manifest in the accretion (or ejection) mechanism.
\item The systems where the mass transfer is stream dominated (\ie proceeds through a Roche lobe overflow like configuration) rather than wind dominated as in most \sgx : Cen X-3, SMC X-1 and LMC X-4 are the main representatives of this category \citep{Boroson:2000du,VanderMeer2007}. Those systems often display spectra indicative of the presence of a disk. The main responsible for the scarcity of those systems is probably the expected instability of this mass transfer when a \ns accretes matter from a Supergiant companion (see section \ref{sec:stab_mass_trans}). 
\end{enumerate}



\section{Other X-ray sources}
\label{sec:others}

Other kinds of extrasolar X-ray sources exist which are not found in X-ray binaries but are believed to be linked to accretion onto a compact object. The brief overview which follows intends to remind the reader about the ubiquity of this wider astrophysical situation.


\subsection{Runaway stellar-mass compact objects}
\label{sec:runaway}

\begin{figure}
\begin{center}
\includegraphics[height=10cm, width=10cm]{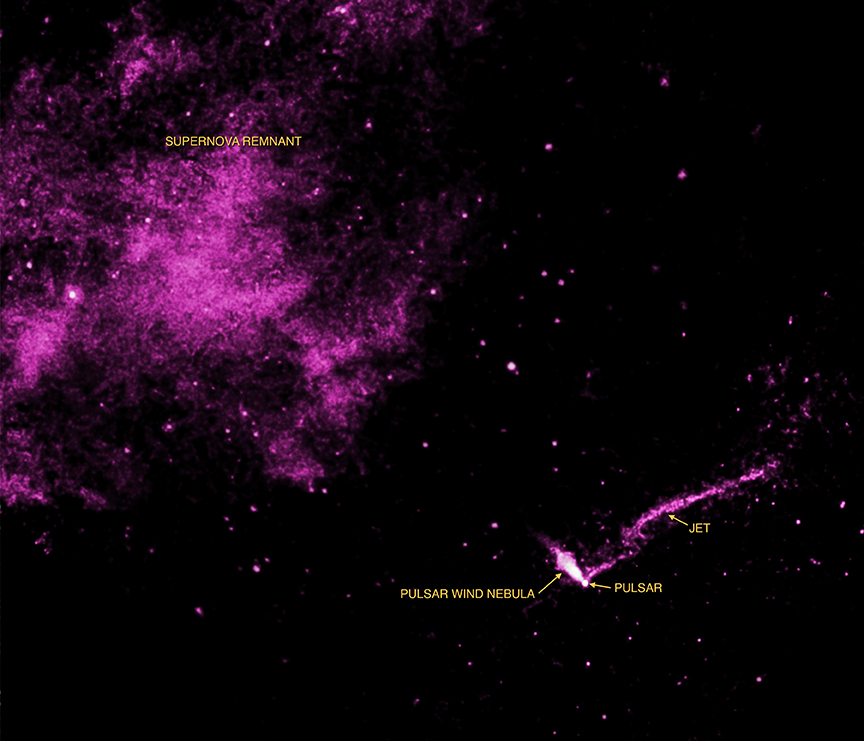}	
\caption{The pulsar IGR J1104-6103 as it speeds through the interstellar medium at over 1,000\kms. It is likely the aftermath of the core-collapse explosion responsible for the supernova remnant in the upper left part of the picture. A jet-like structure is also visible. Credits : \cite{Pavan2014}.}
\label{fig:lighthouse}
\end{center}
\end{figure}

Increasing numbers of hypervelocity stars have been observed, \ie stars with a velocity in the Galactic rest frame larger than 400\kms and doomed to escape the Galactic gravitational potential \citep{Tauris2015}. Those objects support the existence of runaway compact objects whose formation would have disrupted the binary system and provide the hypervelocity star with a large amount of linear momentum\footnote{Note that hypervelocity stars could also gain their large velocities from tidal disruption of tight binary stars by the Galactic central supermassive compact source, SgrA*. As one of the two stars is captured, the other is ejected at high speed via the gravitational slingshot mechanism \citep{Hills1988}.} \citep{Tauris1998}. \bexb with \ns orbiting their high mass stellar companion on a modest eccentricity orbit have been observed \citep{Pfahl2002a}, pointing towards a specific core collapse process hardly providing any kick to the \ns. Nevertheless, we expect most \ns to receive a kick velocity at birth of the order of a few hundreds of kilometers per second \citep{Lyne1994}. Monitoring of the proper motion of statistical numbers of pulsars confirmed those figures and a few of them can be linked to a nearby supernova remnant such as IGR J1104-6103 (Figure\,\ref{fig:lighthouse}). The pulsar then rushes through the interstellar medium at supersonic speeds. In its rest frame, it undergoes a wind accretion mechanism (see Chapter \ref{chap:SgXB}) similar to the one of a \ns travelling in the high speed wind of its stellar companion, though in a more planar fashion. In the second part of this manuscript, we introduce a planar numerical setup of wind accretion which matches the structure of the incoming flow at large distance of a runaway compact object. However, in the case of a young accreting \ns, the pulsar wind, not taken into account in our setup, plays an important role in the shaping of the inflow. To our knowledge, no runaway \bh candidate has been detected up to now, possibly due to the lower formation rate of \bh compared to \ns and to their larger inertia which makes them more difficult to accelerate.

The planarity of this accretion process makes the situation essentially axisymmetric and brings up the question of the longitudinal stability of the accretion wake. The runaway pulsar PSR 2224+65 \citep{Cordes1993} is associated to a prominent nebula (nicknamed the Guitar Nebula) whose evolution over years has been monitored. The shock in the wake of the accretor has a cone-shaped with a pinch or neck which moves along the wake. Measuring its dimensions and its time evolution is a precious asset to understand the instability at stake in this accreting system : are the variations in the wake mere tracers of inhomogeneity in the interstellar medium or are they related to properties of the accretion mechanism?

\begin{figure}
\begin{center}
\includegraphics[height=7cm, width=15cm]{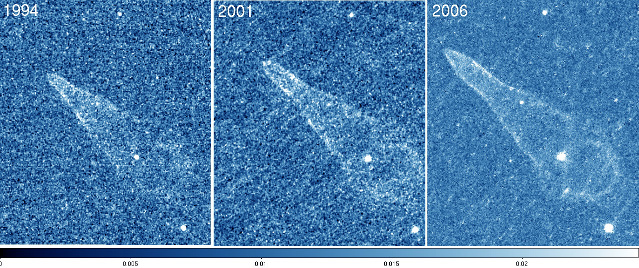}	
\caption{Zoom in on the head of the Guitar nebula where lies the runaway pulsar PSR 2224+65. The change in shape of the wake from year to year is clearly visible and its orientation is compatible with the measured velocity direction. The angular size of each picture is approximately 15 arcseconds and the objects is believed to be located at approximately 2kpc. Credits : HST.}
\label{fig:guitare_nebula}
\end{center}
\end{figure}


\subsection{Tidal disruption events}
\label{sec:tde}

Tidal disruption events (or \tde) correspond to the disruption of a star passing close enough from a supermassive \bh to experience dramatic and irreversible tidal effects \citep{Hills1975,Frank1978}. Because the majority of supermassive \bh are believed to lie dormant and starved of fuel, those transient events could shed some light on those quiet objects \citep{Rees1988a}, in particular in galactic centers where the feeding mechanism remains unclear (see section \ref{sec:sgrA}). Massive stars, which develop a bifurcated structure between a dense core and a tenuous envelope, are more vulnerable to \textsc{tde} ; they preferentially provide fuel to accrete for the superemassive \bh and could significantly contribute to the flaring activity of the latter. Depending on the amount of angular momentum of the inflow, the flare will decay over years \citep{Bonnerot2015}.

\begin{figure}
\begin{center}
\includegraphics[height=6cm, width=12cm]{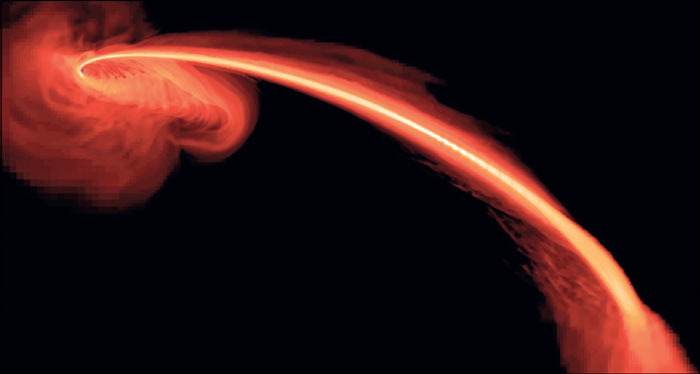}	
\caption{Column density map from a hydrodynamical simulation demonstrating the dynamics of a returning stream on a supermassive \bh produced by the disruption of a giant star. From \cite{Guillochon2014}.}
\label{fig:tde_guillochon14}
\end{center}
\end{figure}


\subsection{Sagittarius A*}
\label{sec:sgrA}

SgrA*, the compact object lying at the center of the Milky Way \citep{Melia2001a,Genzel2010a}, is known to be, nowadays \citep{Clavel2013}, much less luminous than its Eddington limit, possibly pointing to a spherical accretion such as the one portrayed by Bondi (section \ref{sec:Bondi_sph}). A hydrodynamical model of accretion from the wind emitted by distributed point sources around SgrA* has been developed by \cite{Coker1997}.


\subsection{Active galactic nuclei}
\label{sec:agn}

Because Fluid Mechanics is notoriously non trivially scale invariant\footnote{See the non-trivial similitude conversion which leads from an actual ship to its model for example.}, we focused the discussion on stellar-mass accretors. Nevertheless, in terms of spectral range of emission, we saw that the determining parameter was the compactness of the body and it turns out that most if not all galactic centers are believed to host \textsc{smbh}. If some of them wallow in a (well-deserved?) deep rest, like SgrA*, others, called Active Galactic Nuclei (\textsc{agn}), are way more luminous and display features characteristic of accretion at stake. The present work can in no case pretend to any physical relevance in the case of \textsc{agn} but links exist\footnote{As testified by the jets observed in quasars, a specific kind of accreting \textsc{smbh}, but also in microquasars such as GRS1915+105 \citep{Mirabel1994}.} between the two families which might turn profitable. See the book by \cite{Beckmann2012} for more details on \textsc{agn}.



\setlength{\parskip}{0ex} 


\chapter{Numerical tools}
\label{chap:num_tools}
\chaptermark{Numerical tools}
\hypersetup{linkcolor=black}
\minitoc
\hypersetup{linkcolor=red}
\setlength{\parskip}{1ex} 

Fluid Mechanics resolution requires large resolutions to grasp all the relevant spatial and temporal scales\footnote{The coupling between scales is an intrinsic flaw (and strength!) of Continuum Mechanics.} and smart numerical recipes to solve approximately but accurately the partial differential equations which arise. The intervention of a mathematically challenging framework within the complexity of a hydrodynamics environment makes numerical experiments game changers in contemporary Physics. The coming up of High Performance Computing (\aka \textsc{hpc}) architectures has enabled us to envision a yet-to-be-defined breakthrough, numerical epistemology, which could soon refund the very definition of a model at the basis of the scientific method \citep{Ruphy2013,Varenne2013,Varenne2014,Trescases2016}. The numerical modelling step, far from being an additional layer on top of the physical framework, offers an occasion to put in perspective the very notion of a model as a reduced representation of a set of empirically measurable data. Furthermore, it has proven to be a wonderful tool to illustrate, emphasize and reinforce a physical argument.

After a short reminder on the physical laws at the bottom of our work, we introduce the code we used, \textsc{mpi-amrvac}, the numerical scheme we followed and a few notions of High Performance Computing. For a reference describing the mathematical tools used in this chapter (tensor calculus, linear algebra, etc), see \cite{Appel2007}.


\section{Conservation laws}
\label{sec:cons_laws}


\subsection{Context}
\label{sec:context}

Fluid Mechanics is a non formal mesoscopic average of the kinetic theory and its inherited Boltzmann equation \citep{Diu1997}. It introduces a set of variables which tap their epistemological legitimacy in their physical profitability to interpret a wide range of phenomena. With some features inspired by Classical Mechanics, we use the mass density $\rho$, the velocity $\mathbf{v}$ and the total pressure $P$ of a particle of fluid, a subvolume of the system considered whose characteristic size $l$ verifies :

\begin{enumerate}
\item $l$ is large enough to guarantee that any particle of fluid contains a large enough number of actual particles\footnote{\ie in the meaning of the entities covered by kinetic theory.} to wash out the microscopic statistical noise.
\item $l$ is small enough to resolve macroscopic objects of interest, typically the waves of wavelength $\lambda$ which form and develop in the fluid.
\end{enumerate}
To summarize, Fluid Mechanics can prove to be a handy frame of thought provided we can define particles of fluid which verify :
\begin{equation}
n^{-1/3}\ll l\ll \lambda
\end{equation}
with $n$ the density of actual particles.

We can either consider the fluid variables $\rho$, $\mathbf{v}$ and $P$ (called the primitive variables) as :

\begin{enumerate}
\item a collection of values associated to all particles of fluid at a given moment $t$. For example, $\rho(\mathbf{r}(t),t)$ would be the mass density at $t$ of the particle of fluid located with the position vector $\mathbf{r}$, function of $t$. It is the Lagrangian point of view, where we "follow" the particles of fluid.
\item scalar and vector fields defined on an implicitly discrete mesh\footnote{Either due to the physical concept of fluid particle or to the numerical one of cell (see section \ref{sec:architecture}).}. For instance, $\rho(\mathbf{r},t)$ would be the mass density at $t$ of the particle of fluid which happens to find itself within an infinitely small volume around the point located with $\mathbf{r}$ at $t$. It is the Eulerian point of view. 
\end{enumerate}
Both approaches are physically equivalent but as far as the numerical implementation is concerned, we will rely on an Eulerian approach (see section \ref{sec:parall_comm}).

So as to facilitate compact and coordinate free formulations, we introduce the 5-components vector of conservative variables $\mathbf{U}$ defined by :
\begin{equation}
\label{eq:cons_var_vec}
\begin{aligned}
  \mathbf{U} \colon \mathbb{R}^3\times\mathbb{R} &\to \mathbb{R}^5\\
  (\mathbf{r},t) \mapsto &\mathbf{U}(\mathbf{r},t)=\left(
\begin{array}{c}
\rho\\
\rho\mathbf{v}\\
e\\
\end{array}
\right)
\end{aligned}
\end{equation}
where $e$ is the total energy per unit volume given by :
\begin{equation}
e=\frac{1}{2}\rho\mathbf{v}^2+u
\end{equation}
with the first term on the \rhs being the kinetic contribution and $u$, the internal energy per unit volume. Finally, we will refer to the "specific" counterparts of extensive variables as the intensive quantities obtained by dividing by the mass of the system.

\subsection{The Euler equations}
\label{sec:euler_eq}

Fluid Mechanics is based on sets of conservation laws supported by thermodynamical considerations. The simplest form of conservation laws, the Euler equations, is obtained by neglecting external forces, viscosity\footnote{See \cite{Landau1987} for an elegant description of the physical requirements which naturally lead to the mathematical expression of the viscosity term.} and heating/cooling :
\begin{equation}
\label{eq:euler}
\partial _t \mathbf{U}+\mathbf{\nabla}\cdot\mathbb{F}(\mathbf{U})=\mathbf{0}
\end{equation}
with the vector of fluxes :
\begin{equation}
\label{eq:flux}
\begin{aligned}
  \mathbb{F} \colon \mathbb{R}^5 &\to \mathbb{M}_{5,3}\\
  \mathbf{U} \mapsto &\mathbb{F}(\mathbf{U})=\left(
\begin{array}{c}
\rho\mathbf{v}\\
\rho\mathbf{v}\mathop{\otimes}\mathbf{v}+P\mathbb{1}\\
\left(e+P\right)\mathbf{v}\\
\end{array}
\right)
\end{aligned}
\end{equation}
where the divergence operator applied to a tensor does not relate to the one applied to a vector in a trivial way, apart in Cartesian coordinates (see formularies). Beware, in \eqref{eq:flux}, contrary to the expression of $\mathbf{U}$ \eqref{eq:cons_var_vec}, the three-dimensional vector $\mathbf{v}$ unfolds along its own dimension so as to form a tensor in $\mathbb{M}_{5,3}$ rather than a 8-componetns vector in $\mathbb{R}^8$. The dyadic product $\mathop{\otimes}$ is defined by :
\begin{equation}
\mathbf{a}\mathop{\otimes}\mathbf{b}=\left(a_i b_j\right)_{(i,j)\in\llbracket 1,3\rrbracket ^2}
\end{equation}
with $a_i$ and $b_j$ the coordinates of the vectors only in the Cartesian case. The expression \eqref{eq:euler} is a differential formulation of conservation laws, legitimate only in smooth configurations ; this assumption can be relaxed for the integral formulations which serve to derive the jump conditions in appendix \ref{sec:jump}.


\subsection{Closure conditions : the equation-of-state}
\label{sec:eos}

In \eqref{eq:flux}, $\mathbb{F}$ can not be properly considered as a function of\footnote{The system is said to be autonomous if $\mathbb{F}$ is function of $\mathbf{U}$ only (it can be a function of time and space but through $\mathbf{U}$ only).} $\mathbf{U}$ unless we are given a relation between the pressure and the other variables. This additional general requirement to solve the equations of Hydrodynamics is known as the closure condition. Considerations of Statistical Physics we will not remind here guarantee the existence of a canonical equation-of-state\footnote{With $s$ the entropy per unit volume} or \eos, $\Upsilon(s,\rho)$, which contains all the thermodynamical information \citep{Diu2007}. $\Upsilon$ is the internal energy per unit mass. It can be shown that this \eos can be splitted into a caloric and an entropic one, respectively $\Upsilon(P,\rho)$ and $s(P,\rho)$.

All along this manuscript, we will focus on the case of ideal gases. This thermodynamical model corresponds to a gas diluted enough to neglect the covolume associated to each particle\footnote{See the Van Der Waals \eos for a quantification and a physical interpretation of this notion.} and the energy of interaction between particles with respect to their kinetic energy. However, particles are considered thermalized with each other thanks to collisions between infinitely small particles such as a temperature can always be defined. In this case, the caloric\footnote{\label{fn:entropy} The entropic one comes into play in the conservation laws if the temperature is explicitly needed, \eg in a cooling function. For an ideal gas, it is given by : $s=\frac{R}{M}\ln\left(\frac{P}{\rho ^{\gamma}}\right)$ where $R$ is the ideal gas constant and $M$ the mean molar mass of the gas.} \eos is given by :
\begin{equation}
\Upsilon=\frac{P/\rho}{\gamma-1}
\end{equation}
with $\gamma$ the adiabatic index \cite[see][for a definition]{Diu1997,Diu2007}, a constant determined by the number of degrees of freedom per particle and whose value lies between 1 (infinite number of degrees of freedom) and $5/3$ (only 3 translational degrees of freedom) for an ideal gas. The speed of sound in the flow is then given by :
\begin{equation}
\label{eq:sound_speed_def}
c_s^2=\gamma \frac{P}{\rho}
\end{equation}
For a more detailed reminder, see \cite{Toro2009} and for a comprehensive course, refer to \cite{Diu2007}. From now on, unless explicitely stated, we work with ideal gases.

A plethora of fruitful physical submodels can be derived from the Euler equations which have irrigated mindful studies in Geophysics, Climatology \citep{Pedlosky1992}, city planning, etc. 


\subsection{Hyperbolic sets of partial differential equations}
\label{sec:hyp_set_of_PDE}

We now provide some elements of solution for the uni-dimensional Euler equation\footnote{$\mathbf{U}\in\mathbb{R}^3$ and $\mathbb{F}\in\mathbb{M}_{3,1}$ now.} so as to highlight the fundamental mathematical properties of this system at the basis of its computational resolution. Since conservation laws are quasi-linear, they can always be rewritten after using the chain rule as :
\begin{equation}
\partial _t \mathbf{U}+\mathbb{A}(\mathbf{U})\times \partial _x \mathbf{U}=\mathbf{0}
\end{equation}
where the $\times$ symbol is to be understood as a matrix product and $\mathbb{A}(\mathbf{U})$ is the jacobian matrix :
\begin{equation}
\begin{aligned}
  \mathbb{A} \colon \mathbb{R}^3 &\to \mathbb{M}_{3,3}\\
  \mathbf{U} \mapsto &\mathbb{A}(\mathbf{U})=\frac{\partial \mathbb{F}}{\partial \mathbf{U}}
\end{aligned}
\end{equation}
In the case of an ideal gas, we have, after having cautiously expressed the coefficients of $\mathbb{F}$ exclusively as combinations of full coefficients of $\mathbf{U}$ (no $\rho$ of $v$ left alone) :
\begin{equation}
\mathbb{A}(\mathbf{U}) = \begin{bmatrix}
    0 & 1 & 0 \\
    \frac{1}{2}\left(\gamma -3\right)v^2 & (3-\gamma )v & \gamma -1 \\
    \frac{1}{2}\left(\gamma -2\right)v^3-\frac{c_s^2v}{\gamma -1} & \frac{3-2\gamma}{2}v^2+\frac{c_s^2}{\gamma -1} & \gamma v
  \end{bmatrix}
\end{equation}
which can be diagonalized to yield the following eigenvalues $\lambda_i$ and eigenvectors $\mathbf{K}_i$ :
\begin{equation}
\begin{cases}
\lambda _1=v-c_s\\
\lambda _2=v\\
\lambda _3=v+c_s
\end{cases}
\end{equation}
and :
\begin{equation}
\begin{aligned}
\mathbf{K}_1=\left(
\begin{array}{c}
1\\
v-c_s\\
\frac{e+P}{\rho}-vc_s\\
\end{array}
\right) \qquad 
\mathbf{K}_2=\left(
\begin{array}{c}
1\\
v\\
\frac{1}{2}v^2\\
\end{array}
\right) \qquad
\mathbf{K}_3=\left(
\begin{array}{c}
1\\
v+c_s\\
\frac{e+P}{\rho}+vc_s\\
\end{array}
\right)
\end{aligned}
\end{equation}
Since the eigenvalues are all real and distinct, the Euler equations form a set of strictly hyperbolic partial differential equations. This property remains true in three dimensions and for more general \eos. If we write $\mathbb{K}$ the matrix formed with the eigenvectors $\mathbf{K}_i$ as columns and $\mathbb{A}_D$ the diagonalized form of $\mathbb{A}$, we can rewrite te Euler equations as :
\begin{equation}
\mathbb{K}\times\partial _t\mathbf{U}+\mathbb{A}_D\times\mathbb{K}\times\partial _x\mathbf{U}=\mathbf{0}
\end{equation}
If $\mathbb{K}$ is smooth enough with respect to its evolution with space and time, we can write :
\begin{equation}
\mathbb{K}\times\partial _{i}\mathbf{U}\sim \partial _{i}\left(\mathbb{K}\times\mathbf{U}\right) \quad \text{ for} \quad i\in \left\lbrace t;x \right\rbrace
\end{equation}
and we are left\footnote{Within a local region of space and time.} with a set of decoupled uni-dimensional linear advection equations for the variable $\mathbb{K}\times\mathbf{U}$. The use of the so-called characteristics then provides an extraordinary way to transform those partial differential equations into ordinary differential equations\footnote{If the solving of partial differential equations in general remains a tremendous challenge of contemporary Mathematics, the realm of ordinary differential equations has been extensively explored in the past couple of centuries.} \citep{Shu1992}. The eigenvectors\footnote{Associated to discontinuous, rarefaction and shock waves.} form a set of linearly independent vectors from which all solutions - albeit local if $\mathbb{A}$ depends\footnote{Integral curves of the characteristic family can be used as non-linear extensions of the characteristics straight lines associated to linear hyperbolic systems (\ie systems where $\mathbb{A}$ does not depend on $\mathbf{U}$). This approach brings up the notion of Riemann invariants as conserved quantities along the waves.} on $\mathbf{U}$ - can be derived. The selection of a specific solution requires the prior of the initial condition $\mathbf{U}(x,t=0)=\mathbf{U_0}(x)$ and will be investigated in section \ref{sec:riem_prob} where the Riemann problem is addressed.

\section{Numerical scheme}
\label{sec:num_scheme}


\subsection{\texttt{MPI-AMRVAC}}
\label{sec:mpi_amrvac}

The code I have used, \texttt{MPI-AMRVAC}\footnote{For Message Passing Interface - Adaptive Mesh Refinement Versatile Advection Code, sometimes referred simply as \vac in this manuscript (not to be confused with the initial version named alike). Complementary details about the structure of the code can be found in Appendix \ref{sec:amrvac_doc}.} is the latest version of a code whose origins trace back to the mid 90's when G\'abor T\'oth and Rony Keppens first tackled the question \citep{Toth1996a,Toth1998}. It is an explicitely flux-conserving finite volume transport code which can now address hydrodynamical or magneto-hydrodynamical problems, in a classical, a special or a fully relativistic framework, with or without polytropic prescriptions, source terms, etc \citep{Porth:2014wv}. 

The recourse to a coordinate and dimensionality independent syntax\footnote{Thanks to the Loop Annotation SYntax \aka \texttt{LASY} \citep{Toth1997d}. A Perl preprocessor, \texttt{VACPPL}, takes care of converting the .t files from \texttt{LASY} to proper Fortran .f files, using the user defined settings on dimensionality and coordinates but more generally, about the mesh structure.} enables users and developers to freely design algorithms and configurations which apply to a broad range of situations. It is an Eulerian-based code where physical quantities are defined at the center of each cell as averaged values over the whole cell (see section \ref{sec:fin_meth}). The source and user files are designed so as to make a change of coordinates straightforward. Natively, it covers the usual orthogonal curvilinear coordinates \citep{Mignone2014} : cartesian, cylindrical and spherical. They are uniform in the sense that cell centers are equally spaced along each dimension. Thanks to the implementation of Adaptive Mesh Refinement (\textsc{amr}) made by \cite{VanderHolst2007}, different levels of spatial resolution\footnote{Adaptive Time Stepping (\textsc{ats}) has not been yet implemented in \vac but can be found, for instance in the \href{http://www.ics.uzh.ch/~teyssier/ramses/RAMSES.html}{\texttt{RAMSES} code} - see \cite{Commercon2014} for a case where \textsc{ats} proves very profitable once coupled to \textsc{amr}.} can be explored and treated with different numerical schemes. 

Customized data analysis can be carried on internally. Alternatively, many different ouptut formats are available beyond the basic data binary files used by the code : \textsc{vtk} for visualisation with VisIt or Paraview, Native Tecplot, \textsc{dx}, \textsc{idl}. An \textsc{xml}-based module naturally handles the mesh-related information\footnote{Including the informaton relative to \textsc{amr}.} to produce versatile and storage-saving output files. 


\subsection{Algorithmic recipes}
\label{sec:algo_recip}

The following sections are intended to give the reader a sensibility about the main algorithmic issues one can encounter in numerical simulations of fluid dynamics on a grid. More detailed can be found in the related references \citep{LeVeque1992,LeVeque2002,Toro2009,Dullemont}.

\subsubsection{Finite methods}
\label{sec:fin_meth}

Starting from conservation laws such as \eqref{eq:euler}, the spatial and temporal derivatives must be evaluated to advance the computation of an approximate solution at any time from an initial condition. Several approaches can be undertaken to handle the discretisation inherent to the computational tool. Among others can be found :

\begin{enumerate}
\item \underline{the finite difference method} : the most straightforward approach is to associate to a collection of nodes a collection of values. Each variable is a set of point values attached to a set of unstructured positions (\ie the mesh is not necessarily associated to a regular grid). In this way, the spatial derivative of a quantity $u$ in one point $x_i$ is simply given by :
\begin{equation}
\partial _x u=\frac{u_{i+1}^k-u_{i-1}^k}{x_{i+1}-x_{i-1}}
\end{equation}
where we used the super and subscripts accordingly to Figure\,\ref{fig:cells}. There is no cell nor interface in this approach so this equality suffers no ambiguity : it is the only way to have a fair accounting of the contribution of both sides. The system can then be solved by evaluating the fluxes above with $k=n$ or $n+1$, which gives respectively the explicit or the implicit method. The latter requires more computation\footnote{Matrix inversion, Newton-Raphson root finding, etc.} and couples all the points together which makes little sense for hyperbolic system where the information propagates at finite speeds. However, it sometimes has precious assets like stability and convergence, but is not necessarily accurate. The finite difference approach is easier to implement, fast to compute but does not extend to complex geometries nor to \textsc{amr}.
\begin{figure}[!b]
\begin{center}
\def\svgwidth{350pt} 
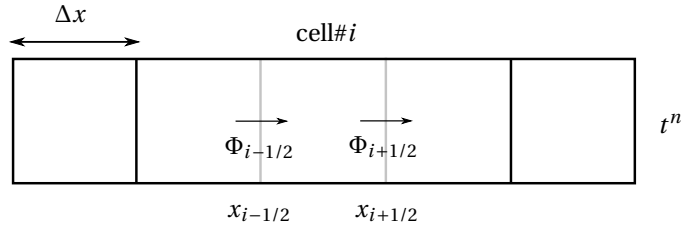
\caption{Sketch to define the different notations in the uni-dimensional finite volume approach where the emphasis is put on edges and fluxes. Subscripts $i$ refer to the position while superscripts $n$ refer to the time iteration. The timestep is written $\Delta t$.}
\label{fig:cells}
\end{center}
\end{figure}
\item \underline{the finite volume method} : in this approach, we solve the integrated form of the conservation laws separately in each cell of the grid. For instance, for the conservation of mass :
\begin{equation}
\frac{\d ~}{\d t}\left[ \int\int\int_{(\mathcal{C}_i)}\rho\left(\mathbf{r_i},t^n\right)\d V \right]+\Phi = 0
\end{equation}
where we used the Green-Ostrogradsky theorem. Also, $\mathbf{r_i}$ refers to all the points within the volume $\left(\mathcal{C}_i\right)$, not to a hypothetical cell center. $\Phi$ is the total flux of mass through the cell edge. There, the discretization can be made on one hand by using an average value $\rho_i^n$ associated to the cell :
\begin{equation}
\rho_i^n=\frac{\int\int\int_{(\mathcal{C}_i)}\rho\left(\mathbf{r_i},t^n\right)\d V}{\Delta V}
\end{equation} 
and on the other hand, with a set of approximated fluxes at the left and right interfaces, respectively $\Phi_{i-1/2}$ and $\Phi_{i+1/2}$ :
\begin{equation}
\label{eq:advec_discrete}
\frac{\Delta x \left(\rho_i^{n+1}-\rho_i^{n}\right)}{\Delta t}+\left( \Phi_{i+1/2} - \Phi_{i-1/2} \right) = 0
\end{equation}
Notice that this method is conservative by construction\footnote{At least up to the computation error level of $10^{-14}$ in double precision.} since :
\begin{equation}
\sum_i \rho_i^{n+1} = \sum_i \rho_i^n - \frac{\Delta x}{\Delta t} \left( \Phi_ {N+1/2} - \Phi_{1+1/2} \right)
\end{equation}
where $\Phi_{1+1/2}$ and $\Phi_ {N+1/2}$ are the fluxes at the edges of the grid considered in this implicitly mesh-based approach. All the secret of the numerical recipe will be contained in the way the fluxes at the interface are determined, a method referred to as reconstruction.
\end{enumerate}


\subsubsection{The Riemann problem}
\label{sec:riem_prob}

The simplest non-trivial initial value problem we can make up for uni-dimensional conservation laws is :
\begin{equation}
\partial _t \mathbf{U}+\mathbf{\nabla}\cdot\mathbb{F}(\mathbf{U})=\mathbf{0} \quad \text{with} \quad
\mathbf{U}(x,t=0)=\mathbf{U}_0(x)=
\begin{cases}
\mathbf{U}_L \quad \text{if} \quad x<0 \\
\mathbf{U}_R \quad \text{if} \quad x>0 
\end{cases}
\end{equation}
with $\mathbf{U}_L$ and $\mathbf{U}_R$ constant and different. It is called the Riemann problem, a mathematical generalization of the physical Sod's shock tube problem. Given the finite volume point of view we adopt, we can already understand the paramount importance of this reduced problem which mimics the situation at each interface : on each side of the interface, the function is constant. Indeed, the solution of the general initial value problem on a grid may be seen as resulting from non-linear superposition of solutions of local Riemann problems. Godunov's methods are based on this principle. At each interface a Riemann problem will be solved such as over a timestep, the propagating characteristic waves do not overlap. 

\subsubsection{Differencing schemes}
\label{sec:diff_schemes}

\begin{figure}[!h]
\centering
   \begin{subfigure}[b]{0.7\textwidth}
	\includegraphics[height=6cm, width=10cm]{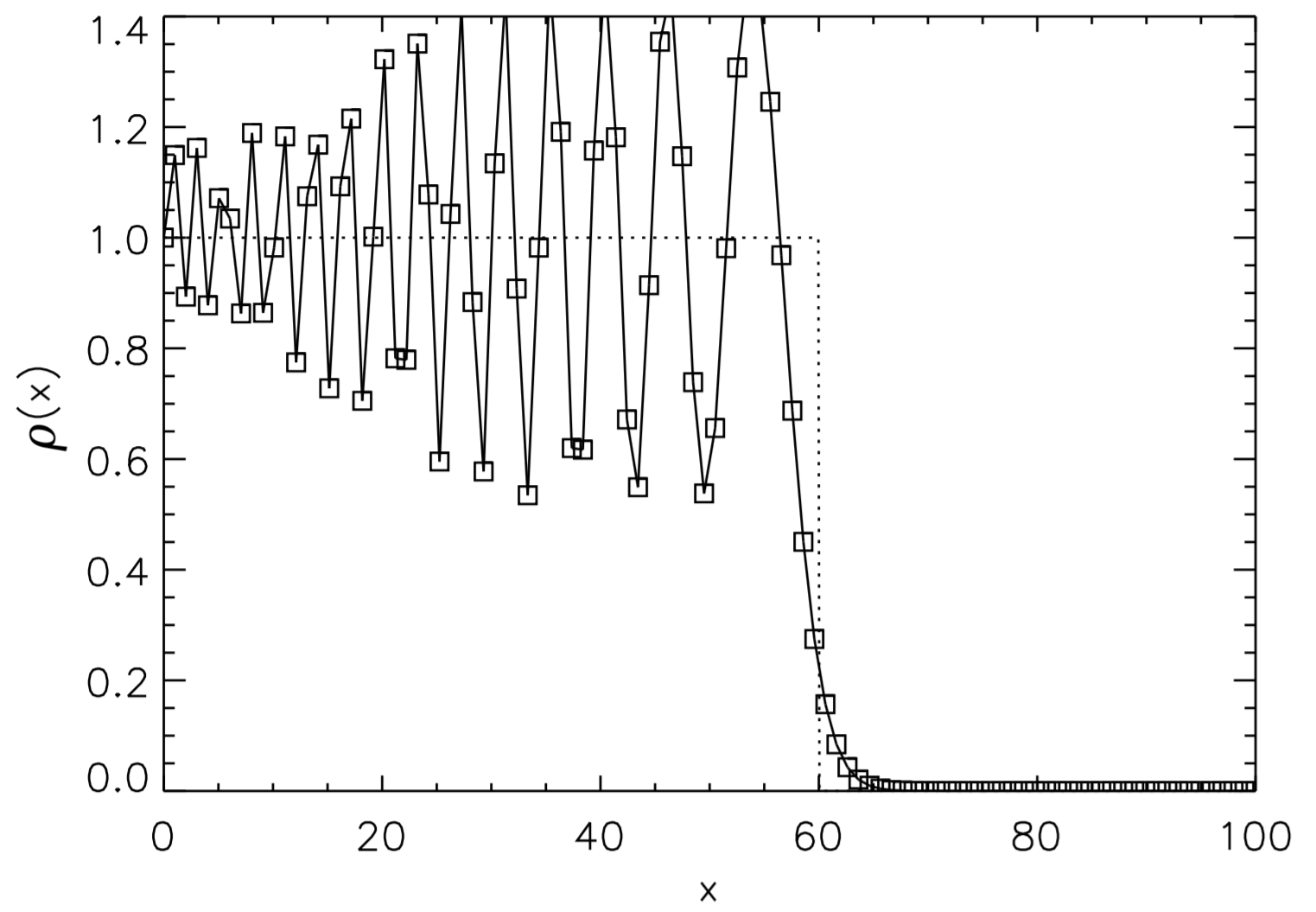}	
   \label{fig:sfig1} 
\end{subfigure}
\begin{subfigure}[b]{0.7\textwidth}
   \includegraphics[height=6cm, width=9.9cm]{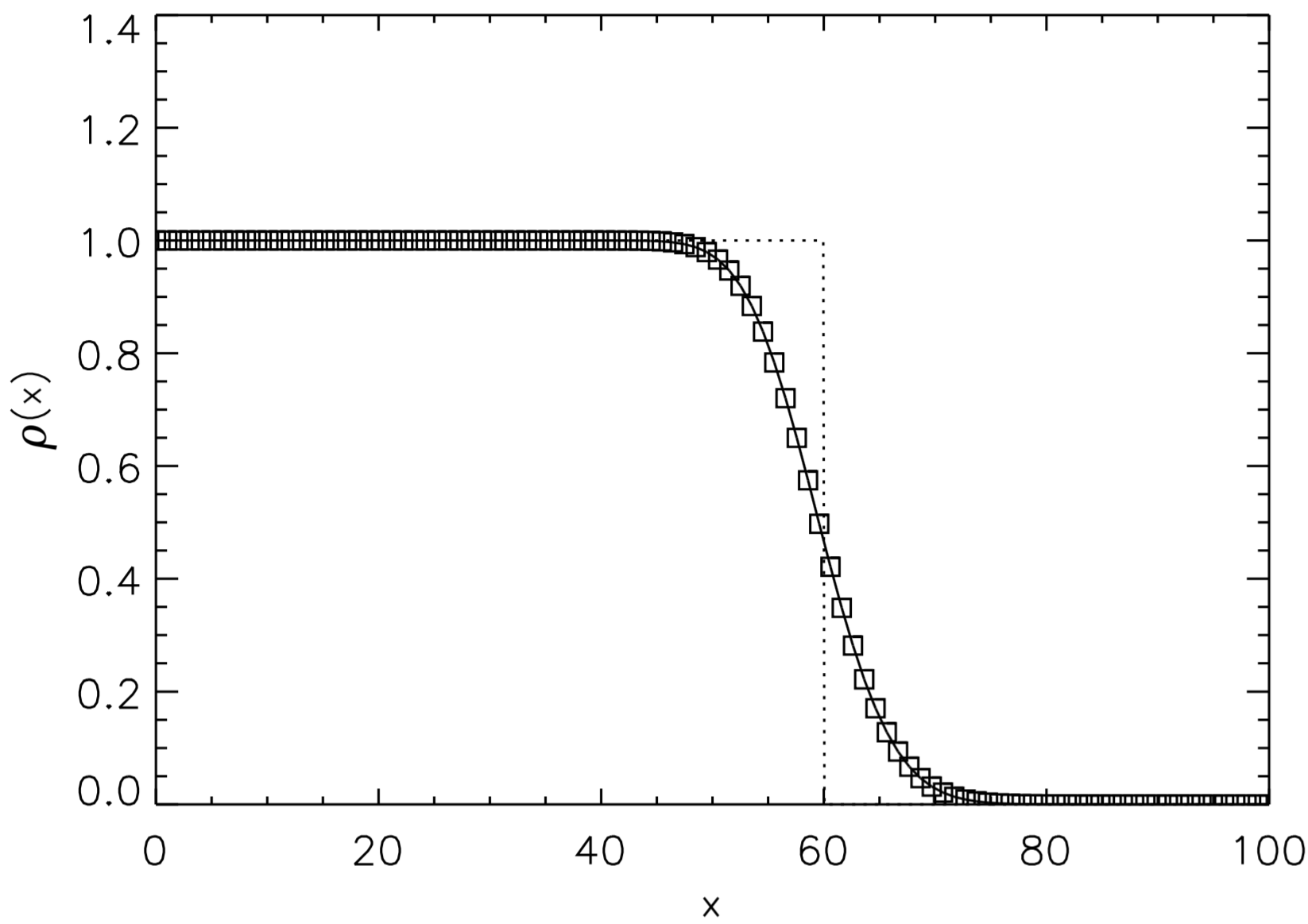}
   \label{fig:sfig2}
\end{subfigure}
\caption[Two numerical solutions]{(\textit{upper panel}) A central differencing numerical solving of the uni-dimensional linear advection equation of a step function which went wrong. The dotted line stands for the analytical solution while the markers and solid line indicate the numerical answer. (\textit{lower panel}) The same numerical problem solved with an upwind differencing scheme. From \cite{Dullemont}.}
\label{fig:oups_centered}
\end{figure}

Accordingly to the finite difference approach, one could want to use a central differencing scheme to resolve, say for the sake of clarity, the uni-dimensional linear advection equation of a step-function with a constant advection velocity $v>0$ :
\begin{equation}
\label{eq:advec_equation}
\begin{cases}
\partial _t\rho + \overbrace{v\partial _x\rho}^{\equiv \Phi}=0\\
\rho\left( x;t=0\right) =
\begin{cases}
1 \quad \text{ for } \quad x<30 \\
0 \quad \text{ for } \quad x>30
\end{cases}
\end{cases}
\end{equation}
Its differentiation in a naive cell-centered approach suggests to replace the fluxes with :
\begin{equation}
\Phi\sim\frac{\Phi_{i+1/2}-\Phi_{i-1/2}}{2\Delta x}\sim v \frac{\rho_{i+1}^n-\rho_{i-1}^n}{x_{i+1}-x_{i-1}}
\end{equation}
In an explicit approach, it means that $\rho_i^{n+1}$ is given by, for an equally-spaced grid :
\begin{equation}
\rho_i^{n+1}=\rho_i^{n}-\frac{\Delta t}{2\Delta x}v\left(\rho_{i+1}^{n}-\rho_{i+1}^{n}\right)
\end{equation}
The numerical result of such a gross approximation is illustrated in the upper panel of Figure\,\ref{fig:oups_centered} where all the wriggles below $x=60$ have little to do with the mathematical solution (the dotted line) and much to do with the numerical scheme. This numerical scheme yields clearly unstable solutions for this problem. The main issue lies in the fact that, to update the value of $u_i^n$, we make use of information from upstream\footnote{\ie the region $x<x_i$ (resp. $x>x_i$) if $v>0$ (resp. if $v<0$).}, but also from downstream, which does not make any physical sense : the information can not go back up\footnote{Note that in this simple situation, there is no other wave speed than $v$ (not even the sound speed).}. Anything which happens in the flow downstream $x_i$ does not belong to its domain of determinacy and can not influence its value at this instant. In hyperbolic systems, information propagates at finite speeds through characteristics. We can use this causality property to design one-sided derivatives in the direction from which information should be coming\footnote{With possibly different directions for different information if there are different characteristic velocities, typically for a system of several equations \ie not for the canonical one-dimensional linear advection one.} : it is the bottom line of the first order upwind scheme we present here. 


\begin{figure}[!hb]
\centering
\includegraphics[height=8cm, width=6cm]{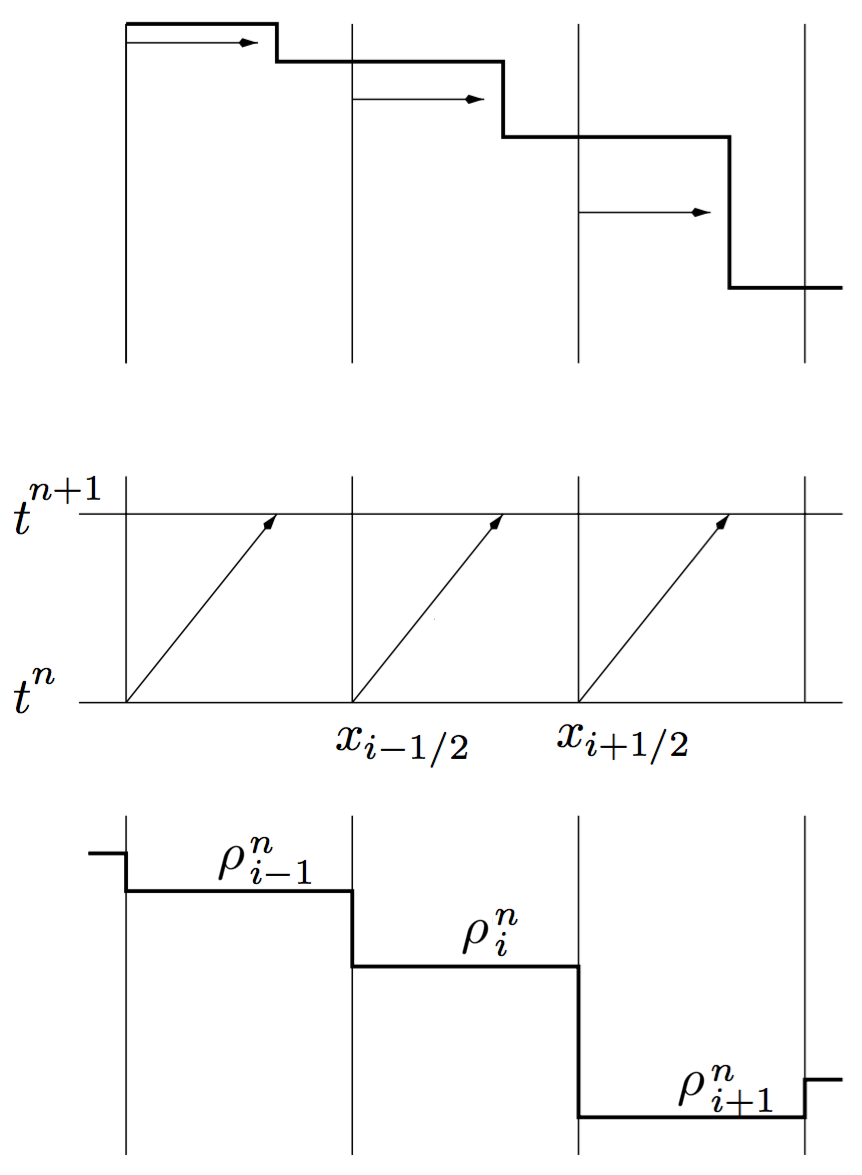}	
\caption{Finite-volume interpretation of an upwind scheme applied to a linear advection equation with $v>0$ (the flow is moving to the right). The bottom graph represents the piecewise constant function at $t^n$ and the upper one yields, if averaged over each cell, the piecewise function at $t^{n+1}$. In-between are represented the characteristics which monitor the advance of the steps. From \cite{LeVeque2002}.}
\label{fig:upwind_advection}
\end{figure}

Since $v>0$, we use the value on the left to define the fluxes at the interface in \eqref{eq:advec_discrete} :
\begin{equation}
\begin{cases}
\Phi_{i-1/2}^n=v\rho_{i-1}^n\\
\Phi_{i+1/2}^n=v\rho_i^n
\end{cases}
\end{equation}
which gives the convex combination of $\rho_i^n$ and $\rho_{i-1}^n$ to get $\rho_i^{n+1}$ :
\begin{equation}
\label{eq:upwind_scheme}
\rho_i^{n+1}=\rho_i^n\left(1-v\frac{\Delta t}{\Delta x}\right)+v\frac{\Delta t}{\Delta x}\rho_{i-1}^n
\end{equation}
where the weights can also be obtained by following the characteristics and interpolating the value of $\rho_i^{n+1}$ between the grid values\footnote{For instance, graphically, on the middle panel of Figure\,\ref{fig:upwind_advection} where we see that $\Delta t/\Delta x\sim0.67/v$ : at $t^{n+1}$, the cell $x_i$ has been overran up to 67\% of its own space by the flow which was in $x_{i-1}$ at $t^n$.} $\rho_{i-1}^n$ and $\rho_i^n$. This simple workaround leads to the results displayed in the bottom panel of Figure\ref{fig:oups_centered}. The scheme proves way more stable and is monotonicity preserving (see the \href{http://www3.nd.edu/~dbalsara/Numerical-PDE-Course/}{comprehensive course} by Dinshaw Balsara on techniques for numerical solving of partial differential equations) but the numerical diffusion of this first order scheme smeared out the step function.

In \eqref{eq:upwind_scheme} appears a necessary condition which turns out to apply to any explicit differencing method : the Courant-Friedrichs-Lewy condition (\aka \cfl condition). Since the value in $x_i$ at $t^{n+1}$ depends only on the value in neighbouring cell at $t^n$, we must make sure that the characteristics from $x_{i-1/2}$ (if $v>0$) does not go beyond $x_{i+1/2}$ within one timestep. More generally, if we write $\lambda_{\text{MAX}}$ the maximum absolute propagation speed for information, we need\footnote{Implicit methods can, to appearances, beat this limit but from experience, they start to lack accuracy when the timestep becomes much larger than $\Delta x / \lambda_{\text{MAX}}$. Besides, their strong stability makes it difficult to point out suspicious behaviour by eye.} :
\begin{equation}
\lambda_{\text{MAX}}\frac{\Delta t}{\Delta x}\leq 1
\end{equation}
Empirically, a Courant number $\lambda_{\text{MAX}}\frac{\Delta t}{\Delta x}$ smaller than 0.5 is usually enough to guarantee the stability and the accuracy of the numerical scheme.

This scheme is said of first order in $\Delta x$ and $\Delta t$ since a Taylor expansion of $\rho_i^{n+1}$ and $\rho_{i-1}^n$ in \eqref{eq:advec_discrete} yields a slightly different expression than the original one, \eqref{eq:advec_equation}, we wanted to solve :
\begin{equation}
\label{eq:twisted_advec_eq}
\partial _t\rho + v\partial _x\rho=-\frac{\Delta t}{2}\partial _{tt}\rho + v\frac{\Delta x}{2}\partial _{xx}\rho + \mathcal{O}(\Delta t^2) + \mathcal{O}(\Delta x^2) 
\end{equation}
where $\Delta t$ and $\Delta x$ take place at the first power, hence the first order. This expression highlights the unavoidable artificial diffusivity introduced by the spatial discretization in the second term of the \rhs. It can be lowered by rising the resolution or using higher-order schemes, for instance with slope limiters or larger stencils\footnote{A mini-grid of points which contribute to the quantities we wish to evaluate.} to compute the derivatives. For time, it means using a predictor to compute the fluxes at mid-timestep to reach a second order precision in time. A final remark : higher-order schemes are more accurate only if there is no strong discontinuities in them according to the two last terms on the \rhs of \eqref{eq:twisted_advec_eq}. When shocks take place, specific shock-capturing schemes must be used.

\subsubsection{Our algorithmic setup}
\label{sec:my_algo_setup}

In most of the simulations presented in this manuscript, we used a second order Lax-Friedrichs method with a Hancock predictor. Thanks to the use of slope limiters\footnote{Mostly the shock-capturing Koren slope limiter, less diffusive than a minmod.}, the scheme is made Total Variation Diminishing which means that no new local extrema are created (it is monotonicity preserving) and the values of local minima (respectively maxima) increase (respectively decrease). The Courant number is typically set to 50\%. Although quite diffusive, the \textsc{tvd-lf} method is robust and not prone to give birth to spurious oscillations.


\section{Splendors \& miseries of \textsc{hpc}}
\label{sec:parall_comm}


\subsection{Architecture of the grid and boundary cells}
\label{sec:architecture}

\begin{figure}[!b]
\begin{center}
\def\svgwidth{350pt} 
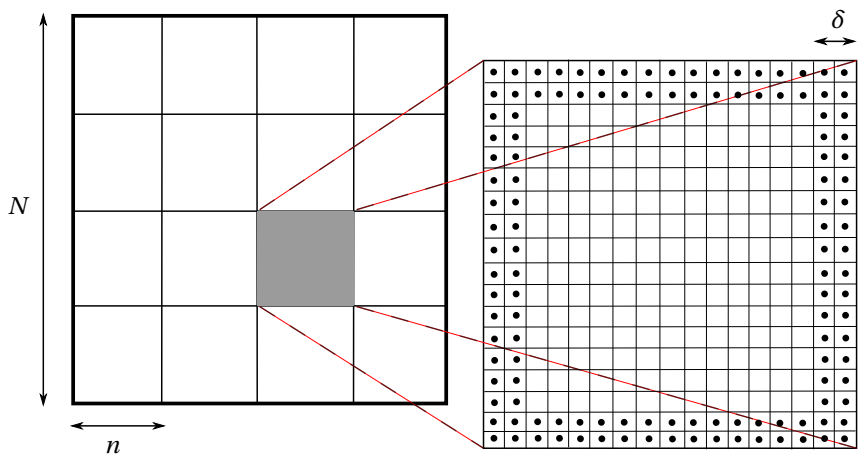
\caption{A representation of a two-dimensional Cartesian grid of size $N$ with blocks of size $n$ on the left. The blocks which are adjacent to a border of the grid (thick black line) have layers of $\delta$ additional ghost cells as neighbours, outside the grid, whose properties are set by the user as boundary conditions. A zoom in on the grey shaded block has been represented on the right. The grid cells are visible and those considered as boundary cells \ie whose values must be communicated to the neighbouring blocks at each iteration are marked with a black dot. The plot is for $\delta=2$.}
\label{fig:grids_subgrids_ghost}
\end{center}
\end{figure}

To make the most of the octree at the basis of the \textsc{amr} structure, the simulation space (\aka grid) is initially subdivided into blocks of equal sizes. We write $N_i$ the number of cells of the grid in the direction\footnote{In the code, the dimensionality of the grid is given by the variable \texttt{ndim} and we have $i\in[1,\texttt{ndim}]$} $i$ and $n_i$ the number of cells of a block in the same direction. A first obvious constraint is that $n_i$ must divide $N_i$. One of this two dimensional cartesian grid has been represented on Figure\,\ref{fig:grids_subgrids_ghost}. The cells on the border of each block\footnote{The border being defined with any "thickness" (\eg 1, 2 or 4 cells), provided it remains small compared to $n_i$ (see section \ref{sec:comm}).} must be communicated to the \cpu\footnote{For Central Processing Unit.} responsible for the computation in the neighbouring blocks (possibly itself) to compute spatial derivatives. Once the run starts, the code can refine\footnote{Or coarsen previously refined ones.} some areas thanks to \textsc{amr}, based on user-defined criteria or standard error estimates, and divide up the existing blocks, down to a specified maximum number of refinement. Without loss of generality, we will now assume that the cells are "squared" which means that the $N_i$ and $n_i$ are the same whatever the direction $i$. The two-dimensionality of the meshes we consider from now on can also easily be relaxed and does not alter qualitatively the following remarks.

On the right side of Figure\,\ref{fig:grids_subgrids_ghost} are displayed the boundary cells, those whose value must be communicated to neighbouring blocks to compute spatial derivatives. The user can specify the "thickness" of this layer with the parameter \texttt{dixB} in \vac, here called $\delta$. One can convince himself that the square configuration is ideal to minimize the fraction of cells at the interface with other blocks, at a fixed total number of cells\footnote{See the classical optimization problem where it is asked to show that the sphere, without any privileged direction like this Cartesian mesh, is the minimal surface of a given volume, or that \href{https://www.youtube.com/watch?v=e0Fgyca6WCw}{the height of a can must be equal to its diameter to minimize the surface of the associated cylinder, at a fixed volume}.}. 

In this section \ref{sec:parall_comm}, we consider static grids which are not destined to be refined. The number of blocks is set initially by the user-defined parameters and does not change during the simulation. The use of \textsc{amr} obviously changes those conclusion but the monitoring of the communication we describe below is not as critical since the number of blocks per \cpu and the total number of \cpus are large enough to allow slight unbalanced working load without dramatic consequences.


\subsection{Parallelization}
\label{sec:parallel}

The philosophy of the code is to strongly compartmentalize the computation on each block so as they can autonomously be monitored by different \cpu, possibly on different nodes, without excessive communication time. Furthermore, one must keep in mind that the number of \cpus selected, $N_{\textsc{cpu}}$, can not be larger than the initial\footnote{In case of \textsc{amr} where the user usually starts with a limited number of blocks, the simulation can be performed on a short number of iterations\footnote{Also called "time step" henceforth.}, enough to get a large number of blocks, and then rerun using the output data files as initial state.} number of blocks given in two-dimensional meshes by $K=\left(N/n\right)^2$ ; we will see that it is the main limitation for strong scalability of the code in the debugging stage of development. We write $k$ the number of blocks per \cpu ; to avoid underloading \cpu which seems to slow them down, we set $k$ to 4 to guarantee that each \cpu has a sufficient number of cells to work on. To assure a good load balancing, we also want each \cpu to work on the same number of blocks so $N_{\textsc{cpu}}$ must divide the total number of blocks $K$. Besides, to guarantee that the "squared cells" assumption remains true at the level of blocks per \cpu\footnote{See the grey shaded group of blocks on Figure\,\ref{fig:load_balancing} for example.}, we require $k$ to be a squared quantity (1, 4, 9, 16...). To maximize the possibilities and make the discussion easier to follow, we work with powers of 2 from now on. The following results can easily be extended to any sizes. The communication between \cpus is monitored within \vac thanks to the subroutines of the \texttt{OpenMPI} library.

\begin{figure}[!h]
\begin{center}
\includegraphics[height=6.5cm, width=10cm]{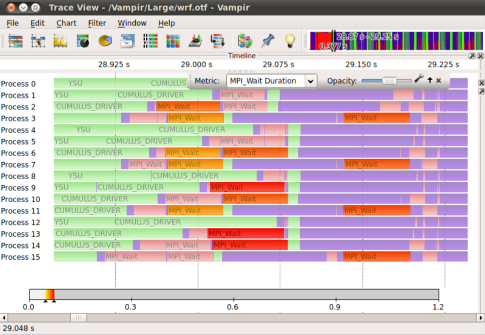}	
\caption{The interface of Vampir enables the user to analyze the output file produced as the job is run. It details the work flow of each \cpu, the time spent in each subroutine and the communication steps.}
\label{fig:vampir}
\end{center}
\end{figure}


\subsection{Computation speed and profiling}

We define $V$ the computation speed per \cpu as : 
\begin{equation}
\label{eq:comp_speed}
V=\frac{N_{\text{op}}\times N_{\text{it}}}{\Delta t \times N_{\textsc{cpu}}}
\end{equation}
where $N_{\text{op}}$ is the number of realized operations per iteration\footnote{Performed at the scale of Arithmetic Logic Units, a subscale within the \cpu.}, $N_{\text{it}}$ is the total number of time iterations required to complete the simulation\footnote{Difficult to evaluate a priori if a laps of physical time -rather than a number of steps - has been specified since the time step can change during the simulation.} and $\Delta t$ is the total elapsed real time in the user frame. \vac offers the possibility to evaluate $V/N_{\text{it}}$ at each time step. From our experience, it can vary a lot for a given numerical setup\footnote{And a given number of blocks and  cells per \cpu.}, from $10^4$ on local university clusters up to a few $10^6$ on national ones. On top of this, it can also vary by a factor up to 20 from a timestep to another during a given simulation. Memory leaks, compilation options and static/dynamic library linking can, among others, be responsible for it. It can also come from the code which must be monitored by using a profiler such as \href{https://www.vampir.eu/news}{Vampir} in association with VampirTrace, its runtime library\footnote{In parallel computing, caveats in the algorithms can not be traced by simple print statements due to the intrinsically non-linear work flow. The very step of writing requires additional precautions when it deals with parallel computing.}. The use of a profiler is mandatory once an anomalously slow computing velocity has been spotted to look for reasons within the code : an error estimate to refine and coarsen the grid where needed, the correction of excessively low density or pressure\footnote{First, there is the physical fallacy to use the hydrodynamics equation (based on the continuous medium assumption) in an environment where the mean distance between particles overcomes the characteristic size or wavelength of the phenomena we are interested in. Second, having large Mach numbers (above a few 10) enlarges the relative numerical error we make on computing the pressure for instance since we get it from the difference between the total and the kinetic energies which become very similar. Large contrasts in density within the simulation space can also lead to artificially high velocities because of errors at the level of the solver (which work with conservative variables) which lead to excessively small timesteps (due to the \textsc{cfl} condition). Those points are the main reasons to set a floor density and pressure.}, a data file being written, a data file being converted into a visualization file, etc.

\begin{figure}
\hspace*{-2.5cm}
\includegraphics[height=5.5cm, width=1.35\textwidth]{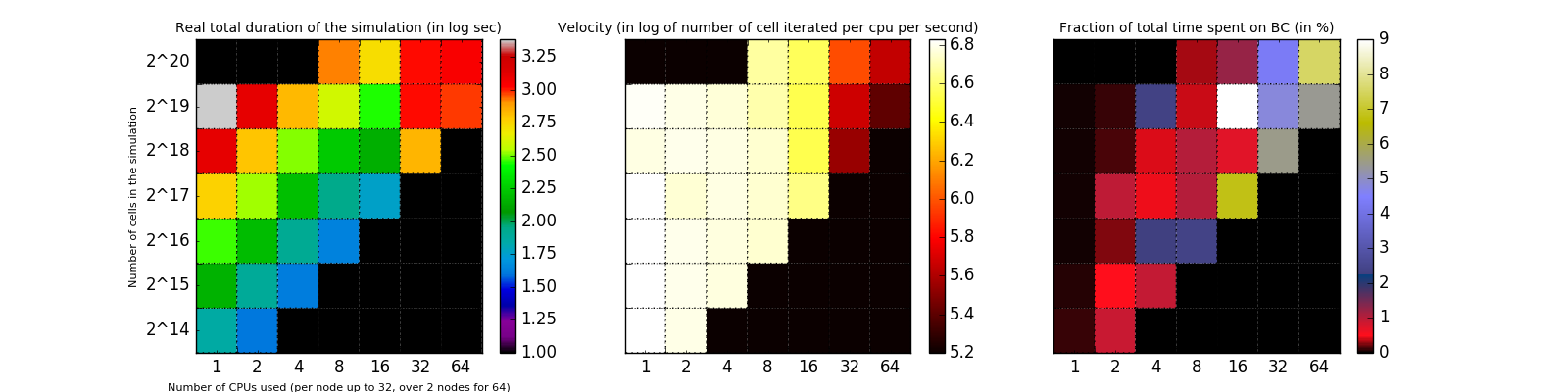}	
\caption{The same simulation has been run on the Arago cluster of the FACe for different $N_{\textsc{cpu}}$ (horizontal axis) and total number of cells (vertical axis). The colormap represents, on the left, the total duration of the simulation, in the middle, the average computation speed $V/N_{\text{it}}$ per unit time step and on the right, the fraction of time spent communicating. The black cells are configurations where the code has not been run.}
\label{fig:scalability}
\end{figure}

We give a few elements of scalability theory to the reader on the basis of \eqref{eq:comp_speed}. It is an essential feature for high performance computing code which certifies its capacity to run efficiently on a large number of \cpus. Two kinds of scalability can be defined :

\begin{enumerate}
\item \underline{Strong scalability} : a code is said to be strongly scalable if $V$ does not decrease excessively as the number of \cpus rise, all other things being kept equal (in particular the number of grid cells, playing a role in $N_{\text{op}}$, must remain the same). It is absolutely necessary for a code to be scalable up to an intermediate number of \cpus (\eg 16 to 32) to be able to debug the code upstream and work out numerical caveats before running it on major facilities and on many \cpus. Indeed, the debugging step, always on a small number of \cpus, requires in priority $\Delta t$ small to appreciate the reactions of the code to modifications, which drives the user into using a low resolution. To decrease $\Delta t$ while debugging, the user then wants to rise $N_{\textsc{cpu}}$ up to 16 or 32 \cpus for instance.
\item \underline{Weak scalability} : in practice, the simulation space can not be subdivided into an infinitely large number of blocks\footnote{For codes compatible with multithreading (and \textsc{gpu} computing), it is theoretically possible to have more processing units than blocks but still, the quick rise in communication time would make the computation speed drop.}, each being monitored by exactly one \cpu as explained in \ref{sec:parallel}. To avoid the caveats which appear when a \cpu is underused or when communication becomes preponderant, one can monitor the evolution of $V$ as the total number of grid cells rises in the same proportions as $N_{\textsc{cpu}}$. Then, the work load per \textsc{cpu} remains approximately similar. A code is said to be weakly scalable if the total elapsed time $\Delta t$ remains fairly constant as the resolution and the number of \textsc{cpu}s rise.
\end{enumerate}
Figure\,\ref{fig:scalability} can be used to illustrate those notions. As one moves along an horizontal line from left to right, the total number of cells remain the same as the number of \cpus rises : we evaluate the strong scalability of the code. The weak scalability corresponds to the bottom-left to upper-right diagonals along which the number of \cpus and cells is multiplied by 2 each time. 

Finally, notice that the core operations to perform when a job is run on a cluster, $N_{\text{op}}$, can be separated into two main categories :

\begin{enumerate}
\item the algorithmic ones, depending on the numerical scheme and its order for instance.
\item the communication steps to transfer the values found in the boundary cells to the neighboring blocks.
\end{enumerate}
The shrinking of the latter is the topic of the last section of this Chapter and the former can be empirically minimized but obeys to mathematical constrains which can not always be bypassed such as the \cfl condition. If communication between nodes is seamlessly fast on modern servers (with matchless InfiniBand\footnote{A computer-networking communications standard which can transfer up to 100Gbit$\cdot$s$^{-1}$.} technology on the best clusters), it is of critical importance on local clusters. It is the user responsibility to solve an optimization problem when $n_i$ and $N_{\textsc{cpu}}$ are chosen for a given $N_i$ so as to minimize the communication time : a larger number of \textsc{cpu}s, all other things being equal, does not necessarily result in a lower amount of total computing time $\Delta t$. It is also the only way to isolate the errors intern to the code from issues due to an incompatibility of the code with the facilities, which requires the interplay of the administrators of the server. To formalize this approach, we will now present a recurrent issue met by \vac on the Arago cluster of the FACe\footnote{On nodes with 32 \cpus each.} namely the use of multithreading. It is a commonly used feature on modern architectures : a single \cpu can run concurrently several processes\footnote{In some ways, it is an internally parallel computing.}. The scheduler of Arago, which is called each time a \textsc{pbs}\footnote{For Portbale Bash Script.} script is submitted, distributes the jobs between the dozen of nodes, each of them being assigned 32 \textit{virtual} \cpus : in reality, there are only 16 of them, each of them being "splitted"in two by multithreading. With \vac, the call to 32 \cpus on 1 node leads to a slowing down so violent ($V$ is divided by approximately 20 from 16 to 32 \cpus) that the rise of $N_{\textsc{cpu}}$ does not compensate it : a similar numerical \vac setup will be 10 times more time-demanding on 32 \cpus than on 16 \cpus on the Arago cluster with multithreading. 

\begin{figure}[!b]
\begin{center}
\includegraphics[height=6cm, width=10cm]{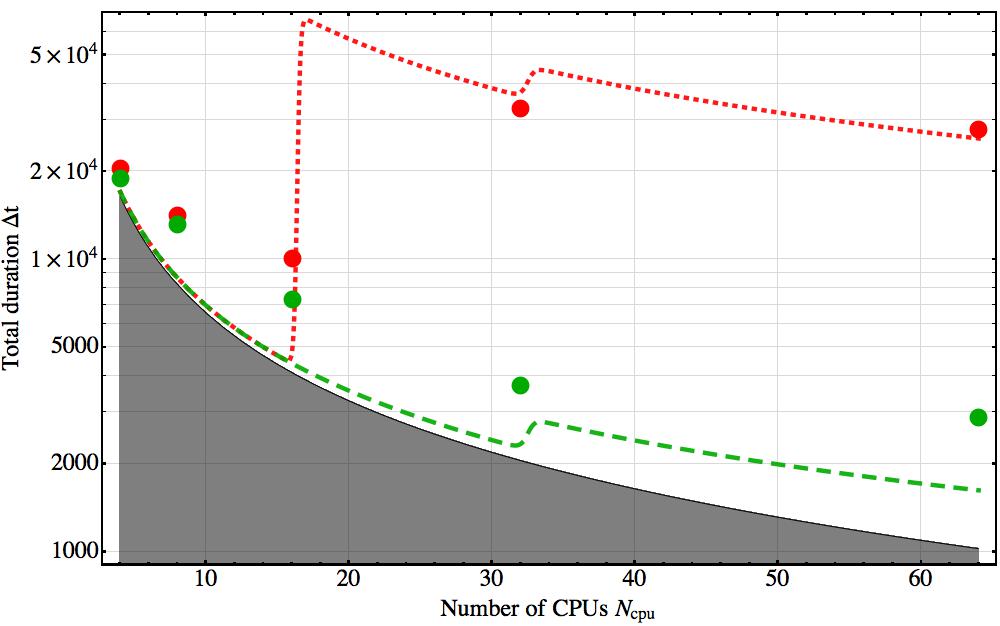}	
\caption{Total duration $\Delta t$ of a given benchmark simulation as a function of the number of \cpus used, $N_{\textsc{cpu}}$. The grey shaded area below the black curve stands for the optimal strong scaling where $\Delta t \propto 1/N_{\textsc{cpu}}$. The red and green points are the observed results (in another timescale than the one represented, but with a similar shape) with and without multithreaded \cpus respectively, taken from a slice of Figure\,\ref{fig:scalability} at a total number of cells of $2^{19}=524288$. The red dotted and green dashed curves are proposed fits to explain those computational behaviour (see equation \eqref{eq:total_time_comp}). The number of cells, the duration scale and the number of timesteps are fiducial ones and do not alter those results, provided they are not excessively small (below 10 seconds).}
\label{fig:optimiz}
\end{center}
\end{figure}


\subsection{Communication}
\label{sec:comm}

\begin{figure}[!t]
\begin{center}
\def\svgwidth{400pt} 
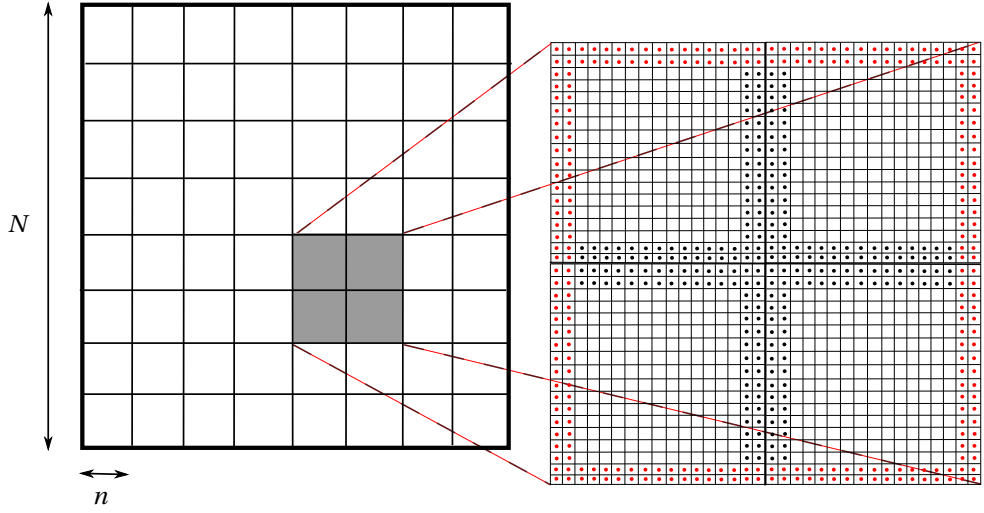
\caption{Same as Figure\,\ref{fig:grids_subgrids_ghost} (except with twice as large resolution in each dimension) but we zoom in on a set of blocks all monitored by the same \cpu, the grey shaded area made of 4 blocks. In red have been represented the boundary cells whose values must be communicated to other \cpus than the one they belong to. $N_{\textsc{cpu}}=16$ and $k=4$}
\label{fig:load_balancing}
\end{center}
\end{figure}

Due to hardware constrains, \cpu in local university clusters are organized within nodes and sockets which typically contain a few dozens of \cpus. A rule of thumb is that communication within a \cpu is always much faster than communication between different \textsc{cpu}s within a node which is itself much faster than communication between different nodes. Furthermore, unfitted hardware tricks such as multithreading\footnote{Which is at the basis of high performance computing methods, not yet compatible with \vac.} can lead to significant decreases of the computation speed per \cpu by a factor of 20 if not bypassed (see previous section). National clusters such as the \textsc{cines} suffer much less flaws and are configured so as to make the most of the facilities\footnote{Identical numerical configurations typically run 20 to 400 times faster on the \textsc{cines} than on local university clusters on a similar number of \cpus. The origin of this discrepancy can not be found in a different fractions of statically and dynamically linked libraries to the executable since the compilation itself is of the order of 10 times faster on the \textsc{cines}.}. To minimize the contact surface between blocks monitored by different \cpus, \vac uses a Morton algorithm to distribute the blocks between \cpus (the load balance step). If we write $N_{\text{bc}}$ the total number of cells within a border (red and black dots in Figure\,\ref{fig:load_balancing}), whose thickness is defined by $\delta$, and $N_{\text{bc,}\neq}$ the cells within a border adjacent to a block monitored by another \cpu (red dots in Figure\,\ref{fig:load_balancing}), we have :
\begin{equation}
N_{\text{bc}}=K\times 
\begin{cases}
2\delta \quad \text{(1D)}\\
2\delta\left(n-2\delta\right)+2n\delta \quad \text{(2D)}\\
4\delta\left(n-\delta\right)\left(n-2\delta\right)+2n^2\delta \quad \text{(3D)}
\end{cases}
\end{equation} 
where the 1D and 3D cases are given for information and where we neglected the edges of the grid and their ghost cells\footnote{Which is acceptable as long as $K\gg1$.}. And we have, for $N_{\text{bc,}\neq}$ in 2D :
\begin{equation}
N_{\text{bc,}\neq}=4\delta\left(\sqrt{k}n-\delta\right)\times N_{\textsc{cpu}}
\end{equation}
which gives, by difference, the number $N_{\text{bc,}=}$ of boundary cells destined to be communicated within a \cpu (black dots in Figure\,\ref{fig:load_balancing}). We can now affirm that the number of operations to be performed by iteration, $N_{\text{op}}$, scales as the sum of the total number of cells plus the number of cells to be communicated within and between \cpus, in a similar fashion as the approach suggested by the Gustafson's law :
\begin{equation}
\label{eq:total_time_comp}
N_{\text{op}}\propto N^2+p_1\times 4\delta N_{\textsc{cpu}} \left( N\sqrt{\frac{k}{N_{\textsc{cpu}}}}-\delta\right)+p_2\times 4\delta N_{\textsc{cpu}} \left[ k \left( \frac{N}{\sqrt{kN_{\textsc{cpu}}}}-\delta \right) - \frac{N}{\sqrt{N_{\textsc{cpu}}}} + \delta \right]
\end{equation}
where we reinjected the expressions $n=N/\sqrt{kN_{\textsc{cpu}}}$ and $K=\left(N/n\right)^2$ to make use of the preexisting dependences and limit the number of variables. Once we set the resolution (\ie $N$), $\delta$ (typically 2) and the number of blocks per \cpu (usually 4), we are left with a function of $N_{\textsc{cpu}}$ only. The weights $p_1$ and $p_2$ each represent the speed of the operation and the relative fraction it occupies within the code\footnote{Due to the philosophy of the code (see section \ref{sec:parallel}), the latter is minimized. Most of the operating time per cell is spent integrating, not communicating.}. We can estimate, for the sake of illustration, $p_1$, the communication between \cpus, to be of the order of 30\% while $p_2$, the communication within a \cpu, is much lower, say 1\%. 

We can now provide an estimate of the total duration of a simulation $\Delta t$ using \eqref{eq:comp_speed} with a fiducial speed $V\sim10^6$ cells per second per \cpu, $N=256$ and the previously given figures. The communication time within a \cpu can be safely neglected since it concerns a comparable number of cells and is an order of magnitude faster than communication between \cpus. To account for the enhanced communication time between \cpus which do not belong to the same node, we modulate $p_1$ with a step function\footnote{Typically $ \propto \frac{1}{2}\left[ 1+\tanh \left( \frac{N_{\textsc{cpu}}-N_{\textsc{cpu,critical}}}{0.5} \right) \right]$.} which rises its value beyond the threshold of 32 \cpus (the maximum number available per node on Arago). To account for the mulithreading issue, we can modulate $V$ similarly with a threshold at 16 \cpus. We see on Figure\,\ref{fig:optimiz} that the experimental benchmarks are in agreement with the model with or without multithreading (respectively in red and green). The small glimpse at 32 \cpus can be smeared out by modern communication devices between nodes such as InfiniBand. The maximum number of \cpus, 64 here, reveals that the values of the number of blocks per \cpu and the total number of blocks in the simulation space can be stringent constrains on low resolution debugging meshes.

This section emphasizes the caution which must be taken when one considers the computational feasibility of a numerical simulation. Empirical investigations to assess the compatibility of the code with the material available must be performed and inherent unbalanced work load within the code must be tracked down with a profiler. Given the complexity of the codes for hydrodynamics, those tests can not be realized on simple benchmarks but must be repeated on each new numerical configuration, or at least on configurations as close as possible from the one destined to yield scientific results.

\subsection{Adaptive Time Stepping}
\label{sec:ats}

\begin{figure}[!h]
\centering
\includegraphics[height=21cm, width=17cm]{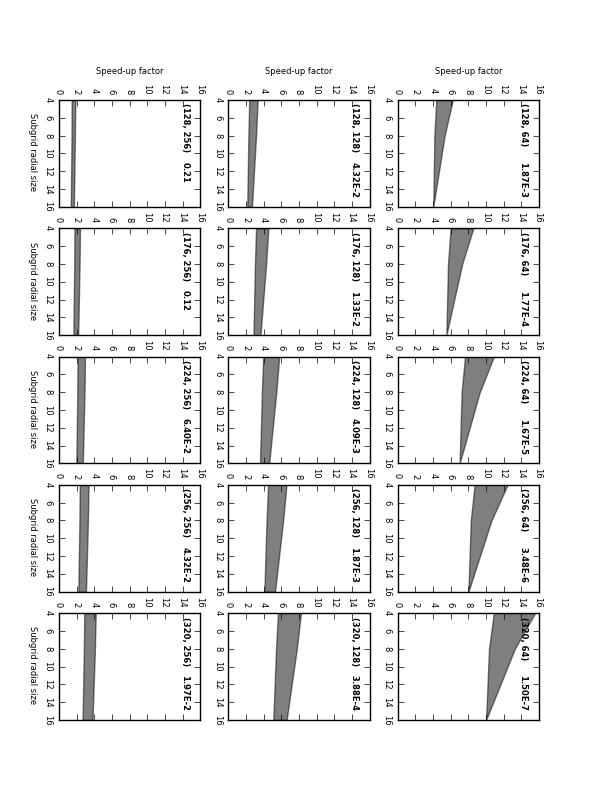}	
\caption{Estimates of the speed-up factor depending on the size of the simulation space and the size of the blocks. The figures between parenthesis in the upper left corner of each panel indicate the radial and angular resolution. In the upper right panel can be found the ratio of the inner boundary size by the outer boundary size.}
\label{fig:ats_speedup}
\end{figure}

The interest of using different time steps for different blocks (\ie doing Adaptive Time Stepping, \ats) is clear when the blocks have very different time steps. It typically occurs when \amr is used, with a spatial dynamics of two to the power of the number of refinement levels. The lower absolute resolution regions are likely to be associated to large time steps and thus, do not require as many numerical iterations as the higher absolute resolution regions. \vac does not include yet the possibility to use \ats but before spending time implementing it, one must carefully wonder about the computing-time it spares. Indeed, if the blocks with the smallest cell size (\ie highest resolution) contain most of the cells (which is likely because it takes more of them to mesh a given simulation space), the speed-up factor is not that high ; the large cells where we save computing iterations are too few to make \ats advantageous in this case. Figure\,\ref{fig:ats_speedup} shows an estimate of the speed-up factor for different total number of cells in the radial (columns) and angular (rows) direction of an axisymmetric 2.5D grid such as the one used in the second part of this manuscript. The size of the block (named here subgrid) is either 4, 8 or 16, which changes the speed-up factor : the smaller the block, the more efficiently we can spot a halve of the time step. On the other hand, blocks with smaller sizes bring more communication time (see section \ref{sec:comm}). The shaded region locates the estimate, with the upper and lower limit of those areas being computed respectively if only gravity or the \cfl condition set the time step. Because \ats requires additional algorithms to deal with the communication of information between blocks at different physical time\footnote{In the same way as in \amr , when it is necessary to communicate the values in the ghost cells of two blocks with different resolutions.}, those speed-up factors are overestimations. Given the significant code development effort \ats represents and the limited interest it represents in our case (with expected speed-up factors being at best of a few), we did not implement it in \vac.

\setlength{\parskip}{0ex} 


\part{Planar \bhl accretion onto a compact object}


\chapter*{Introduction}
\addcontentsline{toc}{chapter}{Introduction}
\adjustmtc

Before addressing the wider topic of wind accretion in X-ray binaries in the third part of this manuscript, we first investigate the accretion process itself. The model of accretion by a moving point-mass initially designed by Bondi, Hoyle \& Lyttleton (henceforth abbreviated with \bhl) offers a quintessential representation of the Physics at stake when a massive astrophysical body rushes through a homogeneous medium\footnote{Equivalently, we will consider a static accretor with a planar supersonic incoming flow.}. The beaming of the flow as it sees the accretor approaching can produce a shock whose properties can not easily be derived from purely analytical considerations ; multi-dimensional partial differential equations from Fluid Mechanics leave little room to explicit solutions. However, the thorough study of the underlying equations revealed fundamental characteristics associated to those flows which can be used as guidelines to separate the numerical artifacts from the physically relevant features. \\
\indent The numerical setup we designed aims at overcoming the main computational obstacle which resides in the scale contrast between the size of the compact accretor and the extension of its gravitational sphere of influence. Although the compact body deflects the streamlines at a large scale, the flow remains free to evolve down to the accretor which lies deep inside the shock. For realistic velocities, we will see that this discrepancy reaches up to 6 orders of magnitude. Characterizing the accretion flow requires to uniformly probe each of those scales. The first attempts to numerically model \bhl flows date back to the eighties and revealed the limitations of the analytical models. Rich structures appeared and were quickly questioned to disentangle the influence of the constraining setup (\eg cylindrical cells and two-dimensional meshes) from the physics behind the scenes. Some features inspired more sophisticated models and the technological progresses of \textsc{hpc} made possible more demanding numerical simulations.\\
\indent In a first Chapter, we remind the reader about the sketch drawn by Bondi, Hoyle \& Lyttleton to follow ballistic test-mass from infinity down to the accretion line in the wake of the accretor. The arguments to derive a first estimate of the mass accretion rate will be of use in the last part of this manuscript to plunge this model into a Roche potential. Unlike this qualitative approach, the spherical case of Bondi accretion offers an occasion to highlight the notion of sonic point. The leading role played by the sonic surface in setting the mass accretion rate will also appear in the second chapter of this part where we numerically address the full hydrodynamical \bhl flow. In this second chapter, our primary goal will be to design a robust axisymmetric numerical setup taking the most of the \textsc{hpc} methods implemented in the \texttt{mpi-amrvac} code presented in Section \ref{sec:mpi_amrvac}. The obtained results are finally discussed and interpreted in order to both evaluate the reliability of the numerical model and to characterize the flow behavior in conditions reminiscent of wind accretion in X-ray binaries.


\chapter{Accretion of zero-angular momentum flows onto an isolated point-mass}
\label{chap:acc_pt-mass}
\chaptermark{Accretion onto a point-mass}
\hypersetup{linkcolor=black}
\minitoc
\hypersetup{linkcolor=red}
\setlength{\parskip}{1ex} 

\section{Ballistic trajectories}


\subsection{Model}

\begin{wrapfigure}{r}{0.3\textwidth}
\begin{center}
\includegraphics[height=7cm, width=0.9\textwidth]{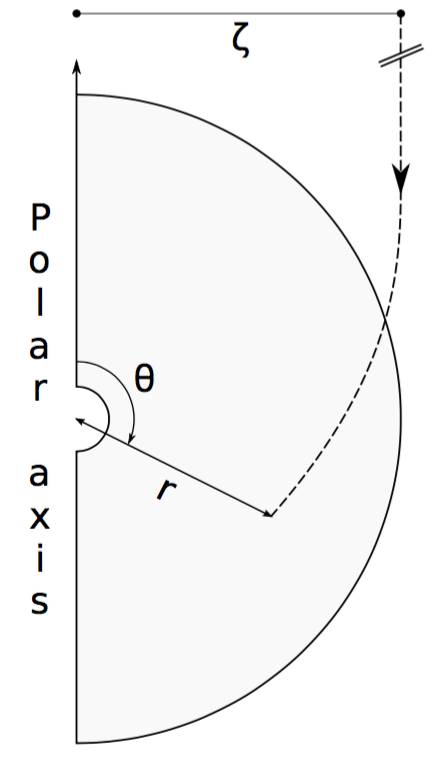}	
\caption{Sketch illustrating the ballistic \bhl flow with the variables and the orientation we consider. A fiducial streamline for a given impact parameter $\zeta$ is plotted in dashed. The light grey area is the simulation space. For the sake of visibility, its inner boundary is highly oversized compared to its outer one.}
\label{fig:context}
\end{center}
\end{wrapfigure}

Let us consider a point-mass of mass $M$ with a relative speed $-\mathbf{v_{\infty}}$ with respect to the ambient medium at infinity, considered as homogeneous. We set this point-mass at the origin of its co-moving frame. This setup is axisymmetric which enables us to work with only 2 coordinates ; we place ourselves in the plane containing the axis passing through the accretor and of direction $\mathbf{v_{\infty}}$. We locate the points in this frame by their distance to the origin, $r$, and the angle $\theta$ (positive) between $-\mathbf{v_{\infty}}$ and the position vector $\mathbf{r}$ as sketched in Figure\,\ref{fig:context}. Finally, we set $v_{\infty}$, the magnitude of $\mathbf{v_{\infty}}$, as the normalisation quantity for velocities and $\zeta_{\textsc{hl}}=2GM/v_{\infty}^2$ as the normalisation quantity for lengths (where $G$ is the gravitational constant). The physical meaning of the latter will appear later on.

In its simplest form \citep{Hoyle:1939fl,Bondi1944b}, the \bhl\footnote{For Bondi, Hoyle and Lyttleton.} incoming flow is supersonic at infinity \ie its Mach number at infinity, $\mathcal{M}_{\infty}=v_{\infty}/c_{\text{s,}\infty}$, is above 1 (where $c_{\text{s,}\infty}$ is the sound speed of the flow at infinity, directly linked to its temperature for an ideal gas). As a consequence, we can characterize the permanent streamlines by the trajectories of test-masses\footnote{Which does not apply to an excessively massive inflow - possibly in a common envelope phase - where the self-gravity of the inflow should be taken into account.} in the gravitational potential of the accretor. 


\subsection{Equations of motion}
\label{sec:motion}

The equation of motion can be straightforwardly solved and gives the dimensionless position and velocity :
\begin{equation}
\label{eq:position}
r=\frac{\zeta ^2}{\sin^2\left(\theta /2\right)+\zeta \sin \theta}
\end{equation}
\begin{equation}
\label{eq:vr}
v_r=-\sqrt{1+\frac{1}{r}-\left( \frac{\zeta}{r} \right)^2 }
\end{equation}
\begin{equation}
\label{eq:vt}
v_{\theta}=\frac{\zeta}{r}
\end{equation}
with $\zeta$ the dimensionless impact parameter of the streamline passing through the point $(r,\theta)$ :
\begin{equation}
\zeta=\frac{r\sin\theta}{2}\left[ 1+\sqrt{1+2\frac{1-\cos\theta}{r\sin ^2\theta}} \right]
\end{equation}
Streamlines with different impact parameters have been plotted using the polar equation \eqref{eq:position} in Figure\,\ref{fig:ball_traj}. Since the Mach number of the flow does not play any role in those ballistic equations, this model of the \bhl flow is often referred as the zero-temperature \bhl flow. However, the equation \eqref{eq:position} shows that streamlines all converge towards a line in the wake of the accretor, $\theta=\pi$, called the accretion line. We will discuss the physical relevance of the ballistic approximation in its vicinity in \ref{sec:limits_ball_model}.


\subsection{Mass and energy density}
\label{sec:mass_and_energy_density}

In this steady-state model, the mass density can be deduced from the frozen-flux of matter. Using the steady state conservation of mass, it implies that $\rho\mathbf{v}$ is a divergence free vector field, which leads to :
\begin{equation}
\label{eq:dens_BK}
\rho= \frac{\zeta ^2}{r\sin\theta \left( 2\zeta-r\sin\theta \right)}
\end{equation}
where we normalized with the homogeneous density of the flow at infinity, $\rho_{\infty}$. If one considers an adiabatic flow (no heating nor cooling on the dynamical timescale), the steady-state conservation of energy (see section \ref{sec:cons_laws}) is given by :
\begin{equation}
\label{eq:cons_energy}
\boldsymbol{\nabla} \cdot \left[ \left( e + P \right) \mathbf{v} \right] = - \rho \mathbf{v} \cdot \boldsymbol {\nabla} \Phi
\end{equation}
with $P$ the pressure, $\Phi$ the gravitational point-mass potential and $e$ the total energy per unit volume (\ie the sum of the kinetic and internal parts). Since the steady-state conservation of mass is given by $\boldsymbol{\nabla} \cdot \left( \rho \mathbf{v} \right) = 0$, we can write the right hand term of \eqref{eq:cons_energy} as a divergence :
\begin{equation}
\label{eq:cont_Bernoulli}
\boldsymbol{\nabla} \cdot \left[ \left( e + P + \rho \Phi \right) \mathbf{v} \right] = 0
\end{equation}
where we recognize the Bernoulli quantity $B$ between parenthesis. Following the expression of the mass density obtained above from $\boldsymbol{\nabla} \cdot \left(\rho\mathbf{v}\right)=0$, we can say that $B$ is given by : 
\begin{equation}
\label{eq:bern_BK}
B= B_{\infty}\frac{\zeta ^2}{r\sin\theta \left( 2\zeta-r\sin\theta \right)}
\end{equation}
where we the normalized $B$ with its value at infinity whose dimensionless expression, $B_{\infty}$, is\footnote{We normalized the energies per unit volume with $\rho_{\infty}v^2_{\infty}$.}, for an ideal gas :
\begin{equation}
B_{\infty} = \frac{1}{2} + \frac{1}{\left(\gamma -1\right)\mathcal{M}_{\infty}^2}
\end{equation}
where $\gamma$ is the adiabatic index\footnote{As defined in section \ref{sec:eos}. See \cite{Horedt2000} for a mindful distinction with the widespread polytropic index.}. The equation above makes use of the first Joule law\footnote{Which states that the internal energy of an ideal gas is function of its temperature only, with the heat capacity at constant volume as a slope.}, of the Mayer relation\footnote{Which relates the heat capacity at constant volume as a function of the adiabatic index for an ideal gas.} and of the law of ideal gases. Combined together, they yield the total energy per unit volume of an ideal gas :
\begin{equation}
u=e-e_K=\frac{P}{\gamma -1}
\end{equation}
where $e_K$ is the kinetic energy per unit volume and $u$ is the internal energy per unit volume. The expression \eqref{eq:cont_Bernoulli} will be of key importance and can not be replaced with a polytropic prescription in the numerical implementation of the problem we consider.



\subsection{Mass accretion rate}
\label{sec:mdot_HL}

\begin{wrapfigure}{r}{8cm}
    \begin{minipage}{8cm}
      \centering
        \includegraphics[height=9cm, width=3.5cm]{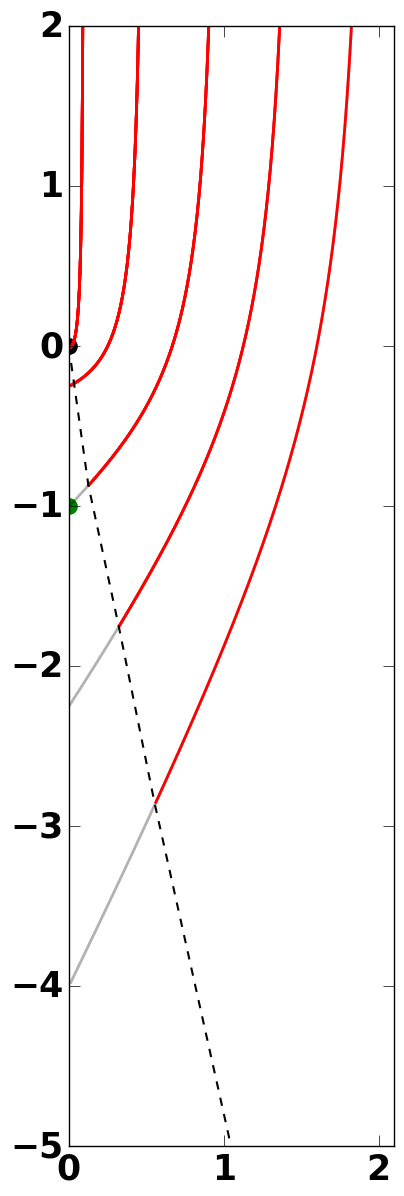}
    \end{minipage}%
    \begin{minipage}{8cm}
      \caption{Set of ballistic trajectories from \eqref{eq:position} (in red) for a Mach number at infinity of 1.1 (mildly supersonic flow) and a dimensionless impact parameter at infinity of 0.1, 0.5, 1, 1.5 and 2 from left to right (in units of $\zeta_{\textsc{hl}}$). The central black dot locates the accretor at the origin while the green one gives the stagnation point. The dotted black line splits the trajectories between their supersonic part (in red) and their subsonic part (represented with a low opacity since they doubtfully correspond to any real trajectory, the ballistic approximation having broke up).}\label{fig:ball_traj}
    \end{minipage}
    \vspace*{-0.2cm}
\end{wrapfigure}

In their seminal paper, Hoyle \& Lyttleton derived a critical impact parameter based on the following arguments. Since the streamlines converge along the accretion line, we can expect dissipative effects to essentially cancel out the normal component of the velocity field. If we write the specific mechanical energy of a test-mass on the accretion line deprived of its $v_{\theta}$ component, we have :
\begin{equation}
\frac{1}{2}v^2_r(\theta=\pi) - \frac{1}{2r} = \frac{1}{2} \left( 1 - \frac{1}{\zeta^2} \right)
\end{equation}
where we used \eqref{eq:position} and \eqref{eq:vr}. We retrieve that for large impact parameters (compared to the soon to be revealed length scale $\zeta_{\textsc{hl}}$), the mechanical energy is left unchanged from its initial value at infinity. This relation indicates that test-masses with an impact parameter below the critical value of $\zeta_{\textsc{hl}}$ will find themselves bound to the accretor (\ie a negative mechanical energy). They are then likely to be accreted. $\zeta_{\textsc{hl}}$ is called the accretion radius though formally, it is a critical impact parameter. The corresponding critical arrival point on the stagnation line which separates the bound and free test-masses is called the stagnation point\footnote{In spite of the fact that the velocity at this point is not zero but just enough to not be able to counterbalance gravity. The test-mass will start to flow away before receding.} (the green dot in Figure\,\ref{fig:ball_traj}) and corresponds to the streamline with an impact parameter $\zeta_{\textsc{hl}}$.

The corresponding physical mass accretion rate $\dot{M}_{\textsc{hl}}$ can be deduced by considering the cylinder of radius $\zeta_{\textsc{hl}}$ and of axis the direction given by the relative speed at infinity :
\begin{equation}
\label{eq:Mdot_HL}
\dot{M}_{\textsc{hl}}=\pi \zeta_{\textsc{hl}}^2 \rho_{\infty} v_{\infty}
\end{equation}
For realistic values of a 1.4\msun runaway neutron stars rushing through the interstellar medium with a relative speed of 200km$\cdot$s$^{-1}$ (see \eg Figure\,\ref{fig:lighthouse}), it gives :
\begin{equation}
\dot{M}_{\textsc{hl}}\sim10^{-18} \left( \frac{M}{1.4M_{\odot}} \right)^2 \left( \frac{\rho_{\infty}}{10^{-24}\text{g}\cdot\text{cm}^{-3}} \right) \left( \frac{v_{\infty}}{200\text{km}\cdot\text{s}^{-1}} \right)^{-3} M_{\odot}\cdot \text{yr}^{-1}
\end{equation}
which is well below any detectable threshold in luminosity given the equation \eqref{eq:Lacc}. Despite its inefficiency, this mass transfer can result in non negligible drags on timescales of the order of the mass doubling timescale \citep{Edgar:2004ip}, always much longer than the timescales we will consider in the present manuscript.

Given the dramatic assumption on the ballistic behaviour of the flow until the normal component of the velocity is suddenly dissipated on the accretion line, equation \eqref{eq:Mdot_HL} provides an order-of-magnitude of the mass accretion rate. The hydrodynamical effects will alter this value, even for high Mach numbers. Attempts to include them are described in the two next sections (\ref{sec:Bondi_sph} and \ref{sec:HD_approaches}).

\subsection{Limits of the ballistic model}
\label{sec:limits_ball_model}

Due to the axisymmetry, the expression of the streamlines \ref{eq:position} indicates a serious convergence of the streamlines in the wake of the accretor, if not a crossing : the velocity field $\mathbf{v}$ can no longer remain divergence-free and the "incompressible flow" assumption\footnote{Indeed, since the ballistic assumption implies that the flow is incompressible - but not the fluid - which means that the Lagrangian derivative is zero along each streamline. Since we are interested in the permanent behaviour, it also means that the streamlines are orthogonal to the isodensity surfaces. The use of the conservation of mass also implies that $\mathbf{v}$ is divergence free. By contraposition, when streamlines cross each other in a steady state framework, it means that the Lagrangian derivative can no longer be considered as zero.} breaks up. It is actually likely that pressure effects must be accounted for before the accretion line. As an example, we plotted in Figure\,\ref{fig:ball_traj} a few streamlines and pinpointed the points beyond which they become subsonic. By this we mean that, if one considers an isentropic flow\footnote{In case of discontinuity along the field line such as a shock, the entropy of the fluid will actually rise - see appendix \ref{sec:jump}. Upstream, the isentropic assumption is essentially validated since the motion is adiabatic (no heating nor cooling term - valid if the thermal timescale is much larger than the dynamical one) and mechanically reversible (\eg if no shock).}, the Laplace relation for an ideal gas gives the expression of the dimensionless sound speed in any point :
\begin{equation}
c^2_{\text{s}}=\frac{1}{\mathcal{M}_{\infty}^2}\rho^{\gamma -1}
\end{equation} 
Then, one can compute the Mach number at any point and detect the "ballistic" front shock (black dotted line in Figure\,\ref{fig:ball_traj}). This surface is well fitted by a cone (\ie constant aspect ratio of the width with respect to the distance to the point-mass along the accretion line) which will motivate the accretion cone model presented in \ref{sec:acc_col}. 


\begin{figure}[!b]
\begin{center}
\includegraphics[height=6.5cm, , width=7cm]{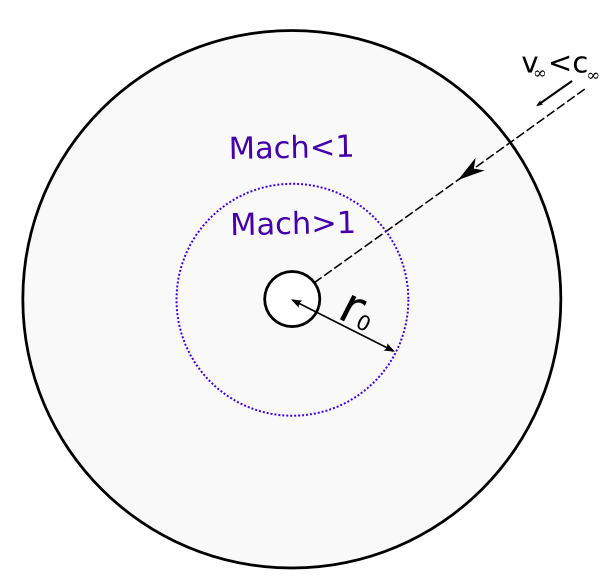}	
\caption{Sketch of the Bondi spherical accretion model. The black dashed line represents a streamline. The purple dotted line represents the sonic surface of an inflowing gas subsonic at infinity.}
\label{fig:bondi0}
\end{center}
\end{figure}


\section{Hydrodynamical intermezzo : the Bondi spherical model}
\label{sec:Bondi_sph}

\subsection{Context}

Before introducing hydrodynamical elements in the \bhl model described in the previous section, we describe another model, the spherical Bondi accretion, where an analytical study can be carried on thanks to the unidimensionality of the problem. Figure\,\ref{fig:bondi0} represents the isotropic situation of a point-mass of mass $M$ at rest with respect to an ambient uniform gas at infinity. The latter has a non-zero but subsonic velocity at infinity, $v_{\infty}$, and a non-zero temperature associated to a sound speed $c_{\text{s}}$ at infinity\footnote{The omission of the $\infty$ symbol will be justified a few lines below by an isothermal assumption.}. We can no longer derive ballistic trajectories since the flow at infinity is subsonic : pressure terms can not be neglected. We undertake a fully hydrodynamical analysis from scratch, with physical quantities for the moment. First, the steady-state conservation of mass in this isotropic configuration yields, if we write $\dot{M}$ the constant and homogeneous mass accretion rate on the central point-mass :
\begin{equation}
\label{eq:cons_mass}
\dot{M}=4\pi r^2 \rho (r) v(r)
\end{equation}
where $v$ stands for the radial component of the velocity. On the other hand, we have the steady-state conservation of the linear momentum :
\begin{equation}
\label{eq:cons_mom}
v\frac{\d v}{\d r}=-\frac{GM}{r^2}-\frac{1}{\rho}\frac{\d P}{\d r}
\end{equation}
For the sake of analytical solvability, we now bypass the equation of energy using a polytropic relation between the pressure and the mass density. It if fully legitimate provided the gas is both ideal and undergoing an isentropic transformation :
\begin{equation}
P\rho^{-\gamma}=\text{cst}
\end{equation}
where $\gamma$ lies between 1 (for an isothermal gas \ie a gas which would instantly radiate away any augmentation of internal energy or would instantly be heated to compensate a loss of internal energy) and $5/3$ (for an adiabatic gas \ie a gas which would not exchange heat with the outside, meaning it would not radiate away the increase in internal energy due to the work done by the pressure force). Then, we have :
\begin{equation}
\label{eq:polytropic_diff}
\frac{\d P}{\d r}=\underbrace{\frac{\d P}{\d \rho}}_{=c_{\text{s}}^2}\frac{\d \rho}{\d r}
\end{equation}
where the identification to the square of the sound speed is valid because of the isentropic assumption. 



\subsection{The isothermal velocity profile}
\label{sec:isothermal_bondi}



Differentiating \eqref{eq:cons_mass}, we get :
\begin{equation}
\frac{\d \rho}{\rho} = - 2 \frac{\d r}{r} - \frac{\d v}{v}
\end{equation}
Injecting this expression in \eqref{eq:polytropic_diff} and then in \eqref{eq:cons_mom} yields :
\begin{equation}
\label{eq:before_main_Bondi_acc}
\frac{r}{v}\frac{\d v}{\d r} = \frac{2}{v^2/c_{\text{s}}^2(r)-1}\left( 1-\frac{GM/2c_{\text{s}}(r)^2}{r}\right) 
\end{equation}
In this section only, we will discuss the isothermal case (\ie $\gamma=1$) so as to analytically derive an implicit expression of the velocity profile : in the expression above, the sound speed $c_{\text{s}}$ no longer depends on the position. The reader is nonetheless invited to refer to \cite{Frank2002}\footnote{And to \cite{Lamers1999} where the emphasis is on the outflowing solutions though.} where the polytropic case in general are treated. Most comments we will make about the structure of the solutions remain essentially unchanged for $\gamma >1$. 

\begin{figure}
\begin{center}
\includegraphics[height=7cm, width=10cm]{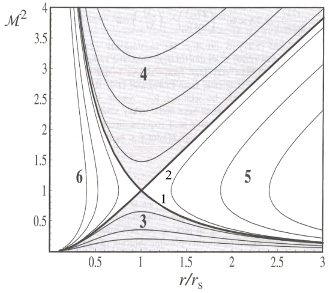}	
\caption{Mach number squared as a function of the ratio of the distance to the stellar center by the sonic radius. Each curve corresponds to a mathematical solution \eqref{eq:vel_prof_Bondi} with a different constant $K$. The 4 different families of solutions are indexed, along with the 2 critical solutions (thick solid lines). From \cite{Frank2002}.}
\label{fig:vel_profiles_Bondi}
\end{center}
\end{figure}

From now on, we can motivate the choice of normalization quantities for the velocity and the length : the sound speed\footnote{Which makes all the dimensionless velocities be Mach numbers.} and a radius we already call sonic, $r_s=(GM)/(2c_{\text{s}}^2)$. Using them, we get the dimensionless equation of motion :
\begin{equation}
\label{eq:main_Bondi_acc}
\frac{\d v}{\d r} = \frac{2v}{r^2} \frac{r-1}{v^2-1} 
\end{equation}
which can, luckily enough, be integrated to give an implicit expression of the velocity profile :
\begin{equation}
\label{eq:vel_prof_Bondi}
\frac{v^2}{2} - \ln (v) = \ln (r^2) + \frac{2}{r} + K
\end{equation}
where $K$ stands for the integration constant. Different values of $K$ set different families of solutions in Figure\,\ref{fig:vel_profiles_Bondi} : for $K<-1.5$, we obtain the double-valued functions on the left and right side of the X-point (indexed 6 and 5) while for $K>1.5$, we get the solutions on the lower and upper sides of the X-point (indexed 3 and 4). A first statement can be made about the asymptotic behaviour of the velocity profile, starting from the lower left panel and going counter-clockwise :
\begin{equation}
\begin{cases}
v\propto e^{-2/r}\quad \text{for} \quad v\ll 1 \quad \text{and} \quad r\ll 1\\
v\propto 1/r^2 \quad \text{for} \quad v\ll 1 \quad \text{and} \quad r\gg 1\\
v\propto \sqrt{\ln (r)}\quad \text{for} \quad v\gg 1 \quad \text{and} \quad r\gg 1\\
v\propto \sqrt{1/r} \quad \text{for} \quad v\gg 1 \quad \text{and} \quad r\ll 1
\end{cases}
\end{equation}
We also notice a decisive feature of those solutions (which will run into again in the different context of radiatively-driven winds later on), the existence of a sonic point - the aforementioned X-point. By sonic point (resp. surface) we mean a point (resp. surface) through which passes a fluid particle which becomes, at this precise point, supersonic\footnote{The reverse, a point where the fluid particle becomes subsonic, is not a sonic point and displays very different properties as we will see in section \ref{sec:bhl_shock} and appendix \ref{sec:jump}.}. Indeed, if one gets a closer look at \eqref{eq:main_Bondi_acc}, it is clear that along the $r=1$ line, the solutions must verify $\d v/\d r = 0$ while along the $v=1$, the solutions must verify $\d r/\d v = 0$. The only exceptions to these results are the critical solutions \ie those who verifies $v(r=1)=1$ ; in this case, the derivative are no longer singular but given by De L'H\^opital's rule : 	
\begin{equation}
\frac{\d v}{\d r} \Bigr|_{r=1} = \pm 1
\end{equation}
The only point which admits two different slopes at the same point is the sonic point $r=1$ and $v=1$ : an inflowing solution, subsonic at infinity (for the negative slope) and an outflowing one\footnote{Indeed, notice that the equation \eqref{eq:main_Bondi_acc} remains unchanged by the transformation $v\rightarrow-v$.}, subsonic in the vicinity of the surface of the object (for the positive slope). The latter, indexed 2 in Figure\,\ref{fig:vel_profiles_Bondi}, corresponds to the Parker wind model \citep{Parker1958} while the former, indexed 1, is the one we are interested in, the Bondi accretion. The other families of solutions will not be discussed here but are detailed in \cite{Lamers1999} and \cite{Frank2002}.


\subsection{The polytropic sonic point}

We now relax the isothermal assumption but not the polytropic one. Intregrating the equation of motion yields the physical Bernoulli invariant :
\begin{equation}
B=\frac{1}{2}v^2+\underbrace{\int \frac{\d P}{\rho}}_{c_{\text{s}}^2/(\gamma -1)} - \frac{GM}{r}=\text{cst}
\end{equation}
where $c_{\text{s}}^2$ is no longer uniform. However, we can get precious insights if we write the equality between the Bernoulli invariant at infinity and at the sonic point\footnote{From here, the isothermal case $\gamma=1$ requires a specific treatment not covered by the equations \eqref{eq:bernoulli_unleashed} and \eqref{eq:son_pt_Bondi}.} :
\begin{equation}
\label{eq:bernoulli_unleashed}
\frac{1}{2}v_{\infty}^2+\frac{c_{\text{s,}\infty}^2}{\gamma -1}=\frac{1}{2}c_{\text{s}}^2+\frac{c_{\text{s}}^2}{\gamma -1}-\frac{GM}{r_{\text{s}}}
\end{equation}
Since we still have, even without the isothermal assumption, a sonic point in $r_{\text{s}}=(GM)/(2c_{\text{s}}^2)$ given the critical point in \eqref{eq:before_main_Bondi_acc}, still valid, we obtain the sonic radius :
\begin{equation}
\label{eq:son_pt_Bondi}
r_{\text{s}}=\frac{5-3\gamma}{4\left(\gamma -1\right)}\frac{2GM/v_{\infty}^2}{1+\frac{2}{\gamma -1}\frac{1}{\mathcal{M}^2_{\infty}}}
\end{equation}
where we recognize the accretion radius $\zeta_{\textsc{hl}}$ at the numerator, though in a very different geometry from the one it has been defined in initially. In the next chapter, this quantity will be referred to as $r_0$ to differentiate it from the distance to the sonic surface in the non spherical case. This expression also appears behind the equation (2.11b) of \cite{Theuns1992} and locates the uniquely determined position where a stationary isotropic subsonic inflowing gas becomes supersonic.

\begin{figure}
\begin{center}
\includegraphics[height=7cm, width=9cm]{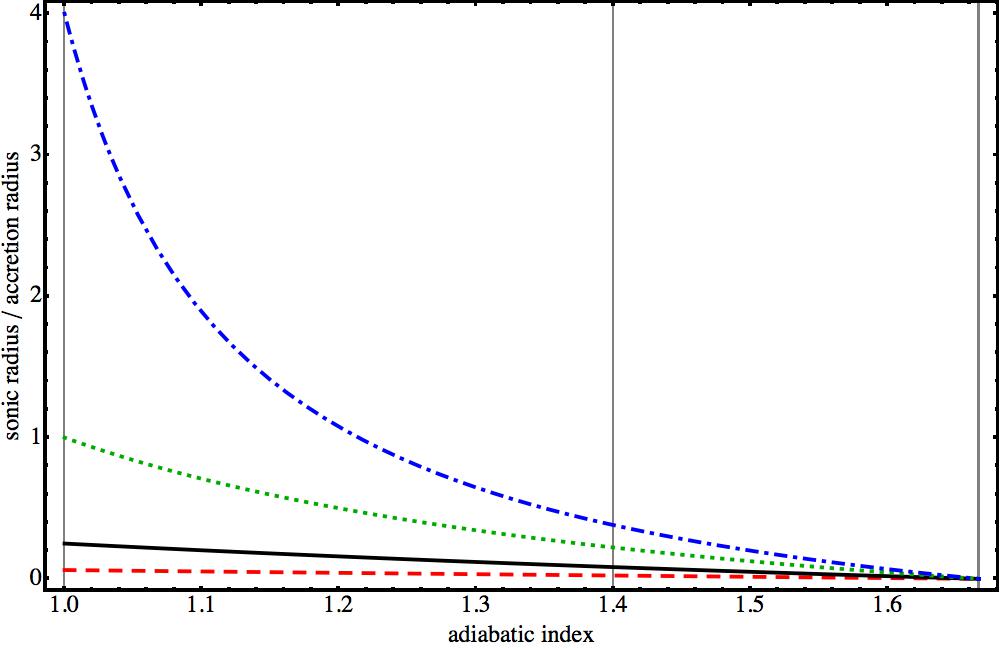}	
\caption{The sonic radius (in units of $\zeta_{\textsc{hl}}$) as a funtion of the adiabatic index for a Mach number at infinity of 0.5, 1, 2 and 4 (from bottom to top). The isothermal flow ($\gamma=1$), the monoatomic adiabatic flow ($\gamma=5/3$) and the diatomic adiabatic flow ($\gamma=7/5$) have been represented with vertical lines.}
\label{fig:sonic_point}
\end{center}
\end{figure}


\subsection{Mass accretion rate}

The only solution which goes from infinity down to the accretor without shock is a transonic one, subsonic at infinity and which becomes supersonic as the flow gets accelerated towards the accretor. More interestingly, the only solution able to do so reaches the sound speed at the sonic radius $r_s$ : we are now in conditions to determine the physical mass accretion rate of Bondi spherical accretion, $\dot{M}_{\textsc{b}}$. Indeed, since $\dot{M}_{\textsc{b}}$ is uniform, it can be evaluated anywhere. The expression \eqref{eq:son_pt_Bondi} suggests to do so at the sonic point.
\begin{equation}
\label{eq:mdot_Bondi_ini}
\dot{M}_{\textsc{b}}=4\pi r_s^2 c_{\text{s}} \rho_{\text{s}}
\end{equation}
Using Laplace relation and the definition of the sound speed, we can get rid of the mass density at the sonic point and are left with the sound speed at the sonic point (and quantities at infinity). But thanks to \eqref{eq:bernoulli_unleashed}, we can express the sound speed at the sonic point as a function of the sound speed at infinity. Thus, we have :
\begin{equation}
\label{eq:mdot_Bondi_sph}
\dot{M}_{\textsc{b}}=\pi \frac{(GM)^2}{c_{\text{s,}\infty}^3} \rho_{\infty} \left[ \left( \frac{\gamma -1}{5-3\gamma} \right) \left( \mathcal{M}_{\infty}^2+\frac{2}{\gamma -1} \right) \right]^{\frac{5-3\gamma}{2(\gamma-1)}}
\end{equation}

For a monoatomic flow\footnote{$\gamma=5/3$.} with a negligible velocity at infinity towards the accretor compared to its sound speed at infinity, we get the following estimate for the mass accretion rate of a putative intermediate mass black hole accreting the gas from a cold interstellar medium at rest at infinity with respect to it :
\begin{equation}
\dot{M}_{\textsc{b}}\sim 10^{-7} \left( \frac{M}{500M_{\odot}} \right)^2 \left( \frac{\rho_{\infty}}{10^{-24}\text{g}\cdot\text{cm}^{-3}} \right) \left( \frac{c_{\text{s,}\infty}}{1\text{km}\cdot\text{s}^{-1}} \right)^{-3} M_{\odot}\cdot\text{yr}^{-1}
\end{equation}
which corresponds to the luminosity levels we observe in X-rays for Ultra-Luminous X-ray sources \citep{Webb2014}. For a stellar mass object in a warmer environment though, typically an isolated neutron star, such a luminosity would be impossible to detect beyond a few kiloparsecs. 

\setlength{\parskip}{0ex} 


\chapter{Numerical simulations of planar accretion onto a compact body}
\label{chap:num_sim_BHL}
\chaptermark{Planar flow accreted onto a compact body}
\hypersetup{linkcolor=black}
\minitoc
\hypersetup{linkcolor=red}
\setlength{\parskip}{1ex} 

\section{Hydrodynamical approaches of \bhl flows}
\label{sec:HD_approaches}

If the proper hydrodynamical treatment of the flow is possible in the case of Bondi spherical accretion, it is in large part due to the unidimensionality of the problem. Once we get interested into an axisymmetric situation like the \bhl flow, things get trickier. We must sacrifice the hydrodynamics to be left with a set of ordinary differential equations ; otherwise, we are doomed to face the much less cooperative realm of partial differential equations. In this section, we introduce two different ways to navigate between the two : Bondi's pragmatic though approximated approach and the accretion cone model. We also prepare the ground for the treatment of the shock by suggesting the reader to refer to appendix \ref{sec:jump} for the jump conditions at planar and oblique shocks.


\subsection{The Bondi empirical formula}
\label{sec:Bondi_empirical}

In an attempt to incorporate thermodynamical considerations in the frame of \bhl flow, \cite{Bondi1952} suggested an intermediate formula\footnote{We prefer not to use the term "interpolation formula" found in the literature since it does not behave as the Bondi spherical mass accretion rate at low Mach number as we shall see in a minute.} which has then been modified by \cite{Shima1985} on the basis of numerical results to verify :
\begin{equation}
\label{eq:Bondi_swindle}
\dot{M}_{\textsc{bh}}=\frac{\dot{M}_{\textsc{hl}}}{\left(1+1/\mathcal{M}_{\infty}^2\right)^{3/2}}
\end{equation}
where $\dot{M}_{\textsc{hl}}$ is given by \eqref{eq:Mdot_HL} and $\mathcal{M}_{\infty}$ is the Mach number of the flow at infinity. At high $\mathcal{M}_{\infty}$, this mass accretion rate does match the one derived from ballistic considerations but we already notice that at low Mach numbers, for any adiabatic index, it can not converge towards the expression \eqref{eq:mdot_Bondi_sph} obtained for spherical mass accretion rates since it does not account for the dependency in $\gamma$. Admittedly, it provides us with a continuous mass accretion rate which empirically extends the \bhl flow to the subsonic regime but does not yet provide a uniform interpolation formula joining the spherical Bondi and the supersonic \bhl flows. We will introduce such an interpolation formula, matching both the low and high Mach number limits, in section \ref{sec:mdot_BHL25D_num}.


\subsection{The accretion cone}
\label{sec:acc_col}

In addition, the matching of $\dot{M}_{\textsc{bh}}$ with $\dot{M}_{\textsc{hl}}$ for high Mach numbers guarantees the self-consistency of the approach but not necessarily its physical relevance. Indeed, $\dot{M}_{\textsc{hl}}$ has been estimated based on ballistic arguments which, strictly speaking, do not apply : whatever the Mach number of the flow at infinity, the accreted particles of fluid always cross a region of space where the ballistic approximation breaks up. So as to refine this model, we follow \cite{Horedt2000} and \cite{Edgar:2004ip} to introduce the accretion cone model (\aka accretion column model). The accretion line is replaced with an infinitely thin accretion cone\footnote{See Figure\,\ref{fig:ball_traj} for an empiric justification of this shape.}. By balancing the mass and linear momentum fluxes crossing a surface of a infinitely short section of this cone, we will be able to estimate a lower limit for the mass accretion rate associated to the \bhl flow.

Let us consider the configuration depicted on Figure\,\ref{fig:acc_column} and adimension the lengths with $\zeta_{\textsc{hl}}$, the densities with $\rho_{\infty}$, the velocities with $v_{\infty}$ and the other quantities with straigthforward products of those quantities\footnote{Without any additional factor, to make sure the geometrical ones do not vanish. For instance, the pressure is simply normalized with $\rho_{\infty}v^2_{\infty}$.}. The flow remains ballistic outside of the accretion cone of half-opening angle $\alpha$. The position along the cone axis is located with $r$ (increasing in the direction of $\mathbf{v_{\infty}}$), its radius at this position with $s$ and we assume $s\sim\alpha r\ll r$ such that, within the cone, the quantities depend only on $r$ and the velocity has no transverse component. We can evaluate the expressions obtained in \ref{sec:motion} as $\theta\rightarrow \pi^-$ to relate the impact parameter $\zeta$ to the longitudinal position $r$ :
\begin{equation}
\label{eq:r_to_zeta}
r \sim \zeta ^2
\end{equation}
Then, we can write the conservation of mass within the red section of infinitesimal length $\d r$ :
\begin{equation}
\label{eq:cons_mass_acc_col}
\d \Phi_{\text{m,lat}} + \d \Phi_{\text{m,}1} + \d \Phi_{\text{m,}2} = 0  
\end{equation}
where the $\d \Phi_{\text{m}}$ correspond to the mass fluxes\footnote{Where the surface elements are oriented outwards.} through, respectively, the lateral surface $2\pi s \d r$, the back surface $\pi s^2\Bigr|_{r}$ and the front surface $\pi s^2\Bigr|_{r+\d r}$. The fluxes are given by :
\begin{equation}
\begin{cases}
\d \Phi_{\text{m,lat}} = - \rho v_{\text{tot}} \cdot 2\pi s \d r = - 2 \pi \zeta \d \zeta = - \pi \d r \\
\d \Phi_{\text{m,}1} + \d \Phi_{\text{m,}2} = - \rho v \pi s^2 \Bigr|_{r} + \rho v \pi s^2 \Bigr|_{r+\d r} = \pi \d r \frac{\d \left( \rho v s^2 \right) }{\d r} 
\end{cases}
\end{equation}
where $v_{\text{tot}}$ is the projection of the velocity on the normal axis to the cone surface in $r$ while $v$ is the velocity in the cone, purely along the cone main axis. For $\d \Phi_{\text{m,lat}}$, we first used $\boldsymbol{\nabla} \cdot \left( \rho \mathbf{v} \right)=0$ to trace back the streamlines up to the corresponding section at infinity and then \eqref{eq:r_to_zeta} for the last equality. In the spirit of this one dimensional model, we now introduce the linear mass density along the cone $\lambda=\d m / \d r$. It enables us to rephrase \eqref{eq:cons_mass_acc_col} using $\rho=\d m / \left( \pi s^2 \d r \right) = \lambda / \pi s^2 $ :
\begin{figure}[!b]
\begin{center}
\def\svgwidth{400pt} 
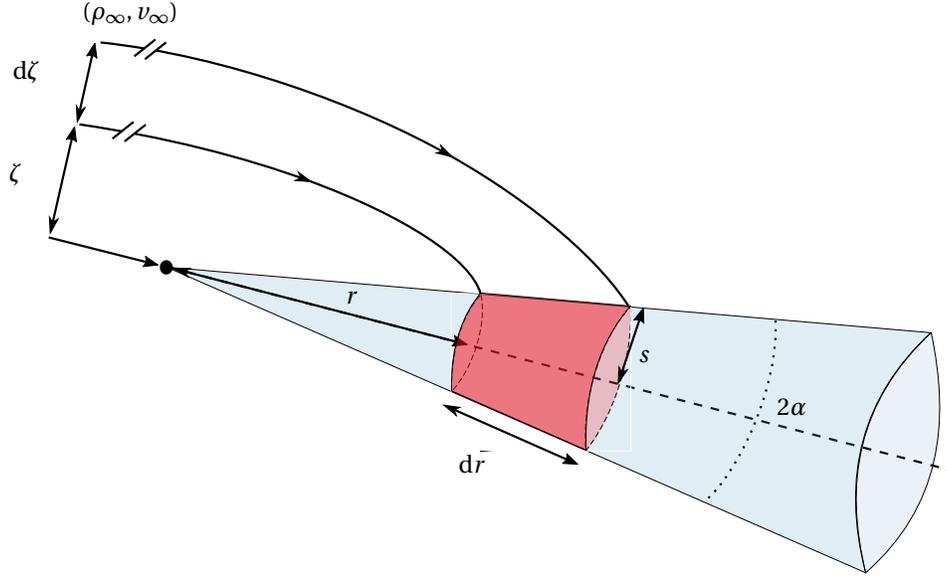
\caption{Model of the accretion cone. The black dot stands for the accretor and the front arrow for the relative speed along the axis. The shock (blueish surface) is represented by a cone of opening angle $2\alpha$. Two streamlines separated by an infinitely small impact parameter $d\zeta$ have been represented. They intersect the shock at a distance to the accretor $r$ and within a range d$r$. At this distance, the width of the cone is given by the first order infinitesimal $s$ (\ie $s\ll r$) and the corresponding surface is shown in red.}
\label{fig:acc_column}
\end{center}
\end{figure}
\begin{equation}
\label{eq:mass_cons_acc_col}
\frac{\d (\lambda v)}{\d r}=\pi
\end{equation}
We now undertake the same balance for linear momentum with the corresponding linear momentum fluxes :
\begin{equation}
\begin{cases}
\d \Phi_{\text{mv$_r$,lat}} = - \rho v_r(\theta\rightarrow\pi ^-) \cdot v_{\text{tot}} 2\pi s \d r = - 2 \pi \zeta \d \zeta = - \pi \d r \\
\d \Phi_{\text{mv,}1} + \d \Phi_{\text{mv,}2} = - \rho v^2 \pi s^2 \Bigr|_{r} + \rho v^2 \pi s^2 \Bigr|_{r+\d r} = \pi \d r \frac{\d \left( \rho v^2 s^2 \right) }{\d r}\\
\d \Phi_{\text{mv,P,}1} + \d \Phi_{\text{mv,P,}2} = - P \pi s^2 \Bigr|_{r} + P \pi s^2 \Bigr|_{r+\d r} = \pi \d r \frac{\d \left( P s^2 \right) }{\d r}
\end{cases}
\end{equation}
where $\d \Phi_{\text{mv$_r$,lat}}$ is the infinitesimal flux of longitudinal linear momentum (the one which is not dissipated at the shock and does enter the cone) and $\d \Phi_{\text{P}}$ are the infinitesimal fluxes due to the work done by the pressure forces. Notice that in the first flux\footnote{Where we used $v_r(\theta\rightarrow\pi^-)=1$}, we used the assumption that the transverse component of the velocity field is instantaneously dissipated. Then, it does not contribute to the balance within the cone. Once we add the dimensionless gravitational source term (\lhs below), the total balance yields :
\begin{equation}
\label{eq:mom_cons_acc_col}
- \frac{1}{2}\frac{\lambda}{r^2} = - \pi + \frac{\d (\lambda v^2)}{\d r} + \frac{\d (P\pi s^2)}{\d r}
\end{equation}
From there, one can either argument in favour of the negligibility of the pressure term \citep{Edgar:2004ip} or make do with it and make assumptions on the thermodynamics \citep{Wolfson1977,Horedt2000}. In the former case, we are left with a set of two non-linear coupled ordinary equations for two variables (the reduced linear mass and velocity along the cone). The first one, \eqref{eq:mass_cons_acc_col}, can be straightforwardly integrated introducing the stagnation point $r_{\textsc{x}}$ :
\begin{equation}
\lambda v = \pi \times \left( r - r_{\textsc{x}} \right)
\end{equation} 
which enables us to get rid of $\lambda$ in \eqref{eq:mom_cons_acc_col} which is now given by :
\begin{equation}
\label{eq:diff_eq_v_acc_col}
v \frac{\d v}{\d r}=\frac{v\left(1-v\right)}{r-r_{\textsc{x}}}-\frac{1}{2r^2}
\end{equation}
A numerical integration with an approximate spatial condition has been represented on Figure\,\ref{fig:vel_prof_acc_col}, where the black dot stands for the position of the stagnation point in the bedrock model of the Bondi-Hoyle-Lyttleton purely ballistic flow. It can be shown that monotonic solutions which match $v\rightarrow 1$ as $r\rightarrow \infty$ can be obtained only for $r_{\textsc{x}}$ above 0.5 ; this value would imply a mass accretion rate twice smaller than $\dot{M}_{\textsc{hl}}$, providing a lower limit we will put on test later on with numerical simulations (section \ref{sec:mdot_BHL25D_num}). 

\begin{figure}[!h]
\begin{center}
\def\svgwidth{250pt} 
\includegraphics[height=6.5cm, width=10cm]{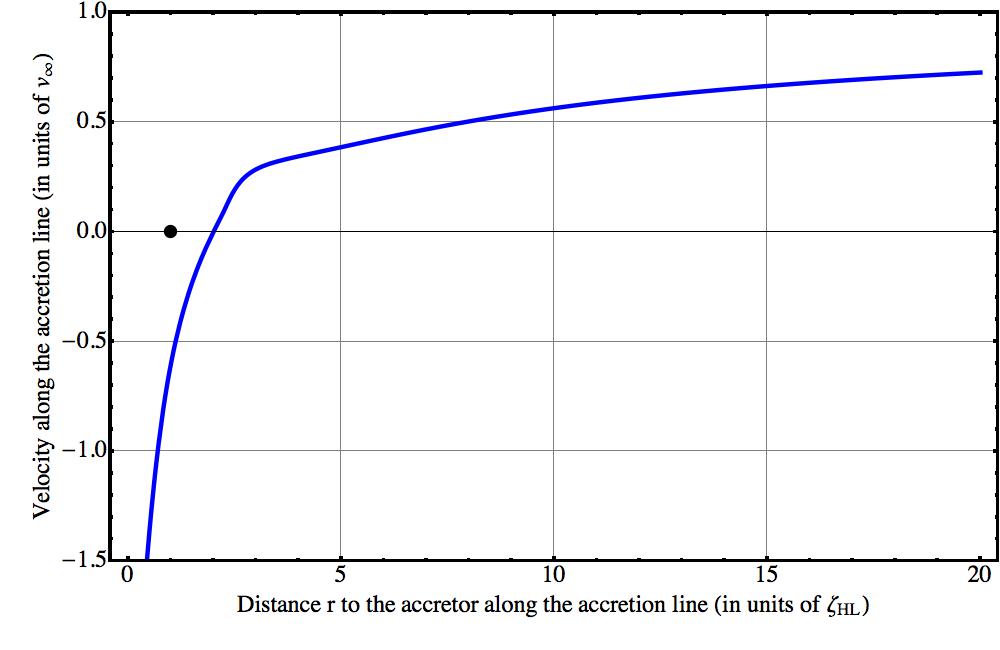}	
\caption{Illustrative velocity profile in the wake of the accretor, along the accretion tail in the accretion cone model. The curve was obtained by numerically integrating \eqref{eq:diff_eq_v_acc_col} with $r_{\textsc{x}}=2$ and the biased condition $v(0.2)=0.0768$ to bypass the stagnation point which is mathematically singular. The black dot indicates the stagnation point through which passes the real solution with $r_{\textsc{x}}=1$. Within a few accretion radii, the velocity retrieves its value at infinity at a few 10\% accuracy level. For a supersonic inflow with a Mach number larger than 2, since the temperature quickly recovers its value at infinity along the accretion line (see text), it means that the flow becomes supersonic again in the cone above a few accretion radii.}
\label{fig:vel_prof_acc_col}
\end{center}
\end{figure}

Finally, the temperature along the accretion line can be estimated with the square of the sound speed which relates the pressure $P$ to the mass density $\rho$ according to \eqref{eq:sound_speed_def}. The latter is simply obtained from the mass per unit length along the tail, $\lambda$, by considering the transverse surface : $\rho=\lambda/\left( \pi s^2 \right)$. Since the ballistic inflow is essentially pressureless compared to the shocked flow, we can estimate that the pressure in the wake origins from the dissipation of the linear momentum transverse to the shock. Balancing the fluxes, we have :
\begin{equation}
P 2\pi s \d r = - \d \Phi_{\text{mv$_{\theta}$,lat}}  = \rho v_{\theta}(\theta\rightarrow\pi ^-) \cdot v_{\text{tot}} 2\pi s \d r
\end{equation}
where $v_{\theta}(\theta\rightarrow\pi ^-)=\sqrt{1/r}$ according to \eqref{eq:r_to_zeta} once injected in \eqref{eq:vt}. The previous remarks yield an estimate of the evolution of the sound speed along the wake :
\begin{equation}
\label{eq:temperature_acc_col}
c_s^2(r)=\gamma \frac{s}{\sqrt{r}}
\end{equation}
In practice, we will see that the accretion tail is not conic and that $s$ evolves slower than $r$ beyond $0.5\zeta_{\textsc{hl}}$ : the tail is concave. As a consequence, $c_s$ does not change much from its value at infinity beyond $0.5\zeta_{\textsc{hl}}$ (section \ref{sec:transv_prof}).

\section{Numerical implementation}
\label{sec:num_impl}

\subsection{Finite size of the accretor}
\label{sec:fin_acc}

\begin{figure}[!h]
\begin{center}
\def\svgwidth{250pt} 
\includegraphics[height=6.5cm, width=10cm]{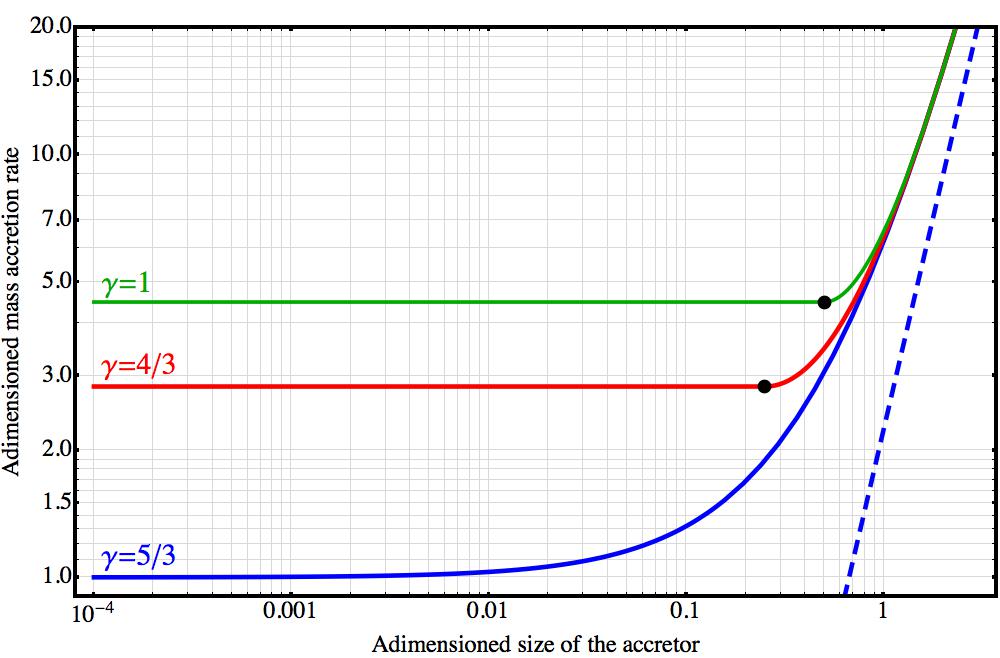}	
\caption{Mass accretion rate as a function of the size of the accretor for 3 different adiabatic indexes. The black dots locate the sonic point, which is expelled to zero for $\gamma=5/3$. The dotted blue line indicates the geometric limit at large radii for $\gamma=5/3$ and depends little on $\gamma$.}
\label{fig:fin_size}
\end{center}
\end{figure}

The mass accretion rate \eqref{eq:mdot_Bondi_sph} we derived in the previous chapter was determined using the existence of a critical point through which the gas could flow, the sonic point. Passing through it was a necessary condition to get a unique transonic solution. However, if the radius $R$ of the accretor is larger than the sonic radius $r_{\text{s}}$, then the mass accretion rate is altered. The influence of the finite spatial extension of the accretor, considered until now as a point-mass, is discussed in detail in \cite{Ruffert1994a} for the Bondi spherical accretion with a zero velocity at infinity. The units of length and mass accretion rates which arise are respectively :
\begin{equation}
\begin{cases}
\left(GM\right)/c_{\text{s,}\infty}^2\\
\pi \frac{\left(GM\right)^2}{c_{\text{s,}\infty}^3}\rho_{\infty}
\end{cases}
\end{equation}  
and when $\mathcal{M}_{\infty}\rightarrow 0$, the dimensionless mass accretion rate $\dot{M}_B$ becomes, if we still write $R$ the dimensionless radius of the accretor \citep{Shapiro1983a} :
\begin{equation}
\dot{M}_B=4\left[ \frac{2}{\gamma +1} R^{\frac{4\left(\gamma-1\right)}{\gamma +1}} + \frac{2\left(\gamma -1\right)}{\gamma +1} R^{\frac{3\gamma -5}{\gamma +1}}\right]^{\frac{\gamma +1}{2\left(\gamma -1\right)}}
\end{equation}
For large accretors, the mass accretion rate behaves as a geometric one\footnote{\ie the one obtained by computing the amount of mass intercepted by a moving disk of radius $R$ per unit time.}, apart from a factor of order unity. We retrieve the formula \eqref{eq:mdot_Bondi_sph} for accretors which lie within the sonic sphere (\ie on the left of the black dots on Figure\,\ref{fig:fin_size}), which never happens for $\gamma=5/3$ because in this case, the sonic radius is null. As a consequence, whatever small the inner boundary is, it alters the numerically measured mass accretion rates if $\gamma=5/3$, such as simulations of a monoatomic gas being accreted require additional precautions concerning the treatment of the inner boundary. On Figure\,\ref{fig:fin_size}, it can be seen that all other things being equal, a gas undergoing accretion in an isothermal way, \ie cooling fast enough to maintain a constant temperature in spite of the work done by pressure forces, yields a larger mass accretion rate than an adiabatic flow, which does not exchange heat with the exterior and sees its temperature rising as it collapses because of the work done by pressure forces.

In the case of a \bhl flow, it will be shown in section \ref{sec:sonic_surf} that this dependence of accretion on the size of the inner boundary of the simulation space still holds. Indeed, the sonic surface of a \bhl flow shares with its spherical sibling tied links which will make the cancelling of the sonic radius for $\gamma=5/3$ decisive.

\subsection{Dynamics}
\label{sec:dyn}

\subsubsection{Spatial dynamics}

Let us start by presenting the most numerically demanding prerequisite of this physical model : the scale dynamics. By scale dynamics, we mean the amplitude of the discrepancy between the smallest and the largest scales in the system. The shape of the shock formed by the incoming wind has not been foreseen analytically. Actually, its very stability, even in the axisymmetric configuration, remains an intensively debated open question \citep[see][Table 1]{Foglizzo2005}. As a consequence, we can not force the shock as a condition exterior to the simulation space and focus on the inner parts only. We can however predict the characteristic lateral size of the shock which corresponds to the scale at which the flow is significantly deflected by the point-mass. Given the arguments described in \ref{sec:mdot_HL}, we can affirm that this scale is given by the accretion radius $\zeta_{\textsc{hl}}$. 

An important conclusion we can draw comes from the analysis of the Mach profile which has been implicitly sketched in \ref{sec:acc_col}. We could see on Figure\,\ref{fig:vel_prof_acc_col} that the velocity at infinity was recovered, in the accretion cone model, beyond a few accretion radii along the accretion line. Since the sound speed also quickly reaches its value at infinity, we can safely affirm that at $8\zeta_{\textsc{hl}}$ in the wake, the flow will have become fully supersonic again \citep[see also][Figure 3]{Blondin2009} : it sets the size of the outer boundary of the simulation space.

\begin{figure}[!h]
\begin{center}
\includegraphics[height=6cm, width=10cm]{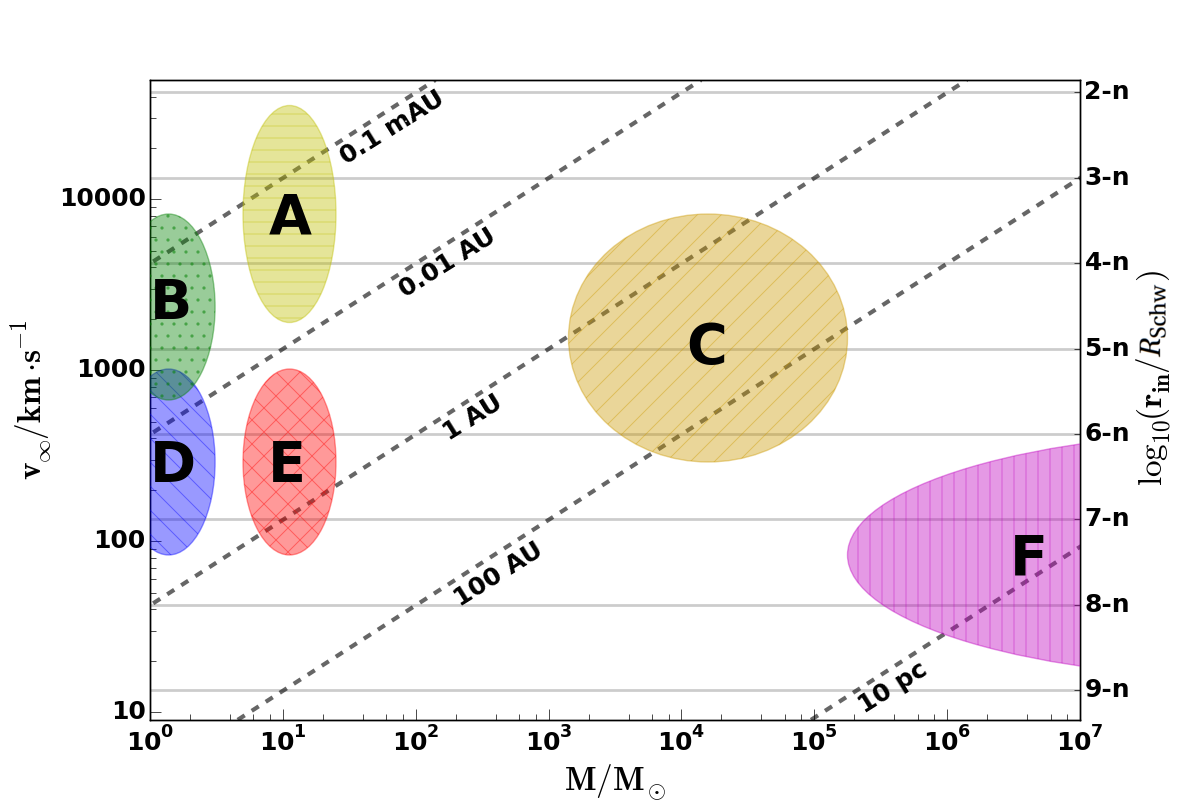}	
\caption{Contour map of the accretion radius $\zeta_{\textsc{hl}}$ of the \bhl flow as a function of the velocity at infinity and of the mass of the accreting body, confronted to an estimate of the computational cost on the right. The latter is represented by the ratio of the inner boundary radius $r_{\text{in}}$ by the Schwarzschild radius of the compact object $R_{\text{Schw}}$, with $n$ being given by $\zeta_{\textsc{hl}} / r_{\text{in}} = 10^n$, such as for n$=$0, the right axis indicates the physical ratio $\zeta_{\textsc{hl}} / R_{\text{Schw}}$. See \cite{ElMellah2015} for a description of the astrophysical objects associated to the different zones.}
\label{fig:dimensions_odm}
\end{center}
\end{figure}

The second scale to grasp is either the sonic surface (if $\gamma<5/3$) or the accretor itself (for $\gamma=5/3$, since the sonic radius cancels out). We set ourselves in the second case since we work with monoatomic gases. If the size of the accretor can be of the order of the accretion radius for an accreting star, it is much smaller when we deal with a compact object, for realistic wind relative speed at infinity, $v_{\infty}$. The discrepancy between the 2 scales depends only on $v_{\infty}$ and the compactness parameter $\Xi$ of the accretor :
\begin{equation}
\frac{\zeta_{\textsc{hl}}}{R}=2\Xi \left(\frac{c}{v_{\infty}}\right)^2
\end{equation}
where $R$ is the size of the accretor and $\Xi$ is of the order of a few 10\% for a neutron star (and is 1 for a black hole with the Schwarzschild radius $R_{\text{Schw}}$ as the characteristic size). At a fixed compactness parameter value, this discrepancy does not depend on the mass of the accretor since the accretion and the Schwarzschild radii both scale with it in the same way. To quantify how computationally demanding a numerical simulation will be, we plotted $\log \left( \zeta_{\textsc{hl}}/R_{\text{Schw}} \right)$ as a function of $v_{\infty}$ respectively on the right (with $n=0$) and left side of Figure\,\ref{fig:dimensions_odm}. We see that, as the relative speed at infinity rises, the discrepancy between scales lowers but for the wind relative velocities expected in X-ray Binaries (zones A and B on the Figure), $\zeta_{\textsc{hl}}\sim 10^5R_{\text{Schw}}$. For a neutron star with $\Xi\sim 15\%$, it implies an outer boundary for the simulation space 240,000 times larger than the inner boundary. As a guideline, the few families of compact objects in astrophysical systems where wind accretion could occur are located :
\begin{itemize}
\item \textbf{A :} black hole high mass X-ray binaries, in particular {\sc lmc x-1} where an Onfp companion star seems to provide a particularly fast wind \citep{Orosz2008}. It could also be the case of lone runners black holes which have been accelerated in a close encounter with other black holes \citep{Sperhake2011,Lora-Clavijo2013}.
\item \textbf{B :} runaway neutron stars and \sgx. The first family might be illustrated by the radio pulsar PSR B2224+65 and its iconic guitar nebula \citep{Cordes1993}, albeit the pulsar wind might play an additional role in the formation of the shock compared to the simple Bondi-Hoyle sketch. A runaway neutron star more suitable to apply the Bondi-Hoyle model might be PSR J0357+3205, where the tail is too long to be accounted for by shocked pulsar wind models \citep{DeLuca2013}. For \sgx, it is believed that we detect the wind of an early spectral type super giant OB star accreted by a slowly-spinning neutron star orbiting on a close-in orbit \citep{Chaty2011a}. Examples of such systems are Vela X-1, 4U 1907+09 \& GX 301-2.   
\item \textbf{C :} intermediate mass black holes, possibly identified as Hyper Luminous X-ray sources ({\sc hlx}) whom the most honourable known member is probably ESO 243-49 HLX-1 \citep{Farrell2009}. It has been suggested, among others, that a wind accretion within a binary system might be responsible for its X-ray luminosity \citep{Miller:2014ti}. An {\sc imbh} could also be present in the Orion Nebula Cluster and undergo wind accretion from a massive stellar companion \citep{Subr2012}.
\item \textbf{D \& E :} neutron star in low mass X-ray binaries, cataclysmic variables and black holes in high mass X-ray binaries where the stellar companion wind is too faint to feed accretion, but might play an indirect role through an interplay with the Roche lobe overflowed formed accretion disc. 
\item \textbf{F :} super massive black holes accreting ambient gas of which the velocity at infinity (deduced from the kinetic energy left if it had fully escaped the super massive black hole gravitational potential) is estimated, here, from the sum of the proper motion of SgrA* and of the characteristic stellar dispersion speed \citep{Reid2003} ; it could be higher if one considers the gas launched from surrounding massive stars \citep[see][for simulations of accretion onto SgrA* along those lines]{Ruffert1994b}. Except for collisions of galaxies, the bulk motion of a super massive black hole compared to the ambient gas is likely to be negligible. 
\end{itemize}

\subsubsection{Time dynamics}

Translated into constrains on the timescale, things get even worse. Indeed, the timestep $\Delta t$ is set by the minimum value between 

\begin{figure}[!h]
\begin{center}
\includegraphics[height=6cm, width=10cm]{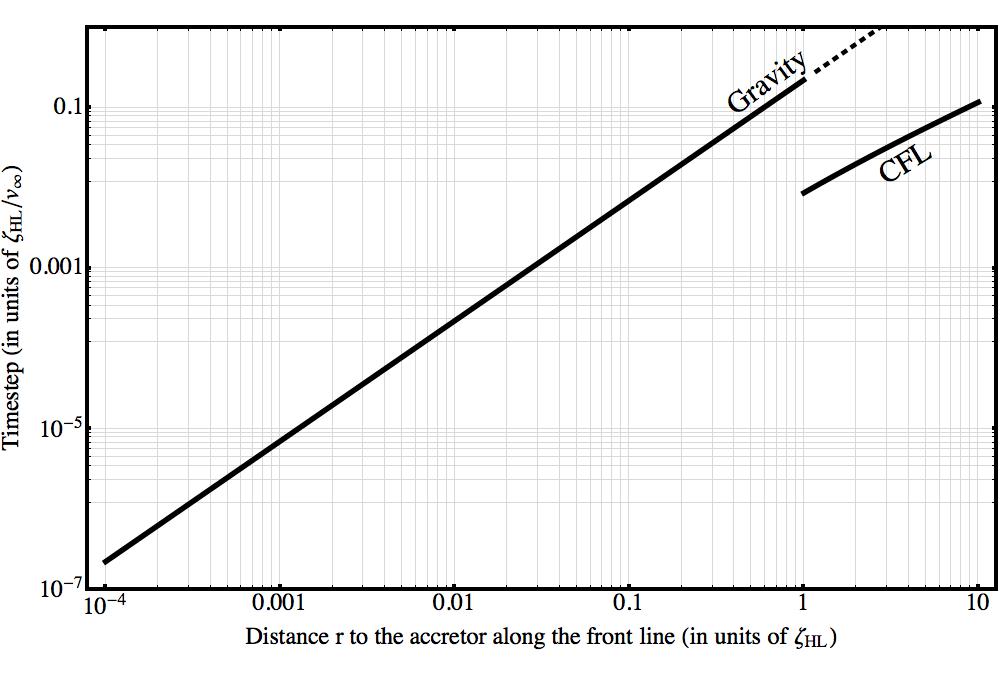}	
\caption{Timestep as a function of the radius along the front line upstream the accretor. From the outer boundary down to the shock ($r\sim 1$), the timestep is set by the \cfl condition. Once the velocity drops downstream the shock, the gravitational timestep takes on and is the constraining one at low radii.}
\label{fig:timesteps}
\end{center}
\end{figure}

\begin{enumerate}
\item the timestep $\Delta t_{\textsc{cfl}}$ set by the \cfl condition (see section \ref{sec:diff_schemes}) :
\begin{equation}
\Delta t_{\textsc{cfl}}=K\text{min}\left(\frac{\Delta r}{V}\right)
\end{equation}
with $K$ the Courant parameter\footnote{Usually set to 0.5 in our simulations.} and $V$ absolute velocity of the flow plus the sound speed\footnote{In the \hd case where the only characteristic wave is the acoustic one.}. Since the cells are approximately squared and the velocities in the radial and orthoradial directions are similar, it is representative in this illustrative argumentation to consider $\Delta r$ only.
\item and the timestep $\Delta t_g$ set by the gravitational acceleration :
\begin{equation}
\Delta t_g = \sqrt{\frac{\text{min}\left(\Delta r\right)}{1/\left[2\text{min}\left(r^2\right)\right]}}
\end{equation}
with, as used in this second part of the manuscript, $\zeta_{\textsc{hl}}$, $v_{\infty}$, $v_{\infty}^2/\zeta_{\textsc{hl}}$ the scales of length, velocity and acceleration. Due to the structure of the mesh we designed (see \ref{sec:mesh}), the cell center radial position is, for each cell and with cell aspect ratios of $\sim$1, of the order of the ratio of the radial by the angular step, $\Delta r/\Delta \theta$. For a resolution good enough, $\text{min}\left(r\right)\sim r_{\text{in}}$
\end{enumerate}
Eventually, the timestep of the whole simulation space is given by :
\begin{equation}
\Delta t=\text{min}\left[ \frac{1}{2}\text{min}\left(\frac{\Delta r}{V}\right) ; \sqrt{2r_{\text{in}}^3\Delta \theta}\right]
\end{equation}
Along the accretion line, upstream the shock, we can use the equations of \ref{sec:motion} with $\zeta=0$ to get the velocity for $r>1$ (\ie before the shock) such as the \cfl timestep at a distance $r$ upstream of the accretor is given by :
\begin{equation}
\Delta t_{\textsc{cfl}}(r)=\frac{\Delta \theta}{2} r\sqrt{\frac{r}{1+r}}
\end{equation}
But this relation is true down to $r=1$. Then, the velocity drops because of the shock and, for the sake of simplicity, we will not consider the reacceleration as large enough to bring $\Delta t_{\textsc{cfl}}$ below $\Delta t_g$. Indeed, given its expression :
\begin{equation*}
\Delta t_g\left( r \right)=\sqrt{2\Delta \theta}r^{3/2},
\end{equation*}
the gravitational timestep sets the simulation timestep $\Delta t$ at low radii. Figure\,\ref{fig:timesteps} summarizes the evolution of this representative timestep along the front line with the radius. At large scales, the bulk motion of the supersonic flow dominates over the influence of the gravitational force while below a certain threshold (fiducially set to the shock here), it is the gravitational force which sets the timestep. As a consequence, we can now estimate the discrepancy between the small and large scales timesteps :
\begin{equation}
\frac{\Delta t_{\textsc{max}}}{\Delta t_{\text{min}}}\sim\frac{\Delta t_g\left(r=1\right)}{\Delta t_g\left(r=r_{\text{in}}\right)}\sim\frac{1}{r_{\text{in}}^{3/2}}
\end{equation}
with $r_{\text{in}}$ the size of the inner boundary in untis of accretion radii. With $r_{\text{in}}=10^{-4}$, the size we have to reach for X-ray binaries according to Figure\,\ref{fig:dimensions_odm}, it means that the motion at large scale requires a timestep 6 orders of magnitude larger than the timestep needed to follow the flow in the vicinity of the inner boundary. To observe the numerically relaxed state which takes place within a few crossing times, it means that we have to plan a few millions of numerical iterations.

\subsubsection{State of the art}

Until now, two approaches have been used to bypass this computational limitation. On one hand, a physically-motivated one which consists in focusing on putative compact objects rushing at almost relativistic speeds through the interstellar medium. If asymmetric Supernova explosions are not expected to account for velocities above 1000\kms \citep{Tauris1998}, some models of encounters between two spinning black holes allow kick velocities to reach 15,000\kms \citep{Sperhake2011}. In this case, the ratio between the accretion radius and the size of the compact object becomes as small as 400, making the simulation way less computational-demanding. Physically, it can be interpreted by saying that the bulk motion of the flow is related to a Mach number so high\footnote{In the comoving frame of the compact object, the flow is rushing towards the accretor but its temperature being invariant by the Galilean change of frame (valid for non relativistic relative velocities and temperatures), the Mach number is extremely high.} that the pressure forces can not come into play until the last minute, when the flow has already arrived very close from the object. The upper front of the shock is then close enough from the object to encompass the two within the same simulation space at an affordable cost. Fully relativistic \hd simulations of accretion onto a superkicked black hole, rotating or not, have been run by \cite{Lora-Clavijo2013} and \cite{Lora-Clavijo2015} who studied the stability of the flow and characterized its structure.

\begin{figure}[!h]
\begin{center}
\includegraphics[height=6cm, width=10cm]{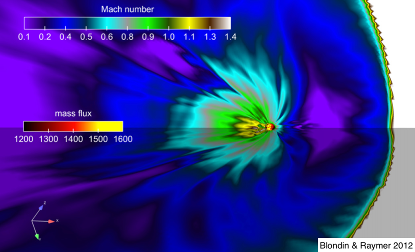}	
\caption{Full three dimensional simulation of a planar flow onto an accretor whose physical size is a few thousands times smaller than the size of the inner boundary represented here as a solid sphere. The yellow to red scale on it represents the local mass accretion rate while the main color scale is the Mach number of the flow. From \cite{Blondin:2012vf}.}
\label{fig:blondin}
\end{center}
\end{figure}

On the other hand, truncating the inner boundary by making it deliberately larger than the physical size of the compact object has been another option, in spite of the potential caveats reminded in \ref{sec:fin_acc}. This option implicitly assumes that the flow which enters the volume around the accretor associated to the oversize inner boundary is trapped and that the enforced boundary condition at the inner boundary is representative enough of the physical answer to prevent the large scale structure from being impacted by it. Simulations of planar \bhl accretion along those lines have been carried on by \cite{Blondin2009}, \cite{Blondin:2012vf} and \cite{Blondin2013a} in two and three-dimensional simulations (see Figure\,\ref{fig:blondin}), with inner boundary sizes down to almost $10^{-3}\zeta_{\textsc{hl}}$ (or $10^{-2}\zeta_{\textsc{hl}}$ in 3D), which corresponds to several hundreds times the actual size of the compact object for wind speeds measured in \sgx. This discrepancy can alter the stability of the flow and bias the mass accretion rates numerically observed. 

\subsection{Mesh}
\label{sec:mesh}

\subsubsection{The geometry}

The planar \bhl flow we address in this second part of the manuscript is a naturally centered problem. The gravitational field produced by the point-mass is isotropic, although the flow at infinity is planar. Preliminary investigations of the flip-flop instability in 2014 (see Figure\,\ref{fig:flip-flop}) led us to give the priority to a numerical scheme which conserves angular momentum up to machine precision. The best way to do so was to use on one hand the axisymmetry to work in a two-dimensional simulation space, and on the other hand a spherical mesh to reproduce the full three-dimensional dynamics of the flow\footnote{In particular the dilution with the distance to the accretor, which plays a role in the explicit form of the divergence operator in spherical coordinates (see \ref{sec:equations_sph}).}. This setup is referred to as "2.5D spherical", where the fractional dimension is a way to say that it reproduces the full three-dimensional behaviour where an axisymmetry would have been enforced ; a two-dimensional slice is then sufficient to follow the motion. The spherical mesh, even once radially stretched as presented below, is associated to orthogonal coordinates, a prerequisite to some of the numerical schemes we rely on, in particular concerning the computation of the fluxes at the interfaces (see section \ref{sec:algo_recip}). 

\subsubsection{The grid}

\begin{figure}[!b]
\begin{center}
\includegraphics[height=1.5cm, width=13cm]{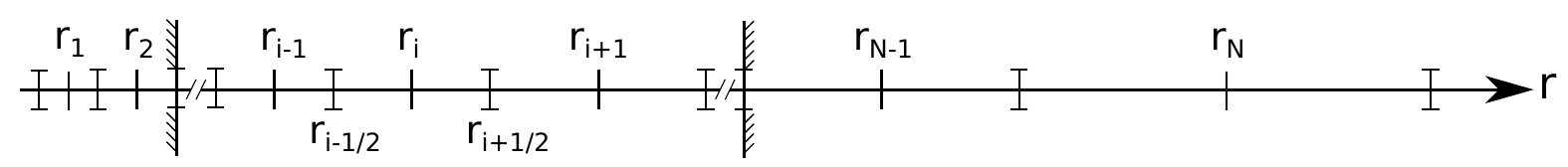}	
\caption{Illustration of the logarithmic grid with the cell centers indexed with integers. The cells indexed 1, 2, N-1 and N are the ghost cells outside of the simulation grid, where the inner and outer boundary conditions are set.}
\label{fig:rule}
\end{center}
\end{figure}
\begin{wrapfigure}{r}{0.25\textwidth}
\begin{center}
\includegraphics[height=8cm, width=0.95\textwidth]{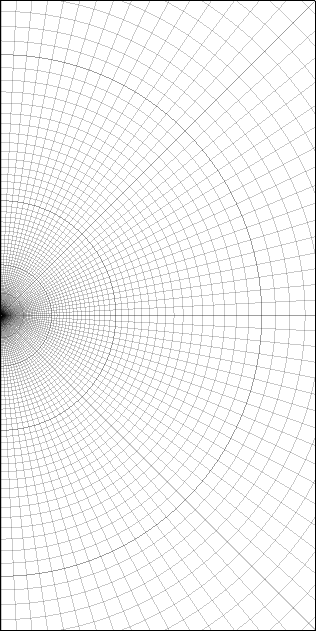}	
\caption{Two dimensional polar self-similar grid with a constant cell aspect ratio at all radii.}
\label{fig:grid}
\end{center}
\end{wrapfigure}
How computationally demanding the time dynamics of the problem is has been estimated in \ref{sec:dyn}. The stake is now to make the number of cells as small as possible while still being able to resolve the characteristic features of the flow. A first naive approach could be to set the radial step such as to be able to properly resolve the inner boundary ; indeed, we need cells at the inner border whose radial extension is small enough with respect to the size of this inner boundary. Due to the presence of two additional ghost cells on the outside border of the simulation space (at the inner edge), we need them at least to fit within the "hole" drawn by the inner boundary. Furthermore, to avoid spurious numerical oscillations due to excessively strong gradients and/or boundary conditions applied in environments too different in terms of relative conditions, we need cells at least an order of magnitude shorter than the radial extension of the inner boundary. Due to the typically 5 orders of magnitude discrepancy between the inner and outer boundary, it means a prohibitively large number of radial cells\footnote{Of the order of a million.}. Another approach would be to make use of the \textsc{amr} option of \vac (see Figure\,\ref{fig:flip-flop}). However, on top of the discontinuous steps of resolution it introduces, the issue of the aspect ratio\footnote{Ratio of the sizes of the cell along the radial and the angular directions.} would not be solved. Indeed, the aspect ratio along a radius changes in the same proportions as the distance to the center, leading to strongly deformed cells, which could turn to be harmful for the accuracy of the numerical schemes\footnote{A numerical scheme such as \textsc{tvd-lf} is better used with similar interface sizes.}.

The best way to homogeneously and continuously resolved all scales is to keep a constant aspect ratio $\xi$ for all the cells within the simulation space :
\begin{equation}
\frac{\Delta r}{r\Delta\theta}=\xi\quad\text{constant}
\end{equation}
with $\theta$ the latitudinal coordinate in the spherical frame\footnote{We also extended it to 3D, in cylindrical and in spherical coordinates. In 3D spherical, $\phi$ is used instead of $\theta$ to compute the aspect ratio $\xi$. Thanks to the help of \href{https://www.researchgate.net/profile/Chun_Xia2}{Chun Xia} at KU Leuven, it was also made compatible with \textsc{amr}.}. Thus, we have a different radial step at each radius given by :
\begin{equation}
\Delta r (r) = \xi \Delta \theta \times r 
\end{equation}
We call such a grid "logarithmic" due to the constant ratio $\Delta r / r$. To minimize the impact on the code, we set the radial center of each cell mid-way between the two edges (see Figure\,\ref{fig:rule}). The stretching is also performed outside of the simulation box, in the ghost cells (those indexed 1, 2, $N-1$ and $N$ on Figure\,\ref{fig:rule}). The position of the cell centers and edges can then be analytically derived. Once the angular resolution and the number of radial cells are fixed, the ratio of sizes between the outer and the inner boundary can be set manually provided we tolerate slight departures\footnote{Below a few percent.} of $\xi$ from its optimal value of 1. Beyond the homogeneity that this implementation brings to the grid by having a constant aspect ratio for any radius, it also saves a lot of computation by enabling the user to have a similar relative resolution $\Delta r /r$ all over the grid without having to oversample the large scale, far away from the center ; the number of cells remains computationally affordable and the only challenge left is the number of timesteps it takes to reach a numerically relaxed flow. 

In the nominal regime, we studied the \bhl flow with 64 latitudinal cells and 128, 176 or 224 radial ones, depending on the size of the inner boundary relatively to the outer one we fix. High angular resolution simulations with an aspect ratio twice as low where also performed as sanity checks to confirm the vanishing influence of the mesh on the observed results.


\subsection{Equations}
\label{sec:equations_sph}

With a null initial East-West longitudinal velocity $v_{\phi}$ and the axisymmetry condition\footnote{Mathematically, the enforcement of the axisymmetry corresponds to the invariance of all quantities with a change of $\phi$, \ie $\partial _{\phi}=0$.}, writing the time evolution of $v_{\phi}$ proves that it must stay to zero during all the simulations, which it numerically does : the entering flow is deprived of any net angular momentum and stays so by numerical construction. Henceforth, we will no longer consider nor this direction nor this component of the velocity field in the rest of this Chapter.

We unfold in spherical coordinates the Euler equations introduced in \ref{sec:euler_eq} to get, with the axisymmetric condition :

\begin{equation}
\label{eq:cons_mass_bhl25D}
\partial _t \rho = - \frac{1}{r^2}\partial_r \left( r^2 \rho v_r \right) - \frac{1}{r \sin\theta}\partial_{\theta}\left( \sin\theta\rho v_{\theta} \right)
\end{equation}
\begin{equation}
\label{eq:cons_mom_bhl25D_1}
\begin{split}
\partial _t \left( \rho v_r \right) = - \frac{1}{r^2}\partial_r \left( r^2 \left( \rho v_r^2 + P \right) \right) -&\frac{1}{r \sin\theta}\partial_{\theta}\left( \sin\theta\rho v_r v_{\theta} \right)\\
& + \frac{\rho v_{\theta}^2}{r} - \rho d_r \Phi
\end{split}
\end{equation}
\begin{equation}
\label{eq:cons_mom_bhl25D_2}
\begin{split}
\partial _t \left( \rho v_{\theta}\right) = - \frac{1}{r^2}\partial_r \left( r^2 \rho v_{\theta} v_r \right) - \frac{1}{r \sin\theta} \partial_{\theta}&\left(\sin\theta \left( \rho v_{\theta}^2 + P \right) \right) \\
& - \frac{ \rho v_{\theta} v_r}{r}
\end{split}
\end{equation}
\begin{equation}
\label{eq:cons_mom_bhl25D_3}
\begin{split}
\partial _t e = - \frac{1}{r^2}\partial_r \left( r^2 \left( e + P \right) v_r \right)  - \frac{1}{r \sin\theta}\partial_{\theta}&\left(\sin\theta \left( e + P \right) v_{\theta} \right) \\
& - \rho v_r d_r \Phi
\end{split}
\end{equation}
where the notations are the same as those introduced in \ref{sec:motion}. The pressure term is computed with the flux at the interface. On the other hand, the source term corresponding to the Newtonian gravitational field of a point-mass is computed at the cell center\footnote{To be understood as the point where medians across each pair of interfaces cross rather than a weighted barycenter. The two do not always correspond \citep{Mignone2014}.}. The geometrical terms \ie those in the divergence which do not write as a derivative are also computed at the cell center and added afterwards. All those operations are performed following the shock-capturing twosteps \textsc{tvd-lf} method with a Koren slope limiter\footnote{While the \textit{superbee} slope limiter would have been too unstable albeit even less diffusive ; the more diffusive \textit{minmod} slope limiter has been used during the 5 first percent of the simulation (in physical time) to perform the initial strong relaxation.} to guarantee robustness and accuracy (see section \ref{sec:algo_recip}).

Many numerical simulations until now have reported on results based on the polytropic assumption (or an isothermal one) to relate the pressure to the density ; if this workaround in analytical studies is an almost compulsory renouncement to go forwards in the computation, it is not as mandatory when it comes to numerical simulations. Furthermore, a major flaw in numerical simulations is to identify the polytropic index $\Gamma=\d \ln P / \d \ln \rho$ to the adiabatic index $\gamma$, determined by the microstructure of the gas. For an ideal gas, the two are the same if and only if the evolution is isentropic\footnote{As clear from its expression reminded in the footnote \ref{fn:entropy} of Chapter \ref{chap:num_tools}.} \citep[as pointed out by][]{Horedt2000}, which is way more constraining than the adiabatic assumption : recall that the entropy not being a conserved quantity, it can be exchanged through heating and cooling, hereby quenched by the adiabatic assumption, but it can also be produced when the gas undergoes a mechanically irreversible transformation (\eg a shock - see appendix \ref{sec:planar_shock}). As we will see in section \ref{sec:bhl_shock}, solving the energy equation albeit with an adiabatic prescription enables us to naturally fulfil the shock jump condition and the associated production of entropy. Accounting for the departure from adiabaticity would require either a full radiative transfer equation solver or strong assumptions about the predominant cooling/heating terms. We chose a more neutral approach which is likely to require refinement as soon as the internal energy density variations due to heat exchange are no longer negligible compared to the ones due to the work done by pressure forces. From now on, since we work with a monoatomic gas, it restrains the comparison of our results to prediction relying on $\gamma=5/3$. Besides, our $\gamma=5/3$ simulations may still depart from theoretical descriptions since we homogeneously describe the flow via an energy equation while writing $P\propto\rho ^{\gamma}$ precludes any conclusive conclusion on the effects of the shock to be drawn.

%


\subsection{Boundary conditions}
\label{sec:bc}

The partial differential equations in the previous section are associated with usual polar symmetric and antisymmetric conditions in $\theta=0$ and $\pi$, using the frame specified on Figure\,\ref{fig:context}. For the outer radial boundary conditions, far ahead the shock, where the flow is supersonic, we prescribe the ballistic solution for $v_r$ and $v_{\theta}$ and the permanent regime solutions deduced for $\rho$ and the Bernoulli quantity $B$ from the mass and energy conservation equations (equations \eqref{eq:vr}, \eqref{eq:vt}, \eqref{eq:dens_BK} and \eqref{eq:bern_BK}). It assures that the gravitational ballistic deflection of the initially planar flow, from infinity to the outer boundary, is taken into account. For the downstream outer boundary condition (\ie where the flow gets out of the simulation space), as specified in section \ref{sec:dyn}, we work with a supersonic flow such as it avoids any spurious reflection of pressure waves. Continuously outflowing conditions then satisfy the requirements.

Concerning the inner boundary conditions, much caution must be taken. Straightforward absorbing conditions (\eg floor density and continuous velocities, provided they leave the simulation space) do alter the stability of the flow without any guarantee of fitting the continuity of the radial fluxes. One has to not prevent the stationary solution of \eqref{eq:cons_mass_bhl25D} to \eqref{eq:cons_mom_bhl25D_3} to be achieved, for example by ensuring the continuity of $\rho v_r r^2$ at the inner boundary : so as to do so, we computed the density in the inner ghost cells with a first order Taylor-Young expansion and deduced the corresponding radial velocities from the continuity of the radial mass flux $\rho v_r r^2$. For the total specific energy $e$, one has to write the gravitational source term as a pure derivative using the steady-state mass conservation equation : 
\begin{equation}
- \rho \mathbf{v} \cdot \boldsymbol{\nabla} \Phi = - \boldsymbol{\nabla} \left( \rho \Phi \mathbf{v} \right)
\end{equation}
to explicit the flux : $(e+P+\rho\Phi)v_r r^2$. The same prescription as above is applied to this flux to derive the value of the pressure in the ghost cells. The value of $v_{\theta}$ is much less critical and is also set via a simple first order expansion. Such inner boundary conditions entitle the flow to reach a permanent regime as the ones described further. Note that those absorbing inner boundary conditions, on top of the gravitational field, make the situation very different from a planar flow deflected by a solid sphere, the archetype we usually rely on to figure out high Reynolds flows.


\subsection{Input parameters}

\subsubsection{Physical parameters}

The input parameters of those simulations can be separated into :

\begin{enumerate}
\item those which can be used as normalization quantities (\aka scale parameters). Setting a velocity at infinity and a mass of the accretor for example sets the velocity and the length scales, respectively $v_{\infty}$ and $\zeta_{\textsc{hl}}=2GM/v_{\infty}^2$.
\item the dimensionless ones which set the shape of the flow (\aka shape parameters), in a similar role as the one played by the Reynolds number for a viscous flow. Here, it is the Mach number at infinity $\mathcal{M}_{\infty}$ which is tuned by changing the temperature of the flow at infinity. It is the only real degree of freedom of those numerical simulations\footnote{Note that the adiabatic index is set to $5/3$, the value corresponding to a monoatomic gas.} since the outputs can be scaled afterwhile to fit any normalization.
\end{enumerate}

\subsubsection{Numerical parameters}

In numerical computation, the unprecision due to the noise\footnote{That is to say by physical phenomena considered as contaminating and polluting since not properly covered by the model nor repeatable enough to be easily monitored and corrected for.} is infinitely smaller than the systematics introduced by the additional layer of numerical modeling (\eg the choice of the solver, the size of the inner boundary or the spatial and temporal resolution). The best way to evaluate the latter is to design a priori sets of acceptable numerical configurations and empirically validate their neutrality by running the same physical configuration on each of them. The similarity of the results obtained with different numerical configurations corroborates a posteriori their physical reliability.

In this spirit,we ran those simulations with sizes of the inner boundary closer and closer from the physical size of the accretor (see sections \ref{sec:fin_acc} and \ref{sec:dyn}), from\footnote{Keeping in mind that each additional decade deeper in space costs an additional factor of $\sim$ 40 in \cpu computational time. Since it is a problem which deals with the number of time iterations and not with the grid resolution, it is not possible to shorten significantly the duration of the simulation by increasing the number of \textsc{cpu}s (see section \ref{sec:parall_comm}).} $r_{\text{in}}=10^{-2}\zeta_{\textsc{hl}}$ to $10^{-4}\zeta_{\textsc{hl}}$. Even with $r_{\text{in}}=10^{-4}\zeta_{\textsc{hl}}$, for a relative wind speed of 1,000\kms, the inner boundary is still ten times larger than the object. In the case of a neutron star, the magnetosphere will anyway comes into play, probably even before, while for a black hole, non Newtonian corrections must be made. In this close-in environment, a new numerical setup, either \textsc{mhd} \citep{Lee:2014uy} with heating/cooling terms to describe the radiation from the neutron star surface or General Relativity to properly describe the black hole vicinity\footnote{In this case, the use of the Eddington-Finkelstein coordinates alleviates the coordinate singularity at the event horizon which can then be fully included within the simulation space. Then, the numerical caveats at the inner boundary described in section \ref{sec:bc} spontaneously vanish.} \citep{Lora-Clavijo2015} must be concatenated to the present one. Finally, the code was also tested with better resolution and different numerical schemes without any significant impact on the results presented in the rest of this Chapter.

\begin{figure*}
\begin{subfigure}{0.48\textwidth}
\begin{center}
\includegraphics[width=7.5cm, height=6.4cm]{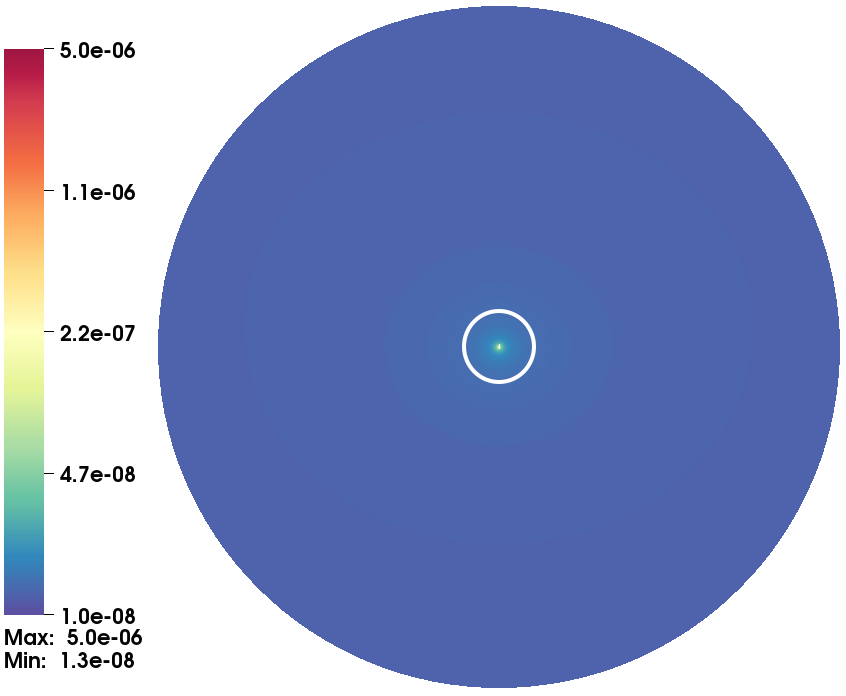} 
\label{fig:subim1}
\end{center}
\end{subfigure}
\begin{subfigure}{0.48\textwidth}
\begin{center}
\hspace*{0.2cm}
\includegraphics[width=7.5cm, height=6.4cm]{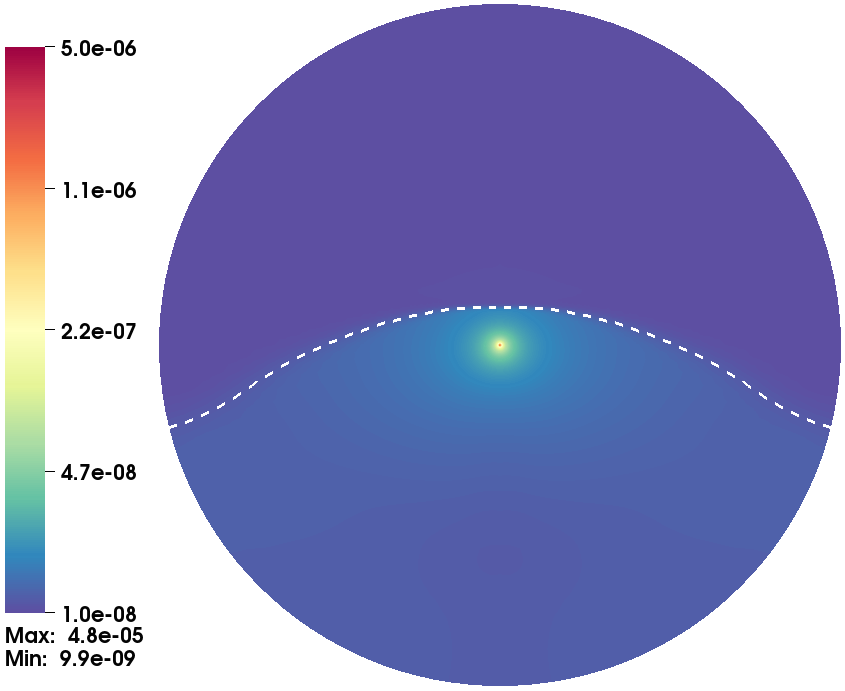}
\label{fig:subim2}
\end{center}
\end{subfigure}
\vspace*{0.8cm}\\
\begin{subfigure}{0.48\textwidth}
\begin{center}
\includegraphics[width=7.5cm, height=6.4cm]{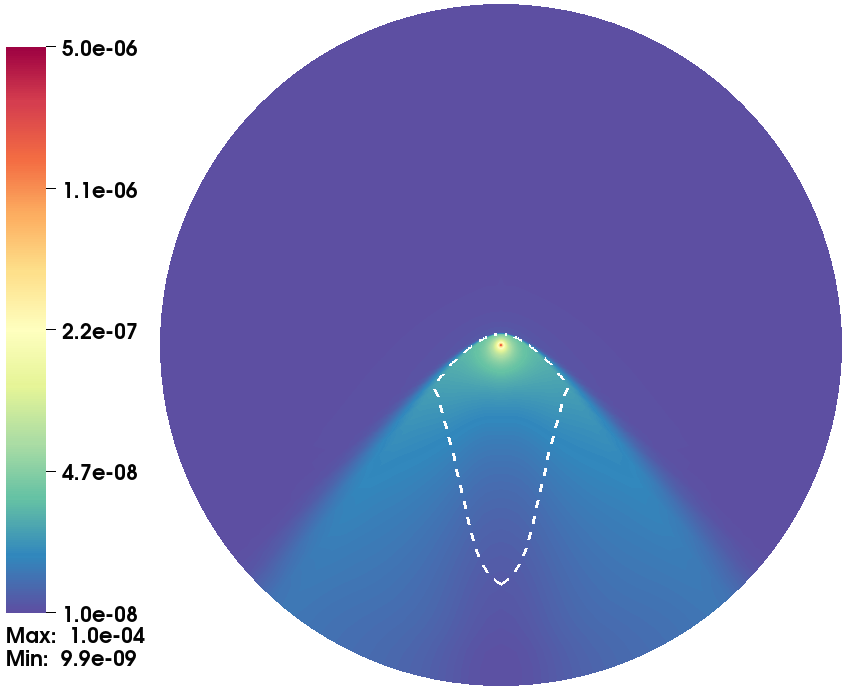}
\label{fig:subim2}
\end{center}
\end{subfigure}
\begin{subfigure}{0.48\textwidth}
\begin{center}
\hspace*{0.2cm}
\includegraphics[width=7.5cm, height=6.4cm]{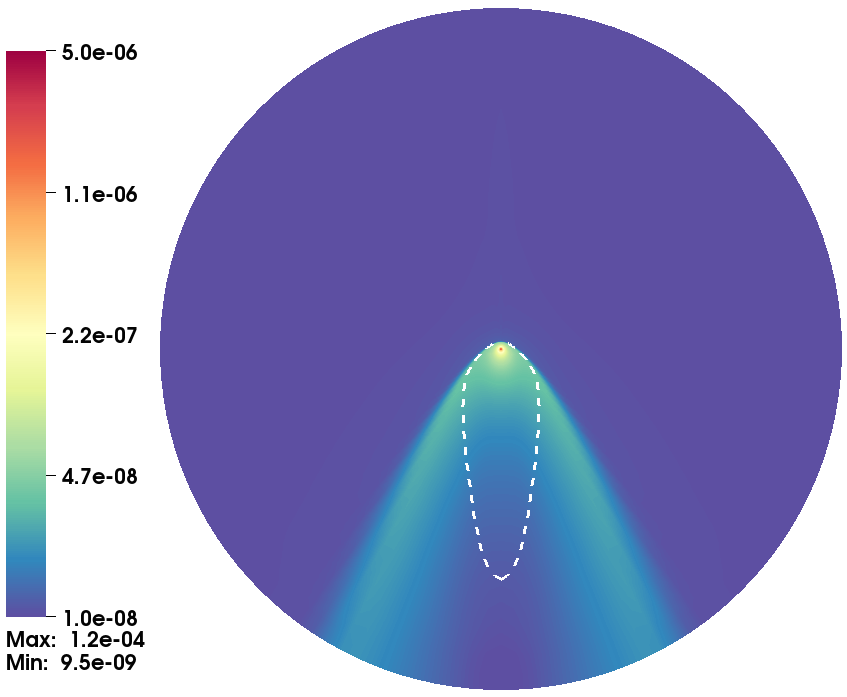} 
\label{fig:subim1}
\end{center}
\end{subfigure}
\vspace*{0.8cm}\\
\begin{subfigure}{0.48\textwidth}
\begin{center}
\includegraphics[width=7.5cm, height=6.4cm]{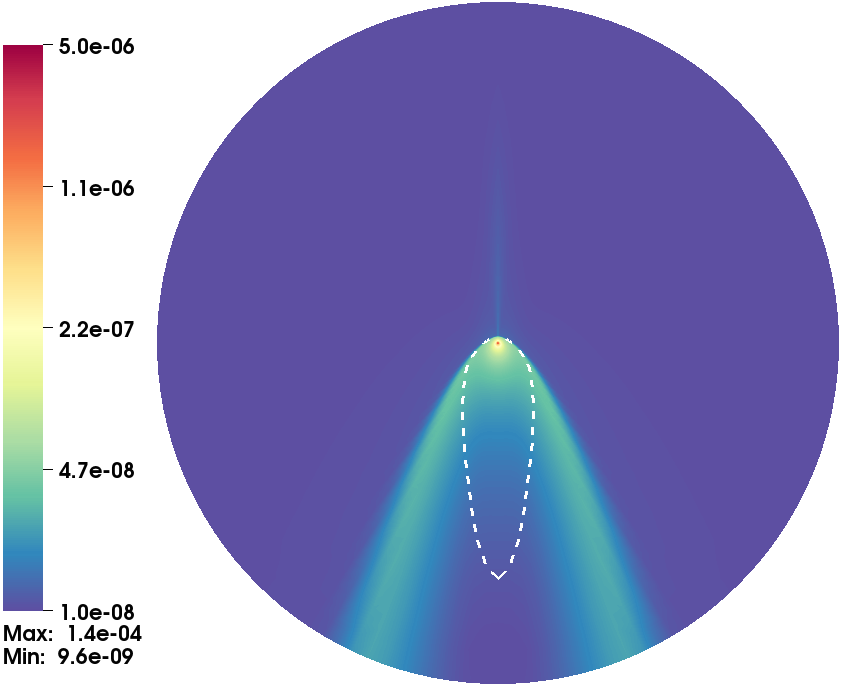}
\label{fig:subim2}
\end{center}
\end{subfigure}
\begin{subfigure}{0.48\textwidth}
\begin{center}
\hspace*{0.2cm}
\includegraphics[width=7.5cm, height=6.4cm]{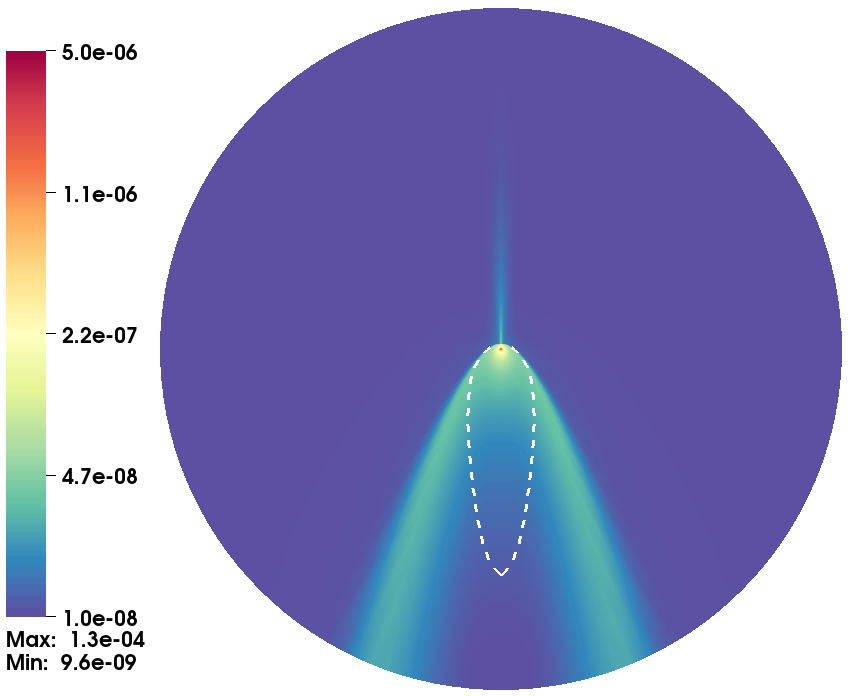}
\label{fig:subim2}
\end{center}
\end{subfigure}
\caption{Logarithmic color maps (arbitrary units) of the stationary density profiles for $r_{\text{in}}=10^{-3}\zeta _{\textsc{hl}}$ and, in reading order, $\mathcal{M_{\infty}}=0.5$, $1.1$, $2$, $4$, $8$ and $16$. The outer Mach 1 surface has been plotted (not the sonic one) as a white dashed line. On the first plot, subsonic, the solid white line stands for the approximate size of the critical impact parameter, $\zeta_{\textsc{hl}}$ ($\zeta_{\textsc{hl}}\sim 0.15 r_{\text{out}}$).}
\label{fig:density}
\end{figure*}


\section{The \bhl shock}
\label{sec:bhl_shock}

In this section, we describe the structure of the flow at the scale of the accretion radius\footnote{The radius of the white circle in the upper left panel on Figure\,\ref{fig:density}.}. For supersonic flows at infinity, a bow shock develops, detached from the inner boundary\footnote{On the contrary, in two-dimensional simulations on a polar grid, assuming a cylindrical vertical extension, the shock is attached (see Figure\,\ref{fig:flip-flop}).}. Its stability is first discussed. Follows a physical-consistency check-up where we rely on the analytical results on adiabatic shocks reminded in the appendix \ref{sec:jump} to validate the robustness of the simulation. Finally, we describe the transverse structure of the shock along the accretion wake and compare it to theoretical predictions. The consistency of those results, reinforced in the next section \ref{sec:acc_prop_BHL}, enables us to affirm that the observed flow in those numerical simulations does result from the gravitational beaming of the flow without spurious influence of the inner boundary.


%
%
%


\subsection{Opening angle}

As reminded in the appendix \ref{sec:jump}, as we go from the planar shock at the front to further downstream along the shock surface, the shock becomes more and more oblique. Figure\,\ref{fig:deflection_angle} emphasizes that, for a given Mach number at infinity, there is a maximum obliqueness\footnote{Obliqueness to be understood here as synonym for "departure from planarity", so $\pi-\beta$ with the notations of appendix \ref{sec:jump}.} beyond which the shock vanishes : the downstream and upstream variables are equal. The corresponding theoretical minimum angles have been plotted in red in Figure\,\ref{fig:opening_angle} for the supersonic Mach numbers we considered. 
\begin{figure}[!t]
\begin{center}
\includegraphics[height=6cm, width=12.5cm]{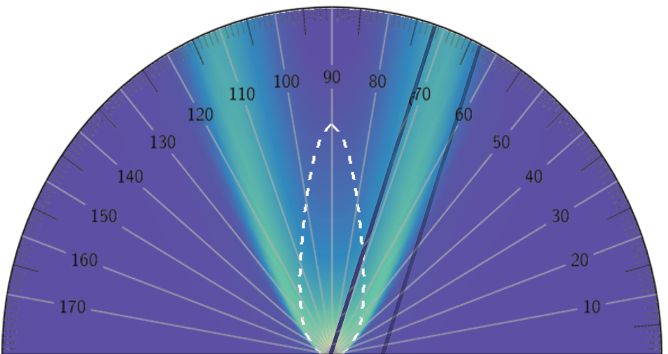}	
\caption{Measuring protocol to evaluate the opening angle on a fiducial relaxed configuration $\mathcal{M}_{\infty}=16$. The outermost opaque stripe is the estimated envelope and the innermost one is its parallel passing through the accretor. We read the opening angle with the latter one and the overlaid protractor.}
\label{fig:opening_angle_16}
\end{center}
\end{figure}

We measured the opening angle on each of the four relaxed supersonic simulations with $\mathcal{M}_{\infty}\ge 2$ following the protocol illustrated in Figure\,\ref{fig:opening_angle_16}. We drew the straight line tangent to the shock at the outer edge of the simulation space, in the wake, and measured its inclination relative to the direction of the flow at infinity. Since we do not have access to the whole tail and have to truncate it, it only provides an upper limit on this inclination which overestimates the actual opening angle. in Figure\,\ref{fig:opening_angle} have been represented in black the measured angles for the different Mach numbers ; the associated uncertainties represented with error bars come from the reading of the angle on the overlaid protractor but are, like always with numerical simulations, widely dominated by systematics. The opening angles we measure decrease with increasing Mach numbers, as intuitively expected. 

The first systematic shift one can think about is the fact that the theoretical opening angle is not necessarily reached within the simulation box, which explains that the measured one is always larger due to the concave shape of the shock. However, one must also notice that the oblique shock theory we remind in the appendix \ref{sec:jump} assumes a homogeneous planar inflow free of any exterior influence while we work on a gravitationally beamed configuration : when they reach the shock, most streamlines do so with a velocity vector which is no longer aligned with the velocity at infinity we took as a reference\footnote{Although for impact parameter above a critical non trivial value, they start to tend to hit the shock without having being significantly deflected.}. In some way, we can say that they are more normal than they would have been without the gravitational field. It makes the meaning of the opening angle we measured different from the $\beta$ introduced for oblique shocks from a planar flow in \ref{sec:oblique_shock}.
\begin{figure}[!h]
\begin{center}
\includegraphics[height=6.5cm, width=10cm]{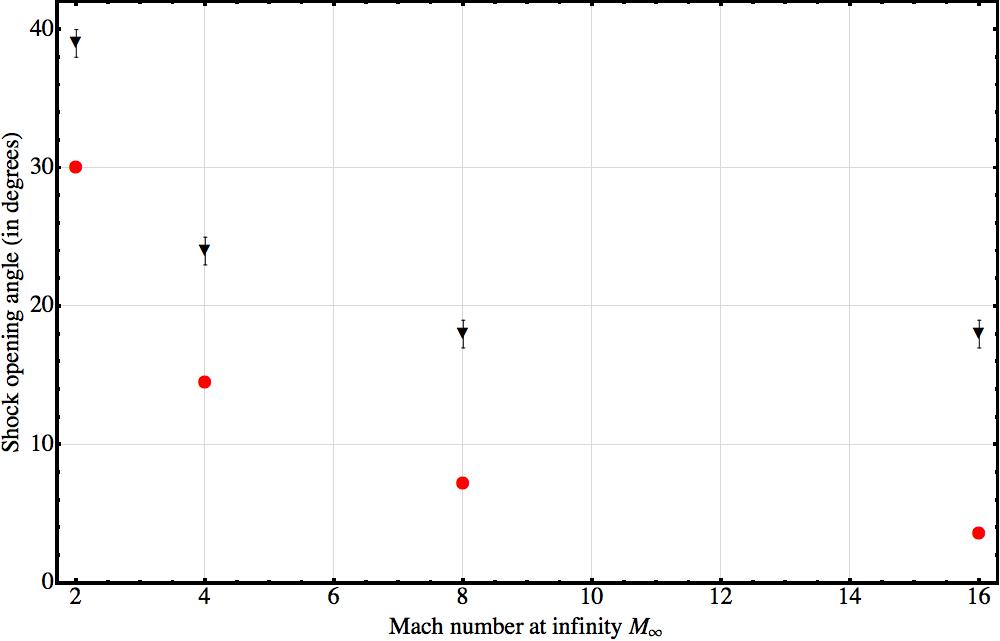}	
\caption{The black triangles represent the observed minimum opening angles $\beta$ of the relaxed shock within the simulation space as a function of the Mach number of the flow at infinity. In red is the theoretical minimum opening angle $\beta$ below which there is no longer any deflection of the flow (\ie $\theta=0$ in \eqref{eq:deflection_angle}).}
\label{fig:opening_angle}
\end{center}
\end{figure}


\subsection{Jump conditions}
\label{sec:jump_measured}

To study the jump conditions related to the subsequently formed bow shock, we focus on the front shock, locally planar. The extension of the results below to the whole shock might however be of prime importance for incoming follow-ups of the inhomogeneous entropy downstream the shock due to different obliquenesses of the shock as one goes from the front to the tail \citep{Foglizzo2000}.

\begin{figure}[!h]
\begin{center}
\includegraphics[height=8.5cm, width=13cm]{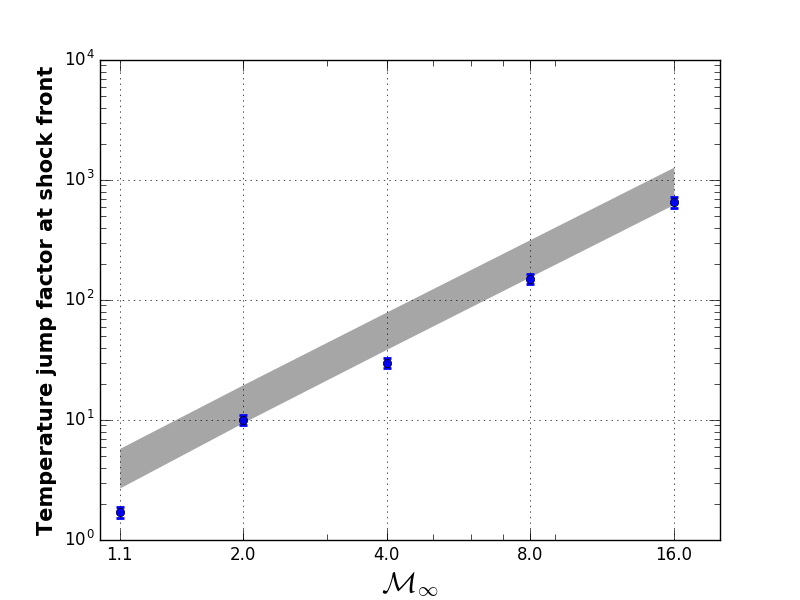}	
\caption{Ratio of the upstream to the downstream temperature as a function of the Mach number at infinity. The blue points correspond to the numerically observed ratios (with a fiducial 10\% spreading) while the gray area is where lies the theoretically expected ones.}
\label{fig:Temperature_jump_at_front_front_shock_at_10_8K}
\end{center}
\end{figure}

Rather than describing the jump conditions for all variables like in the appendix section \ref{sec:jump} devoted to it, we undertake a precise study of the thermal jump condition in particular. The shock is highly non isothermal and the corresponding measured jumps in temperature are shown in blue square markers in Figure\,\ref{fig:Temperature_jump_at_front_front_shock_at_10_8K}. One can understand the evolution of those discontinuities with the Mach number at infinity by figuring out the corresponding Rankine-Hugoniot jump condition at the shock front \eqref{eq:thermal_jump} which is given by, with subscripts 1 and 2 corresponding respectively to upstream and downstream quantities :
\begin{equation}
\frac{T_2}{T_1}=1+\frac{2\gamma}{\gamma +1}\left( \mathcal{M}_1^2-1 \right)
\end{equation}
To relate $\mathcal{M}_1$, the Mach number just before the shock, to $\mathcal{M}_{\infty}$, we use the fact that the density immediately before the front shock does not depart much from its value at infinity. Indeed, with $r\sin\theta \xrightarrow[\theta\rightarrow 0]{}\zeta$ in \eqref{eq:dens_BK}, we retrieve $\rho\sim\rho_{\infty}$ for low $\theta$. Since the evolution is isentropic before the shock\footnote{Due to an adiabatic energy equation and a ballistic motion without discontinuity.}, we can conclude that the pressure\footnote{Via the Lagrange law for ideal gases or, equivalently, by the formula for the entropy of an ideal gas in \ref{fn:entropy} of chapter \ref{chap:num_tools}.} and then, the temperature itself did not change by much. The sound speed at infinity is then similar to its value before the front shock. For the velocity, it can straightforwardly be deduced from \eqref{eq:vr} and gives, if we write $r_{\text{sh}}$ the position of the shock front :
\begin{equation}
\mathcal{M}_1\sim\sqrt{1+\frac{\zeta_{\textsc{hl}}}{r_{\text{sh}}}}\mathcal{M}_{\infty}
\end{equation}
With a shock front ranging between $\zeta_{\textsc{hl}}$ and a third of $\zeta_{\textsc{hl}}$, we expect a temperature jump lying within the gray area in Figure\,\ref{fig:Temperature_jump_at_front_front_shock_at_10_8K} (with the upper line corresponding to $r_{\text{sh}}=\zeta_{\textsc{hl}}/3$). This Figure shows that the numerical shock remains more dissipative than the real one, due to the addition of numerical viscosity. Higher resolutions and higher order schemes can help to lower this discrepancy. Also, we notice a suspicious beaming along the polar axis (visible in Figure\,\ref{fig:density} at high Mach number) which might lead to underestimated values for $r_{\text{sh}}$.
Eventually, we notice that, as intended, the Mach-1 surface corresponding to the flow going from being supersonic to being subsonic\footnote{The white dashed line in Figure\,\ref{fig:density} - which is not the sonic surface since it corresponds to a surface of decreasing Mach number following the flow.} always stops before the outer edge for any $\mathcal{M}_{\infty}\ge 2$. However, we can not help noticing an interesting feature of oblique shocks : since the tangential component of the velocity field does not change as the flow crosses the shock and because the requirement for making the velocity subsonic downstream applies only to the normal component, the total Mach number does not necessarily goes below 1 as the flow crosses what is still a shock. Further away along the tail\footnote{Outside of the simulation space presently covered.}, we could even witness the vanishing of the shock, when downstream and upstream conditions become the same.

\subsection{Transverse profiles}
\label{sec:transv_prof}

\begin{figure}[!h]
\begin{center}
\includegraphics[height=7.5cm, width=12cm]{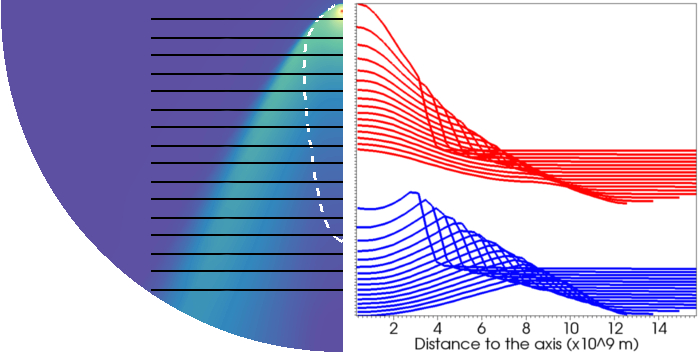}	
\caption{(right) Vertically shifted transverse profiles of density (blue) and temperature (red) sampled along the tail of the steady-state. The scales are linear and the parameters are $\mathcal{M_{\infty}}=4$, $r_{\text{in}}=10^{-3}\zeta _{\textsc{hl}}$ and $\zeta _{\textsc{hl}}\sim 2.7\cdot 10^9$m. (left) Corresponding slices in black.}
\label{fig:transverse_profiles}
\end{center}
\end{figure}

The transverse profiles along the tail have also been the object of theoretical predictions. In \cite{Bisnovatyi-Kogan1979}, the regime with the largest adiabatic index displays bow shocks with "a cone-shape cavity". in Figure\,\ref{fig:transverse_profiles} have been displayed in red the transverse temperature profiles and in blue the transverse density profiles. To give an idea of the hull of those profiles along the tail, we sampled the profiles along the accretion line, from 0.5 to $6\zeta _{\textsc{hl}}$ (resp. upper and lower curves), and shifted them by a constant vertical offset on this diagram. We clearly see that the shock is stronger close to the accretor than far downstream in the wake and that its position is obviously the same for mass densities and temperatures. More interestingly, the mass density reaches a maximum at the shock and decreases slowly within the cone shock, even down to its value outside the shock beyond $6\zeta _{\textsc{hl}}$. This feature has also been glimpsed in numerical simulations of relativistic \bhl flows onto spinning black holes \cite[see][Figure 3]{Gracia-Linares2015} following the publication of \cite{ElMellah2015}.

We can also remark how the position of the successive maxima of temperature along the accretion line confirm a posteriori the assumption we made in section \ref{sec:acc_col} that the sound speed does not evolve much in the wake of the accretor which tends to favour a concave shape for the shock. From 6 to $0.5\zeta _{\textsc{hl}}$, the temperature approximately doubles which makes the sound speed only rise by a factor of $\sqrt{2}$.


\section{Accretion properties of the \bhl flow}
\label{sec:acc_prop_BHL}


\subsection{Sonic surface}
\label{sec:sonic_surf}
\subsubsection{For the Bondi spherical model}

As seen in Figure\,\ref{fig:sonic_point}, whatever the Mach number at infinity, all flows feature a sonic radius which converges to zero for monoatomic gas ($\gamma=5/3$). Indeed, according to \eqref{eq:son_pt_Bondi}, the sonic radius of a monoatomic gas being accreted adiabatically is strictly zero\footnote{In practice, as the flow gets closer from the accretor, it cannot remain adiabatic and will, at some point, radiate away a fraction of its energy. The cooling terms are then no longer negligible.}. We made use of this critical point to determine the mass accretion rate in \eqref{eq:mdot_Bondi_ini} and want now to export this approach to the \bhl planar flow.

\subsubsection{Analytical insights in the case of the \bhl flow}

In the case of a subsonic inflow, the equation of the sonic surface has been studied by \cite{Beskin1995}. Yet, most of the astrophysical objects of interest we focus on undergo supersonic accretion. We can make the case for ubiquitous supersonic flows using Figure\,\ref{fig:dimensions_odm} where the scale of relative velocities of the gas with respect to the accretor at infinity can be related to the temperature with the following rule of thumb : to convert the left velocity scale into a temperature one in Kelvins for a Mach-1 flow of Hydrogen particles with $\gamma=5/3$, one can multiply 72 by the square of the velocity at infinity in km$\cdot$s$^{-1}$. Thus, for a bulk motion at a few 100\kms, common if not underestimated in the systems we consider, a temperature above almost one millions Kelvins would be required to make this flow subsonic\footnote{The reader might want to keep in mind that the temperature of the interstellar medium ranges from a few 10 to 10$^6$K \citep{Ferriere2001}, the latter corresponding to a sound speed of the order of 100 km$\cdot$s$^{-1}$.}.

\begin{wrapfigure}{r}{0.3\textwidth}
\begin{center}
\includegraphics[height=1.8\textwidth, width=1\textwidth]{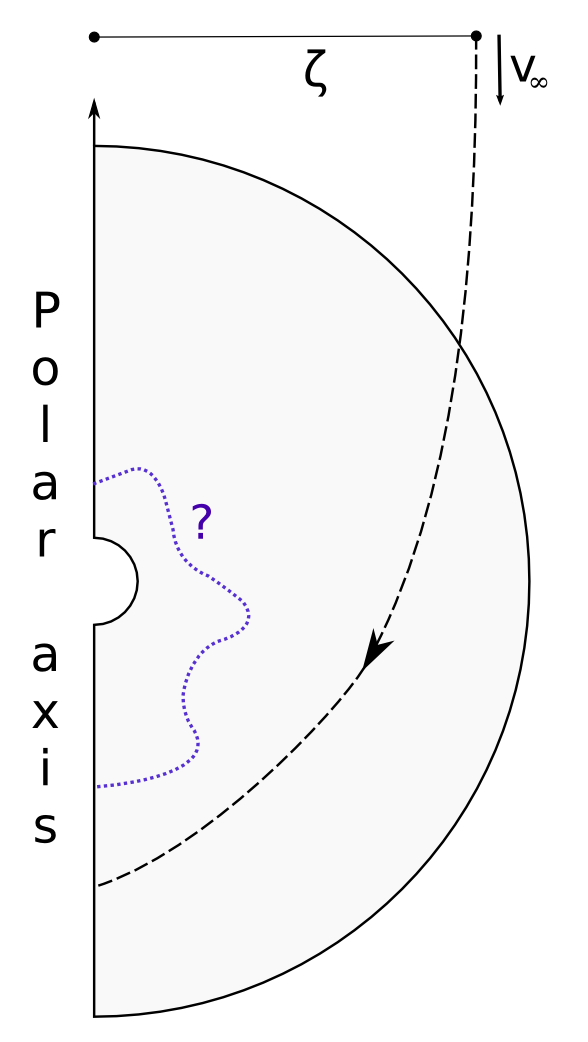}	
\caption{Sketch of the BHL flow where the unknown sonic surface has been represented in purple. It turns out that it is actually anchored into the accretor for $\gamma=5/3$ (see text).}
\label{fig:BHL_sonic_surface}
\end{center}
\end{wrapfigure}

If the symmetry of the problem guarantees the axisymmetry of the sonic surface in the steady-state, it is no longer simply described by one quantity (like the sonic radius $r_0$), as illustrated by Figure\,\ref{fig:BHL_sonic_surface}. The derivation of the polar curve $r(\theta)$ of the sonic surface is beyond the scope of our mathematical capacities but a topological property determined by \cite{Foglizzo1996} guarantees that the sonic surface intersects the corresponding sonic sphere whose radius is given by \eqref{eq:son_pt_Bondi} for the same conditions at infinity. Indeed, if one denotes $S$ the cross-section of a tube made of infinitesimally close streamlines in a specified direction and $r$ the distance to the accretor, \cite{Foglizzo1996} pinpointed that the increase rate of $S$ as a function of $r$ verifies :
\begin{equation}
\begin{cases}
\frac{r}{S}\frac{\d S}{\d r}>2 \quad \text{if} \quad r_{\text{s}}<r_0\\
\frac{r}{S}\frac{\d S}{\d r}<2 \quad \text{if} \quad r_{\text{s}}>r_0
\end{cases}
\end{equation}

\begin{figure}
\hspace*{0.05\columnwidth}
\begin{subfigure}{0.9\textwidth}
 \includegraphics[width=0.55\columnwidth]{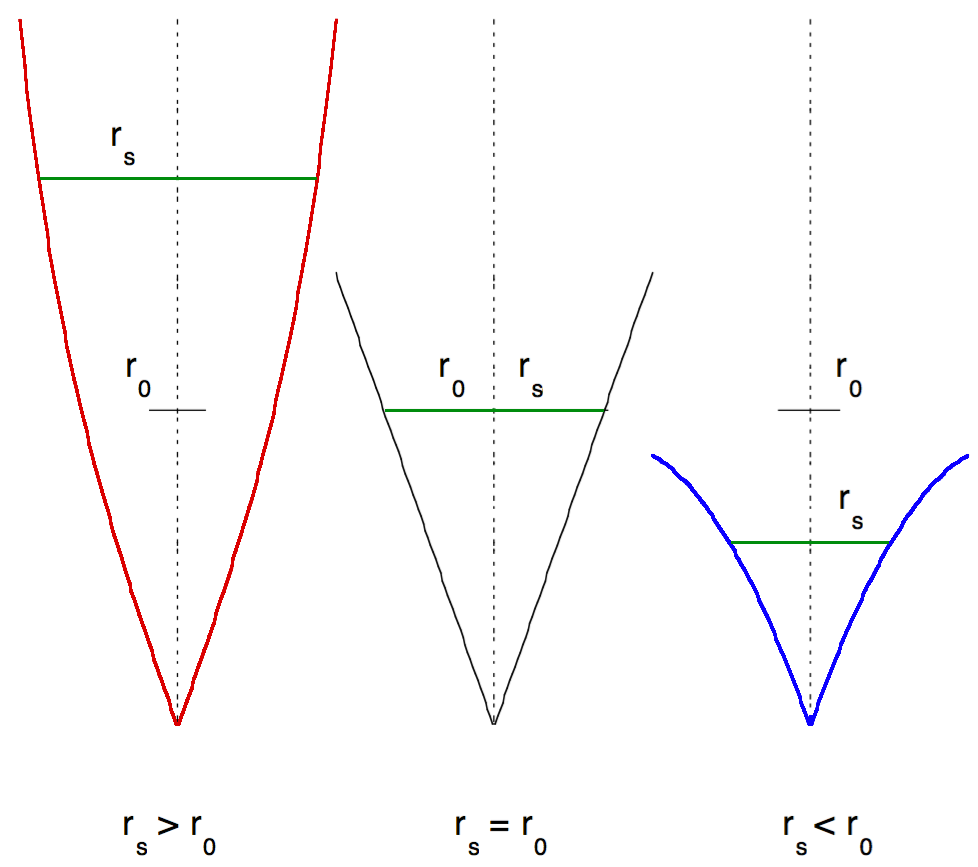}
  \label{fig:sfig1}
\end{subfigure}%
\hspace*{-0.3\columnwidth}
\begin{subfigure}{0.9\textwidth}
 \includegraphics[width=0.35\columnwidth]{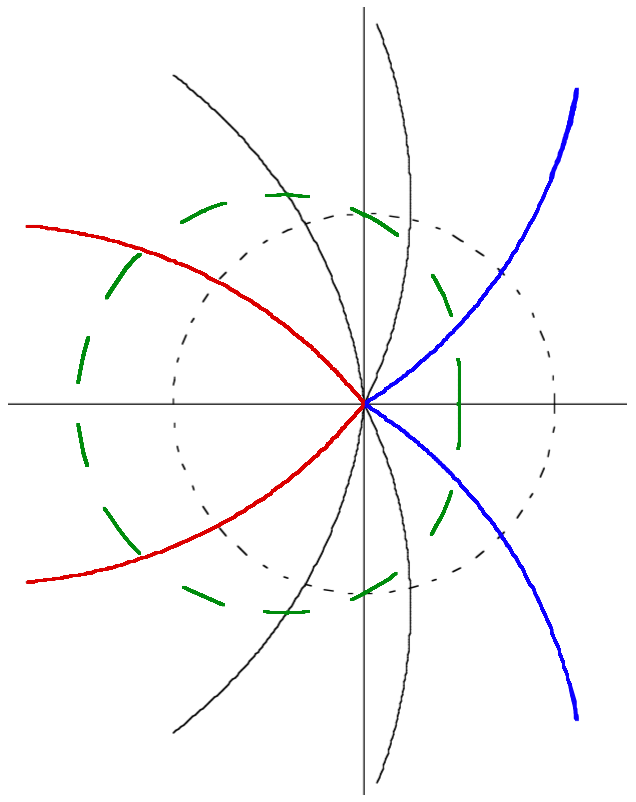}
  \label{fig:sfig2}
\end{subfigure}
\caption{\textit{(left)} 3 different configurations for the streamlines in a direction, depending on the distance between the central accretor and the sonic surface in this direction $r_{\text{s}}$ (green) with respect to the sonic radius $r_0$ (always at the same level on those 3 sketches). The streamlines are concave for $r_{\text{s}}>r_0$ (red), radial for $r_{\text{s}}=r_0$ and convex for $r_{\text{s}}<r_0$ (blue).\\
 \textit{(right)} Global sketch of the streamlines around the accretor with sections where they are concave (red) and others where they are convex (blue), which means that in some direction, the distance to the sonic surface (green dashed line) is larger than the one to the sonic sphere (dotted dashed line) while in others, it is smaller. From \cite{Foglizzo1996}.}
\label{fig:thierry}
\end{figure}

where $r_{\text{s}}$ is the distance from the accretor to the sonic surface in the $\theta$ direction considered and $r_0$ is the sonic radius set by the Bernoulli invariant at infinity\footnote{Which is the same for all streamlines due to the homogeneity of the conditions at infinity.}. For $r_0=r_{\text{s}}$, we retrieve the spherical case with a cross-section which goes as the square of the distance to the center. However, for $r_{\text{s}}>r_0$, it evolves slower, hence concave streamlines, while for $r_{\text{s}}<r_0$, it evolves faster hence convex streamlines (left panel of Figure\,\ref{fig:thierry}). For orthoradial velocities $v_{\theta}$ of constant sign along the sonic surface on each side of the axis\footnote{In Appendix C of \cite{Foglizzo1996} can be found a demonstration which shows that this assumption is actually not required to guarantee the intersection between the sonic surface and the sonic sphere.}, we can rely on the right panel of Figure\,\ref{fig:thierry}, where the sonic sphere has been plotted in dotted dashed\footnote{Since the configuration is not spherical though, the spherical sonic surface of radius $r_0$ has no proper physical sense in this situation. Its interest lies in its link with the actual sonic surface of this axisymmetric configuration.}, to understand the implications of this result. If some streamlines converge to the accretor in a concave way on one side (red streamlines), they will "occupy" a larger angular fraction than if they had done so in a radial way and will necessarily force other streamlines to converge to the accretor in a convex way on the other side (blue streamlines). Due to the equivalence between this convexity property and the position of the sonic surface with respect to the sonic sphere, it means that the sonic surface will be above the sonic sphere in some directions and below in other ones. By continuity, the sonic surface does intersect the sonic sphere which implies, for $\gamma=5/3$, that the sonic surface is anchored into the accretor.

\subsubsection{Numerical results}

\begin{figure}[!h]
\begin{center}
\hspace*{-1cm}
\includegraphics[height=6cm, width=18cm]{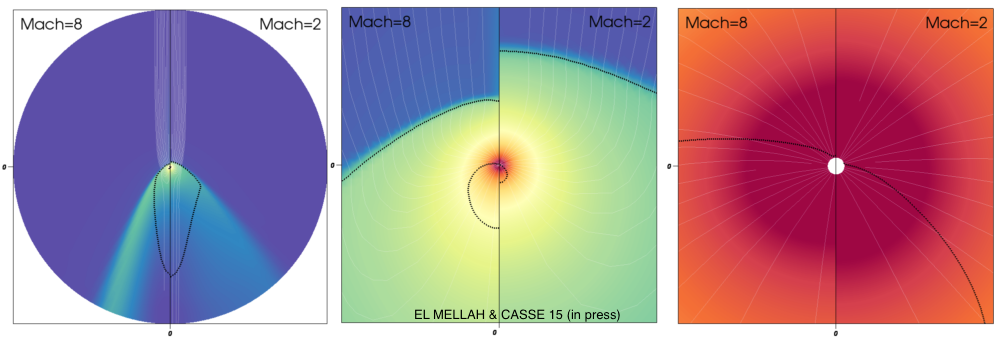}	
\caption{Logarithmic colormap of the density (purple stands for low, red for high). Successive zoom in (by a factor 20 each time) on the innermost parts of a planar flow (coming from the top) being deflected by a central accretor for different Mach numbers at infinity. In white are represented the streamlines while the the dotted black lines represent the Mach=1 surfaces (the inner one being the sonic surface).}
\label{fig:sonic_surface}
\end{center}
\end{figure}

\begin{figure}[!b]
\hspace*{-1.5cm}
\begin{subfigure}{0.45\textwidth}
 \includegraphics[height=7.9cm, width=7.4cm]{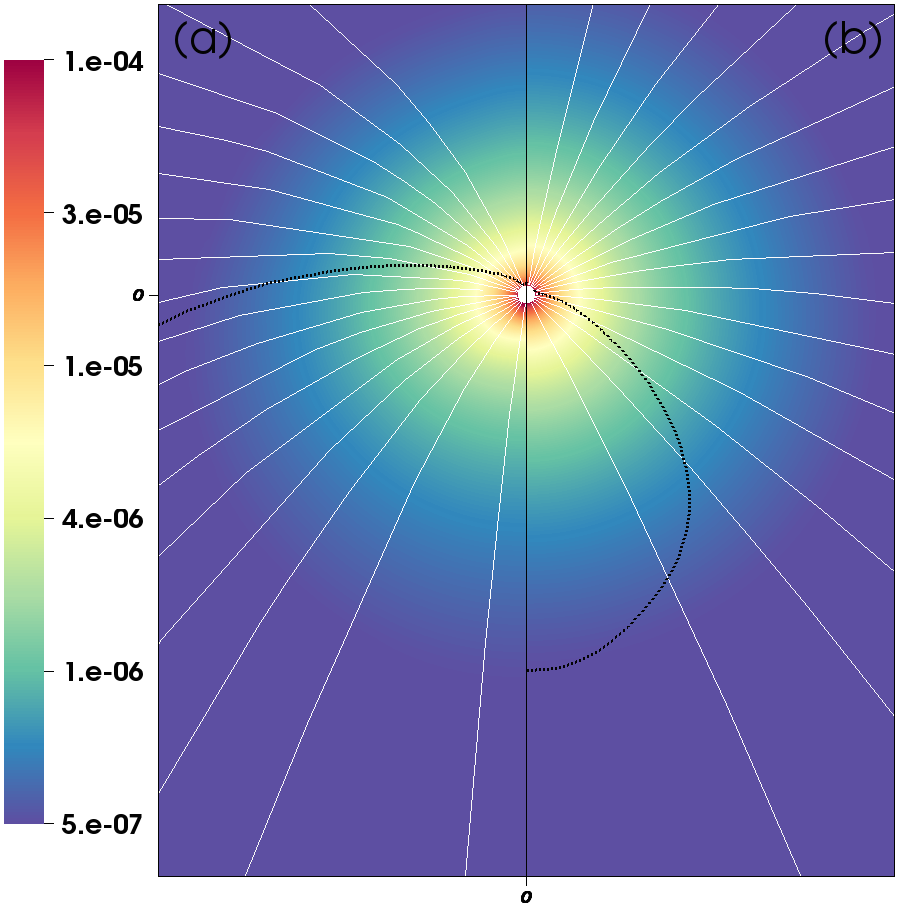}
  \label{fig:sonic_surface_different_Mach}
\end{subfigure}
\hspace*{0.8cm}
\begin{subfigure}{0.45\textwidth}
\includegraphics[height=7.9cm, width=7.4cm]{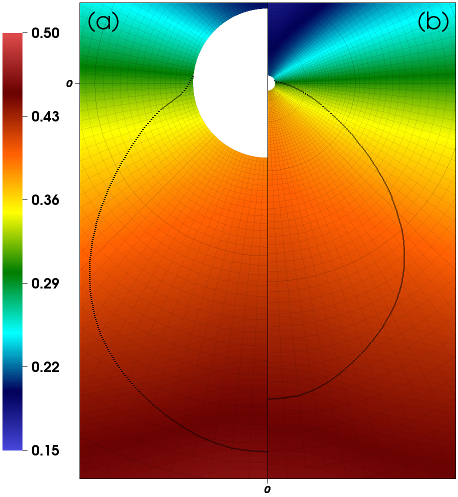}	
  \label{fig:sonic_surface_different_inner_boundary_size}
\end{subfigure}
\caption{(\textit{left}) Logarithmic color map of the density (arbitrary units), with $r_{\text{in}}=10^{-3}\zeta _{\textsc{hl}}$, for (a) $\mathcal{M_{\infty}}=8$ and (b) $\mathcal{M_{\infty}}=2$. The white lines are the streamlines and the thick dotted black line is the sonic surface, anchored into the inner boundary in both cases.  Zoom in by a factor 300 on the central area of the simulation. (\textit{right}) Zoom in by a factor 500 on the central area of the simulation. Colour map of the local mass accretion rate $\rho v_r r^2$, with $\mathcal{M_{\infty}}=2$, for (a) $r_{\text{in}}=10^{-2}\zeta _{\textsc{hl}}$ and (b) $r_{\text{in}}=10^{-3}\zeta _{\textsc{hl}}$. The scale is linear (arbitrary units) and the inflowing mass is counted positively. The thick dotted black line is the sonic surface, anchored into the inner boundary in both cases. The radial logarithmic mesh is over plotted.}
\label{fig:sonic_surfaces_2_plots}
\end{figure}

An important conclusion of those simulations is the confirmation of Foglizzo and Ruffert's analytical prediction about the topology of the sonic surface for an adiabatic flow with $\gamma=5/3$ : whatever the size of the inner boundary or the Mach number of the supersonic flow (Figures\,\ref{fig:sonic_surface} and \ref{fig:sonic_surfaces_2_plots}), the sonic surface is always anchored into the inner boundary and it extends along the wake of the accretor. For supersonic flows, the density distribution we found is mostly isotropic and the streamlines, radial, in the vicinity of the inner boundary. 

\subsection{Mass accretion rate}
\label{sec:mdot_BHL25D_num}

\subsubsection{Analytical predictions}

\cite{Foglizzo1996} used the sonic surface and its relations to its spherical counterpart to estimate a lower limit on the mass accretion rate by computing it through the sonic surface\footnote{For $\gamma<5/3$, the sonic surface needs not to be anchored and all the mass accreted pass through the sonic surface so it is not a lower limit but a direct estimate if it is indeed not anchored. For $\gamma=5/3$, most of the mass is still accreted through the sonic surface.}. The result they obtained \citep[equation (108) in][]{Foglizzo1996} depends on the Mach number at infinity and on a free parameter $\lambda$ that we will constrain below with our numerical simulations.

To summarize, they split the mass flux in 3 different components, but only two of them are defined in a detached shock configuration as ours : the mass flowing through the sonic surface and the mass flux through the angular sector where the flow is still subsonic (right panel of Figure\,\ref{fig:sonic_surfaces_2_plots} for a zoom in on the vicinity of the inner boundary). Neglecting the latter, they derived an interpolation formula to determine the former :
\begin{equation}
\label{eq:interpolFR}
\dot{M}_{\textsc{fr}}=\dot{M}_\textsc{b} \left[ \frac{\left(\gamma +1\right)\mathcal{M_{\text{eff}}}^2}{2+\left(\gamma -1\right)\mathcal{M_{\text{eff}}}^2} \right]^{\frac{\gamma}{\gamma -1}} \left[ \frac{\gamma +1}{2\gamma \mathcal{M_{\text{eff}}}^2-\gamma +1}\right]^{\frac{1}{\gamma -1}} \times \left[ 1+\frac{\gamma -1}{2} \mathcal{M_{\infty}}^2 \right]^{\frac{5-3\gamma}{2\left(\gamma -1\right)}}
\end{equation}
with the last factor being equal to 1 for $\gamma=5/3$ and with $\mathcal{M_{\text{eff}}}$ the effective Mach number which depends on a free parameter $\lambda$ assessing the aforementioned discrepancy between $\dot{M}_{\textsc{hl}}$ and the observed mass accretion rate at high $\mathcal{M_{\infty}}$ :
\begin{equation}
\label{eq:lambda}
\begin{split}
\dot{M}_{\textsc{fr}}\xrightarrow[\mathcal{M_{\infty}} \to \infty]{}\lambda \dot{M}_{\textsc{hl}}
\end{split}
\end{equation}
\begin{equation}
\label{eq:Meff}
\frac{\mathcal{M_{\text{eff}}}}{\mathcal{M_{\infty}}}\sim \frac{1}{2^{\gamma}\lambda ^{\frac{\gamma -1}{2}}} \sqrt{\frac{2}{\gamma}} \frac{\left( \gamma +1 \right)^{\frac{\gamma +1}{2}}}{ \left( \gamma -1 \right)^{\frac{5\left(\gamma -1\right)}{4}} \left( 5-3\gamma \right)^{\frac{5-3\gamma}{4}} }
\end{equation}
Physically, $\lambda$ accounts for an essential non-ballistic feature of the flow, whatever high the Mach number is. As a consequence, for $\gamma=5/3$, Foglizzo and Ruffert's formula must be understood as a lower limit since it neglects the matter being accreted from a subsonic region whom angular extension becomes larger as $\mathcal{M_{\infty}}$ decreases.

\subsubsection{Homogeneous normalization}

In the left panel of Figure\,\ref{fig:mdot_num_2_plots} is represented the evolution of the spatially averaged mass accretion rate in the vicinity of the inner boundary, well below the shock front radius, as a function of time. It takes the system approximately 10 crossing times (at most) to relax the upstream outer boundary conditions specified in \ref{sec:motion} of chapter \ref{chap:acc_pt-mass} and to reach a plateau. The normalisation of the mass accretion rates, set by $\dot{M}_{\textsc{hl}}$, is the same for all points : $\dot{M}_{\textsc{hl}}$ does not depend on the Mach number at infinity which was evolved by changing $c_{\infty}$, not $v_{\infty}$. Then, we do observe larger mass accretion rates for colder ambient media, at $v_{\infty}$ and $\mathcal{M_{\infty}}$ fixed, and with a saturation level, measured by $\lambda$, which is lower than the one prescribed by the qualitative model of Hoyle \& Lyttleton.

For each Mach number, we computed a time-averaged mass accretion rate but beforehand, we construct a non-physically motivated reference mass accretion rate, $\dot{M}_0$, whom only requirement is to verify :
\begin{equation}
\label{eq:M0_lim}
  \left\{
      \begin{array}{l}
        \dot{M}_0 \xrightarrow[\mathcal{M_{\infty}} \to \infty]{} \lambda \dot{M}_{\textsc{hl}} \\
        \dot{M}_0 \xrightarrow[\mathcal{M_{\infty}} \to 0]{} \dot{M}_B
        \end{array}
    \right.
\end{equation}
with $\lambda$ the constant further discussed above. Indeed, since Bondi's derivation \eqref{eq:mdot_Bondi_sph} for $v_{\infty}\ll c_{\infty}$ is based on the same physical laws as the ones we implemented, the numerically computed mass accretion rates must tend towards $\dot{M}_{\textsc{b}}$ for low $\mathcal{M_{\infty}}$. On the contrary, Hoyle \& Lyttleton's qualitative approach does not make $\dot{M}_{\textsc{hl}}$ more than an order of magnitude. Thus, $\dot{M}_0$ verifies the necessary asymptotic behaviors, both at low and high Mach numbers, for any acceptable interpolation formula. In no case $\dot{M}_0$ should be taken as a physically meaningful formula by itself.

The structure requirement of $\dot{M}_0$ relies on a quadratic average velocity $\tilde{v}$ between $v_{\infty}$ and $c_{\infty}$, in the same spirit as Bondi's interpolation formula albeit introducing a $\gamma$-dependent weighting factor for $c_{\infty}$ and a $\lambda$ one for $v_{\infty}$ :
\begin{equation}
\label{eq:vnorm}
\tilde{v}=\sqrt{\lambda v_{\infty}^2+\frac{1}{\left(\gamma -1 \right)^4}\left( \frac{2}{5-3\gamma}\right)^{\frac{5-3\gamma}{\gamma -1}}c_{\infty}^2}
\end{equation}
We then introduce the modified accretion radius, $R_E$, which compares the gravitational influence of the accreting body to the joint action of kinetic and enthalpic terms at infinity :
\begin{equation}
\label{eq:RE}
R_E=\frac{GM}{\displaystyle\frac{v_{\infty}^2}{2}+\frac{c_{s,\infty}^2}{\gamma -1}}
\end{equation}
In this way, it continuously pinpoints the characteristic length scale from high to low Mach numbers :
\begin{equation}
\label{eq:RE_lim}
  \left\{
      \begin{array}{l}
        R_E \xrightarrow[\mathcal{M_{\infty}} \to \infty]{} \zeta_{\textsc{hl}} \\
        R_E \xrightarrow[\mathcal{M_{\infty}} \to 0]{} \frac{GM}{\displaystyle c_{s,\infty}^2/\left(\gamma -1 \right)}
        \end{array}
    \right.
\end{equation}
In the first asymptotic case, $R_E$ can be seen as the maximum impact parameter below which an amount of gravitational energy larger than the kinetic one at infinity will be converted into kinetic energy, at constant pressure and entropy (no shock). In the second, it is the radius below which thermodynamical properties of the gas have been significantly altered from their undisturbed state at infinity, without major change of velocities.

Eventually, by writing $\dot{M}_0\propto R_{\text{E}}^2 \rho_{\infty} \tilde{v}$, a neutral formula not privileging a spherical nor an axisymmetric point of view, we make $\dot{M}_0$ verify the empirical condition \eqref{eq:M0_lim} by setting the proportionality constant to $\pi$ :
\begin{equation}
\label{eq:norm}
\dot{M}_0=\pi R_{\text{E}}^2 \rho_{\infty} \tilde{v}
\end{equation}
Thanks to $\dot{M}_0$, we have a homogeneous normalisation variable to compare the mass accretion rates to at any $\mathcal{M_{\infty}}$.

\subsubsection{Numerical results and discussion}

\begin{figure}[!b]
\hspace*{-1.5cm}
\begin{subfigure}{0.45\textwidth}
 \includegraphics[height=7.cm, width=8.5cm]{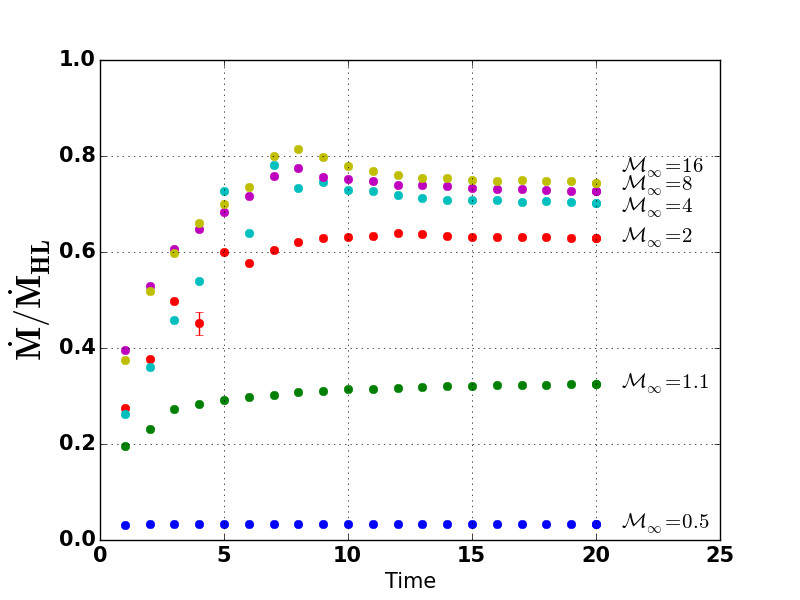}
  \label{fig:Mdot_time}
\end{subfigure}
\hspace*{0.8cm}
\begin{subfigure}{0.45\textwidth}
\includegraphics[height=7.cm, width=8.5cm]{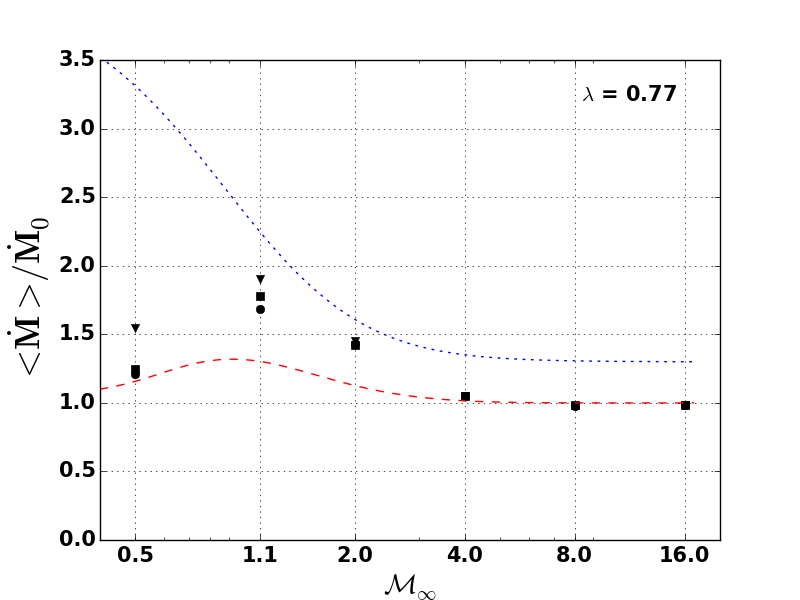}	
  \label{fig:Mdot_Mach}
\end{subfigure}
\caption{(\textit{left}) Mass accretion rate as a function of time for different Mach numbers and $r_{\text{in}}=10^{-3}\zeta _{\textsc{hl}}$. The time unit is set by the crossing time $\zeta_{\textsc{hl}}/v_{\infty}$ and the mass accretion rate unit, by \eqref{eq:Mdot_HL}. (\textit{right}) Mass accretion rates $\dot{M}$ as a function of the Mach number $\mathcal{M_{\infty}}$, normalized with the empirical mass accretion rate given by \eqref{eq:norm}. In black are the numerically computed ones for $\zeta_{\textsc{hl}} / r_{\text{in}} = 10^2$ (triangles), $\zeta_{\textsc{hl}} / r_{\text{in}} = 10^3$ (squares) and $\zeta_{\textsc{hl}} / r_{\text{in}} = 10^4$ (circles, only computed for the three lowest Mach numbers). The blue dotted line is the Bondi interpolation formula and the red dashed is the one from \cite{Foglizzo1996}.}
\label{fig:mdot_num_2_plots}
\end{figure}

The steady-state mass accretion rates $\langle\dot{M}\rangle$ are computed by averaging the instantaneous ones, and the error bars from the dispersion, both for $t>10$ crossing times, once the steady state is reached. The obtained $\langle\dot{M}\rangle$ are presented in the right panel of Figure\,\ref{fig:mdot_num_2_plots}. The error bars are smaller than markers extension but at low $\mathcal{M_{\infty}}$, the data dispersion is dominated by systematics whom influence is given by the three sets of points from the different inner boundary sizes. As expected, the traditional Bondi intermediate formula \eqref{eq:Bondi_swindle}, in blue dotted in Figure\,\ref{fig:mdot_num_2_plots}, matches the Hoyle-Lyttleton mass accretion rate at high Mach numbers, which itself turns out to be an overestimate of the actual mass accretion rate by approximately 30\% : it is consistent with the previous reports of $\dot{M}_{\textsc{hl}}$ being a slight overestimate of the numerically observed mass accretion rates in the asymptotically supersonic regime - see, \eg, Figure\,7 of \cite{Edgar:2004ip}. Yet, since \eqref{eq:Bondi_swindle} does not properly include thermodynamics (\eg it does not depend on the adiabatic index), it is a poor estimate in the subsonic regimes, overestimating $\dot{M}$ by a factor up to 4 for $\gamma=5/3$ - which corresponds to the analytically derived limit of $\dot{M}_{\textsc{bh}}/\dot{M}_{\textsc{b}}$ at low Mach numbers given in the consistent and comprehensive review of \cite{Ruffert1994a}. But surprisingly enough, $\dot{M}_{\textsc{bh}}$ is not too far off in the mildly supersonic one, where the Hoyle-Lyttleton formula is invoked with the velocity at infinity replaced by its quadratic average with the sound speed at infinity.

Although pretty similar to Ruffert's numerical conclusions \citep{Ruffert1994c,Ruffert1994}, our results do not show any decreasing trend at high Mach number and match the interpolation formula $\dot{M}_{\textsc{fr}}$ (in dashed red in the right panel of Figure\,\ref{fig:mdot_num_2_plots}) in the asymptotically supersonic regime for $\lambda\sim 0.77$, down to $\mathcal{M_{\infty}}=4$, at a few percent precision level. However, the precision on $\lambda$ could be limited by the numerical beaming along the $\theta=0$ axis\footnote{Visible in Figure\,\ref{fig:density} for $\mathcal{M_{\infty}}\ge 8$.} mentioned in the section \ref{sec:jump_measured}, although compensated by a similar depletion along the $\theta=\pi$ axis. Concerning the resolution dependency of our simulations, with the higher resolution simulations, we did not observe any significant change of the mass accretion rate beyond a 2\% level in the case $\mathcal{M_{\infty}}=4$. Yet, it turned out that the $\mathcal{M_{\infty}}=1.1$ case gave accretion rates 5 to 10\% lower in the high resolution runs, which emphasizes the larger systematic uncertainties for mildly supersonic flows. It also enables the reader to figure out an estimate of the systematic uncertainties in Figure\,\ref{fig:mdot_num_2_plots}.

We also grasp the main feature of the axisymmetric \bhl flow that is to say an amplification by a few 10\% of $\dot{M}$ around $\mathcal{M_{\infty}}=1$ compared to the interpolation formulae \eqref{eq:interpolFR} and \eqref{eq:norm} verifying $\dot{M}\xrightarrow[]{\mathcal{M_{\infty}} \to 0}\dot{M}_{\textsc{b}}$ and $\dot{M}\xrightarrow[]{\mathcal{M_{\infty}} \to \infty}\sim\dot{M}_{\textsc{hl}}$. It must be noticed that for $\mathcal{M_{\infty}} = 0.5$, $1.1$ \& $2$, the mass accretion rates measured converge towards the interpolation formula derived by \cite{Foglizzo1996} as $r_{\text{in}}/\zeta_{\textsc{hl}}$ drops, without fully settling down (which is in agreement with the zero value of the sonic radius) : since the sonic surface is anchored into the accretor, there are always directions of accretion where the flow is not supersonic. The smaller the inner boundary size, the closer the simulation from the model drawn in \cite{Foglizzo1996}. It can also be seen from direct visualization  of the sonic surface (in the right panel of Figure\,\ref{fig:sonic_surfaces_2_plots}) which tends to occupy an angular region around the inner boundary larger for smaller inner boundary radii. Given those elements, those numerical $\dot{M}$ must be seen as upper limits of the ones one would get for a smaller absorbing\footnote{In the meaning of the inner boundary conditions described in \ref{sec:bc}.} inner boundary. \\

Due to the mostly radial streamlines in the vicinity of the inner boundary, we can affirm that the accretion is regular in the sense that there is no infinite mass accretion rate direction, although the local mass accretion rate is enhanced by a factor of a few units in the back hemisphere compared to the front hemisphere. As a consequence, the non isotropy of the mass accretion rates around the inner boundary is mostly due to the non isotropy of the velocity field ; along the mock accretion line, in the wake of the accretor, the flow has been more reaccelerated than upstream after the shock.


\section{Discussion}
\label{sec:discussion_BHL_2}

We designed a numerical setup optimized to embrace the whole dynamics of a 2.5D planar hydrodynamical Bondi-Hoyle accretion flow onto a body whom size is 3 to 5 orders of magnitude smaller than its accretion radius, reproducing typical weakly or non-magnetized compact objects. Despite the challenging dynamics, we are able to characterise the mass accretion rates for Mach numbers at infinity ranging from 0.5 to 16, along with the geometrical properties of the flow : the bow shock forming for supersonic flows, the internal depletion of mass along the accretion line (beyond the stagnation point) and the temperature jump at the shock are all in agreement with previous analytic solutions \citep{Bisnovatyi-Kogan1979}. Those simulations also emphasized the relevance of the sonic surface in the derivation of the mass accretion rates. As first suggested by \cite{Foglizzo1996}, anchoring of the latter to the accretor surface allows one to set an analytical lower limit on the mass accretion rates and to understand, at least qualitatively, its dependence on the Mach number for mildly supersonic flows. We evaluated the underestimation factor induced by such an approach by accounting for the whole mass flow, not only through the supersonic surface but also in the subsonic angular section. It is worth mentioning that the anchoring of the sonic surface, shown here for the first time, is a strong consistency check which brings robustness to the physical accuracy of those simulations. Furthermore, we observed the relaxation towards a numerically stationary regime, which does not present any significant oscillatory behaviour, discarding the growth of axisymmetric instabilities with this level of numerical dissipation. Concerning non axisymmetric instabilities like the so-called flip-flop instability, unfolding those 2.5D setups in full 3D would enable us to investigate whether its amplitude is similar to the 2D cases. 

The use of a proper energy equation alleviated the introduction of a polytropic index and ensured the anchoring of the sonic surface for any monoatomic ideal gas ; it also naturally handled the jump conditions at the shock. Yet, we relied on an adiabatic assumption which might overestimate the gas heating as matter flows in. Besides, we neglect radiative feedbacks which are likely to play a role through, for instance, the radiative pressure. Indeed, comparing the Eddington luminosity to the Hoyle-Lyttleton mass accretion rate with a fiducial efficiency of 10\% gives :

\begin{equation}
\frac{10\% \dot{M}_{\textsc{hl}}c^2}{L_{\text{Edd}}} \sim 4\% \left( \frac{M}{20M_{\odot}} \right) \left( \frac{v_{\infty}}{10^6\text{m}\cdot\text{s}^{-1}} \right)^{-3} \left( \frac{\rho _{\infty}}{10^{-11} \text{kg}\cdot \text{m}^{-3}} \right) 
\end{equation} 

Taking into account this radiative term in the energy equation is necessary for slower winds, {\sc imbh} \citep{Park2013} or {\sc smbh} \citep{Novak2011}. Persistent radiative feedback of accreting stellar-mass black holes on their host primordial galaxies is also under current investigation, along with the central supermassive black hole influence \citep{Wheeler2011}. Concerning orbital effects in binaries \citep[see][for {\sc sph} simulations]{Theuns1993,Theuns1996d}, the present study has not included them but serves as a test case for a forthcoming work accounting for their effects. Yet, if the orbitally induced torque remains small enough, those axisymmetric simulations are not expected to depart much from the actual configuration of \sgx where the mass transfer is believed to occur mainly through fast stellar winds. Thanks to our numerical setup, which reconciles the requirement for physical size of the accretor together with the necessity to include the accretion radius within the simulation space, we stepped through a numerical threshold beyond which we expect the wind accretion process at stake in binaries to be simulated comprehensively.

%

\setlength{\parskip}{0ex} 


\chapter*{Transition}
\addcontentsline{toc}{chapter}{Transition}
\adjustmtc

The present numerical setup serves as much as a model to study the evolution of a \bhl flow depending on its Mach number as a prerequisite to characterize the wind accretion process. We constantly monitored our results to compare them to firmly established analytical constraints, derived accurate numerical mass accretion rates and investigated the evolution of the properties of the flow with its Mach number at infinity, from slightly subsonic to highly supersonic setups : the bow shock forming for supersonic flows, the internal depletion of mass along the accretion line and the temperature jump at the shock for instance. It turns out that those simulations confirm the analytical prediction by \cite{Foglizzo1996} concerning the topology of the inner sonic surface. Using a continuously nested logarithmic mesh with a shock-capturing algorithm and specific boundary conditions, we can resolve the flow from the vicinity of the central compact object up to the significant deflection scale, at an affordable computational price and with realistic wind velocities. In agreement with the properties of \bhl accretion on finite size objects, we can finally reach numerical regimes where the size of the inner boundary no longer significantly alters the flow properties. We solved the full set of hydrodynamical Eulerian equations, including the energy one, with the less restrictive adiabatic assumption instead of considering a polytropic flow and identifying the adiabatic and the polytropic indexes. \\
\indent With this first numerical test-case of a \bhl flow, we paved the way to multi-dimensional follow-ups of fast flows encountering the gravitational well produced by a compact object, much smaller than its accretion radius. The topology of the numerically derived sonic surface serves as a strong sanity check which drives us into adapting this setup to match the configuration of wind accreting X-ray binaries. In particular, the orbital effects must be included (Chapter \ref{chap:roche}) along with the specificities of the radiatively-driven wind emitted by the Supergiant stellar companion (Chapter \ref{chap:wind}). Indeed, we will see that wind accretion occurs preferentially in Supergiant X-ray binaries which host, in addition of the compact accretor, an early type evolved star with a large luminosity and mass loss rate. In a binary system, the evolutionary path of a massive star is drastically altered by the presence of a nearby companion. If the nature of the binary interaction is largely determined by the initial orbital period and mass ratio of the system, the role of mass transfer in shaping the evolution of a high mass star remains uncertain, partly because of the paucity of measurements of the intrinsic distributions of orbital and stellar parameters.\\
\indent To constrain the latter, the approach we advocate is to use the coupling between the stellar, orbital, wind and accretion parameters in Supergiant X-ray binaries. In a ballistic framework, we designed a toy-model to follow the evolution of streamlines and the subsequent observables as a function of a reduced set of characteristic numbers and scales. We show that the shape of the permanent flow is entirely determined by the mass ratio, the filling factor, the Eddington factor and the $\alpha$ force multiplier. Provided scales such as the orbital period are known, we can trace back the observables to evaluate the mass accretion rates, the accretion mechanism (wind or stream-dominated), the shearing of the inflow and the stellar parameters. This setup also provides us with physically-motivated outer boundary conditions to investigate the vicinity of the accretor and focus on the bottom line problematic of this PhD manuscript : can a wind accretion regime give birth to disc-like structures around a compact object in \sgx ? 


\part{Wind accretion in persistent Supergiant X-ray Binaries}


\chapter{Winds of isolated hot massive stars}
\label{chap:wind}
\chaptermark{Winds of hot stars}
\hypersetup{linkcolor=black}
\minitoc
\hypersetup{linkcolor=red}
\setlength{\parskip}{1ex} 

Since the seminal papers of the 70's \citep{Lucy1970,Castor1975b}, the winds of massive stars have been thoroughly studied. Unlike intermediate and low mass stars (\ie below a few solar masses) where winds are believed to be sustained by thermal pressure gradient\footnote{Which requires a hot corona with temperatures high enough to reproduce the terminal speed observed for the Sun. The heating mechanism responsible for this corona is an intense object of research \citep[see \eg][]{Xia2011} but it is believed that the magnetic field produced by convective motions in the outer layers of the star plays a major role. Yet, beyond a few solar masses, the outer layers are no longer convective but radiative \citep[see][for a first glimpse of the question with the Schwarzschild buoyancy criterion.]{Shu1992}.}, massive stars harbour photospheres hot enough to emit a significant fraction of the light above the ionization frequency of metals. Their interaction with the partly ionized atoms (section \ref{sec:compton_scatt}) in the outer layers of the star provides a significant amount of net linear momentum to the gas (\ref{sec:MdotV}), much more important than the one from continuous radiative pressure on free electrons (\ref{sec:thom_scatt}). A wind develops (\ref{sec:motion_CAK}) and reaches terminal speeds of the order of a few times the escape velocity at the stellar surface (\ref{sec:mdot_CAK}). In this section, we only tackle the case of isolated stars. The coupling with the Roche potential described in Chapter \ref{chap:roche} will be done later on in Chapter \ref{chap:SgXB}. This section settles for a short reminder of some features we will use later on to study \sgx. For more detailed and comprehensive reviews, the reader can refer to the chapter 7 of \cite{Lamers1999} and to the reviews by \cite{Kudritzki2000} and \cite{Puls2008}.


\section{Continuous radiative pressure}
\label{sec:cont_rad_press}


\subsection{Thomson scattering on free electrons}
\label{sec:thom_scatt}

Thomson scattering describes the interaction of light with free charged particles. In the outer stellar layers, free electrons from fully ionized Hydrogen will be the privileged partners due to their low inertia\footnote{Bound electrons can also interact as we will see in section \ref{sec:compton_scatt}.}. We thereby undertake a corpuscular interpretation of this model but a wave one is also possible and would give strictly the same results. The incoming photon\footnote{An electromagnetic wave in the wave approach, hence the need for a charged particle.} excites a free electron which oscillates and isotropically\footnote{Given the fact that we do not work with relativistic flows here, this term does not suffer any ambiguity and can be understood in the frame of the star or the accelerated particle.} emits a photon at the same frequency as the absorbed one, following Larmor's result on radiation emitted by accelerated charged particles\footnote{Inelastic scattering is to be expected if the rest mass energy of the particle ($\sim$511keV for an electron) is lower than the energy of the interacting photon.}. The absorption does not depend on the wavelength which justifies the name of "continuous radiative pressure" we will sometimes use to refer to this process, as opposed to line absorption in section \ref{sec:compton_scatt}. 

Since the initially absorbed photons come from the photosphere below, there is a net gain of linear momentum for the electrons which then redistribute it to the ambient flow via Coulomb dragging. The Thomson cross-section of a free particle of charge $q$ and of mass $m$ can be shown to be, in \textsc{cgs} :
\begin{equation}
\sigma_e = \frac{8\pi}{3} \left( \frac{q^2}{m c^2} \right)^2
\end{equation}
where $c$ is the speed of light. This expression shows why free electrons have more influence than free protons if in equal quantity, given their smaller mass. For an assembly of free electrons in a stellar atmosphere, if we write $I_{He}$ the number of electrons provided per Helium nucleus\footnote{Typically 2 for an O-star.} and $Y$ the number fraction of Helium with respect to Hydrogen\footnote{Set to 20\% in this manuscript.}, we get the total opacity\footnote{Cross-section per unit mass of proton.} $\kappa_e$ of the stellar material \citep{Kudritzki1989a} :
\begin{equation}
\kappa_e \sim 0.398 \frac{1+I_{He} Y}{1+4Y} \text{cm}^2\cdot\text{g}^{-1}
\end{equation}
Since the radiation pressure is given by the ratio of the stellar flux $F$ by the speed of light, once we account for the cross-section, we get the outwards acceleration exerted on the flow\footnote{Remarkably enough, the opacity does not depend on the light frequency which makes the integration of the flux per frequency over the whole frequency range straightforward and immediately yields the total stellar flux $F$.} :
\begin{equation}
g_{\text{Thomson}}=\kappa_e\frac{F}{c}
\end{equation}

The opacity $\kappa_e$ varies little in the wind and will be considered in this manuscript to be constant and homogeneous. As a consequence, the radial dependence of this acceleration term is the same as the one of the stellar flux : it decreases as the square of the distance to the stellar center. This statement invites to compare this acceleration to the gravitational one which follows the same radial evolution.


\subsection{The Eddington limit}
\label{sec:eddington_lim}

The outer layers are bounded to the star by gravitation but subjected to continuous radiative pressure\footnote{And, to a lesser extent for massive stars, to thermal pressure and centrifugal force.}. If the latter overcomes the former, the star undergoes consequent mass losses much more important than the outflows driven by line-absorption we will introduce below. The luminosity it takes to reach this limit is called the Eddington luminosity and is given by :
\begin{equation}
\label{eq:LEdd}
L_{\text{Edd}}=\frac{GM/r^2}{g_{\text{Thomson}}}=\frac{4\pi GMc}{\kappa_e}\sim42,000L_{\odot} \left( \frac{M}{1M_{\odot}} \right) \left( \frac{0.31\text{cm}^2\cdot\text{g}^{-1}}{\kappa_e} \right)
\end{equation} 
where $M$ is the mass of the star, $G$ the gravitational constant and the value of $\kappa_e$ has been obtained with the formula and the figures mentioned in the previous section. For a star like the Sun, it shows that the continuous radiative pressure does little to counterbalance gravity but since luminosity evolves faster than mass, its importance is no longer negligible for massive OB-stars. So as to quantify the latter, we introduce the dimensionless Eddington factor $\Gamma$, the ratio of the stellar luminosity $L$ by its Eddington luminosity :
\begin{equation}
\label{eq:GEdd}
\Gamma=L/L_{\text{Edd}}
\end{equation}
It evaluates the relative importance of continuous radiative pressure with respect to gravitation. For OB-star, it can reach 30\%\footnote{For Wolf-Rayet stars and Luminous Blue Variables, it even gets higher.}.

Due to the similar evolution of the gravitational and continuous radiative accelerations, $\Gamma$ is a constant set by the stellar parameters and one can define an effective gravity and, by then, an effective escape speed, by lowering the stellar gravity with a factor $1-\Gamma$. However, this phenomenon can not be responsible for the launching of a wind due to this very reason : either the continuous radiative pressure overcomes the stellar gravity at every radius and additional limiting effects come into play \footnote{An altered coupling between matter and radiation in clumpy winds (since photons tend to escape through the more tenuous gas between optically thick clumps) can allow a star with $\Gamma>1$ to remain stable \citep{VanMarle2009}.}, or it does not and will never be large enough, whatever the distance to the stellar center, to drive a wind.


\section{Global properties of a line-driven wind}
\label{sec:glob_prop_CAK}


\subsection{Compton scattering on metal ions}
\label{sec:compton_scatt}

\begin{figure}[!b]
\begin{center}
\includegraphics[height=8cm, width=7cm]{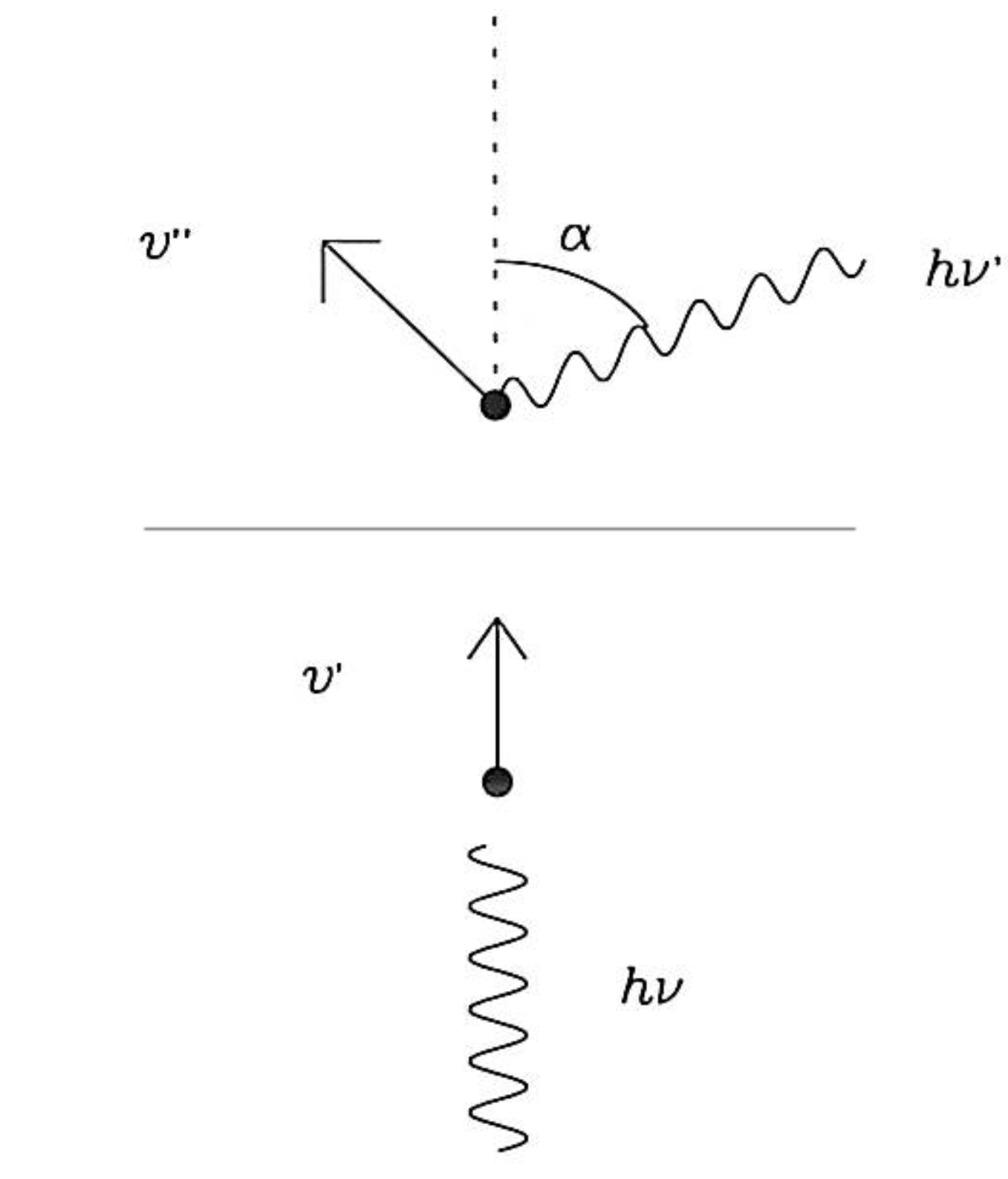}	
\caption{\textit{(lower panel)} Absorption of a photon with frequency $\nu$ by a particle with a velocity $v_r$ along the direction of incident light. \textit{(upper panel)} Spontaneous emission of a photon with frequency $\nu'$ in a direction identified with the angle $\alpha$ with respect to the initial direction of incident light. As it does so, the velocity of the particle along the direction of incident light goes from $v_r'$ to $v_r''$. From \cite{Lamers1999}.}
\label{fig:compton_scatt}
\end{center}
\end{figure}

Compton scattering covers the wide scope of interactions between a photon and free charged particles. It encompasses the Thomson case described above but is more general since the scattering can be inelastic : the energy of the photon reemiteed by the particle of mass $m$ has a lower energy than the incident photon. A quick computation can be carried on to evaluate the linear momentum brought to a particle as it absorbs and reemits a photon in a direction $\alpha$ with respect to the incident one (Figure\,\ref{fig:compton_scatt}). Without loosing much of the essential results, we can consider the original emitting source as a point such as, if we consider only pristine photons from the photosphere which have not been scattered yet, this initial direction is actually a radius. We also write $v_r$ the radial components of the velocity, the only one we are interested in here. After the absorption of a photon at $\nu$ in the frame initially comoving with the absorber, the latter is provided an additional velocity along the direction of the incident photon :
\begin{equation}
\label{eq:compton1}
mv_{r,2}-mv_{r,1}=\frac{h\nu}{c}
\end{equation}
where 1 refers to before the absorption and 2 to after the absorption. The absorber now finds itself in an excited state, as we will see in a minute when we will tackle the nature of the interaction itself rather than its dynamical effects on the particle. But in the blink of an eye, it relaxes by emitting a photon at $\nu'$ in a random direction $\alpha$ in its reference frame, which alters once again its momentum accordingly :
\begin{equation}
mv_{r,3}-mv_{r,2}=-\frac{h\nu'}{c}\cos\alpha
\end{equation}
If we express the gain of momentum between the initial and final states as a function of the rest frequency corresponding to $\nu$ \ie $\nu_0=\nu/\left(1+v_{r,1}/c\right)$, we have :
\begin{equation}
\label{eq:compton2}
mv_{r,3}-mv_{r,1}=-\frac{h\nu'}{c}\cos\alpha+\frac{h\nu}{c}=-\frac{h\nu_0}{c}\left(1+\frac{v_{r,2}}{c}\right)\cos\alpha+\frac{h\nu_0}{c}\left(1+\frac{v_{r,1}}{c}\right)
\end{equation}
For non relativistic velocities and photon energies well below the rest mass of the scattering particle, we have, injecting \eqref{eq:compton1} in \eqref{eq:compton2} :
\begin{equation}
mv_{r,3}-mv_{r,1}=\frac{h\nu_0}{c}\left(1-\cos\alpha\right)
\end{equation} 
Since $\alpha$ is uniformly sampled between 0 and $\pi$, we have the following net gain of linear momentum :
\begin{equation}
\langle mv_{r,3}-mv_{r,1}\rangle=\frac{h\nu_0}{c}\frac{\int_{\alpha=0}^{\pi}\int_{\phi=0}^{2\pi}(1-\cos\alpha)\sin\alpha\d \alpha \d \phi}{4\pi} = \frac{h\nu_0}{c}
\end{equation}
where $\phi$ is the East-West coordinate of the spherical coordinates. Eventually, it shows that the momentum increase by isotropic scattering is the same as for the case of pure absorption of a photon at $\nu_0$.

Concerning the phenomenon responsible for absorption and spontaneous emission itself, it is mostly because of the excitation of electrons in partly ionized metal atoms (C, N, O, Ne, Si, P, S and Fe-group elements) : they are excited as they absorb the light (provided they have still bound electrons left). The energy it takes to excite one of the electrons of the atom to a higher energy level is typically an electron-Volt, in agreement with the UV / X photons one can find from the photosphere of massive stars. The relaxation is triggered either by collisions if the plasma is dense enough or by the Heisenberg characteristic timescale deduced from the ratio of the Planck constant by the energy between levels, hence the "blink of an eye" above. The sharp and magnified response of resonant atomic lines to photons with the right frequency $\nu_0$ quickly makes those lines optically thick : they absorb all the photons available at this frequency and are accelerated\footnote{Optically thin absorption lines also decrease as $r^{-2}$ and can not account for the launching of the wind \citep{Kudritzki2000}.}. As they do so, they see the stellar surface recede and are able to absorb photons of higher energy energy previously untouched. They keep accelerating and if the metal atoms transfer their momentum to their plasma mates faster than they are accelerated (see next section, \ref{sec:coulomb_drag}), a bulk motion appears : the wind is launched. 


\subsection{Momentum redistribution}
\label{sec:coulomb_drag}

The ambient gas in the stellar atmosphere contains ions which are coupled with each other and in particular with the ones among them which are the most involved in the line-acceleration process due to their high resonant opacity. Is this coupling strong enough to redistribute the gained linear momentum to the surrounding particle and induce a bulk motion? The coupling is done through Coulomb interactions between the metal ions and the field particles : as the latter exerts a drag on the former, momentum is redistributed. If this redistribution occurs on a timescale smaller than the acceleration timescale $t_{\text{acc}}$, a bulk motion appears. The latter can be evaluated by monitoring the time it takes, given an acceleration representative of the one exerted on metal ions, to reach a velocity of the order of the sound speed. Once compared to the Coulomb dragging timescale, we can affirm that the coupling is strong enough to drive a global motion of the plasma is the absolute mass loss rate $\dot{M}$ is above a certain limit set by the stellar luminosity $L$ and the velocity of the flow $v$ \citep{Lamers1999} :
\begin{equation}
\dot{M} > 2\cdot 10^{-9} \left( \frac{L}{10^5L_{\odot}} \right) \left( \frac{v}{1000\text{km}\cdot\text{s}^{-1}} \right) M_{\odot}\cdot\text{yr}^{-1}
\end{equation}
This result is an additional explanation to the absence of significant line driven winds in later than B spectral type stars\footnote{In addition to the scarcity of photons energetic enough.}, where the mass outflows observed are below this threshold. The Sun for instance, displays a mass loss rate through its wind which is of the order of 10$^{-14}M_{\odot}\cdot$yr$^{-1}$, which is slightly below the corresponding threshold while OB-star displays mass loss rates 3 orders-of-magnitude above this threshold as we will see later on.


\subsection{Wind mass rate}
\label{sec:MdotV}

We now evaluate the total energy and momentum deposited in the wind by stellar radiation. To do so, we consider only one absorption line at a frequency $\nu_0$, optically thick enough to absorb every photon available. We write $F_{\nu}$ the stellar flux at frequency $\nu$ and the light rays are supposed to be purely radial (point-source, see section \ref{sec:fd} for a refinement). As the flow accelerates from negligible speeds up to the terminal speed $v_{\infty}$, it absorbs higher energy photons than the ones at $\nu_0$ due to their Doppler-reddening. All the photons between $\nu_0$ and $\nu_0\left( 1+v_{\infty}/c \right)$ will then be absorbed. The total luminosity transferred to the wind momentum is given by, for $v_{\infty}\ll c$ :
\begin{equation}
L_{\text{abs}}=\int_{\nu_0}^{\nu_0\left( 1+v_{\infty}/c \right)} 4\pi R^2 F_{\nu} \d \nu
\end{equation}
where $R$ is the stellar radius. This luminosity accounts for the wind momentum which is given by, if one assumes a blackbody spectrum for the flux $F_{\nu}$ :
\begin{equation}
\dot{M}v_{\infty}=\frac{L_{\text{abs}}}{c}=0.6\frac{L}{c^2}v_{\infty}
\end{equation}
This expression highlights an important feature which still holds if we go through a more rigorous derivation : the mass accretion rate can be determined independently from the wind velocity at infinity. In particular, for $N_{\text{eff}}$ effective optically thick absorption line, we have :
\begin{equation}
\dot{M} \sim N_{\text{eff}}L/c^2 \sim 7\cdot 10^{-7} \left( \frac{L}{10^5L_{\odot}} \right) \left( \frac{N_{\text{eff}}}{100} \right)
\end{equation}
where $N_{\text{eff}}$ has been set so as to approximately yield mass loss rates in agreement with the ones observed for OB-stars. 

If all photons of the blackbody stellar spectrum are absorbed once and the contribution of reemitted photons is neglected (single-scattering assumption), $L_{\text{abs}}=L$. We then get an upper limit on the wind momentum at infinity. Comparing the actual one to this maximum amount can be accomplished through the wind efficiency number $\eta$ :
\begin{equation}
\label{eq:eta_CAK}
\eta=N_{\text{eff}}\frac{v_{\infty}}{c}
\end{equation}
In this section, we implicitly neglected the contribution of reemitted photons. It is the single-scattering assumption\footnote{In practice, as long as Sobolev's framework holds, it is meaningful to neglect the role of scattered photons whose energy is quickly (\ie after at most a few scattering) too low anyway, especially with respect to the receding ions, to be reabsorbed.}. This is a reasonable hypothesis if there are few enough absorption lines in the wind or if they are spaced enough such as the reemitted photon has little chance to be absorbed. Otherwise, it will contribute by being absorbed by another atomic transition. In this case, $\eta$ can not go over 1. If one wants to relax the single-scattering assumption and exceeds $\eta=1$, proper Monte-Carlo techniques must resorted to in order to follow the path of a photon and its successive scatterings \citep{Vink2000}. Multiple scatterings have little influence on the winds of OB-Supergiants but are important for the winds of Wolf-Rayet stars or Luminous Blue Variables.

For OB-Supergiants, the previous comments enable us to predict a mass outflow between the Coulomb coupling lower-limit and the single-scattering limit :
\begin{equation}
2\cdot 10^{-9} \left( \frac{L}{10^5L_{\odot}} \right) \left( \frac{v}{1000\text{km}\cdot\text{s}^{-1}} \right) < \frac{\dot{M}}{1M_{\odot}\cdot \text{yr}^{-1}} < 2\cdot 10^{-6} \left( \frac{L}{10^5L_{\odot}} \right) \left( \frac{1000\text{km}\cdot\text{s}^{-1}}{v_{\infty}} \right) 
\end{equation}
in solar masses per year. Yet, to get a precise value within this range, we now need to consider the explicit expression of the line-absorption acceleration. We will see that the mass outflows generated by this process are rather on the high limit on the range above.

\begin{figure}
\begin{center}
\includegraphics[height=7cm, width=10cm]{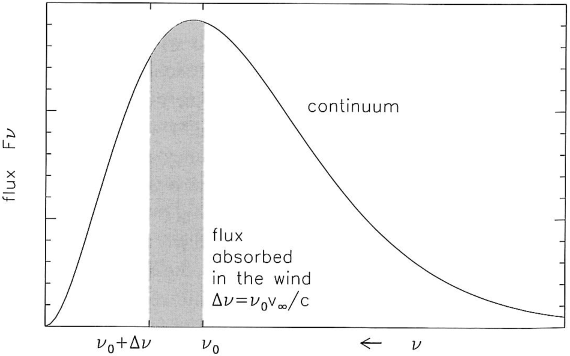}	
\caption{Stellar black body spectrum. The grey area represents the band fully absorbed by the wind given the absorption line at $\nu_0$ at rest and the velocity at infinity $v_{\infty}$ we considered in section \ref{sec:MdotV}. From \cite{Lamers1999}.}
\label{fig:stellar_flux_abs}
\end{center}
\end{figure}


\section{Dynamics of the wind}
\label{sec:dynamics_CAK}


\subsection{Acceleration term}
\label{sec:gCAK}

The derivation of the acceleration term is out of the scope of the present manuscript but can be found either in the seminal papers \citep{Castor1975b} or in more recent publications \citep{Lamers1999}. The Sobolev approximation \citep{Sobolev1960} enables to write the optical depth as a local quantity\footnote{Strictly speaking, integrations along ray tracks should be performed to evaluate the acceleration term - see the computations realized in the comoving frame without the Sobolev approximation.} modulated by the radial velocity gradient\footnote{Due to the aforementioned Doppler shift.}. In the \cak model (for Castor, Abbott \& Klein), after summation over all contributing atomic lines using the convenient force multipliers, we obtain the following force per unit mass \citep{Ud-Doula2002} :
\begin{equation}
\label{eq:gCAK}
g_{\textsc{cak}}=\frac{1}{1-\alpha}\frac{\kappa_e Q L}{4\pi r^2 c} \left( \frac{1}{\rho(r)\kappa_ecQ} \frac{\d v}{\d r} \right)^{\alpha} = \frac{\Gamma GM}{r^2}\frac{Q}{1-\alpha} \left( \frac{1}{\rho(r)\kappa_ecQ} \frac{\d v}{\d r} \right)^{\alpha}
\end{equation}
where $\rho$ is the mass density, $Q$ and $\alpha$ are the force multipliers and the factor in parenthesis is a dimensionless optical depth. We recognize the continuous radiative pressure force $\Gamma GM / r^2$ in the \rhs. The $\alpha$ force multiplier evaluates the relative contribution of strong and weak lines to the acceleration of the wind : for $\alpha=0$, we recover the optically thin case (where absorption decreases as $r^{-2}$) while for $\alpha=1$, all lines are optically thick. The original works substituted to $Q$ the $k$ force multiplier\footnote{$k=\frac{Q^{1-\alpha}}{1-\alpha}\left( \frac{v_{\text{th}}}{c} \right)^{\alpha}\sim\frac{Q^{1-\alpha}}{1-\alpha}\left( \frac{T}{5.5\cdot 10^{12}\text{K}}\right)^{\alpha /2}$.} but the former, introduced by \cite{Gayley1995}, has proved to be more satisfying both for computational and physical reasons. Indeed, $k$ depends on the stellar surface temperature because of the use of the thermal speed at the surface in \cite{Castor1975b} instead of a constant such as the speed of light above. On the other hand, $Q$ is fairly independent of $\alpha$ and the effective stellar temperature. It is a dimensionless effective quality factor which evaluates the global efficiency level of the absorption lines wind acceleration with respect to the continuous radiative pressure on free electrons ; though the latter are way more numerous than electrons bound in metallic ions whose transitions intercept the stellar continuum, $Q$ typically reaches 10$^3$ in OB-Supergiants. There exists an additional force multiplier, $\delta$, introduced by \cite{Abbott1982} to take the non homogeneity of the ionization level in the wind into account. Including it consistently in our binary model would require to also account for the self-ionization of the accretion flow, an effect we discarded for reasons explained in more details in Chapter \ref{chap:SgXB}. Thus, in our steady-state framework, we make the assumption of a frozen-in ionization state ($\delta=0$) whose consequences are discussed in greater details in section 4.1 of \cite{Kudritzki1989a}.

The proper calculation of the force multipliers (like for $N_{\text{eff}}$ in section \ref{sec:MdotV}) requires stellar atmospheric models we will not address here. Their values are provided in tables such as Table 1 of \cite{Gayley1995} or Table 2 of \cite{Abbott1982} (for $\delta$) as a function of the effective stellar temperature, the surface gravity, the stellar metallicity\footnote{The acceleration term is believed to be approximately proportional to the metallicity - since the launching of the wind implies metal ions. See the Figure 10 of \cite{Kudritzki2000} for a visualisation of the systematic bias on the terminal speed of the wind for lower metallicity stars in the Small and Large Magellanic Clouds.} and the photosphere electronic density. For the OB-Supergiant we will focus on in this work, $\alpha$ ranges from 0.45 to 0.65 while $Q$ stays approximately constant around 1,000 \citep{Shimada1994}.


\subsection{Ballistic equation of motion in the point source limit}
\label{sec:motion_CAK}

In a very analogue way as the approach we undertake to study Bondi spherical accretion (section \ref{sec:Bondi_sph}), we write the steady-state hydrodynamical equation of conservation of the mass and linear momentum along the radial axis\footnote{The former is the integrated form while the latter is in its local form.}, but with the two additional radiative acceleration terms\footnote{$g_{\text{Thomson}}$ for the continuous and $g_{\textsc{cak}}$ for the line absorption.} and with $v>0$ :
\begin{equation}
\begin{cases}
4\pi r^2 \rho v = \dot{M} = \text{cst}\\
v\frac{\d v}{\d r} = - \frac{GM}{r^2} + \underbrace{g_{\text{Thomson}}}_{\kappa_e\frac{L/4\pi r^2}{c}} + g_{\textsc{cak}} - \frac{1}{\rho}\frac{\d P}{\d r} = - \frac{GM(1-\Gamma)}{r^2} + g_{\textsc{cak}} - \frac{1}{\rho}\frac{\d P}{\d r}
\end{cases}
\label{eq:CAK_cons_eq}
\end{equation}
where $P$ stands for the thermal pressure\footnote{We saw in section \ref{sec:eddington_lim} that the radiative pressure can be included in the gravity term via the Eddington factor $\Gamma$.}, $M$ for the stellar mass and where we used the expressions of the Eddington luminosity \eqref{eq:LEdd} and the Eddington factor \eqref{eq:GEdd} to rewrite the Thomson scattering radiative pressure force. In an OB-Supergiant, we can safely neglect the thermal pressure since the envelope is radiatively-dominated\footnote{See \cite{Lamers1999} for a quantification of the small influence of the thermal pressure.}. It is equivalent to ignoring the discrepancy between the stellar radius and the sonic point, slightly above. Given the velocity profiles we will obtain, it is clear that the flow becomes supersonic quickly enough after it leaves the stellar surface to not significantly alter the whole velocity profile and the mass loss rate, something we will verify a posteriori. 

\begin{figure}
\begin{center}
\includegraphics[height=5cm, width=10cm]{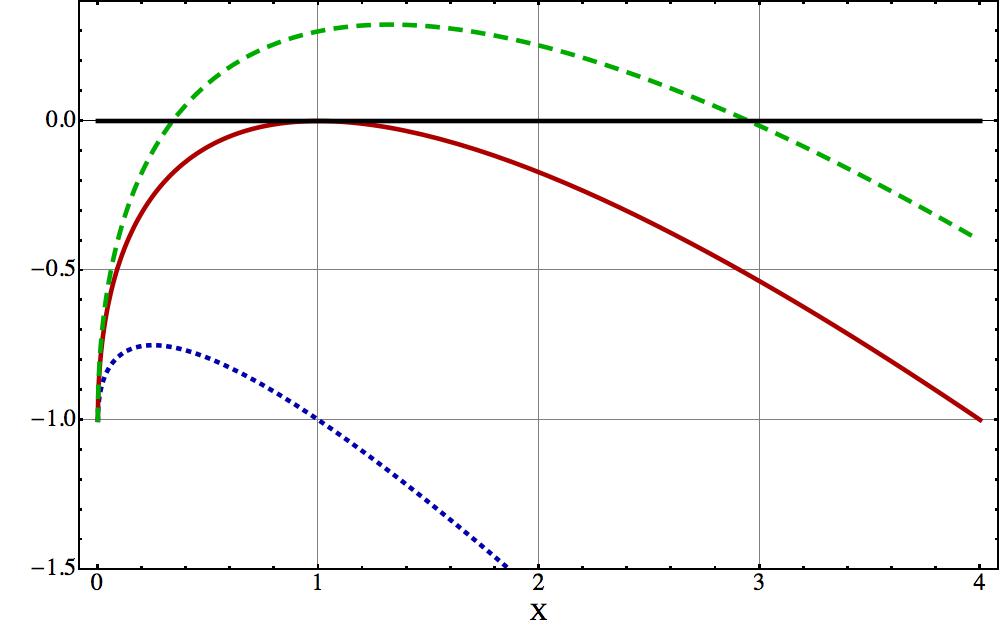}	
\caption{Function given by \eqref{eq:adim_CAK} as a function of the dimensionless quantity $X=vr^2\frac{\d v}{\d r}$. In blue dotted, red solid and green dashed are represented the functions for $\aleph=1$, $2$ and $2.3$ respectively, with $\alpha$ set to 0.5. The thick black horizontal line stands for $y=0$ and its intersection with the 3 curves indicate the roots of \eqref{eq:adim_CAK}.}
\label{fig:critical_point}
\end{center}
\end{figure}

Let us make these equations dimensionless to highlight the decisive parameters at stake in this problem. The natural dimension quantities one can think about are the stellar radius $R$ for length and $\sqrt{GM(1-\Gamma)/R}$ for the velocities. Then, multiplying by $r^2$ and using the conservation of mass above to express the mass density as a function of the mass outflow rate in the conservation of linear momentum, we have the scaled equation of motion :
\begin{equation}
\label{eq:adim_CAK}
X = -1 + \aleph X^{\alpha} \quad \text{with} \quad X=vr^2\frac{\d v}{\d r}
\end{equation}
where the dimensionless parameter $\aleph$ is a constant given by :
\begin{equation}
\label{eq:aleph}
\aleph=\frac{1}{1-\alpha}\left(\frac{\Gamma Q}{1-\Gamma}\right)^{1-\alpha} \underbrace{\left( \frac{L}{\dot{M}c^2} \right)^{\alpha}}_{\sim N_{\text{eff}}^{\alpha}}
\end{equation}
Equation \eqref{eq:adim_CAK} is an equality which can not be verified for any $X$ but one in particular : $X$ is a constant. If we plot the function $f\left( X \right) = \aleph X^{\alpha}-X-1$ as a function of $X$ for a fiducial $\alpha$ and different values of $\aleph$ (Figure\,\ref{fig:critical_point}), it appears that there exists a critical value of $\aleph$ below which there is no $X$ solution for the equation $f\left( X \right)=0 $, above which there are two solutions and for which there is a unique one. Since the mass loss rate in \eqref{eq:aleph} can not be degenerated (\ie bi-valued), it means that $\aleph$ is given by this precise critical value and the corresponding value of $X$ can be found by looking for the extremum of the expression \eqref{eq:adim_CAK} with respect to $X$ :
\begin{equation}
\label{eq:critical_X}
X=\left(\alpha\aleph\right)^{\frac{1}{1-\alpha}}
\end{equation}
This mathematical singularity plays a similar role as the one played by the sonic point in the case of Bondi accretion, and will naturally lead to an evaluation of the mass loss rate in the incoming section \ref{sec:mdot_CAK}. Meanwhile, we head towards an analytic derivation of the velocity profile.

%
%


\subsection{Velocity profile}
\label{sec:vel_prof_CAK}

The velocity profile summoned by observers to model the wind of hot stars, the so-called "$\beta$-wind" \citep[see \eg][]{Wen:1999wg} finds its theoretical justification in the use of the previous results. Indeed, if we inject the expression \eqref{eq:critical_X} of the critical $X$ in \eqref{eq:adim_CAK}, we have :
\begin{equation}
\label{eq:X_crit_al}
X=\frac{\alpha}{1-\alpha}
\end{equation}
which can be integrated from the stellar surface where the velocity can be taken to be zero up to infinity to first determine the dimensioned terminal speed :
\begin{equation}
\label{eq:vel_inf_CAK}
v_{\infty}=v_{\text{esc,m}}\sqrt{\frac{\alpha}{1-\alpha}}
\end{equation}
where $v_{\text{esc,m}}$ is the modified escape velocity, $\sqrt{2GM(1-\Gamma)/R}$. We then determine the velocity profile :
\begin{equation}
\label{eq:vel_prof}
v(r)=v_{\infty}\left( 1-\frac{R}{r} \right)^{0.5}
\end{equation}
The ratio of the modified escape speed by the thermal speed of the flow at the stellar surface is given approximately by :
\begin{equation}
\left(\frac{v_{\text{esc,m}}}{c_{\text{s}}}\right)^2\sim 1,000 \left( \frac{M}{15M_{\odot}} \right) \left( \frac{\mu}{1} \right) \left( \frac{1-\Gamma}{1-0.1} \right) \left( \frac{R}{18R_{\odot}} \right)^{-1} \left( \frac{T}{20\text{kK}} \right)^{-1} 
\end{equation}
which shows that, given the fiducial velocity profile plotted in Figure\,\ref{fig:vel_prof}, the flow quickly becomes supersonic as it rises above the stellar surface, which washes out the influence of the thermal pressure in \eqref{eq:CAK_cons_eq}. More generally, the exponent $0.5$ which appears in \eqref{eq:vel_prof} along with the terminal speed $v_{\infty}$ are usually fitted to observational data\footnote{See \cite{Kudritzki2000} and \cite{Puls2008} for more information about the way we derive the "observed" wind properties.} as degrees of freedom - the former being written $\beta$, hence the $\beta$-wind profile. The resulting values and trends are $\beta$ exponents above 0.5 and larger terminal speeds and evolving with $\alpha/(1-\alpha)$ instead - see \eg Table 3 in \cite{Crowther2006} or Table 8 in \cite{Searle2008} for B Supergiants. Refinements such as the one presented in section \ref{sec:fd} help to bridge the gap between the predicted and observed values.

\begin{figure}
\begin{center}
\includegraphics[height=7cm, width=10cm]{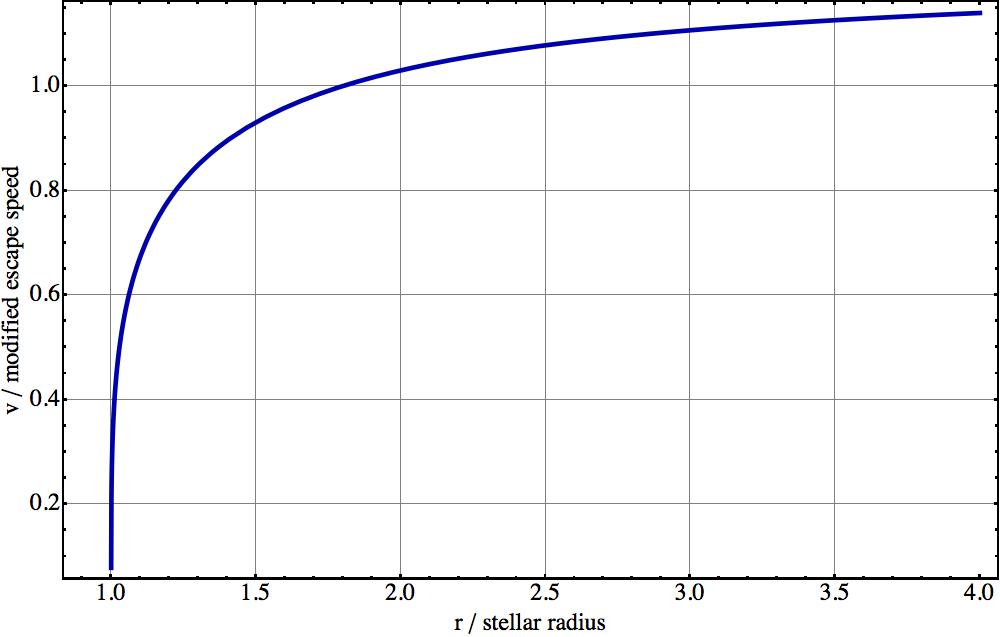}	
\caption{$\beta$-wind velocity profile for $\alpha=0.6$ and $\beta=0.5$ (\ie the theoretical expectation for the simple most \cak model).}
\label{fig:vel_prof}
\end{center}
\end{figure}


\subsection{Mass loss rate and terminal velocity}
\label{sec:mdot_CAK}

\begin{figure}
\begin{center}
\includegraphics[height=8cm, width=12cm]{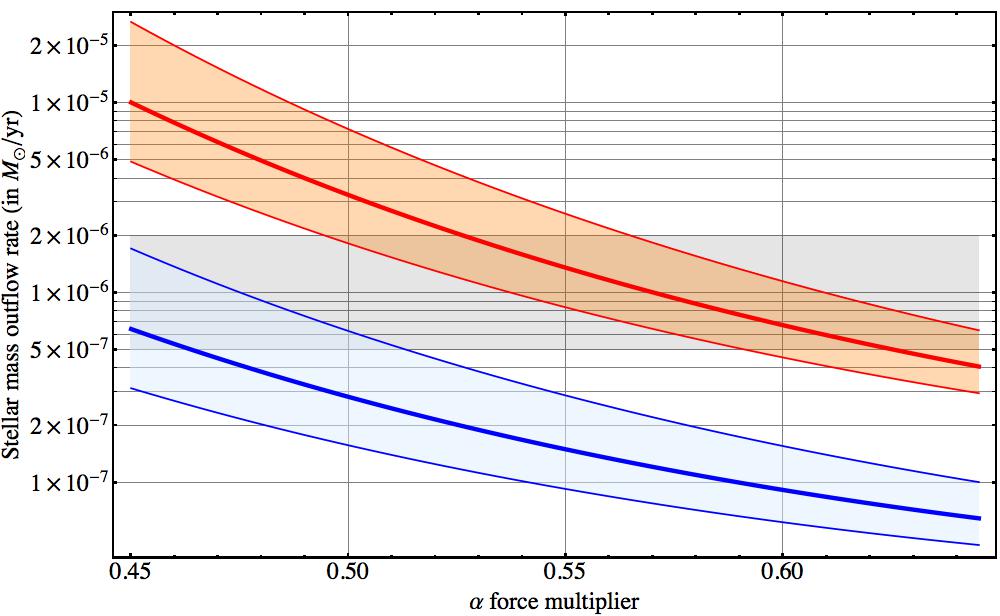}	
\caption{Stellar mass outflow as a function of the $\alpha$ force multiplier. In thick red (on top) is represented the case for $\Gamma=0.3$ and $Q=900$ ; the light red filled area around is for $Q$ ranging from 500 (lower thin red curve) to 2000 (upper thin red curve). Same thing in blue in the lower part of the diagram but for $\Gamma=0.1$. The stellar mass (to compute the scale, the Eddington luminosity) is set to 20\msun and we used \eqref{eq:LEdd}. The grey area locates the usual range of mass loss rates observed for the stellar companion in \sgx such as Vela X-1, XTE J1855-026 or IGR J18027-2016 \citep{VanLoon2001,Searle2008,Gimenez-Garcia2016}.}
\label{fig:Mdot}
\end{center}
\end{figure}

Accurate mass-loss rates are decisive elements to better appreciate the evolutionary tracks of massive stars. It is also tells us more about the feedback on the interstellar medium \citep{Coleiro2013a}. We make use of the singularity which forces the $\aleph$ parameter given by \eqref{eq:aleph} to a specific value to compute the mass loss rate of the \cak model in the point source limit. It gives the mass loss rate $\dot{M}$ as a function of the stellar luminosity $L$ and of several dimensionless parameters : its Eddington factor $\Gamma$, the $\alpha$ force multiplier and the $Q$ force multiplier.
\begin{equation}
\label{eq:Mdot_MCAK}
\dot{M} = \frac{L_{\text{Edd}}}{c^2} \frac{\alpha}{1-\alpha} \Gamma \left( \frac{\Gamma Q}{1-\Gamma} \right)^{\frac{1-\alpha}{\alpha}} 
\end{equation}
where $c$ is the speed of light and $L_{\text{Edd}}$ is the Eddington luminosity ; the latter is preferred to the stellar luminosity as a normalization quantity since within a given range of stellar masses, it varies more slowly than the former (typically by a factor at most 3 given the range of masses we consider). A slightly more realistic mass loss rate but with similar dependences, $\dot{M}'$, is represented in Figure\,\ref{fig:Mdot} and given by 50\% of \eqref{eq:Mdot_MCAK} (see the incoming section \ref{sec:fd}). It shows that optically thicker winds gives much smaller mass loss rates, almost a decade smaller for $\alpha$ going from $0.45$ to $0.55$. Higher stellar luminosities for a given stellar mass also favour larger mass loss rates, with mass outflows approximately 10 times more important for $\Gamma=0.3$ than for $\Gamma=0.1$ (for $\alpha=0.5$). Eventually, the $Q$ force multiplier, likely to lie between 500 and 2000, accounts for a variation of a factor $4^{\frac{1-\alpha}{\alpha}}\sim 4$. Given the range of mass outflows we observe for stellar companions in \sgx, this sketch is consistent with the force multipliers determined by \cite{Shimada1994} or \cite{Gayley1995} for OB-Supergiants.

\subsection{Correction for finite cone angle effect}
\label{sec:fd}

\begin{figure}
\begin{center}
\def\svgwidth{400pt} 
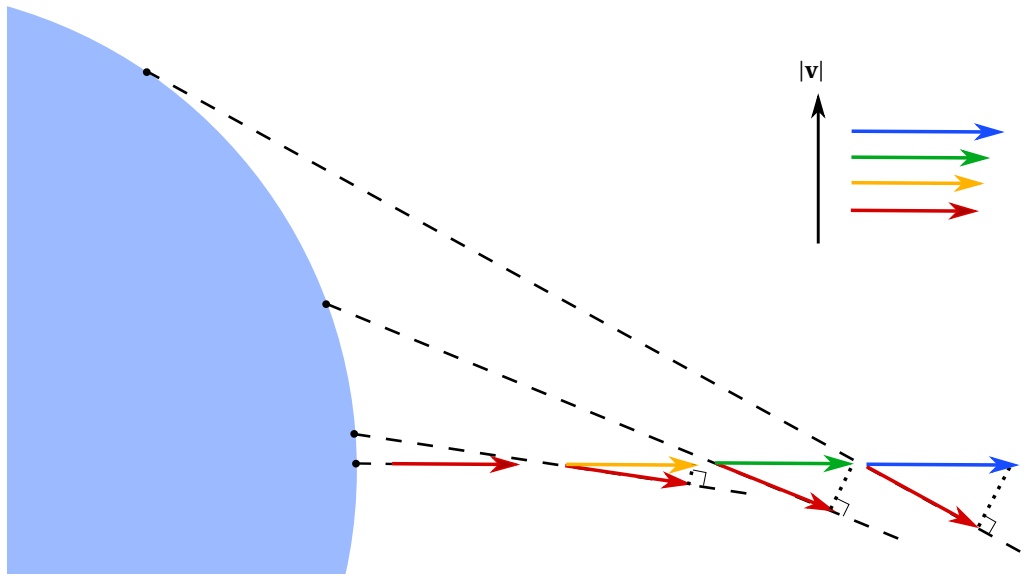
\caption{Sketch to illustrate the capacity of non-radial light rays to contribute to the wind acceleration. Photons of larger energy (associated to the velocities in red) than the one being absorbed radially at a given radius (successive colors) can still be absorbed provided they come from higher inclinations. Only the upper half of the star has been represented but the situation is obviously symmetric so there is no deviation from the radial direction.}
\label{fig:fd}
\end{center}
\end{figure}

\begin{figure}
\begin{center}
\includegraphics[height=8cm, width=10cm]{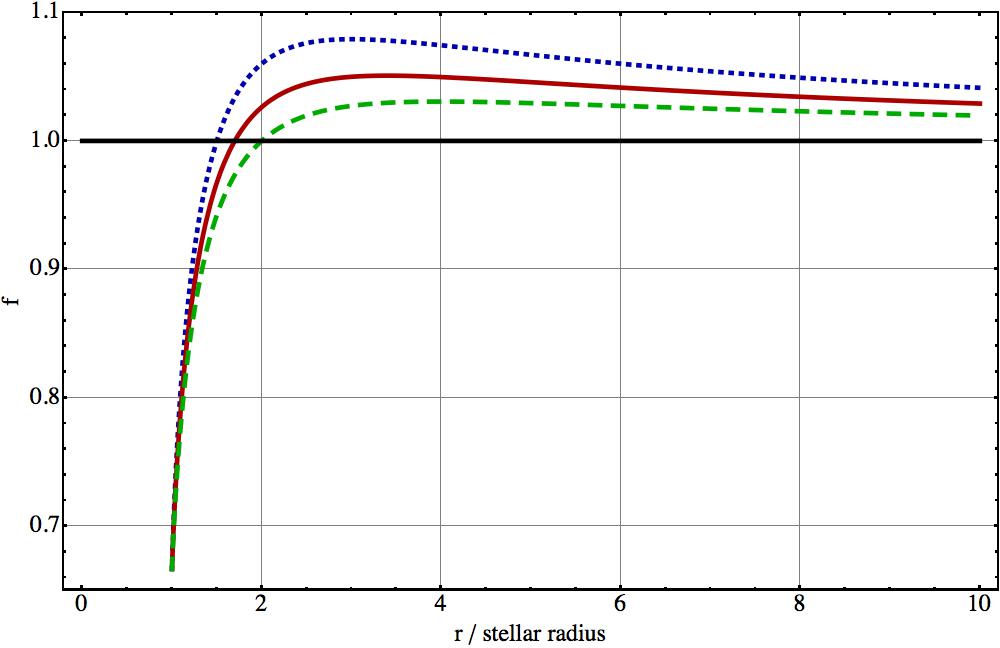}	
\caption{Finite cone angle factor $D$ computed from a $\beta$-wind velocity profile as a function of the reduced radius. From top to bottom, the exponent of the $\beta$-wind law is 0.5, 0.7 and 1. The $\alpha$ force multiplier is set to 0.5 and determines the initial value of the curves at the stellar surface, $1/(1+\alpha)$.}
\label{fig:fd_factor}
\end{center}
\end{figure}

The previously summarized results can be conveniently analytically derived because we worked with the simplified form \eqref{eq:gCAK} of the line acceleration term ensued from the point-source approximation : all the photons absorbed were moving radially outwards. However, in the vicinity of the stellar surface where most of the acceleration takes place, the star can hardly be approximated by a point and a significant fraction of the photons absorbed do not move radially. In Figure\,\ref{fig:fd}, one can see that photons at a given frequency can be absorbed at different distances provided they come from more inclined regions and thus, sees the same projected relative velocity (the vectors in red in Figure\,\ref{fig:fd}) ; it is another feature not accounted for in the bedrock model and which does alter the velocity profile. A precise quantification of this cone angle effect (\aka finite disk effect) shows that the new line acceleration term $g_{\text{fd}}$ is given by \citep{Pauldrach1986,Friend1986} :
\begin{equation}
g_{\text{fd}}=g_{\textsc{cak}}\times D\left( \sigma, \mu ; \alpha \right)
\end{equation}
where the correction factor $D$ is given by \citep{Lamers1999} :
\begin{equation}
\label{eq:fd_factor}
D\left( \sigma, \mu ; \alpha \right) = \frac{\left(1+\sigma\right)^{\alpha+1}-\left( 1+\sigma \mu^2\right)^{\alpha+1}}{\left( 1-\mu^2 \right) \left( 1+\alpha \right) \sigma \left( 1+\sigma \right)^{\alpha} }
\end{equation}
with $\sigma$ and $\mu$ defined as follow :
\begin{equation}
\begin{cases}
\sigma\left( r,v,\frac{\d v}{\d r}\right)=\frac{r}{v}\frac{\d v}{\d r}-1\\
\mu^2\left( r \right)=1-\left(\frac{1}{r}\right)^2
\end{cases}
\end{equation}
It turns out that the inclusion of this refinement does not alter the shape of the velocity profile too much, which is now a $\beta$-law profile with an exponent $\beta$ not necessarily equal to 0.5. The evolution of this factor $D$ as a function of the radius for $\beta$-law velocity profiles with different exponents is shown in Figure\,\ref{fig:fd_factor}. We retrieve $D=1$ at large distance, where the star does look like a point-source, and $D$ amounts to $1/(1+\alpha)<1$ at the stellar surface.

Semi-analytical models and numerical computations have been developed to fully take this effect into account \citep{Friend1986,Villata1992,Muller2008,Araya2014a,Noebauer2015} and provide self-consistent estimates of the mass loss rate and the terminal speed of the wind. They all agree within a level of precision sufficient for our model of \sgx on a halved mass outflow $\dot{M}' \sim0.5\dot{M}$ and a terminal speed given by :
\begin{equation}
\label{eq:vel_inf_mod}
v_{\infty}\sim 2.5 \frac{\alpha}{1-\alpha} v_{\text{esc,m}}
\end{equation}
instead of \eqref{eq:vel_inf_CAK}. It must be noticed that this result holds for stellar effective temperatures above the bi-stability jump threshold, of the order of 20,000K \citep{Vink1999,Kudritzki2000}, typically below the temperatures we observe \sgx. More precisely, we will see later on, in section \ref{sec:num_scheme}, that in models where the sonic surface and the stellar surface are at the same level, the mass accretion rate can be analytically corrected for the finite cone effect by writing :
\begin{equation}
\dot{M}'=\left( \frac{1}{1+\alpha} \right)^{1/\alpha} \dot{M}
\end{equation}
For the values of $\alpha$ we explore, it does correspond to a factor of almost 50\%. It is the mass outflow we consider by default from now on. 

\subsection{Wind momentum - Luminosity relationship}
\label{sec:WLR}
In the 90's, \cite{Kudritzki1994} and \cite{Puls1996} emphasized a relation between the modified wind momentum and the stellar luminosity, the so-called Wind momentum Luminosity Relationship (\wlr). Once properly calibrated, this relation could prove to be a reliable extra-galactic distance indicator by using purely spectroscopic information \citep{Puls2008}. The modified wind momentum $D_{\text{mom}}$ is given by :
\begin{equation}
\label{eq:WLR}
D_{\text{mom}}=\dot{M}'v_{\infty}\sqrt{\frac{R}{R_{\odot}}}\sim \left( \frac{1}{1+\alpha} \right)^{\frac{1}{\alpha}} \frac{2.5\sqrt{2}}{c^2\sqrt{R_{\odot}}} \left( \frac{\kappa_e}{4\pi c} \right)^{\frac{1-\alpha}{\alpha}} \left( \frac{\alpha}{1-\alpha} \right)^2 \left[GM\left(1-\Gamma\right)\right]^{\frac{3}{2}-\frac{1}{\alpha}} Q^{\frac{1-\alpha}{\alpha}} L^{\frac{1}{\alpha}} 
\end{equation}
where we used the modified expression of the mass loss rate given by \eqref{eq:Mdot_MCAK} and \eqref{eq:vel_inf_mod} for the terminal speed. We wrote $D_{\text{mom}}$ so as to highlight the primary role of the $\alpha$ force multiplier to set the strength of the dependence between $D_{\text{mom}}$ and the different parameters. For realistic values of $\alpha$, $D_{\text{mom}}$ depends first on the stellar luminosity and then, on the $Q$ force multiplier ; the mass of the star and its Eddington parameter have little influence on the modified wind momentum.

As an illustration, we compared this relationship for different values of $\alpha$ to observational data for early-type B Supergiants from \cite{Searle2008}, believed to be isolated analogues of the stellar companions we will study in more details in \sgx - see Figure 15 by \cite{Searle2008} for similar plots. A dependence on the spectral type of the \wlr has been shown \citep[][Figure 8]{Kudritzki1999} but we will focus our study on the more numerous massive Supergiants, the early B-type ones. Including the $\delta$ force multiplier results in an effective $\alpha$ which is given by $\alpha-\delta$, where $\delta$ is approximately equal to 0.1. The values of $\alpha$ found in the literature by fitting observational data - Table 6 in \cite{Kudritzki1999} or Table 2 in \cite{Kudritzki2000} - then match the force multipliers derived by \cite{Shimada1994}, albeit marginally\footnote{The large relative uncertainties on the fitted $\alpha$ for early B-type Supergiants make any conclusive statement hazardous but it seems that the $\alpha$ derived from the \wlr are somewhat larger than the ones found from atmospheric models, although accounting for the $\delta$ force multiplier lowers this discrepancy.} : for early type B Supergiants, $\alpha$ ranges from 0.45 to 0.55 with $Q\sim 900$.

\begin{figure}
\begin{center}
\includegraphics[height=8cm, width=12cm]{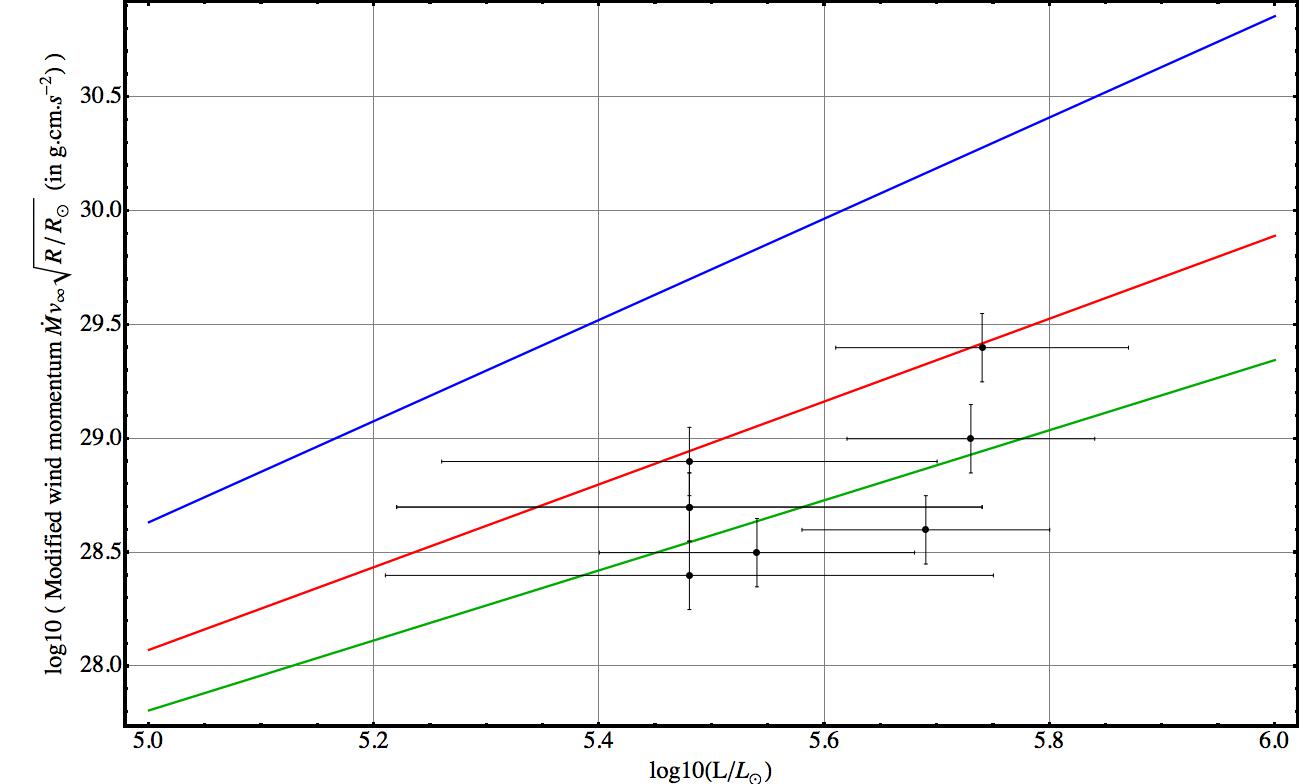}	
\caption{Wind momentum - Luminosity Relationship (WLR). The logarithm of the modified wind momentum rate $D_{\text{mom}}$ has been represented as a function of the luminosity. The black dots stand for 8 Galactic B-type supergiants with a stellar temperature above the bi-stability jump \citep{Vink1999} \ie the early-type B Supergiants observed by \cite{Searle2008}. The error bar on the modified wind momentum are set to 0.15dec, following the prescription by \cite{Kudritzki1999}. The lines indicate the predicted modified wind momentum \eqref{eq:WLR} with, from top to bottom, $\alpha=0.45$, 0.55 and 0.65. $Q$ was set to 900, the stellar mass to 15\msun and the Eddington factor to 20\% - with little if any influence of the two lattest parameters.}
\label{fig:WLR}
\end{center}
\end{figure}


\section{Conclusion}

This section has been devoted to the introduction of a simple model to represent the wind of an isolated hot supergiant star. First, we presented the simplest form of the \cak theory so as to guide the reader in the realm of analytically solvable equations and derived the qualitative trends which still hold for more sophisticated versions. Then, we succinctly introduced an important refinement to bridge the quantitative gap between observed and predicted terminal speeds with the cone angle effect taken into account. We must finally acknowledge the abundance of works concerning more realistic wind models. In short, the following questions are still under investigation within the community devoted to a better understanding of those winds albeit not necessarily in relation with \sgx :

\begin{enumerate}
\item \underline{Multiple-scattering}. In a configuration where the mass outflow exceeds the single-scattering limit, photons are scattered more than once and provide an additional momentum to the wind \citep{Vink1999}. The subsequent enhancement of the mass loss rate is significant for Wolf-Rayet and possibly for Hypergiant stars such Wray 977, the stellar companion in the hypertrophied \sgx GX301-2 \citep{Leahy2008,Servillat2014}.
\item \underline{Wind-blanketing}. When multiple-scattering comes into play, it becomes possible for the radiative energy to be scattered back to the star and heat the photosphere. The wind acts as a blanket then and profoundly alters the thermal structure of the upper photospheric layers. 
\item \underline{Clumpiness}. Line-driven instabilities have been revealed since the very first papers on winds of hot stars \citep{Lucy1970}. In the 80's, the effect has been quantified by \cite{Owocki1984} and \cite{Owocki1985}. Numerical simulations have brought some insight on the structure of a clumpy wind \citep{Dessart2005} and models have been developed, including in the case of \sgx \citep{Ducci2009}.
\item \underline{Porosity}. It is possible that the clumpiness is large enough in some winds that the clumps are optically thick and most of the mass lost is contained in the clumps. In this case, since observations are mostly sensible to the clumps, assuming a homogeneous wind would lead to overestimated mass loss rates. The altered coupling between matter and radiation can be quantified relying on porosity approaches \citep{Owocki2004}.
\item \underline{$\delta$ force multiplier}. The non homogeneity of the ionization state of the flow can be modelled in a first approximation with the inclusion of an additional force multiplier, $\delta$. In the case of \sgx where the ionizing flux from the compact object will dominate, it is likely to not be a good approximation to account for $\delta$ and discard the radiative feedback of the accreted flow. 
\item \underline{Stellar rotation}. Massive stars start their life as fast rotators which breaks up the assumption of a purely radial and isotropic flow around the star \citep{Puls2008}. In Be-\textsc{xb} where the donor star is fastly rotating, this has to be taken into account to justify the formation of a decretion disk around the star.
\item \underline{Stellar magnetic field}. Since it plays a role in the structure of the properties of the stellar surface, critical for the launching of the wind, the stellar magnetic field must have an influence on the properties of the wind.
\end{enumerate}

Our intention is not to provide a thorough description of winds of isolated hot stars but to encompass enough of their essential properties to explain the observed trends in \sgx. In this respect, we did not include those effects in the current model we introduce in Chapter \ref{chap:SgXB}. We can now rely on this semi-analytical approach to implement a numerical configuration where a low mass compact object orbits a massive Supergiant star without altering the launching mechanism of its wind.

\setlength{\parskip}{0ex} 


\chapter{Binary systems}
\label{chap:roche}
\chaptermark{Binary systems}
\hypersetup{linkcolor=black}
\minitoc
\hypersetup{linkcolor=red}
\setlength{\parskip}{1ex} 

Following the Jeans criterion for gravitational collapse of gas cloud in the interstellar medium and naively applying a conservation of angular momentum argument such as in \ref{sec:obj} leads to excessively large radii for the subsequently formed proto-star. A workaround which has been suggested to overcome the centrifugal barrier responsible for these radii is to split the initial angular momentum between different bodies orbiting each other ; most of the angular momentum goes into the orbital one, with, in addition, the spins of the fragments. This process could explain the large if not major fraction of systems composed of two (or more) stellar mass objects \citep[see][for a recent review on stellar multiplicity frequency and dependences]{Duchene2013}, making the solar system an outlier. In addition, it is now seriously considered that the multiplicity is positively correlated to the initial mass of the stars : the more massive the more pressing the need for a partner to distribute angular momentum into two spins and an orbital one \citep{Mason2009,Sana2012}. Outnumbering the isolated stars, the binary systems also play a decisive role in shaping their host Galaxy, providing the energy, the metal-enriched gas and sometimes the trigger for the upcoming generations of stars. We will not tackle the question of the formation of multiple star systems but more information can be found in the section 3.1 of A. Coleiro's PhD \citep{Coleiro2013a} and references therein.

\begin{figure}
\begin{center}
\includegraphics[height=12cm, width=10cm]{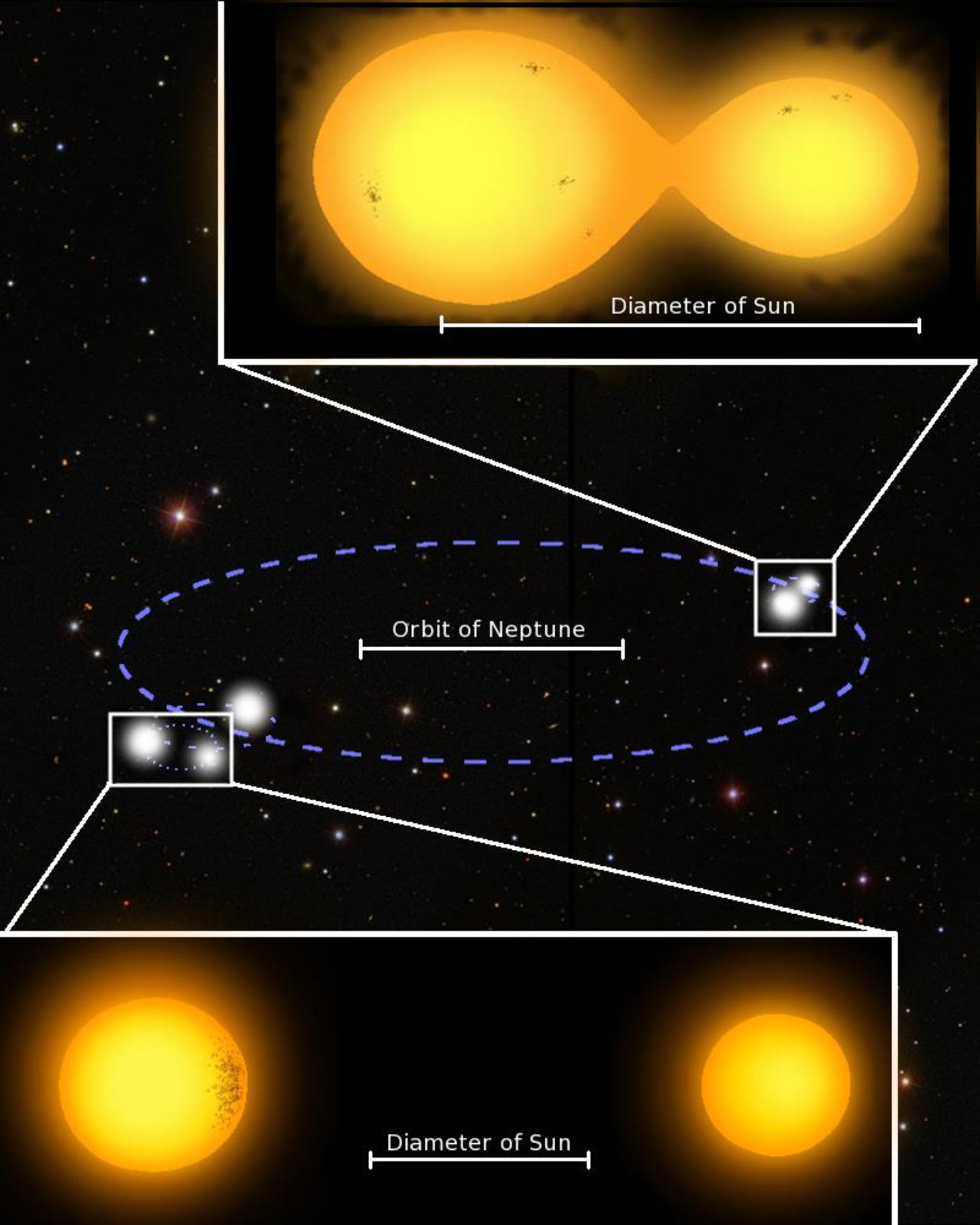}	
\caption{An artist’s impression of the quintuple star system J093010 \citep{Lohr2015}. The blue dotted line marks the orbital path of the centers of mass of the two pairs of stars. The fifth star, whose position is uncertain, is to the right of the left pair. If the two binaries, with an angular separation of a few arcseconds, could be resolved \citep[based on the resolution reached by][Figure 1]{Schutz2011}, each of the two orbits (not to scale relative to the large one on the main caption) are too small to be imaged.}
\label{fig:artist_pic_mult}
\end{center}
\end{figure}

If the time evolution of isolated stars is now well constrained, their combined evolution in binary systems remain largely speculative given the plentiful possible evolutionary tracks. However, the interplay between the two bodies we witness in binary systems gives us the occasion to reach extraordinary levels of precisions when it comes to stellar parameters\footnote{Asteroseismology has also proven to be a game changer over the last decade, in particular thanks to the \textsc{CoRoT} satellite (for Convection, Rotation and planetary Transits).} \citep[see][for an example of impressive precision on the stellar parameters]{Bakos2010}. The wobbling motion of the two stars as they orbit each other translates into periodic time-modulation not only of the spectral lines but also of the light curve. The more huddled and edge-on the system, the larger the amplitude of this modulation. Eclipsing binaries\footnote{See paragraph 10.1.1. and Figure 12 in \cite{Kirk2016} for a roadmap to compute the correction completeness factor due to the geometrical probability of transit and a comparison to the available catalogs.} enable us to constrain inclinations angles and thus, orbital elements \citep[see \eg][]{Heminiak2009}. Coupled to the precise study of the transit shapes, this knowledge provides at least lower limits on the stellar radius, at best the granulation \citep{Bastien2013} or the spot activity \citep{Sanchis-Ojeda2012} at the surface. The stellar spins themselves can be figured out using spectroscopic high time resolution monitoring of the most luminous star as its companion passes by \citep{Albrecht2007}, opening the door to statistical data on spin-orbit misalignment and providing insightful constrains on the tidal interactions between the two bodies. The previously unseen bolometric precision of instruments such as the Kepler satellite also makes possible photometric measures which were, up to now, only feasible in a spectroscopic approach \citep{Faigler2011} : the mass functions deduced from spectroscopic and photometric measures can now be compared so as to tell us more about and correct for the specific biases of each method. A handful of other techniques (transit-timing variations, Roemer delay, etc) has been successfully applied to the study of binary systems and now provides a harvest of data which, brought together, outline possible trends about stellar evolution in binary systems.

\begin{wrapfigure}{r}{0.5\textwidth}
\begin{center}
\includegraphics[height=6.75cm, width=0.95\textwidth]{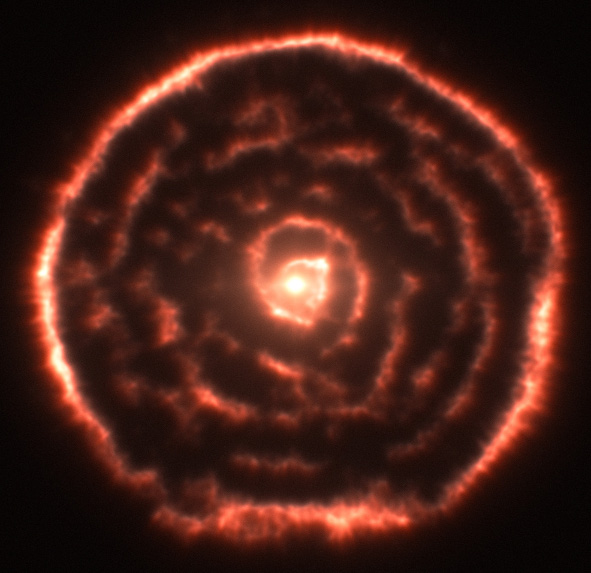}	
\caption{Spiral nebula of gas and dust winding outward from the giant red star R Sculptoris. The spiral shock, clearly visible at the center, points towards an orbital motion of the asymptotic giant branch star due to an underlying companion \citep{Maercker2012}. An empirical model to fit the spiral structure  can be found in \cite{Homan2015}. Credits : Alma/ESO/NAOJ/NRAO.}
\label{fig:scluptoris}
\end{center}
\end{wrapfigure}

But the picture is still incomplete. If our statistical knowledge of those systems has been considerably broadened by large scale surveys\footnote{The Kepler satellite has been continually monitoring more than 145,000 point sources during the last 7 years.}, we remain unable to fully appreciate the evolutionary singularities which may account for the different spectral and time-variability found in X-ray binaries, an evolved stage of binary systems. Sophisticated stellar models draw conclusions for the evolution of isolated stars which still not fit the bill for binaries : for example, discrepancies between evolutionary and spectroscopic masses remain. Historically, the Algol paradox illustrates the specificities of stellar evolution in a binary system, which sometimes involves mass transfer (see section \ref{sec:massTransfer}). The mutual irradiation of the two stars\footnote{More generally, for a founding reference on light curve distortions due to the binarity (mutual illumination, Doppler boosting and ellipsoidal light variations essentially), the reader may consult the comprehensive book by \cite{Kopal1978}. It shows how the three aforementioned modulations can be disentangled using the basis property of the sinusoidal functions with different frequencies.}, clearly observed on bolometric light curves, can also alter the properties of the outer layers and the stellar mass-loss rates. In this sense, X-ray binaries offer a snapshot of a short-lived phase of binary systems where the intertwining of the two bodies displays a rich and complex behavior we have to investigate. The understanding of stellar evolution will not be satisfying as long as this ephemerous yet determining phase remains unclear. Conversely, a better appreciation of this stage is a cornerstone to understand one of the following ones, involving two close-in orbiting compact objects ; the observational masterstroke accomplished (twice!) earlier this year by the \textsc{ligo}\footnote{For Laser Interferometer Gravitational-Wave Observatory.} and Virgo collaborations \citep{Abbott2016,Abbott2016a} has just ushered in a gold rush to study those systems from a privileged and previously unseen angle.


In a first part, we remind the reader about the theoretical framework of binary systems and in particular the Roche potential as a guideline to appreciate the trajectory of test-masses when the Coriolis force can be neglected. Then, we go on with a few reminders about the effects of mass transfer over secular time scales\footnote{Which refers to stellar evolutionary time scales in this manuscript, of the order of the nuclear timescale.} and use them to justify the bimodality in the observed distribution of stellar mass in X-ray binaries. 


\section{The Roche model}
\label{sec:roche_model}

\subsection{Context}
The Roche model has been first designed by the XIX$^{\text{th}}$ century astronomer Edouard Roche. It takes place in a rotating framework at constant angular speed  vector $\boldsymbol{\Omega}$ where two point masses, henceforth indexed 1 and 2, produce a gravitational field (see Figure\,\ref{fig:roche_config}). The center of mass CM is used as the origin of the framework. So as to relate this model to celestial mechanics of an isolated system of two bodies, the distance between those points is not left at the physicist's discretion but is set by Kepler's third law. The latter is deduced from the conservation of the angular momentum vector which guarantees to the system a constant orbital plane and an orbital separation $a$ set by :
\begin{equation}
a^3\Omega ^2 = G(M_1+M_2)
\end{equation}
where $M_{\text{i}}$ refers to the mass of the point i. 


\begin{figure*}
\begin{subfigure}{0.55\columnwidth}
  \centering
  \hspace*{-1cm}
 \includegraphics[width=0.95\columnwidth]{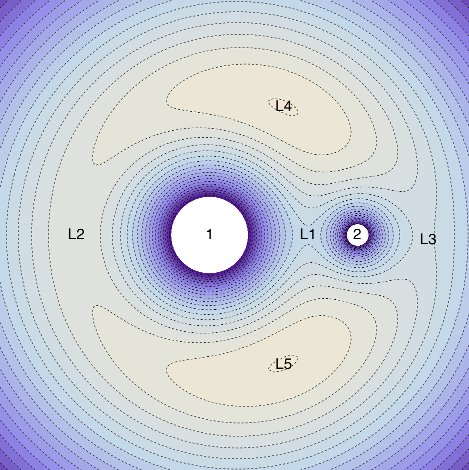}
\end{subfigure}%
\begin{subfigure}{0.55\columnwidth}
  \centering
    \hspace*{-1.5cm}
 \includegraphics[width=0.95\columnwidth]{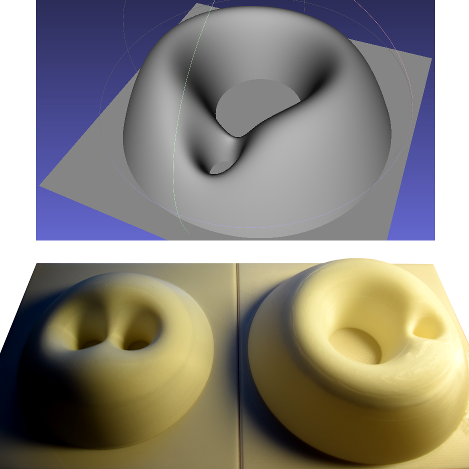}
\end{subfigure}
\caption{(\textit{left}) Logarithmic colormap of the Roche model with the two point-masses 1 and 2. The mass ratio $q=M_1/M_2$ is set to 5. The white parts have been truncated. From dark purple to light beige, the algebraic value of the potential rises (and so do the altitude of the associated surface represented on the right). The 8 shaped critical isopotential is clearly visible and delimits the two regions of gravitational influence, the Roche lobes. (\textit{upper right}) 3D surface ($q=10$). This .stl file, which has been produced from a fine mesh sampling with Mathematica and has been processed with the \href{http://meshlab.sourceforge.net}{Meshlab software}, is the standard input format for 3D printers. (\textit{lower right}) Two 3D printed surface for $q=1$ (left) and $q=10$ (right). Each 3D printed surface is made of successively disposed layers of plastic which naturally traces the contours of the potential and emphasizes the equilibrium points.}
\label{fig:roche_map}
\end{figure*}

\begin{figure}
\begin{center}
\includegraphics[height=6cm, width=10cm]{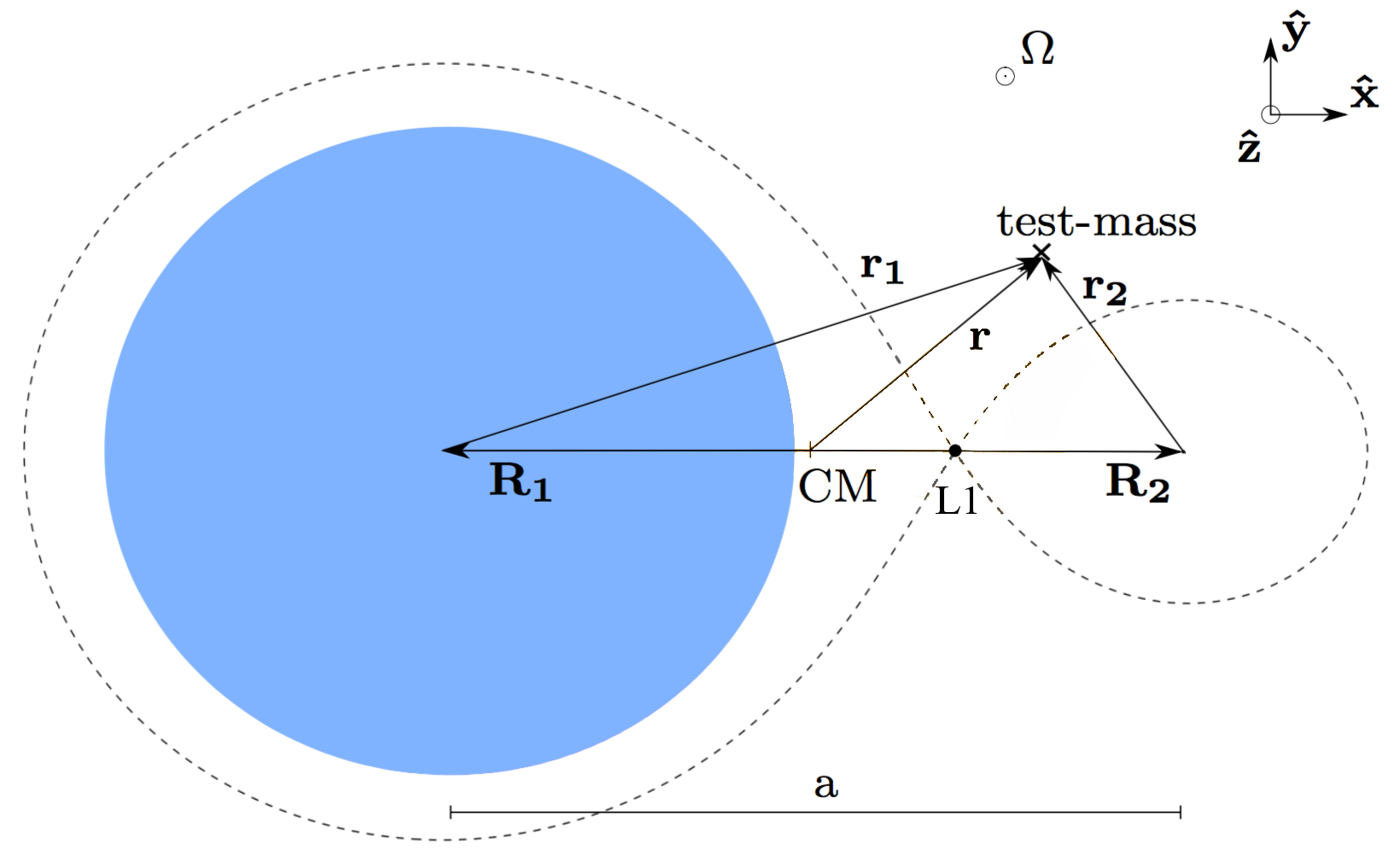}	
\caption{Orbital slice of the \sgx configuration we consider with, in blue, the supergiant stellar companion and on the right, the compact object. The eight-shaped black dotted line is the Roche surface passing through the first Lagrangian point, L$_1$. The latter has been represented with the black dot at the crossing of the critical isopotential. $a$ is the orbital separation. The different vectors used to express the Roche potential have been represented, with the center of mass CM as the origin of the framework whose basis can be found in the upper right corner. For the sake of clarity, the test-mass belongs to the orbital plane here.}
\label{fig:roche_config}
\end{center}
\end{figure}


\subsection{The Roche potential}

\subsubsection{Expression}

Now, let us consider a test-mass \ie a point of mass $m\ll M_1, M_2$ so it does not produce any significant gravitational field. It experiences the two gravitational forces from 1 and 2, along with the two non-inertial forces\footnote{Notice that if $1$ and $2$ are on non circular orbits (eccentric ones or even on free trajectories), this formalism still applies provided we relax the constant angular speed assumption, adding a new fictitious force, the so-called Euler force. The latter case is of particular interest when it comes \eg to tidal disruption events (see \ref{sec:tde}).} :
\begin{equation}
\label{eq:basicBall}
m\frac{d\mathbf{v}}{dt}=-\frac{GM_1m}{|\mathbf{r}-\mathbf{r_1}|^3}\left(\mathbf{r}-\mathbf{r_1}\right) - \frac{GM_2m}{|\mathbf{r}-\mathbf{r_2}|^3}\left(\mathbf{r}-\mathbf{r_2}\right) + m\Omega^2 \mathbf{r_{\bot}} - 2m\boldsymbol{\Omega}\wedge\mathbf{v}
\end{equation}
where $\mathbf{v}$ is the velocity vector of the test mass in the co-rotating frame and $\mathbf{r_{\bot}}$ is the position vector of the projection of the test-mass on the orbital plane : $\mathbf{r_{\bot}}=\mathbf{r}-\left(\mathbf{r}\cdot \mathbf{\hat{z}}\right)\mathbf{\hat{z}}$. The vectors $\mathbf{r_1}$ and $\mathbf{r_2}$ locate the two masses with respect to the center of mass. It clearly appears in this equation that the motion is independent of the mass of the test-mass (provided $m\ll M_1, M_2$). It turns out that the 3 first forces are conservative \ie they can be written via a potential\footnote{One can always write the force as a spatial gradient if there is a first integral of the motion, which is guaranteed if the time coordinate does not show up by other means than through the spatial component.} called the Roche potential :
\begin{equation}
\label{eq:roche}
\Phi_{\text{R}}\left( \mathbf{r} \right)=-\frac{GM_1}{|\mathbf{r}-\mathbf{r_1}|} - \frac{GM_2}{|\mathbf{r}-\mathbf{r_2}|} - \frac{1}{2} \Omega^2 r_{\bot}^2
\end{equation}
which is not the case of the Coriolis force, even if the latter can not be called dissipative either since it does no work. An important point to make concerning this potential is that it no longer displays the convenient isotropy of an isolated point mass. As a consequence, the angular momentum of the test-mass has no reason to be constant, which can sound paradoxical since the system composed of 1 \& 2 keeps the same angular momentum and the whole system is isolated. This physical artefact is to be attributed to the assumption $m\ll M_1, M_2$ : when the test-mass separates from 1, strictly speaking, it modifies $M_1$ and, in doing so, the orbital separation and the distribution of angular momentum. This point can no longer be put aside when one considers the secular evolution of a binary system and a dynamical Roche potential must be considered (see section \ref{sec:massTransfer}).
 

If the norm of the angular momentum vector of a test-mass can change, its direction remains colinear to the angular speed vector $\boldsymbol{\Omega}$ provided its initial position and velocity vectors belongs to the orbital plane. The motion of such a test-mass, trapped in the orbital plane, can be appreciated with the analogy of a rolling marble on a 2D surface in a homogeneous gravity field, as long as the Coriolis force plays a minor role. The vertical motion of the marble then traces the potential while its horizontal motion follows the actual motion of the test-mass in the orbital plane. Let us discuss the shape and the amplitude of this surface, represented in Figure\,\ref{fig:roche_map}.

\subsubsection{The scale and the shape}

This restricted three body problem\footnote{Concerning the actual three body problem, its formal resolution has made a major leap forward in the early XX$^{\text{th}}$ century with Sundman's work but unfortunately, the serial expression of the solution converges so slowly that it is of little use for computational purpose.} first seems to require the prior of three parameters, typically the two masses and the orbital separation. Yet, equation \eqref{eq:roche} can be adimensioned in order to differentiate the shape from the scale parameters :
\begin{equation}
\Phi_{\text{R}}\left( \mathbf{\tilde{r}} \right)=-\Phi_{\text{R,0}} \cdot \left[ \frac{q}{\left| \mathbf{\tilde{r}} - \mathbf{\tilde{r}_1} \right|} + \frac{1}{\left| \mathbf{\tilde{r}} - \mathbf{\tilde{r}_2} \right|} + \frac{1}{2} (1+q) \tilde{r}^2 \right]  
\end{equation}
where the shape of second factor on the \rhs is entirely determined by the mass ratio and with :
\begin{equation}
\begin{cases}
\Phi_{\text{R,0}}=\frac{GM_2}{a}\\
\mathbf{\tilde{r}_1}=-\frac{1}{1+q}\mathbf{\hat{x}}\\
\mathbf{\tilde{r}_2}=\frac{q}{1+q}\mathbf{\hat{x}}
\end{cases}
\end{equation}
The tildes stand for the length normalized to the orbital separation $a$. A change in the orbital separation then only leads to a stretching of the potential but not to a different shape : the shape is scale invariant. Once the length scale and the potential scale have been set at $a$ and $\Phi_{\text{R,0}}$ respectively, the corresponding dimensionless equation of motion is given by :
\begin{equation}
\frac{d\mathbf{v}}{dt} = - \boldsymbol{\nabla}\Phi_{\text{R}} - 2\sqrt{1+q}\cdot\mathbf{\hat{z}}\wedge\mathbf{v}
\end{equation}
where the tildes have been omitted and the velocity has been scaled with the square root of $\Phi_{\text{R,0}}$. We notice that even the Coriolis force depends only on the mass ratio such as in the end, the shape of the trajectories remains unchanged by a modification of the scale. It only depends on the dimensionless quantity $q$. 

Such a distinction between scale and shape parameters look straightforward on this academic example but will be extended in the more complicated case where the wind launching of the Supergiant star is included in the next Chapter. The key element to grasp with this separation is that it is an essential prerequisite before running any numerical computation. Indeed, the scale parameters intervene only as a posteriori multiplicative factors of numerical outputs since the latter do not depend on the physical units. Though, what is physically a problem with 3 degrees of freedom is numerically a problem with only one degree of freedom, the scaling being performed afterwhile in a much less computationally-demanding stage. A similar reduction of the parameter space dimensionality will not only be a decisive computational time-saving feature but also a gateway to physical interpretations rather than plain descriptions of the results.


\subsection{Trajectories}

\subsubsection{Numerical solutions}
We limit our analysis to the motion in the orbital plane. The Roche potential now serves as a background to investigate the possible trajectories followed by a test-mass launched from any initial position $\mathbf{r_0}=(x,y)$ with the initial speed $\mathbf{v_0}$ The equation of motion for the test-mass once projected is now given by :
\begin{equation}
\label{eq:RocheSystemBallEq}
\begin{cases}
\ddot{x}=-\frac{\partial V}{\partial x}+2\Omega y\\
\ddot{y}=-\frac{\partial V}{\partial y}-2\Omega x
\end{cases}
\end{equation}
This set of non-linear coupled equations can be integrated using a 4$^{\text{th}}$ order Runge-Kutta\footnote{Since the mechanical energy of the test-mass is conserved in this non-dissipative problem, it can be used as a probe to evaluate the accuracy of the numerical integration. See the famous Lotka-Volterra prey-predator model for a very similar mathematical and numerical framework where a first integral of the motion can also be determined.} or a symplectic integrator. A \href{http://demonstrations.wolfram.com/TrajectoryOfATestMassInARochePotential/}{Mathematica applet} has been developed, in association with 3D printed Roche potentials\footnote{Those two models, for a mass ratio of 1 and 10, have been printed thanks to a 3D printer funded by the \textit{r\'egion Ile-de-France} via a \textit{Domaine d'Int\'er\^et Majeur en Astrophysique et Conditions d'Apparition de la Vie} (DIM ACAV) project. Hubert Halloin validated the printing, Jean-Luc Robert provided a funding for the use and material and Marco Agnan was responsible for the technical support and realization.}, to illustrate the profusion of trajectories one can obtain in the orbital plane from this very simple model.

\subsubsection{The Lagrangian points}
\label{sec:lagrange}
The Roche potential has 5 equilibrium points, the Lagrangian points. Along the line joining the two bodies ($y=0$), it can be shown that there are 3 equilibrium points : one between the two bodies, L$_1$, and one on each side, L$_2$ and L$_3$. Once the appropriate change of variables are made in \eqref{eq:RocheSystemBallEq}, it can also be shown that this system admits two lateral points of equilibrium, L$_4$ and L$_5$. Those 5 points have been represented in Figure\,\ref{fig:roche_map} : L$_4$ and L$_5$ are maxima of potential while L$_1$, L$_2$ and L$_3$ are saddle points. The equilibrium study of those points requires the proper consideration of the Coriolis force. The linearly perturbed expression of \eqref{eq:RocheSystemBallEq} around the equilibrium points leads to coupled systems of equations whose Jacobian contains the eigenvalues and the eigenvectors of the problem\footnote{\cite{Murray1999} provides more details about this computation but see \cite{Strogatz} for a pedagogical introduction to this kind of analysis - with the Figure A.2 in \cite{Murray2002} as a crystal clear summary.}. The latter provide information on the shape and period of the linearly perturbed orbits. Sign considerations of the eigenvalues for L$_1$, L$_2$ and L$_3$ show that they are unstable equilibria but L$_4$ and L$_5$ turn out to be stable. Actually, a whole arch from a point to another is stable and admits confined trajectories, the horseshoe orbits. This statement led people to consider the possibility to form and maintain a circumbinary ring of matter in common envelope symbiotic binaries - the quasi-dynamical mass transfer described in \cite{Podsiadlowski2010}.

An important feature to keep in mind and illustrated by the applet mentioned above, since the system is Hamiltonian, the mechanical energy of the test-mass is conserved ; in other words, the isopotentials are also isovelocity curves. Thus, if a test-mass is left within a Roche lobe (where the isopotentials are closed around lower potential regions) with an initial velocity too low to exit the Roche lobe (\ie with an initial kinetic energy below the absolute value of the potential at the first Lagrangian point), the test-mass is trapped within its Roche lobe. The singular case where a test-mass is left at the first Lagrangian point with a negligible velocity will be the guideline of the incoming discussion on Roche-lobe overflow. On the contrary, if a test-mass has (or acquires, see previous Chapter) an initial velocity large enough, it can not remain bounded to the system and will necessarily run away. In case of a flow, shocks can dissipate energy and modify this analysis.


\subsection{The spatially extended star in the ideal Roche potential}

In this section, unless explicitly stated, the tidal forces are supposed to have fully circularized the orbits and brought the stellar component we consider into sychronous co-rotation (see \ref{sec:tidal_interactions}).

\subsubsection{The Roche lobes}

\begin{figure}
\begin{center}
\includegraphics[height=7cm, width=10cm]{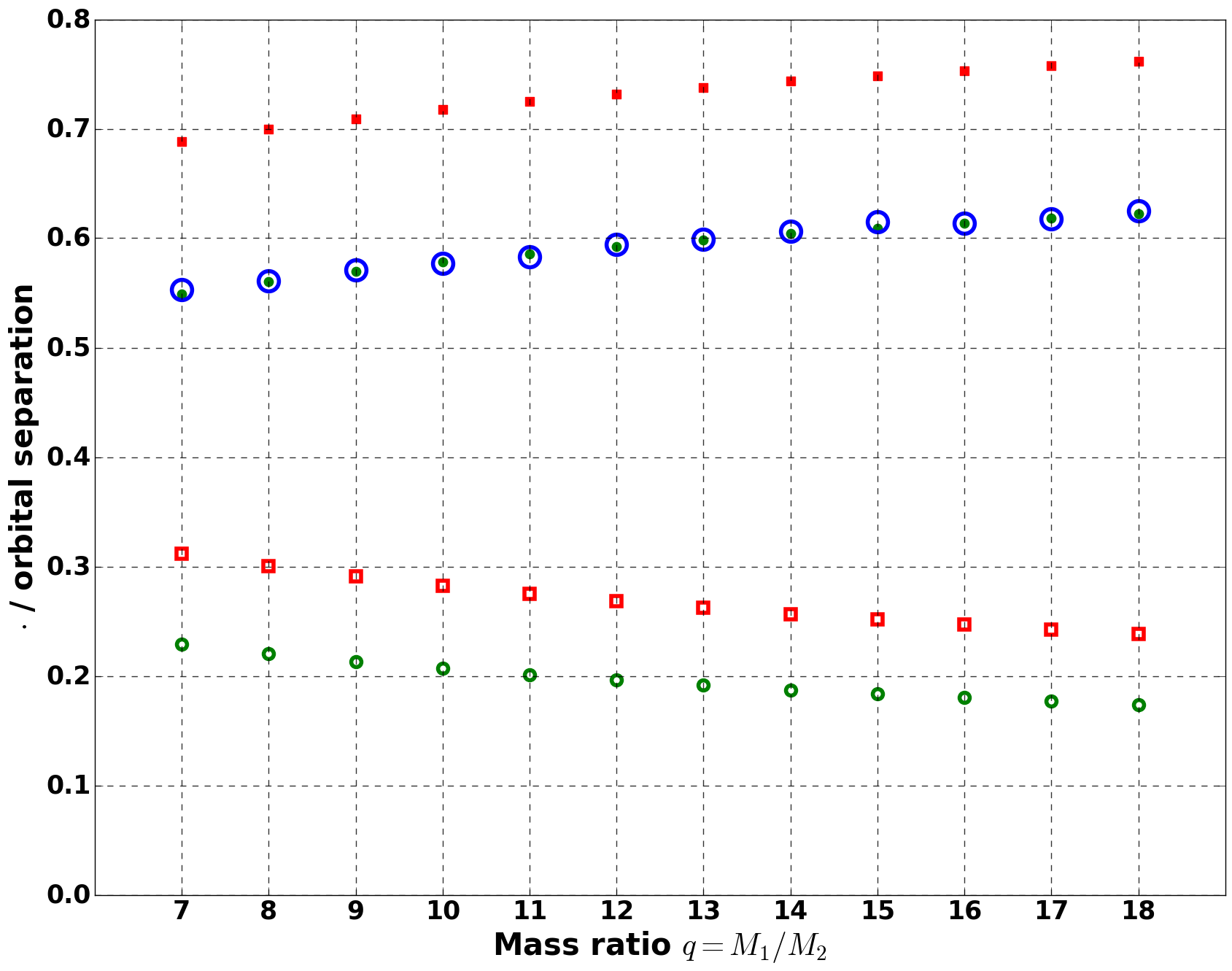}	
\caption{The red filled squares are the computed distances from the stellar center to the first Lagrangian point, $D_{1,L_1}$, as a function of the mass ratio (1 indexing the star and 2 the compact object). They have been computed with a Newton-Raphson algortihm which provided results such as the relation at stake matched with a relative precision of $10^{-10}$. The green discs stand for the radius of the Roche lobe according to Eggleton's formula \eqref{eq:egg} while the blue circles around them are Monte-Carlo computed radii of the Roche lobe. The two sets match within less than a percent. The red empty squares and the green empty circles are the corresponding values for the accreting body. All quantities have been normalized to the orbital separation.}
\label{fig:roche_sizes}
\end{center}
\end{figure}

The first Lagrangian point, L$_1$, belongs to a critical 8-shaped isopotential which splits the inner region in two closed sections : the Roche lobes. We define the radius of the Roche lobe on the side of the body $i$, $R_{\text{R,}i}$ as the radius of the sphere with the corresponding volume. Due to the aforementioned exclusive dependence of the Roche potential shape on $q$ when the orbital separation $a$ is used as a length scale, we already know that the function $\mathcal{E}=R_{\text{R,}i}/a$ depends only on the mass ratio. The expression of $\mathcal{E}$ is unknown but can be approximated to better than 1\% over the whole range of $q$ values with \citep{Eggleton1983} :
\begin{equation}
\mathcal{E}(q)=\frac{0.49q^{2/3}}{0.6q^{2/3}+\ln{(1+q^{1/3})} }
\label{eq:egg}
\end{equation}
This function, along with the normalized distance of the stellar center to the first Lagrangian point, has been represented in Figure\,\ref{fig:roche_sizes} for the range of mass ratios we will focus on in this work. The reader might find an equivalent Figure spanning several orders-of-magnitude in $q$ in Figure 1 of \cite{Tout1991}. When one has to account for non-synchronous stellar rotation, see \cite{Avni1976} for a modified formula. An extension of this expression has also been suggested to account for the radiation pressure which lowers the effective gravity of high luminosity stars \citep{Dermine2009}.

\subsubsection{The hydrostatic Roche problem}
In order to explain the departure of the stellar surface from spherical shape, we write the hydrostatic equivalent of \eqref{eq:basicBall} :
\begin{equation}
\mathbf{0} = -\frac{1}{\rho}\boldsymbol{\nabla}P -\boldsymbol{\nabla}\Phi_{\text{R}} 
\end{equation}
where the first \rhs term is the pressure force and the velocity of the flow is zero since the flow is hydrostatic within the co-rotating frame (which presupposes synchronous rotation, see section \ref{sec:tidal_interactions}). According to this equation, isobares, which define the fluid extension and by then, the stellar extension, coincide with isopotentials ; any closed contour delimiting a domain of lower potential are possible stellar surfaces and vice versa.

\subsubsection{The filling factor}

\begin{figure}
\begin{center}
\includegraphics[height=7cm, width=10cm]{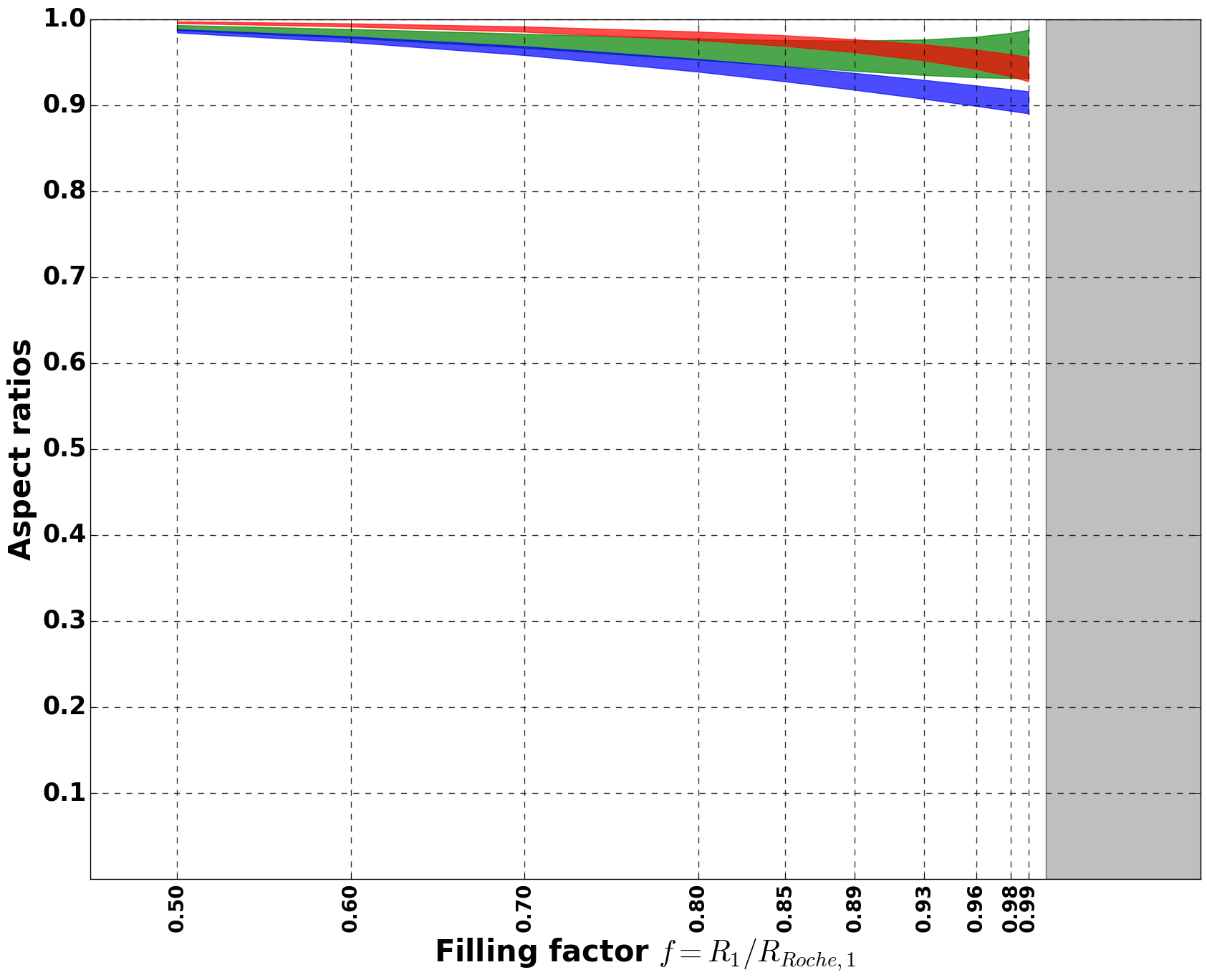}	
\caption{Three aspect ratios ($\Delta_1$ in red, $\Delta_2$ in green and $\Delta_3$ in blue) quantifying the departure of the stellar surface from the spherical shape as the filling factor rises (with 1 as upper value). The vertical extension of each curve represents the range of value for an aspect ratio within the range we consider \ie from 7 to 18 (with 7 corresponding to the lower bound of the red stripe and to the upper bounds of the blue and green ones).}
\label{fig:aspect_ratios}
\end{center}
\end{figure}

   The filling factor $f$ is then the ratio of the radius of the star by the one of its Roche lobe :
\begin{equation}
\label{eq:fillingFactor}
f=\frac{R_1}{a\mathcal{E}(q)}
\end{equation}   
It will be a key dimensionless quantity later on to study the motion of the wind at the orbital scale. Already, an interesting feature is the deformation of the stellar surface with respect to the spherical shape one expects in the case of an isolated non-rotating star, since the isopotentials are not spherically-shaped. We quantify it by introducing three aspect ratios $\Delta_1$, $\Delta_2$ and $\Delta_3$ computed as follow. For $\Delta_1$, we evaluate the potential at the point $(x_1,R_1,0)$ with $x_1$ the position of the star with respect to the center of mass (with $x_1<0$) and $R_1$ the stellar radius. We then look along the x-axis for the 2 points within the stellar Roche lobe with the same potential (using a Newton-Raphson solver). Since the extension of the sphere of radius $R_1$ along the y-axis is always enclosed within the Roche lobe, even for large filling factors, those 2 points always exist. The ratio of twice the stellar radius by the distance between those two points gives $\Delta_1$. It represents the egg-shape deformation of the isopotentials in the orbital plane. Since the extension of the sphere of radius $R_1$ along the z-axis can go beyond the Roche lobe for large filling factors, we can no longer rely on the same method to directly compare the extensions along the z and x axis. Consequently, we first compare the extensions along the y and z axis using the same method as above but looking for a point along the z axis passing through the stellar center instead of the x axis. The ratio of this coordinate by the stellar radius provides $\Delta_2$. Eventually, $\Delta_3$ which compares the extensions along the x and z axis is obtained by the ratio of $\Delta_2$ and $\Delta_1$. All of them are below one and are represented in Figure\,\ref{fig:aspect_ratios} in red, blue and green in the order of rising index as a function of the filling factor and the mass ratio. 
   
\begin{figure}
\begin{center}
\includegraphics[height=5cm, width=10cm]{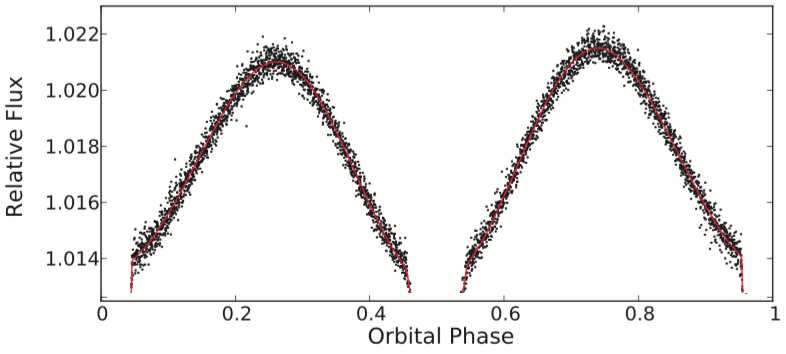}	
\caption{Folded bolometric light curve of KOI (for Kepler Object of Interest) 1224 at the orbital period of approximately 2.7 days (black dots). The out-of-eclipse behavior of this short-period system hosting a distorted low-mass star orbited by a hot white dwarf, displays a clear enhancement of the luminosity, the ellipsoidal light variation. It marks the non-sphericity of the stellar surface. In this waveband, the star is approximately 25 times more luminous than its compact companion. From \cite{Breton2012}.}
\label{fig:ELV}
\end{center}
\end{figure}

\begin{figure}
\begin{center}
\includegraphics[height=5cm, width=10cm]{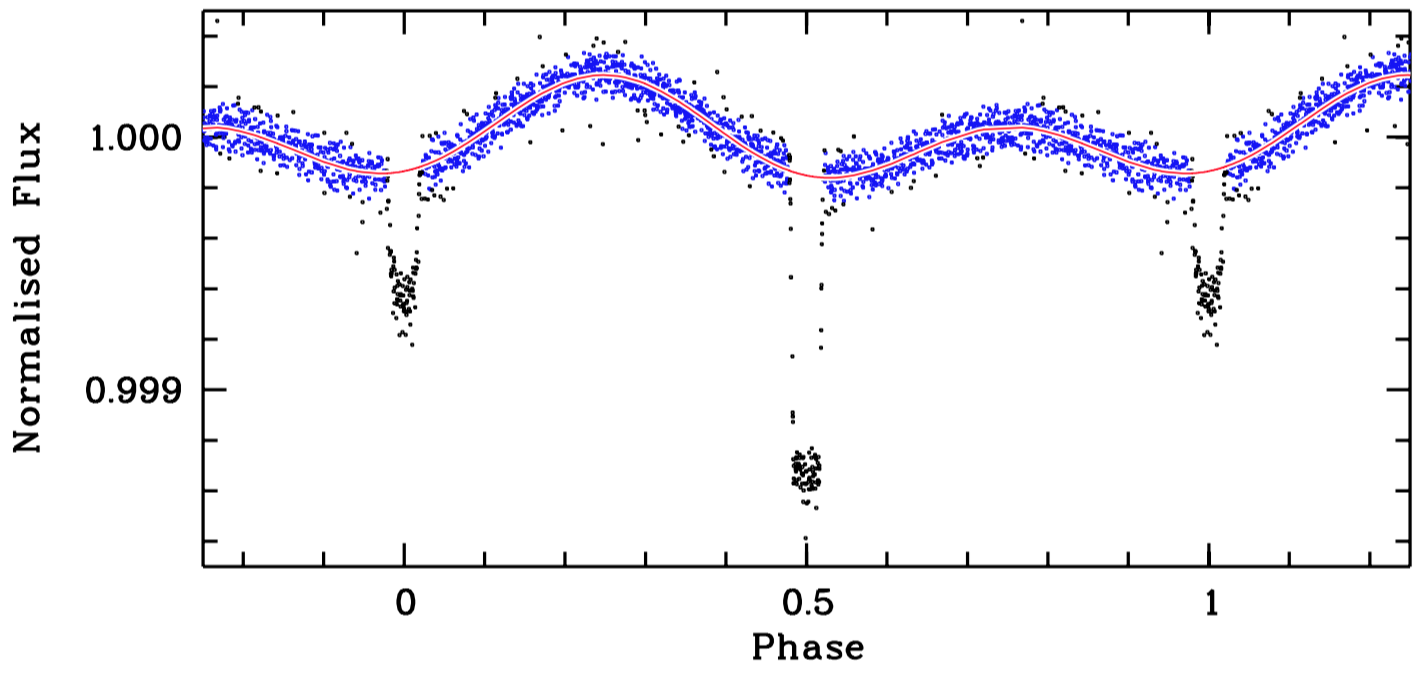}	
\caption{Folded bolometric light curve of KOI 74 at the orbital period of approximately 5.2 days (blue dots). The out-of-eclipse behaviour not only displays the ellipsoidal light variation at twice the orbital frequency but also an asymmetry between the light enhancement before and after the primary eclipse. This additional light is to be understood as the Doppler beaming of the stellar light due to the special relativistic geometrical Doppler effect (\aka Doppler boosting). In this waveband, the star is expected to be 500 times more luminous than its white dwarf companion. From \cite{VanKerkwijk2010}.}
\label{fig:DB}
\end{center}
\end{figure}

   Such an egg-shape deformation entails a variation in the light observed from the system due to the enlarged cross-section\footnote{Strictly speaking, to quantify this effect, one also wants to account for the non uniform temperature distribution at the surface, with a colder photosphere further away from the stellar center.} when the system is in quadrature with respect to the eclipsing (or transiting) configuration\footnote{See \cite{Wilson1994} for three dimensional representations of the stellar surface at different orbital phases.} (Figure\,\ref{fig:ELV}). Another classical observational feature of close-in binaries is the Doppler beaming - when the star approaches - and fainting - when it recedes - of the stellar light due to the special relativistic light aberration, \aka Doppler boosting (Figure\,\ref{fig:DB}). In the case of high luminosity ratios of the two bodies within a given waveband, the instrumental precision we now reach in some ranges opens the door to photometric measures of radial velocity. 

\section{Mass transfer}
\label{sec:massTransfer}

The need for a theory of mass transfer in binary systems was first triggered by the puzzling observation of the Algol binary\footnote{It is actually a hierarchical triple system.} (\aka $\beta$ Persei), an eclipsing system historically called "The Demon Star" due to its light variability. Algol consists of a 3.7$M_{\odot}$ main sequence star and a 0.8$M_{\odot}$ red giant. According to the theory of stellar evolution, the more massive a star the faster its evolution, but the red giant stage comes after the main sequence. The Algol paradox can be resolved provided we account for a possible mass transfer in the past, which brings up the burning question of the specificities of stellar evolution when they are not isolated. To better appreciate the role played by mass transfer, we now describe it in more details, from its triggering to the different regimes we must expect, with a special emphasis on the case where the accretor is a compact object. To keep things simple, we assume that the two bodies are on circular orbits and remain so as matter is transferred (see section \ref{sec:circ_orb} for a discussion of circularisation in binary systems).


\subsection{Secular evolution of the filling factor}

Through this section, we study the reasons which can lead an initially detached binary system \footnote{\ie two bodies underfilling their Roche lobes (that is to say their filling factors are below 1).} to become the stage of a mass transfer. According to \eqref{eq:fillingFactor}, it takes either an expansion of the stellar radius or a shrinking of the orbital separation, the Eggleton factor being essentially constant for slow changes of the mass ratio which will thus be considered as constant in this section. As a consequence of the instability of the first Lagrangian point, midway between the two bodies, mass transfer will take place once the filling factor reaches unity ; this regime of mass transfer is called Roche lobe overflow (\rlof) but keep in mind that it tells something about the geometry of the star relatively to its Roche lobe more than about the mass transfer per se\footnote{See section \ref{sec:struct_acc_flow} where Roche lobe underfilling Supergiant stars can transfer mass in a way which is a reminiscent of \rlof, the stream-dominated regime.}.

The proper treatment of the growth of the stellar radius as times flow on secular scales require sophisticated models out of the scope of the present manuscript\footnote{The \href{http://mesa.sourceforge.net}{\texttt{mesa} code} provides a very comprehensive toolkits to handle stellar evolution. It has become the main scientific stellar evolution code and its proactive community brings maintenance, collaborative improvements and technical/scientific support. A \href{http://mesa-web.asu.edu}{Web-base interface} also exists for immediate hands-on use.}. Let us first frame a binary system hosting two Roche lobe underfilling stars and whose orbital period would be constant. As the stars evolve, their radii grow such as the filling factor of, \eg, the star indexed 1 and called the primary can reach unity. If it happens, stellar matter gets close enough from the first Lagrangian point to leak into the Roche lobe of the secondary star (section \ref{sec:RLOF}). We will see in the next section \ref{sec:consMassTrans} that this regime can be stable if the mass of the donor is lower than the one of the accretor (usually a neutron star in the Supergiant X-ray binaries we will focus on in the next Chapter). 




The filling factor can also reach unity due to a shrinking of the orbital separation and not to the evolution of the stellar radius. On the Corbet diagram of Figure\,\ref{fig:corbet_diag}, one of the systems represented, 4U 2206+543, hosts a main sequence star. In absence of \rlof mass transfer, we will mention in the next section how the orbital separation can still shrink as angular momentum is lost.


\subsection{Stability of the mass transfer}
\label{sec:stab_mass_trans}

The reader will find a comprehensive and detailed discussion of the secular evolution of X-ray binaries in the review by \cite{VandenHeuvel2009}. We remind here the main results which lead to two separated populations of X-ray binaries. To do so, we first investigate the stability of the mass transfer and then go on to deduce the main evolutionary tracks which drives a binary system into emitting large amount of X-rays. 

\subsubsection{Conservative mass transfer}
\label{sec:consMassTrans}

The archetype of the conservative mass transfer is the \rlof described in section \ref{sec:RLOF} : the mass lost by the donor 1 (with a rate $\dot{M}_1<0$) is entirely captured by the accretor 2. The angular momentum $\mathbf{L}$ of the system composed of the two orbiting bodies in the inertial frame is given by :

\begin{equation}
\mathbf{L}=\underbrace{\mathbf{L_1^{\text{orb}}}+\mathbf{L_2^{\text{orb}}}}_{M_1M_2\sqrt{Ga/(M_1+M_2)}}+\mathbf{L_1^{\text{spin}}}+\mathbf{L_2^{\text{spin}}}
\end{equation}
where "orb" refers to the orbital angular momentum whose magnitude has been expressed while "spin" refers to the self-rotation of the two bodies. If we neglect the latter ones\footnote{For a star in synchronous rotation, it can be justified by the small stellar radius compared to the orbital separation (this ratio being squared). The neutron star rotates much faster than its orbital speed but with a spin period larger than the millisecond and nominal parameters of a \sgx, one can easily show that its spin angular momentum can also be neglected compared to its orbital one, mostly due to its tiny size.}, we can write the conservation of the total mass and angular momentum of the system and get the rate of change of the orbital separation \citep[see \eg][and references therein for more details]{DSouza:2005jx} :
\begin{equation}
\label{eq:adot}
\frac{\dot{a}}{a}=-2\frac{(-\dot{L})}{L}+2\frac{(-\dot{M}_1)}{M_1}(1-q)
\end{equation}
where the loss of angular momentum ($\dot{L}<0$) of the system can be due to stellar winds or magnetic torques\footnote{Or to an emission of gravitational waves, significant for short-period compact binaries such as the historical PSR 1913+16.} \citep{Schatzman1962a}. During a stage of significant conservative mass transfer between the two bodies such as a \rlof, the change in the orbital separation due to the loss of angular momentum can be neglected. In this case, this expression shows that a transfer of matter from the lighter component to the heavier body ($q<1$) results, on secular time scales, in a widening of the orbital separation. If the transfer of matter occurs from the heavier to the lighter component ($q>1$), the orbital separation shrinks until possibly the mass ratio goes below 1. The local relation above can be integrated to get the analytical net change of the orbital separation for a given set of initial and final masses. We can then differentiate \eqref{eq:egg} to get the rate of change of the radius of the Roche lobe as a function of $q$, $\dot{M}_1$ and $M_1$. On the other hand, the rate of change of the donor radius can also be deduced, provided a mass-radius relation\footnote{The classical polytrope model suggests different values for $n=d(\log R)/d(\log M)$ depending on the dominant physical processes at stake : for instance, 0.5 to 1 for a high (radiative envelope) to low (convective envelope) mass star on the main sequence and $-1/3$ for an evolved fully convective star.}. By comparing the evolution of the Roche lobe and stellar radius as the mass is transferred, we can distinguish two main regimes :

\begin{enumerate}
\item either the filling factor lowers as the mass transfer happens. This negative feedback loop guarantees a slow evolution since mass is supplied on a rate controlled by the secular expansion of the donor.
\item or the filling factor (which was already of the order of unity) gets higher as the mass is transferred and then, the system evolves on the short timescale associated to the stellar equilibrium.
\end{enumerate} 

One can derive a critical value of the mass ratio, $q_0$, function of the polytropic index of the donor, of its mass and of its mass loss rate\footnote{And of the orbital angular momentum and its loss rate if it has not been neglected.}, which sets a threshold between those two regimes : for values of $q$ below (resp. above) $q_0$, the mass transfer is stable (resp. unstable). The exact value of $q_0$ depends on the model but is of the order of 1.

A first conclusion we can draw from this analysis is that large mass ratio systems are not likely to be observed in a \rlof configuration given the instability of the mass transfer : such a system would quickly find itself in a common envelope phase\footnote{This scenario is favoured if the donor has a convective envelope at the onset of the mass transfer. It possibly leads to the formation of the spectacular albeit speculative Thorne–Żytkow objects \citep{Thorne1977}.}, unless the mass ratio reaches $q_0$ before the secondary component fills its own Roche lobe. Consequently, in an X-ray binary hosting a low mass accretor with respect to the donor mass, any extended stellar atmosphere (probably present around large Eddington factor stars such as Blue Luminous Variables, hypergiant or Wolf-Rayet stars) is likely to be quickly transferred to the accretor. In X-ray binaries where the large X-ray luminosity points towards a \rlof mass transfer and an accreting compact object of mass generally larger than one solar mass, it means that the stellar companion is of low mass : it is the case of \textsc{lmxb}. We do not expect proper \rlof mass transfer for high mass star orbited by a neutron star. The few systems which have been observed in a \rlof -like regime while their mass ratio is likely larger than unity have been described in section \ref{sec:BH_RLOF_SgXB}. 


\subsubsection{Common envelope}

For common envelope systems\footnote{Also called contact binaries for stellar binaries, with W Ursae Majoris as the observational reference.}, the stellar matter fills the whole space below the lowest isopotential between the second and the third Lagrangian point. This phase brings up the question of the legitimacy of the point-mass assumption for the star. Indeed, the gravitational field we considered in section \ref{sec:roche_model} is only accurate outside the star (\ie in any point such as a sphere with the stellar center as an origin and which does not intersect the surface can be drawn). Indeed, the Green-Ostrogradsky theorem, coupled to the Poisson equation to derive the gravitational field from the mass distribution\footnote{Which becomes the Einstein equation in a general relativistic framework.} guarantees the right to replace the spatially extended mass distribution with a weighted Dirac distribution. In systems in a common envelope configuration, self-gravity must be properly included to represent the dynamics of the system.

\subsubsection{Birth of an X-ray binary}
\label{sec:birth}

In a binary where both stars underflow their Roche lobe, mass transfer can be triggered. Its fate depends mostly on its initial orbital period and stellar masses, along with the kind of mass transfers at stake (conservative or not). It also depends on whether the envelope of the donor is convective or radiative when the mass transfer starts. The different scenarii are described in detail in the compendious references \cite{Verbunt1995} and \cite{VandenHeuvel2009} (see in particular Figure 10 and the 3 cases A, B and C). Though, as long as none of the two bodies has collapsed into a compact object, the mass transfer will go unnoticed in X-rays.

A first scenario accounts for the \textsc{lmxb}. In brief, it involves a high mass ratio system where the less massive body is quickly engulfed into the envelope of its massive companion as the latter evolves. It leads to a dramatic reduction of the orbital separation and, provided the binary survives the supernova explosion associated to the formation of the compact object\footnote{See also section \ref{sec:runaway} on the possible formation of runaway compact objects.}, to a stable \rlof transfer of matter from the bullied low mass star to the freshly formed compact remnant : a \lmxb is born. During this process, the low mass star has not necessarily evolved enough to leave the main sequence.

\begin{figure}
\begin{center}
\includegraphics[height=10cm, width=6cm]{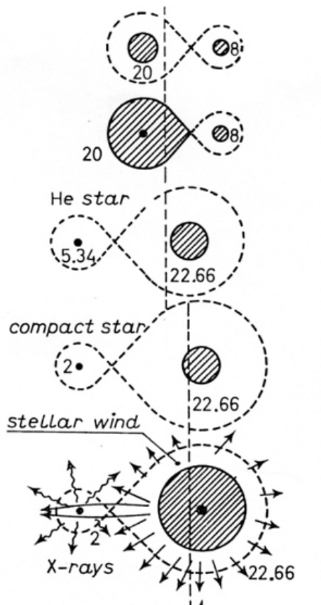}	
\caption{A scenario with the successive stages to account for the formation of \sgx. The stellar masses are indicated and the vertical dashed line traces the position of the center of mass. From \cite{VandenHeuvel2009}.}
\label{fig:formation_sgxb}
\end{center}
\end{figure}

Concerning the \textsc{hmxb}, two main scenarii account for the two main observed families (see section \ref{sec:HMLMXB} for a description of their observational properties) : the \sgx and the Be-\textsc{xb}. In both cases, the initial mass of each star is above 8\msun but the progenitors of \sgx start with somewhat larger masses for the two stars, as depicted on the first sketch of Figure\,\ref{fig:formation_sgxb}. As the more massive star evolves, it fills its Roche lobe and due to the unstable \rlof, quickly transfers its envelope to its companion which becomes more massive. The naked Helium core keeps evolving, not much affected by the absence of its erstwhile envelope, and collapses into a neutron star. Later on, the remaining massive star, in spite of the initial rejuvenating stage, is on its last legs : it has become a supergiant and displays strong radiatively-driven winds. A fraction of it is captured by its compact companion. A \sgx is born.

Most of the difference in the evolution of Be and \sgx lies in the initial masses at stake and the fraction of the mass deposited in the Helium core. Since only the envelope surrounding the core will be lost in case of unstable \rlof, it implies different amounts of mass being transferred and by then, a different evolution of the orbital separation. \cite{VandenHeuvel2009} suggests an empirical relation between the initial stellar mass and the mass of the Helium core which, once injected in the integrated expression of \eqref{eq:adot} for \rlof (\ie without loss of angular momentum) gives the final period $P_\text{f}$ after the envelope has been given away to the secondary star as a function of the initial one $P_\text{ini}$ :
\begin{equation}
P_{\text{f}} \sim \frac{100 P_{\text{ini}}} { \left(\frac{M_{1,\text{ini}}}{13M_{\odot}}\right)^{1.3} \left[ 1 + q_{\text{ini}} \left(1-\frac{2}{9}\frac{M_{1,\text{ini}}}{14M_{\odot}}\right)^{0.42} \right]^3 }
\end{equation}
where $q_{\text{ini}}$ is the initial mass ratio and $M_{1,\text{ini}}$ is the initial mass of the donor. The term at the denominator ranges from 15 to 50, for increasing mass ratios and initial mass of the donor from $(1.5,10M_{\odot})$ to $(2.5,20M_{\odot})$ ; the first configuration corresponds more to the progenitors of Be-\textsc{xb} while the second is for the ones of \sgx. Thus, we expect orbital periods (at the moment when the Helium core is unveiled) a few times larger for the progenitors of Be-\textsc{xb} compared to the ones of \sgx. According to the distribution of orbital periods of the two groups (see the Corbet diagram in Figure\,\ref{fig:corbet_diag}), this trend seems to be preserved as the system evolves.

Because the orbital separation is wider, the angular momentum provided to the stellar accretor during the first phase of \rlof is larger. The secondary star is accelerated as it accretes the envelope of its companion and reaches spins of the order of the disruption limit\footnote{By balancing the gravitation and the centrifugal force at the equator, one gets the following minimum spin period for a star of mass $M$ and radius $R$ below which the outer layers are no longer bounded : $1.5\text{ day } \left( \frac{R}{15R_{\odot}} \right)^{3/2} \left( \frac{M}{20M_{\odot}} \right)^{-1/2}$}. An excretion disk forms around the newly formed Be star and will be tapped by the compact object once formed (see Figure\,\ref{fig:BeXB}).  

\begin{figure}
\begin{center}
\includegraphics[height=10cm, width=10cm]{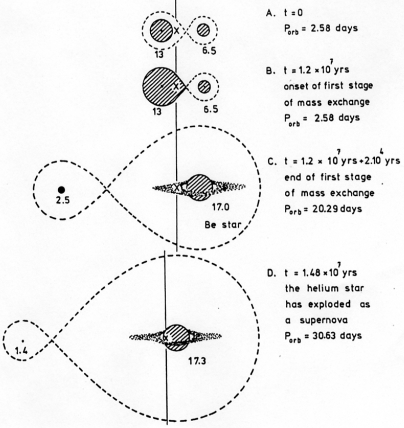}	
\caption{A scenario with the successive stages to account for the formation of \bexb. The stellar masses are indicated and the vertical line traces the position of the center of mass. From \cite{VandenHeuvel2009}.}
\label{fig:formation_Bexb}
\end{center}
\end{figure}

\subsubsection{The scarcity of Intermediate Mass X-ray Binaries}

The observed distribution of stellar companion masses is bimodal, hence the two separated families of X-ray binary systems : \textsc{lmxb} and \textsc{hmxb}, with generally different kinds of mass transfer. If \rlof is triggered in a \textsc{hmxb} hosting a neutron star, we have seen that the system is doomed : it will evolve towards a common envelope system on a short (radiative) timescale.

Let us consider the evolution of a system initially alike the progenitor of the \lmxb we considered in the previous section \ref{sec:birth} but where the mass ratio is lower such as the secondary star is of a few solar masses. In this case, it’s easier for the intermediate mass star to eject the common envelope of the primary which collapsed into a neutron star. The secondary is not massive enough to display strong winds - it does not form a \sgx - and once \rlof mass transfer starts, it finds itself in the unstable regime \citep[see the purple U-shaped beyond 1$M_{\odot}$ in Figure 1 of][]{Lin2011} ; the low mass neutron star is not able to accommodate the super-Eddington incoming accretion flow, most of the mass is quickly lost from the binary. The mass transfer slows down once the mass ratio is below one and the secondary star is now a low mass companion. It is therefore possible that most of the currently observed donors in \lmxb started their evolution as intermediate mass stars. The reader may find more information on this Intermediate Mass X-ray Binaries desert in \cite{Podsiadlowski2010}. See also \cite{King1999} and \cite{Podsiadlowski2000} for an application to the case of Cyg X-2.

\begin{figure}
\begin{center}
\includegraphics[height=9cm, width=10cm]{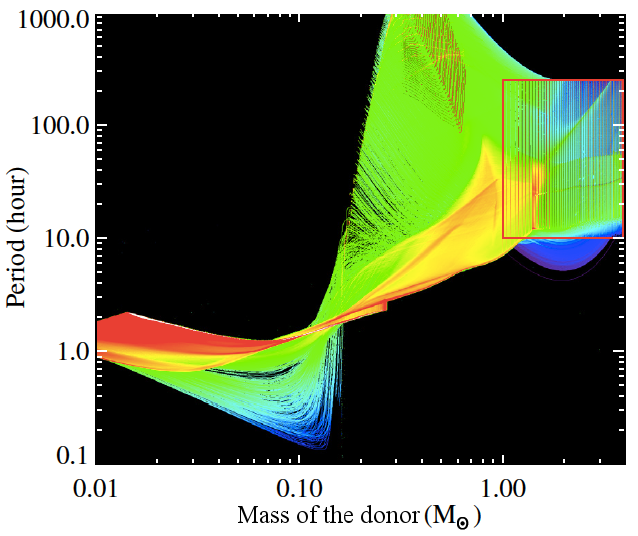}	
\caption{Evolution of 38,000 binary systems uniformly sampled with a mass of the donor and orbital period within the red frame. The colormap indicates the accumulated time spent in each pixel by all the binaries, from 10$^3$ years (purple) to 10$^{10}$ years (red). Shortly, few systems are left with a mass of the donor above 1\msun. From \cite{Lin2011}.}
\label{fig:evolution_IMBH}
\end{center}
\end{figure}


\subsection{Roche lobe overflow}
\label{sec:RLOF}
In this part, we evaluate the X-ray luminosity produced by this configuration when the accretor is a compact object and question the existence of an accretion disk. 

\begin{figure}
\begin{center}
\includegraphics[height=9cm, width=9cm]{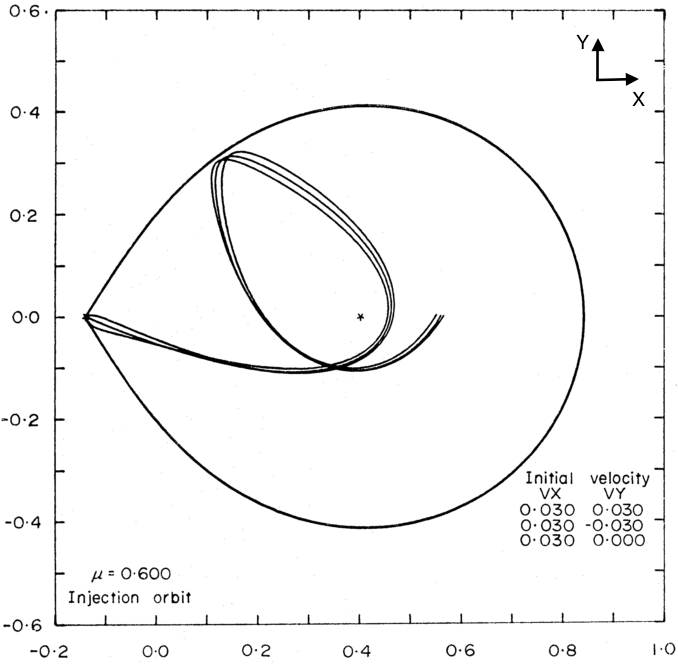}	
\caption{Ballistic trajectories of test-masses left at the first Lagrangian point with kinetic specific energies negligible compared to the absolute Roche potential at $L_1$. The Roche lobe represented is the one of the accretor. From \cite{Flannery1974}.}
\label{fig:flannery}
\end{center}
\end{figure}

\subsubsection{Mass accretion rate}
The reader might find a detailed semi-analytical ballistic model to describe in more details the formation of the stream flowing through the first Lagrangian point in \cite{Lubow1975}.

In the \rlof configuration, the stellar donor is strongly distorted and starts to pour matter into the Roche lobe of its companion\footnote{When one of the two bodies fills its Roche lobe, the system is said to be "semi-detached".}. The mechanical specific energy of the gas on the edge of the first Lagrangian point is given by the Roche potential and the thermal specific energy at the stellar surface, with the latter being negligible. Then, according to the final remark of section \ref{sec:lagrange}, the flow is likely to be trapped within the Roche lobe of the companion, especially if the latter is more massive\footnote{In this case, the ratio of the potentials at $L_1$ and at the Lagrangian point behind the companion is larger which leaves less room to a leakage through the latter point.}. The dimensionless\footnote{With the orbital separation $a$ and $\Phi_0=GM_2/a$ as references for the length and the potential (homogeneous to a squared velocity).} width $\delta$ of the nozzle which forms at the first Lagrangian point can be estimated by comparing this thermal specific energy to an expansion of the Roche potential along the $y$ transverse axis in $L_1$ :
\begin{equation}
\Phi_{\text{R}}\sim\Phi_{\text{R,}L_1} + \frac{1}{2} \left[ q \left( \frac{1-X_{\text{L,1}}^3}{X_{\text{L,1}}^3}\right) + \left( \frac{1-X_{\text{L,2}}^3}{X_{\text{L,2}}^3} \right) \right] y^2
\end{equation}
where $y$ is the small dimensionless displacement in the y-direction. $X_{\text{L,i}}$ stands for the dimensionless distance of the center i to the first Lagrangian point. Since we are interested in an estimate of the mass accretion rate, we did not consider in the expression above the expansion along the $z$ transverse axis. Because the net restoring gravitational force due to a departure from $L_1$ is partly compensated by the increase in the outwards centrifugal force along the $y$ direction but not along the $z$ one, the cross section of the flow flowing through the first Lagrangian point will be ellipsoidal (flattened in the $z$ direction, with the ratio of the second order derivatives of the potential along the two directions giving the aspect ratio). From now on, we neglect this flattening and rely on the expansion along the $y$ direction only. A usual approximated expression found in the literature \citep[see \eg][]{Frank2002} gives : 
\begin{equation}
X_{\text{L,i}}=0.5+0.227\log q_\text{i} \quad \text{with} \quad i\in\left\{1;2\right\}
\end{equation}
with :
\begin{equation}
\begin{cases}
q_{\text{i}}=M_\text{i}/M_\text{j} \quad \text{with} \quad j\neq i\\
X_{\text{L,1}}+X_{\text{L,2}}=1
\end{cases}
\end{equation}
Within the range of mass ratios we consider, this factor, which is the one in front of $y^2$ above, is of the order of $3\left(1+q\right)$. The flow is able to overcome the potential barrier around $L_1$ provided its specific thermal energy is higher than the additional amount of potential compared to $\Phi_{\text{R,}L_1}$ due to the extension of the nozzle along the y-axis :
\begin{equation}
3(1+q)\delta^2\sim \frac{c_s^2}{2}
\end{equation}
where $c_s$ is the sound speed. The mindful reader will have noticed at this point that $\delta$ is essentially the small parameter $\epsilon$ \cite{Lubow1975} based their analysis on. The mass outflow through the first Lagrangian point is then given by, in physical units :
\begin{equation}
-\dot{M}_1\sim \pi \delta ^2 \rho c_s = \pi\rho \left( \frac{c_s^3}{6\Omega ^2} \right) \sim 10^{-8} M_{\odot}/\text{yr} \left( \frac{\rho}{10^{-7}\text{g}\cdot\text{cm}^{-3}} \right) \left( \frac{P}{10\text{hr}} \right)^2 \left( \frac{c_s}{7\text{km}\cdot\text{s}^{-1}} \right)^3
\end{equation}
where $\dot{M}_1<0$, $\rho$ is the mass density at the first Lagrangian point and we used Kepler's third law. Classical values for low-mass evolved stars\footnote{The mass density in the photosphere is lower than the mean density and the mean density drops quickly as a star evolves. The Sun sets an order-of-magnitude for the initial value of this mean density on the main sequence with $\rho_{\odot}\sim1.4$g$\cdot$cm$^{-3}$} have been used. Those mass accretion rates are much higher than the ones we can derive from wind accretion considerations as we will see. Let us now study the fate of this stream dominated flow.

\subsubsection{The accretion disk}
Onc can show that the thermal specific energy is much smaller than the absolute potential at the first Lagrangian point which means that the flow quickly acquires a supersonic velocity (in the co-rotating frame) as it flows. In this way, the ballistic preliminary work we realized in the previous section can be summoned and applied to draw the streamlines : the angular momentum in the co-rotating frame initially rises due to the Coriolis force\footnote{And, to a lesser extent, to the combined action of the two gravitational forces.} which accounts for the deviation of the streamline with a zero initial transverse velocity in Figure\,\ref{fig:flannery}. Particles are then doomed to travel within the Roche lobe and the trajectories, which depend weakly on the initial conditions, quickly cross which breaks up the ballistic approximation\footnote{A necessary condition to be ballistic for a flow is to be incompressible - not too be confused with an incompressible fluid. Yet, a flow is said to be incompressible if and only if its Lagrangian derivative is $\mathbf{0}$ \ie $\mathbf{\nabla}\cdot\mathbf{v}=0$ (and matter is conserved). As for the magnetic field, in terms of field lines, it means that the streamlines do not cross. If they do, the flow is no longer incompressible and a proper hydrodynamical treatment must be carried out.}. Dissipative effects come into play and the flow arranges so as to minimize its energy. Since the potential is approximately isotropic once the flow comes into the Roche lobe of the accretor, after the first phase of addition of angular momentum by the Coriolis force, the angular momentum with respect to the accretor is essentially conserved and the flow forms a ring\footnote{It is the orbit which minimizes the energy, at a constant angular momentum.} of radius $R_{\text{c}}$, the circularization radius. If one neglects the angular momentum change induced by the combined action of the two gravitational forces from $L_1$ to the circular orbit, we can find $R_{\text{c}}$ by noting that the Coriolis force vanishes in a frame which would not be co-rotating and where the system would rotate at its orbital angular speed. Then, the specific angular momentum with respect to the accretor in $L_1$ would be given by $D_{2,L_1}^2\Omega$, where $D_{2,L_1}$ is the distance from the accretor center to the first Lagrangian point. By definition of the circularization radius around the secondary body (\ie the accretor) :
\begin{equation}
\frac{GM_2}{R_{\text{c}}^2}=\frac{\left( D_{2,L_1}^2\Omega \right)^2}{R_{\text{c}}^3}
\end{equation}
According to Figure\,\ref{fig:roche_sizes} seen from the point of view of the secondary object, $D_{2,L_1}$ is slightly larger than the radius of the Roche lobe, by a factor we write $1+\epsilon$. We then have, using Kepler's third law :
\begin{equation}
\label{eq:circ_rad_rlof}
\frac{R_{\text{c}}}{a}\sim \frac{1}{a} \left\lbrace \frac{\Omega ^2}{GM_2} \left[ \underbrace{(1+\epsilon)a\mathcal{E}(1/q)}_{\sim D_{2,L_1}} \right]^4 \right\rbrace \sim (1+\epsilon)^4(1+q)\mathcal{E}^4(1/q)
\end{equation}
This ratio shows that, for realistic orbital separations and for compact accretors (white dwarfs included), the circular orbit will be well above the surface\footnote{Which is not necessary the case if the accretor is a star \citep[see][for numerical simulations of \rlof in short-period Algols]{Raymer2012}.}.

Due to viscous effects we will not detail in this manuscript, the ring spreads into a disk \citep[see the section 5.2 of][for instance]{Frank2002} where matter spirals in towards the accretor. As it does, gravitational energy is converted into heat which is radiated away following Shakura and Sunyaev's multi-color black body $\alpha$ disk model \citep{Shakura1973b}, as detailed in \ref{sec:lum_spec}. If the efficiency of this conversion is of the order of 10\%\footnote{It is an estimate of the product of the compactness of the accretor by an efficiency factor which evaluates the fraction of the gravitational potential energy which is released as free radiation.}, the corresponding X-ray luminosity for the nominal mass accretion rate computed in the previous section is :
\begin{equation}
L_X = 0.1 (-\dot{M}_1) c^2 \sim 6\cdot 10^{37}\text{erg}\cdot s^{-1}
\end{equation}
We will see in the next Chapter that this luminosity is an order of magnitude above the very upper edge of the range of luminosities expected for a wind accretion mechanism. As a consequence, observers tend to infer the presence of a disk from an X-ray luminosity above 10$^{37}$erg$\cdot$s$^{-1}$. In many intrinsically obscured systems such as \sgx, the emission is strongly absorbed in the soft X-rays, where the disk emits most of its light : its direct observation is unthinkable. It is also an energy range strongly absorbed by the interstellar medium.

\section{Tidal interactions}
\label{sec:tidal_interactions}


Evidence for circulatization in stellar binaries have been witnessed in spectroscopic and eclipsing binaries. The former observed sample is short of early-type binaries\footnote{Early-type stars have few spectral lines in their spectra.} and is mostly made of stars with convective envelopes ; since we are interested in the evolution of the progenitors of \sgx, we will only focus on the case of stars hosting a convective core surrounded by a radiative envelope \ie, observationally speaking, on the eclipsing binaries.

\subsection{Theoretical arguments in favour of circularized orbits}
\label{sec:circ_orb}

Following Zahn's seminal papers on tidal interactions in binaries \citep{Zahn1975,Zahn1977}, \cite{Claret1997} suggested an expression (eq. (18)) for the circularization\footnote{Not to be confused with the circularization radius of an accreted flow.} timescale $\tau_{\text{circ}}$ of a stellar orbit in a binary system. Normalized by the stellar lifetime for high-mass stars\footnote{Beyond 1.5\msun, approximately those with a radiative envelope.} $\tau_0\sim2.25(M/M_{\odot})^{-3}$ \citep{Daigne2015}, we see that this circularization timescale is given by : 

\begin{equation}
\label{eq:tides_circ}
\frac{\tau_{\text{circ}}}{\tau_0}=\underbrace{\frac{q}{\left(1+1/q\right)^{11/6}}}_{\sim 1 \quad \text{for} \quad q\sim 0.5} \left( \frac{10^{-6}}{E_2} \right) \left( \frac{M}{15M_{\odot}} \right)^{5/2} \left( \frac{R}{14R_{\odot}} \right)^{3/2} \left( \frac{1/3}{r} \right)^{21/2} \left( \frac{10\text{Gyr}}{\tau_{0\text{,}\odot}} \right)
\end{equation}
where $\tau_{0\text{,}\odot}$ is the lifetime of the Sun, $M$ and $R$ are the mass and stellar radius respectively and $q$ is the ratio of the mass of the star we are focusing on by the mass of the companion. The structural parameter $E_2$ can be computed and is of the order of 10$^{-6}$ \citep{Claret1997}. 

\begin{figure}
\begin{center}
\includegraphics[height=8cm, width=11cm]{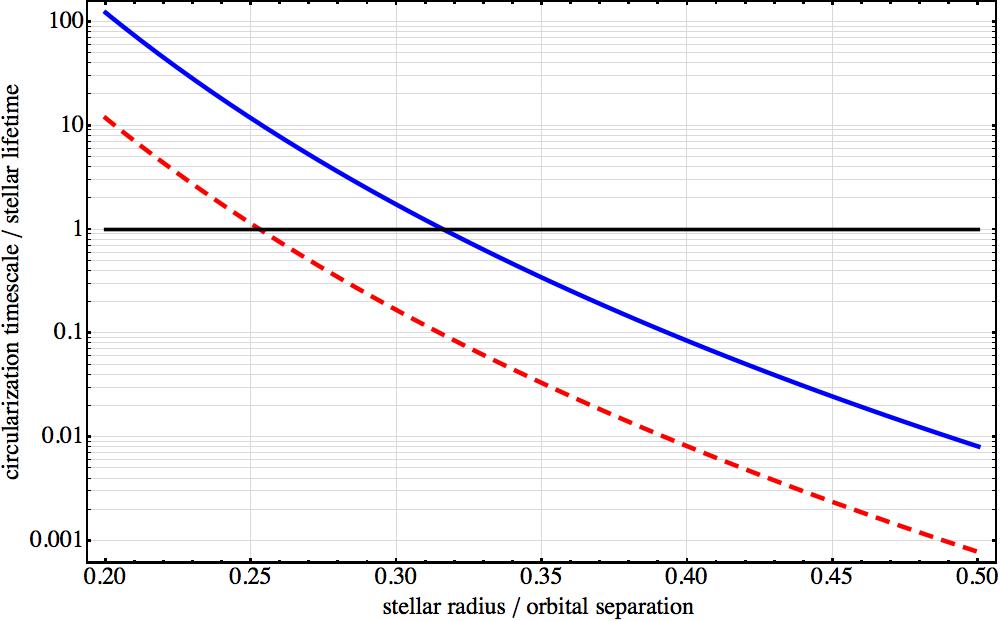}	
\caption{Ratio of the circularization timescale by the stellar lifetime for stars with radiative envelopes as a function of the fractional radius. The blue continuous line is for a 15\msun star with a radius of 10\rsun and the red dashed line corresponds to a 8\msun star with a radius of 6\rsun. Below the horizontal thick black line, the orbit can be circularized within the stellar lifetime. We set $E_2=10^{-6}$ and $q=2$ (\ie the star whose orbit is being circularized is twice as massive as the companion, typically the initial situations of Figure\,\ref{fig:formation_sgxb} and \ref{fig:formation_Bexb}).}
\label{fig:timescales}
\end{center}
\end{figure}

This expression shows that the relative speed at which a stellar orbit circularizes in a binary system depends mostly on a key quantity, the fractional radius $r$, which is the ratio of the stellar radius by the orbital separation. Provided it is large enough, the circularization of the orbit will occur on a timescale smaller than the stellar lifetime (see Figure\,\ref{fig:timescales}). Given the strong dependency, we expect a cutoff, a quick drop in the eccentricity measured for fractional radii below a threshold\footnote{Different thresholds are expected depending on the structure of the envelope ; eccentric orbits in-between could then be conclusively considered to be associated to a star with a convective envelope.}. in Figure\,\ref{fig:timescales}, we see that for different types of massive stars, the circularization is possible within the stellar lifetime for fractional radii above 0.25 ; above $1/3$, it takes less than $10$\% of the stellar lifetime to circularize.

\subsection{Confrontation to observations}

Recent observational analysis \citep{Mazeh2008,VanEylen2016} have undergone measures of this threshold relying on lowly biased samples such as eclipsing binaries. Due to the unknown value of the longitude at periastron $\omega$, the eccentricity itself cannot be accessed directly from the analysis of the light curves ; observers need to introduce an additional layer of assumptions to deduce the eccentricity from the value of $e\cos \omega$, determined from the phase of the secondary eclipse relative to the primary \citep{VanEylen2016}. However, since we only look for a threshold in fractional radius below which the eccentricity is essentially zero, plotting $e\cos \omega$ instead of $e$ will not alter the value of the threshold\footnote{Beyond the threshold, we just expect the distribution of eccentricities to be modulated by the cosine of a uniform distribution angle $\omega$.} \citep{North2004}.

\begin{figure}
\begin{center}
\includegraphics[height=6cm, width=13cm]{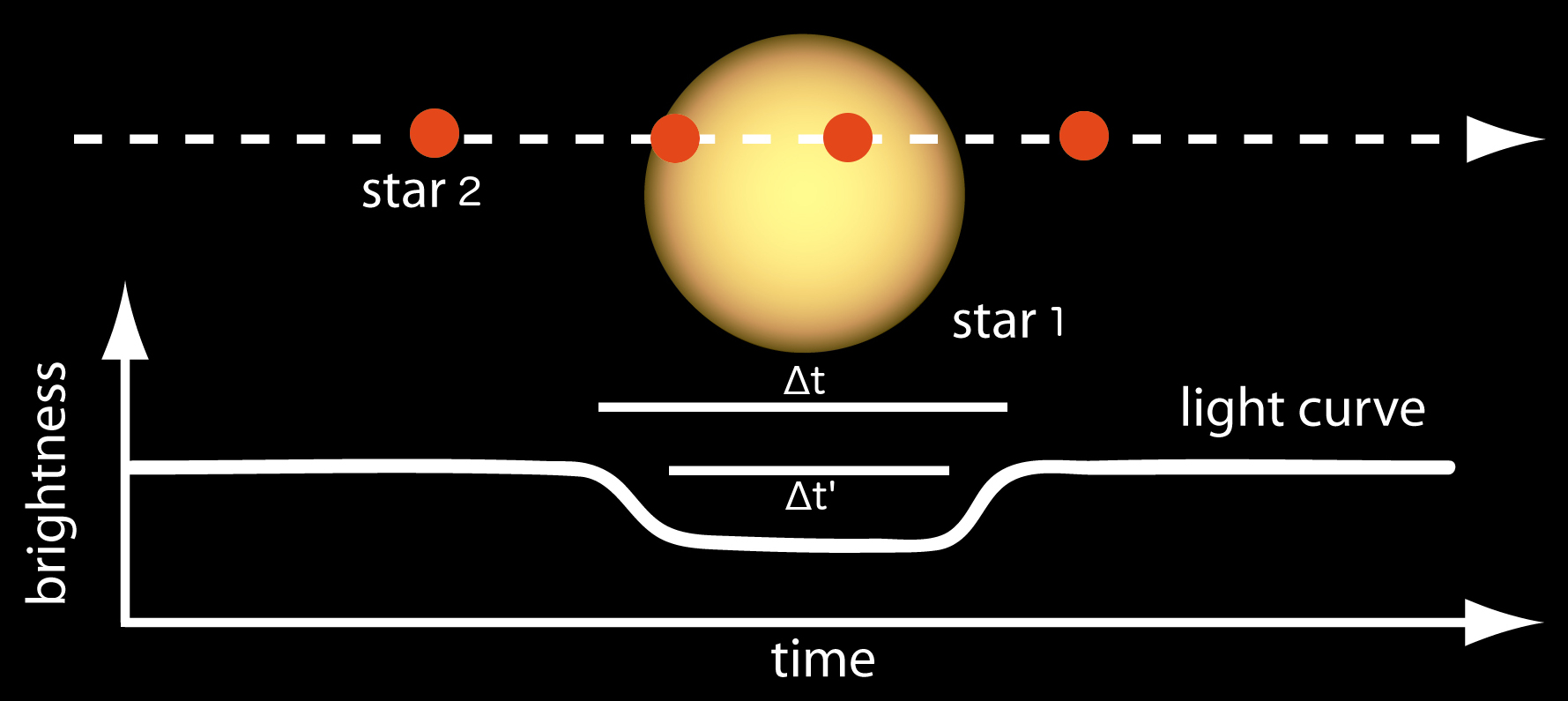}	
\caption{Illustration of the evolution of the brightness as a star transits its larger companion. Credits : NASA Ames.}
\label{fig:transit}
\end{center}
\end{figure}

\begin{figure}
\begin{center}
\includegraphics[height=9cm, width=10cm]{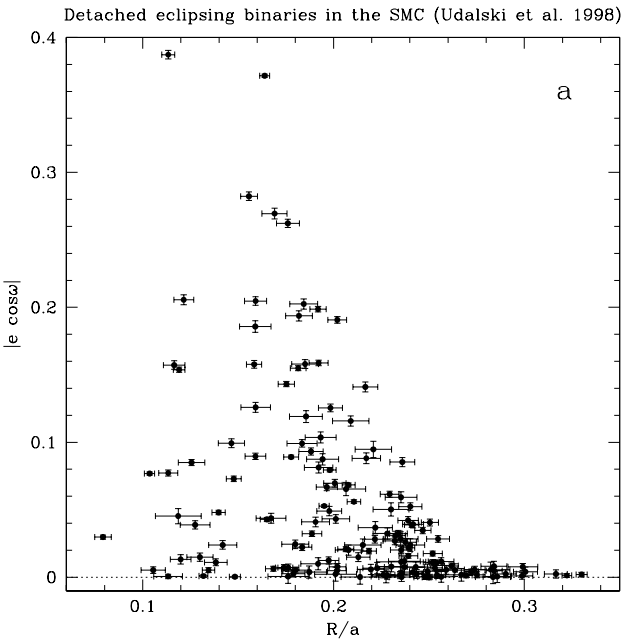}	
\caption{Value of the product of the eccentricity by the cosine of the longitude at periastron, as a function of the fractional radius. Sample of detached eclipsing binary systems from Small Magellanic Cloud \citep[data from][]{Udalski1998}. From \cite{North2004}.}
\label{fig:obs_circularization}
\end{center}
\end{figure}

Concerning the fractional radius, it can be straightforwardly measured in eclipsing binaries using considerations on $\Delta t$, the duration between the first and last time of dimming of the light, and $\Delta t'<\Delta t$, the duration of the total eclipse (when the foreground star is entirely within the disk of the background star). If one of the two eclipses admits a total eclipse, one can show that (see Figure\,\ref{fig:transit}), \eg, when star 2 passes in front of star 1 :
\begin{equation}
\begin{cases}
\Delta t \sim\frac{R_1+R_2}{\pi a}P\\
\Delta t'\sim\frac{R_1}{\pi a}P
\end{cases}
\end{equation}
where $P$ is the orbital period, precisely measured\footnote{An uncertainty we neglect in this equality is the curvature of the orbit. As the fractional radius shrinks, it becomes more important but remains largely negligible.}. Using a high photometric precision instrument such as Kepler enables us to measure with incredible levels of precision those time lapses and, by then, the corresponding fractional radii of the two bodies (without any additional assumption on their masses for instance). Plotting $e\cos \omega$ as a function of $r$ is then possible in Figure\,\ref{fig:obs_circularization}. The threshold deduced from this Figure ($\sim$0.25) is slightly lower than the one predicted in the previous section ($\sim 1/3$). A mass ratio of 1 or later-type stars than the O-stars (more rare) considered above are enough to correct this discrepancy.


Then, we can safely assume that short-period stellar binaries hosting early-type stars are circularized. What about their offspring, X-ray binaries? 

Since \bexb progenitors host lower masses objects than the progenitors of \sgx, the orbital separations involved are larger (see \ref{sec:stab_mass_trans}). Hence, the orbit is less likely to be circularized. Be that as it may, due to the strong non conservation of the mass of the system when the Supernova explosion takes place, the orbit will be highly altered\footnote{Going through the computation suggests that the binary is disrupted by the explosion if it occurs in the more massive of the two objects, making an additional case for preliminary mass transfer - see \ref{sec:stab_mass_trans}.}. In addition, we expect the explosion to be non-spherical such as the neutron star is left with a kick velocity \citep{Tauris2003}. Those two effects statistically contribute to rise the eccentricity of the orbit of the freshly formed compact object \citep[see][for detailed calculations]{Tauris1998}. However, the quantification of those effects show that they will have fewer impacts, all other things equal, for lower ratios of the kick velocity with respect to the orbital speed. Since the progenitors of \sgx host more massive stars on closer orbits, we can then suggest a first explanation about the bimodal distribution in eccentricity between \sgx and \bexb : the combined evolution of the phase preceding the Supernova explosion and the way the system reacts to the latter makes \sgx much less eccentric than \bexb. In the next chapter, we can then afford to consider the \sgx we study as fully circularized.


\subsection{Synchronization}
\label{sec:sync_tides}

The theory of tidal evolution predicts synchronization timescales smaller than circularization timescales. Given the above considerations, we will then safely suppose that synchronization has been reached in all the \sgx we study.



\setlength{\parskip}{0ex} 




\chapter{Persistent Supergiant X-ray Binaries}
\label{chap:SgXB}
\chaptermark{Persistent \sgx}
\hypersetup{linkcolor=black}
\minitoc
\hypersetup{linkcolor=red}
\setlength{\parskip}{1ex} 

\section{Modelling wind accretion}
\label{sec:models}


\subsection{The inspiring precedents of \agb donors}
\label{sec:symb_bin}

Barium stars (\aka Ba stars) are population\footnote{The stars in the Milky Way are separated between population III (the very first generation, relics of the early stage of stellar formation in our Galaxy, with very low metallicity levels), II (late type stars, intermediate metallicity levels and dispersed around the Galactic plane) and I (early type stars, high metallicity and gathered in the spiral arms).} III G and K spectral-type giants displaying excesses of Barium\footnote{Above a threshold of [Ba/Fe]$\hat{=}\log\left( \frac{N_{\text{Ba}}/N_{\text{Ba},\odot}}{N_{\text{Fe}}/N_{\text{Fe},\odot}} \right)>$0.5 where $N_{\text{X}}$ refers to the number density of element X and $\odot$ refers to the values measured for the Sun.}, Carbon and s-process elements\footnote{For slow-neutron capture.} in general compared to usual red giants. Indeed, they are not massive enough (below 5\msun) to reach the Asymptotic Giant Branch (\agb) where they would have been able to produce heavy elements, subsequently brought to the surface by third dredge-up \citep{Iben1983}. Ba stars build up the main fraction of a wider family, the Carbon-Enhanced Metal-Poor stars (\cemp stars) where we observe [C/Fe]>1. See \cite{Beers2005} and \cite{Masseron2009} for a more detailed review of the different kinds of \cemp stars. Those objects have turned out to have oscillating radial velocity profiles in agreement with the presence of a white dwarf companion \citep{Lucatello2005}, suggesting an enrichment by mass transfer rather than intrinsic scenarii. The progenitor of the white dwarf can have been an \agb whose massive stellar wind enabled it to loose enough mass to escape the Supernova ending \citep{Gawryszczak2002}, producing spectacular planetary and proto-planetary nebulae. The less massive progenitor of the Ba star would have then accreted a fraction of this metal-enriched wind\footnote{Contrary to what initially suggested \cite{McClure1983} the photosphere of the \agb does not need to fill its Roche lobe during this phase to make the mass transfer effective enough in those long-period ($>$ a few thousands days) systems.}. The evaluation of this fraction is one of the key quantities to assess the viability of this scenario.  

Symbiotic binaries consist of an \agb or a red giant star and a hot accreting companion, usually a white dwarf. The orbital period lies between hundreds of days and a hundreds of years. The white dwarf taps the wind of its companion whose mass loss rates $\dot{M}_{\text{out}}$ ranges from 10$^{-8}$ to 10$^{-6}M_{\odot}\cdot$yr$^{-1}$. A first approach could be to model the incoming wind onto the white dwarf with a planar supersonic flow at a velocity $v$ of a few 10\kms (the usual terminal velocity for red giant winds) without taking into account the orbital speed\footnote{Which is of the same order at most, not changing too much this result.}, the non spherical dilution of the wind or the eccentricity of the orbit. The \bhl theory reminded in Chapter \ref{chap:acc_pt-mass} would yield a fraction $\beta$ of stellar wind being accreted of the order of :
\begin{equation}
\label{eq:betaBHL_SgXB}
\beta\hat{=}\left|\frac{\dot{M}}{\dot{M}_{\text{out}}}\right|=\frac{1}{4}\left(\frac{\zeta_{\textsc{hl}}}{a}\right)^2=\left(\frac{GM_{\bullet}/v^2}{a}\right)^2 
\end{equation}
with $M_{\bullet}$ the mass of the white dwarf and $a$ the orbital separation. Once estimated, the 10\% efficiency associated accretion luminosity $L$ for a system where the \agb star is twice heavier than the white dwarf is :
\begin{equation}
\label{eq:x-ray_lum_wind_acc}
L=0.1\times\left( \beta\dot{M}_{\text{out}}\right)\times c^2\sim 100\text{L}_{\odot} \left( \frac{v}{40\text{\kms}} \right)^{-4} \left( \frac{M}{0.5M_{\odot}} \right)^{4/3} \left( \frac{P}{10^4\text{days}} \right)^{-4/3} \left( \frac{\dot{M}_{\text{out}}}{10^{-7}M_{\odot}\cdot\text{s}^{-1}} \right) 
\end{equation} 
which is more than an order of magnitude too low to account for the observed luminosity but also for the metal enrichment problem in \cemp stars \citep{Boffin1994}. A more sophisticated approach which includes the orbital velocity and accounts for the eccentricity\footnote{Although it does consider $v_0=v$.} can be found in \cite{Boffin1988} and was used by \cite{Abate2013} in symbiotic binaries. They concluded that larger accretion fractions than the ones given by the formulae above could be reached once two additional aspects, widely explored by \cite{Mohamed}, were taken into account in a mass transfer regime called "wind-\rlof" :
\begin{enumerate}
\item The fact that the wind is not instantaneously accelerated but on the contrary, admits a finite characteristic acceleration length scale which can be larger than the Roche lobe of the donor star or even larger than the orbital separation. 
\item The related fact that the wind can be highly non spherical but beamed towards the accretor.
\end{enumerate}
Many different semi-analytical approaches and several numerical setups have been developed to address this question, among which \cite{Theuns1993}, \cite{Theuns1996d}, \cite{Jahanara2005}, \cite{Nagae2005}, \cite{deValBorro:2009gk}, \cite{HuarteEspinosa:2012wq} or \cite{Blackman:2013td}. This idea of a possible enhancement of mass transfer once the launching of the wind was properly accounted for and the fruitful results it brought up drove us into adapting this model to persistent \sgx in section \ref{sec:mass_acc_rate}. Note that the main difficulty for \agb donors is that the launching mechanism\footnote{Believed to be initiated with pulsations which bring matter far enough to condensate and then being accelerated by radiation pressure on the subsequently formed dust grains \citep{Hofner2007}.} is much less well constrained than the launching of the wind of hot stars.


\subsection{Mass accretion rates \& X-ray luminosity}
\label{sec:X-ray_lum_wind_acc}

\subsubsection{Using the \bhl formalism}

In X-ray binaries, the spectral properties of the emission are very different. As indicated by its name, it emits most of its radiation in the X-ray range\footnote{\ie above a few 0.1keV and below a few 100keV.} while the accreting white dwarfs of symbiotic binaries emit in a softer range, typically UV light\footnote{\ie above a few eV and below a few 100eV.}. This discrepancy reflects the higher compactness parameter of the accretors in X-ray binaries which are mostly neutron stars and sometimes black holes\footnote{Note that the waverange of the emission also depends on the emitting environment being optically thin or thick - see \ref{sec:lum_spec}.}. However, the similar accretion luminosities can be explained by the formula \eqref{eq:x-ray_lum_wind_acc} used with nominal values for \sgx and a similar efficiency for the conversion of potential energy into luminosity :
\begin{equation}
\label{eq:estim_X-ray_wind}
L\sim 3.7\cdot 10^{35}\text{erg}\cdot\text{s}^{-1} \left( \frac{v}{1,000\text{\kms}} \right)^{-4} \left( \frac{M}{1.3M_{\odot}} \right)^{4/3} \left( \frac{P}{10\text{days}} \right)^{-4/3} \left( \frac{\dot{M}_{\text{out}}}{10^{-6}M_{\odot}\cdot\text{s}^{-1}} \right)
\end{equation}
Although this estimated luminosity is the same within a few percent as the one derived in the previous section for symbiotic binaries\footnote{No mysterious coincidence here, the mass of the accretor has been tuned so that it guarantees this while still corresponding to a possible mass - see \eg the values measured in \sgx in Figure 7 of \cite{Clark2002a} or Figure 12 of \cite{VanderMeer2007}.}, the parameters it deals with are very different. The wind speed when it reaches the accretor is much faster which contributes to significantly lower the accretion radius, but the period is much smaller which almost compensates the previous drop and makes the angular size of the accretion radius as seen from the donor star similar. This value lies on the lower end of the range of observed time-median X-ray luminosities in persistent \sgx (see Figure\,\ref{fig:wind_accretion}). There are two additional and somewhat more fundamental reasons why this approach can in no case be considered as satisfying. First, the theory of radiatively-driven winds (see chapter \ref{chap:wind}) predicts significant inhomogeneities due to internal shocks provoked by the intrinsic non-linearity of the acceleration term - see \cite{Dessart2005} for numerical simulations. It means that overdense clumps with possibly lower velocities due to their larger inertia are present in the wind\footnote{Especially within two stellar radii, where the neutron star usually lies.}. The formula above indicates how sensitive the level of luminosity is to the kinetic energy of the flow it accretes, which brings up a large possible spreading around this order of magnitude, if not a shifted value \citep{Ducci2009}. Second, the comparison with symbiotic binaries does not hold due to a strong scale non-invariance of the problem. The characteristic scale quantities do order in the same way (\eg, for the length, the accretion radius is smaller than the stellar radius which is smaller than the acceleration radius itself smaller than the orbital separation, for usual values) but the spatial scale dynamics in symbiotic binaries, limited to a few tens at most, can reach a few hundreds in \sgx between the smallest and the largest characteristic scale. Such an enlargement of the adimensioned flow numbers formed by ratio of the characteristic scales can have dramatic effects on the structure of the flow, such as the well-known triggering of turbulence for large Reynolds numbers in other contexts. As a consequence, the actual structure of the flow and its time variability are likely to depart a lot from the results obtained for \agb donors in binary systems, in particular concerning the likelihood of the formation of a wind-capture disk \citep{Jahanara2005,deValBorro:2009gk,HuarteEspinosa:2012wq}. 

\subsubsection{The $\beta$-wind velocity profile}

In those low eccentricity systems where the wind speed is much larger than the orbital one, the aforementioned formula by \cite{Boffin1988} does little to solve the discrepancy if one keeps using the terminal speed of the wind. More sophisticated attempts to account for the X-ray luminosity observed rely on the actual velocity profile of the wind. Indeed, when it reaches the close-in neutron star, it is still at a lower speed than the terminal one (see also the upper panel of the Figure\,\ref{fig:vel_prof} in section \ref{sec:res_1D}). It is common to fit the wind velocity profile with a formula inspired by \eqref{eq:vel_prof} derived in Chapter \ref{chap:wind}, the so-called $\beta$-law \citep{Lamers1999} :
\begin{equation}
\label{eq:beta_wind}
v(r)=v_{\infty}\left(1-\frac{R}{r}\right)^{\beta}
\end{equation}
where $v_{\infty}$ is the measured terminal speed, $R$ is an estimate of the stellar radius\footnote{To enforce the velocity to match the speed of sound at the stellar surface, $R$ is usually multiplied by a coefficient slightly smaller than 1 \cite[see \eg][]{Ducci2009}.} and $\beta$ is a quantification of the smoothness of the velocity profile, believed to lie between 0.7 and 1 \citep{Muller2008}. Using this profile with $\beta=0.8$ and a terminal speed of 1,000\kms, one can derive the velocity of the flow once it reaches the compact object at the orbital separation\footnote{Or at the peri and apastron for a non circular orbit.}. This value can be used instead of the fiducial 1,000\kms in \eqref{eq:estim_X-ray_wind} to compute a more realistic X-ray luminosity in function of, for instance, the orbital period. The latter no longer only set the orbital separation but take part in the estimate of the wind velocity when it reaches the accretor.

\begin{figure}[!b]
\begin{center} %
\includegraphics[height=10cm, width=14cm]{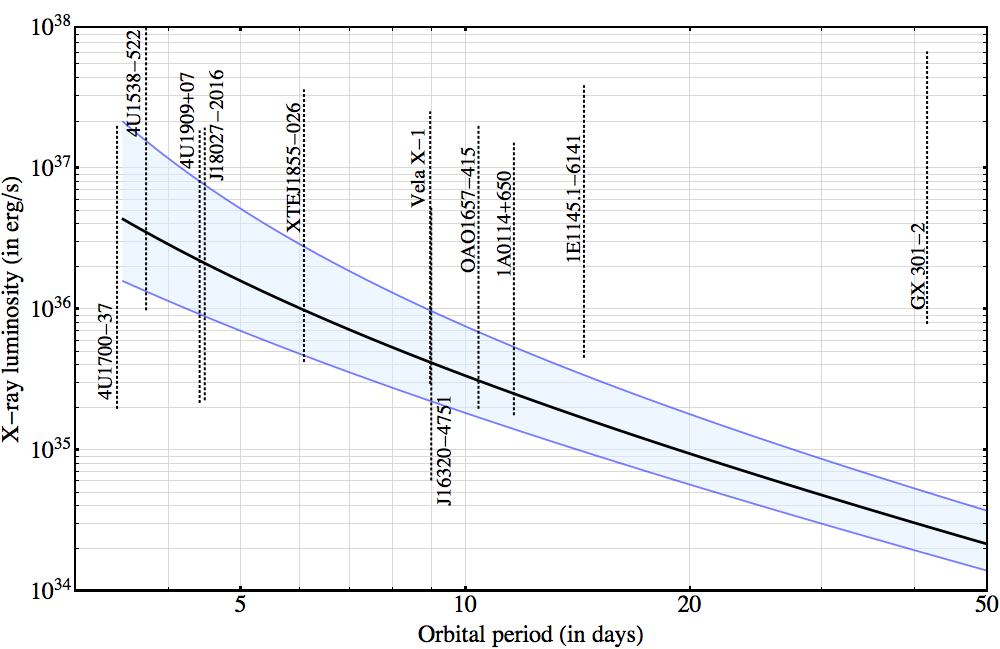}	
\caption{Range of X-ray luminosities observed for a sample of \sgx with, overlaid, an estimate of the X-ray luminosity as a function of the period (see text for more details). We use a mass for the neutron star and the donor star of 1.4\msun and 20\msun respectively. The stellar radius is set to 15\rsun, the terminal speed to 1,000\kms, the $\beta$ velocity profile exponent to 0.8 and the mass loss rate to 10$^{-6}M_{\odot}\cdot$yr$^{-1}$. The central thick black line corresponds to a circular orbit while the upper and lower case are the values at the peri and apastron respectively, for an eccentricity of 0.2.}
\label{fig:wind_accretion}
\end{center}
\end{figure}

The orbital period, when measured, is extremely well known compared to the other quantities of the system, which enables us to neglect the uncertainty on its value and motivates its use. Using formula (6) from \cite{Boffin1988} (a refinement where the orbital speed is also accounted for) with the velocity to compute the accretion radius given by the $\beta$-wind profile\footnote{Evaluated at the the orbital separation or at the peri and apastron if the orbit is eccentricity.} yielded the solid black curve and the filled blueish region around displayed in Figure\,\ref{fig:wind_accretion}. Overplotted is the domain of variation for a subset of Sg\textsc{xb} : we selected the persistent \sgx which have a measure of orbital period and a relatively precise measure of distance to derive the associated luminosity, where both quantities are taken from\footnote{We also multiply the result by 3 to extrapolate the 17 to 60keV fluxes given by \cite{Walter15} to broadband X-ray fluxes.} \cite{Walter15}. The range of X-ray luminosities for 4U1700-37, 4U1538-522, Vela X-1 and OAO1657-415 is taken from \cite{Walter15}. For the other systems, we enforced the same luminosity variation relatively to their median X-ray luminosity which gives a total dynamic factor between the lowest and highest level of luminosity of approximately 100, consistent with their persistent \sgx nature.

What we observe is that apart from their values at low orbital period (\ie below 5 days, when the underfilling of the Roche lobe becomes a strong constraint on the stellar radius), the median X-ray luminosities are also consistent with a non dependence with respect to the orbital period. The spreading, over several orders of magnitude of the X-ray luminosity can not be explained by the departure from circularity of the neutron star orbit since the eccentricity hardly reaches 0.2 in those systems. The discrepancy which remains emphasizes the need for a self-consistent framework where all the elements of the problem are coupled together : for example, where the terminal speed and the mass loss rate are not artificially set by hand but derived from the stellar properties and where the velocity profile unfolds itself in the Roche potential of the two bodies. In this way, the predicted X-ray luminosities will be able to self-consistently handle the wobble around the black curve in Figure\,\ref{fig:wind_accretion} due to the mass-loss rate, the mass of the accretor, the stellar mass, the wind terminal speed, the stellar radius, etc.



\subsection{Wind-captured discs}
\label{sec:wind_formation}

\begin{figure}[!h]
\begin{center}%
\includegraphics[height=8cm, width=8cm]{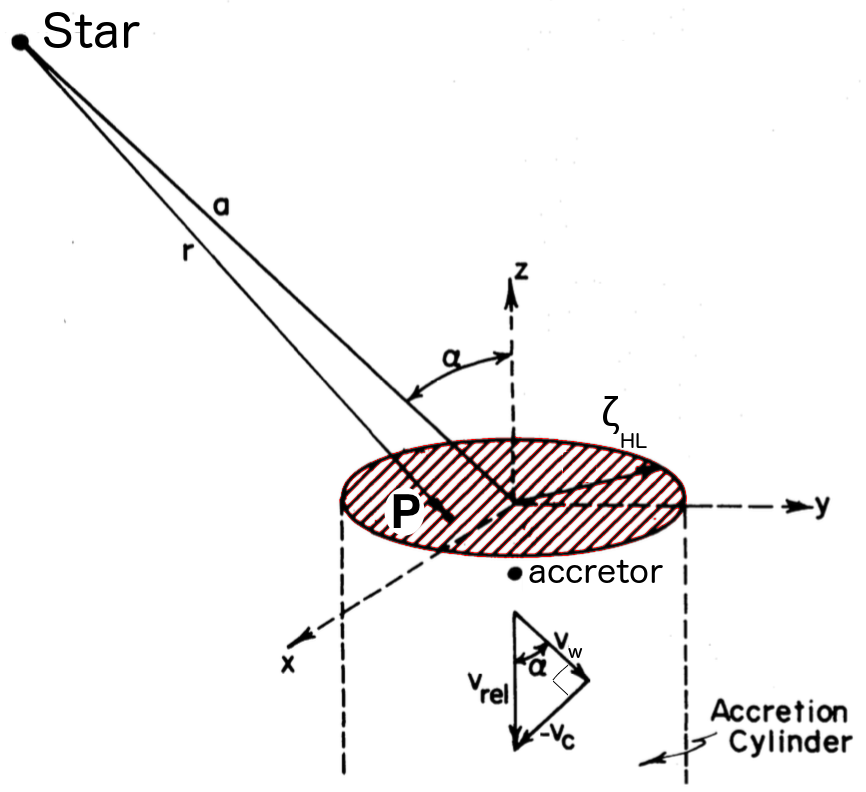}	
\caption{Sketch to represent the tilted accretion cylinder due to the circular motion at a velocity $\mathbf{v_c}$ of the accretor around a static star which emits a radial wind. The red hatched surface is a three-dimensional representation of the cross-section of the accretion cylinder which has the z-axis as a normal axis. The x-axis is normal to the orbital plane. The upper right panel represents the configuration in the orbital plane From \cite{Shapiro1976}.}
\label{fig:sketch_Shapiro}
\end{center}
\end{figure}

For the sake of simplicity, we make a few additional assumptions with respect to the more realistic model presented in the next sections. We follow the model introduced by \cite{Shapiro1976} to estimate the net angular momentum carried across the accretion cylinder face because of the variation of speed and density of the incident flow across its cross-section. First, we consider that the mass ratio is large enough to make the star lie at the center of the inertial (\ie non-corotating) frame. The low-mass\footnote{Typically, not a \bh.} accretor is on a circular orbit at $\mathbf{v_c}$ around the star and captures wind material as it flows radially outwards from the star. The matter is intercepted in a co-orbiting tube, the aforementioned accretion cylinder. It is not aimed directly along the line joining the two bodies but pointed ahead of it by an angle $\alpha$ due to the relative velocity $\mathbf{v_{\text{rel}}}$ of the wind with respect to the compact object. It is the latter which defines the main axis of the accretion cylinder, tilted with respect to the radial direction by the angle $\alpha$ given by :
\begin{equation}
\alpha=\arctan\left(\frac{v_c}{v_{w}}\right)
\end{equation}
where $v_{w}$ is the velocity of the wind at the orbital separation. The configuration is represented in Figure\,\ref{fig:sketch_Shapiro}. We go on with a second assumption : the accretion radius of the hatched cross-section in Figure\,\ref{fig:sketch_Shapiro} is small enough compared to the orbital separation (and to the acceleration radius of the wind) to consider the incident angle $\alpha$ on the cross-section as uniform. As a consequence, we can safely neglect the variation of the velocity of the wind relative to the star across the cross-section\footnote{Which is less constraining than assuming the orbital separation small compared to the acceleration radius such as, once the wind reaches the accretor, it would already be at a constant terminal speed $v_{\infty}$} and write the relative speed of the wind with respect to the accretor $\mathbf{v_{\text{rel}}}=\mathbf{v_{w}}-\mathbf{v_c}$. We also neglect any variation along the x-axis, normal to the orbital plane. 

\begin{figure}[!h]
\begin{center}%
\def\svgwidth{340pt} 
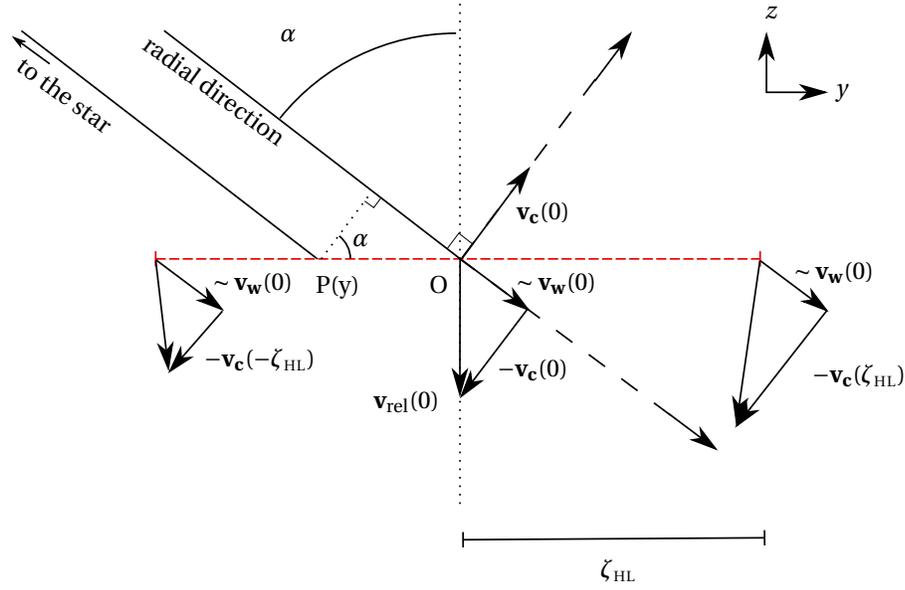 
\caption{Section in the orbital plane of the accretion cylinder. The vectorial difference of the wind speed $v_w$ by the orbital speed $v_c$ yields the relative velocity $\mathbf{v_{\text{rel}}}$ we use to compute the flow of angular momentum through the transverse cross section (short red dashed horizontal line). From the closest to the furthest side of the transverse cross section, the orbital speed rises due to the solid rotation at the angular speed $\Omega$ around the star : $\left|v_c(-\zeta_{\textsc{hl}})\right|<\left|v_c(0)\right|<\left|v_c(\zeta_{\textsc{hl}})\right|$. The coordinates between parenthesis are given relatively to the point O, center of the disk corresponding to the transverse cross section, along the y-axis. A similar sketch can be drawn with slices at $x\neq 0$ by replacing $\zeta_{\textsc{hl}}$ with $\zeta_{\textsc{hl}}\cos(\theta)$, where $\theta$ is the angular position of a point on the disk aforementioned, with $\theta=0\left[\pi\right]$ in the orbital plane.}
\label{fig:sketch_Shapiro_perso}
\end{center}
\end{figure}

The inflow through the transverse cross-section does contain a net angular momentum due to :
\begin{enumerate}
\item a lower density (since the flow is isotropically diluting) on the side with $y>0$, further away from the star than the side with $y<0$.
\item a different relative velocity for different segments of the cross-section\footnote{Due to, as explained above, a different orbital velocity for each segment ; we neglect the wind speed variation across the cross-section, an assumption we will discuss in the end of this section.}. Indeed, the cross-section is a disk in solid rotation around the star, which makes each of its surface element at a variable distance $r$ from the star orbits it with a local velocity $r\Omega$, where $\Omega$ is the orbital pulsation. We do neglect the direction shift (visible on the two extremities in Figure\,\ref{fig:sketch_Shapiro_perso}) but we do not neglect the change of magnitude it induces.
\end{enumerate}
The angular momentum per unit time crossing an element of surface centered in a fiducial point P of the cross-section is given by :
\begin{equation}
\d ^3\mathbf{L}=\mathbf{OP}\wedge \mathbf{v}(P)(\d ^3m)
\end{equation}
By symmetry, the net angular momentum relative to the y-axis will yield zero after integration. We then focus on the component along x (omitting the subscript notation) :
\begin{equation}
d^3 L=-y\cdot v_{\text{rel}}(y)\times\underbrace{\rho (y) \cdot \d x \cdot \d y \cdot v_{\text{rel}}(y) \d t}_{\d ^3m}
\end{equation}
We now proceed to a first order Taylor-Young expansion of $v_{\text{rel}}$ and $\rho$ around the zero value :
\begin{equation}
\begin{cases}
\rho (y) \sim \rho (y=0) + y \frac{\d \rho}{\d y} \Big| _{y=0}  = \rho (r=a) + y \sin \alpha \frac{\d \rho}{\d r} \Big| _{r=a}  \\
v_{\text{rel}} (y) \sim v_{\text{rel}} (y=0) + y \frac{\d v_{\text{rel}}}{\d y} \Big| _{y=0} = v_{\text{rel}} (r=a) + y\sin \alpha \frac{\d v_{\text{rel}}}{\d r} \Big| _{r=a} 
\end{cases}
\end{equation}
where we used that $\d r/\d y=\sin \alpha$. We now assume that the variation of the wind speed along the cross-section is negligible compared to the variation of the orbital speed. Using the magnitude relation $v^2_{\text{rel}}=v^2_{w}+v^2_c$ , we get : 
\begin{equation}
\label{eq:sys_shapiro_lightman}
\begin{cases}
\frac{\d \rho}{\d r} \Big| _{r=a} \sim -2\frac{\rho}{a}\\
\frac{\d v_{\text{rel}}}{\d r} \Big| _{r=a} \sim \frac{v_{\text{rel}}}{a}\sin ^2\alpha
\end{cases}
\end{equation}
where we used, in the second expression, $v_c(a)/v_{\text{rel}}(a)=\sin \alpha$. The asymmetries in density and relative velocity across the cylinder face contribute with opposite signs to the net angular momentum flux : the fastest portion of the flow with respect to the cross-section is also the least dense. The computation of the net specific angular momentum is thus non trivial. That being said, we are now left with the following integral which yields the net specific angular momentum entering the accretion cylinder :
\begin{equation}
\frac{\d L}{\d m}=\frac{1}{\dot{M}_{\textsc{hl}}}\int \d \left( \frac{\d ^3L}{\d t} \right) \sim a \frac{v_{\text{rel}}(a)}{\pi}\left(\frac{a}{\zeta_{\textsc{hl}}}\right)^2 \times I\left(\frac{\zeta_{\textsc{hl}}}{a}\right)
\end{equation}
where $\dot{M}_{\textsc{hl}}=\pi \zeta_{\textsc{hl}}^2 \rho (a) v_{\text{rel}}(a)$ and $I(u_0)$ is the following integral :
\begin{equation}
I(u_0)=\int_{u=0}^{u=u_0} \int_0^{2\pi} u^3 \cos ^2\theta \d u \d \theta
\end{equation}
where $\theta$ and $u$ refers to the polar coordinate on the cross-section, with $y=u\cos\theta$. Once carried on, this calculus provides the following analytic expression for the estimate of the net specific angular momentum :
\begin{equation}
\label{eq:spec_ang_mom}
\frac{\d L}{\d m} = \frac{1}{2} \Omega \zeta_{\textsc{hl}}^2
\end{equation}

\begin{figure}[!b]
\begin{center}%
\includegraphics[height=10cm, width=14cm]{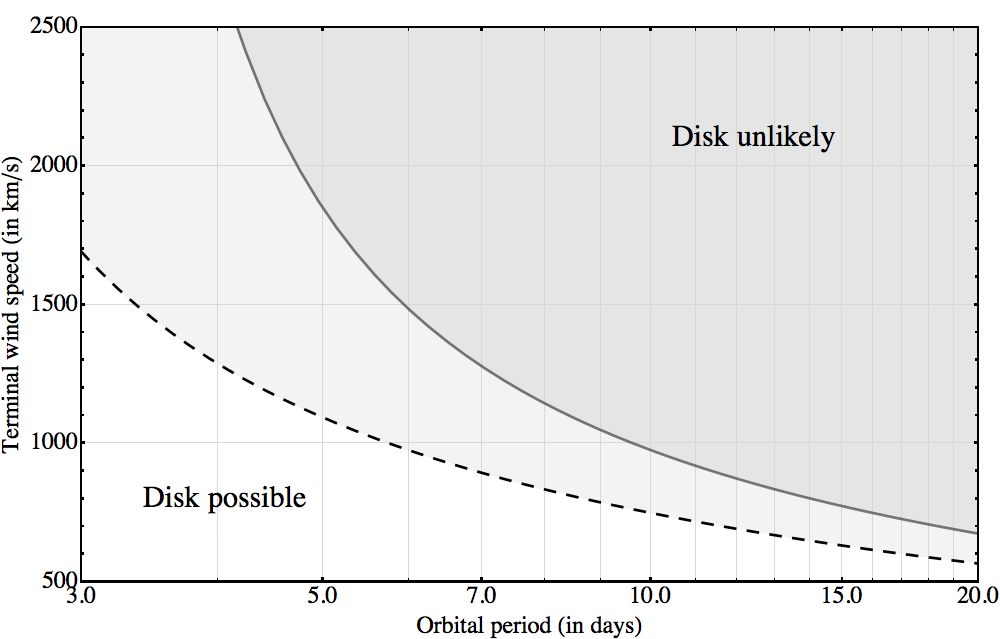}	
\caption{Likelihood of the formation of a wind-capture disk as a function of the orbital period and the terminal wind speed. The dashed line is the maximal envelope which matches the criterion \eqref{eq:circ_rad_needed} for a stellar mass and radius of respectively 20\msun and 15\rsun and a neutron star of 1.4\msun while the solid one is for a stellar mass and radius of respectively 25\msun and 25\rsun and a neutron star of 1.9\msun. The size of the neutron star is set to three times its Schwarzschild radius and the $\beta$-exponent at 0.8.}
\label{fig:ang_mom_for_disk}
\end{center}
\end{figure}

In practice, we expect deviations from this value since it relies on the accretion radius which is only an order-of-magnitude deduced from a qualitative approach. Numerical simulations by \cite{Livio1986} and \cite{Ruffert1999} suggest that the actual accreted specific angular momentum for high Mach number flows undergoing adiabatic evolution is actually three times smaller. Indeed, the main flaw of this demonstration is the neglecting of the variation of the wind velocity across the accretion cylinder cross-section. The domain of validation of this assumption, if we take the $\beta$-velocity profile to evaluate the wind speed variation, is indeed hardly met since we would need :
\begin{equation}
1+\beta \left( \frac{v_w}{v_c} \right)^2 \ll \frac{a}{R}
\end{equation}
which is generally not verified in a \sgx, where the wind speed at the position of the accretor is already several times larger than the accretor orbital speed, where the filling factor is not small ($>$ 60\%) and where the mass ratio is large. The breaking up of this hypothesis has the respective minor and major effects :
\begin{enumerate}
\item in the second equation of the system \eqref{eq:sys_shapiro_lightman}, in the \rhs, a second term, dominant, should be added following, for instance, the $\beta$-velocity profile as a guideline. Since the wind is faster further away from the star, it enhances the relative velocity asymmetry in \eqref{eq:sys_shapiro_lightman}. 
\item the model of the accretion cylinder itself is jeopardized since the accretion radius is not the same on both sides of the accretor. In spite of the enhanced relative velocity asymmetry pinpointed above, the influence of the relaxation of this assumption is not obvious because the cross-section over which is integrated the angular momentum (no longer a disk) becomes larger on the side closer from the star and smaller on the other one.
\end{enumerate}
Aware of this limitation, let us use the approximate evaluation \eqref{eq:spec_ang_mom} with the correction factor $1/3$ mentioned above to compare the circularization radius associated to this specific angular momentum to the size of the accretor. With $R$ and $M$ the radius and mass of the accretor, and $M_*$ the mass of the star, the necessary condition of existence of a disk-like structure around the compact object deduced from \eqref{eq:spec_ang_mom} is the following :
\begin{equation}
\label{eq:circ_rad_wind}
\left(\frac{\zeta_{\textsc{hl}}}{a}\right)^4>\underbrace{\frac{36}{1+M_*/M}}_{\sim 3 \text{ for } M_*=10M}\frac{R}{a}
\end{equation}
It is a necessary condition to assure that the circularization radius is larger than the accretor. To compute the accretion radius $\zeta_{\textsc{hl}}$, we rely on the $\beta$-velocity profile \eqref{eq:beta_wind} and translate the condition above into a trend in the "orbital period - terminal speed" parameter space :
\begin{equation}
\text{disk } \Rightarrow v_{\infty} < v_{\infty\text{,\textsc{max}}} \hat{=} \frac{1}{\left(1-R_*/a\right)^{\beta}} \left( \frac{2GM}{a} \right)^{1/2} \left( \frac{1+M_*/M}{36} \frac{a}{R} \right)^{1/8}
\label{eq:circ_rad_needed}
\end{equation}
with $R_*$ the stellar radius and $a$ the orbital separation linked to the period through Kepler's third law. At a fixed terminal speed, the dependence on the orbital separation can be understood through the dependence of the velocity profile (the shorter the orbital separation, the larger the accretion radius and the specific angular momentum) and the larger shear in the accreted flow for lower orbital separation. The Figure\,\ref{fig:ang_mom_for_disk} summarizes this discussion by emphasizing the need for short periods and low terminal speeds to favour the formation of a wind-capture disk. The expression \eqref{eq:circ_rad_wind} provides an expression\footnote{When the inequality is replaced with an equality, the radius $R$ is the circularization radius.} for the circularization radius of a wind accretion flow, to be compared to the one we had for a \rlof mass transfer, given by \eqref{eq:circ_rad_rlof}. We see that the likelihood of the formation of a disk is much smaller for a wind mass transfer, as long as the accretion radius is smaller than the distance from the compact object to the first Lagrangian point. One can see that it is the case for realistic wind terminal speeds but it takes slightly lower velocities\footnote{Due to an accretor lying within the acceleration zone of the donor star or an aborted wind acceleration due to an ionizing flux from the compact object high enough to strip the metals from all their electrons.} to make the two scales comparable ; in this case, the existence of a disk ceases to be a chimera.

\section*{Conclusion}

If we have seen that the formation of a disk was a possibility to keep in mind even for a wind mass transfer, its actual existence and its structure are difficult to foresee. In particular, the disk can extend well beyond the circularization radius derived above because of angular momentum by viscous torques. However, if the central object is a neutron star, its magnetosphere will interact with the disk, possibly disrupting its inner parts. Taking those effects into account requires to go beyond the simplified model above and actually compute the specific angular momentum entering the vicinity of the accretor, what we perform in the next sections of this Chapter.

\section{Numerical launching of the stellar wind}
\label{sec:num_launching}

Since we work with \sgx, systems where the mass ratio is large, we can afford to neglect the influence of the neutron star on the launching of the wind. Indeed, the latter is mostly determined in the very vicinity of the stellar surface, far from the gravitational influence of the tiny compact companion. To validate our computation of the streamlines in the \sgx configuration, we first confront the numerically integrated one-dimensional velocity profile for an isolated star to the theoretical predictions. 


\subsection{Ballistic approximation}
\label{sec:ball_app}

As pinpointed by \cite{Shapiro1976}, the stellar wind is highly supersonic and this approximation holds until the gravitational deflection induced by the compact object leads to the crossing of the ballistic streamlines. Numerical simulations of the wind such as the ones performed by \cite{Walder2014} and \cite{Manousakis2015c} suggest that from the stellar surface to the vicinity\footnote{Within a volume whose typical size is a few times the accretion radius.} of the accretor, the flow is essentially ballistic. Indeed, as validated a posteriori by the one-dimensional numerical benchmark in Figure\,\ref{fig:vel_prof} described in section \ref{sec:res_1D}, the line acceleration peaks close from the stellar surface and the wind, typically at a the temperature of the photosphere, reaches supersonic speeds shortly after it leaves the surface : we amalgamate the sonic surface to the stellar one, which prevents us from investigating intrinsically time variable behaviors related to internal shocks \citep{Cohen2014a} or instabilities (\eg clumps) within the wind. For a steady-state investigation of the outflow which accounts for the pressure, see \cite{Lamers1999}. It must also be mentioned that including the ionizing feedback from the close-in of the compact object on the flow upstream leads to departure from ballistic configurations \citep[see the photoionization wake in][though at values of the mass ratio too low to fit \sgx]{Blondin1990}.

However, starting the wind requires initial conditions not only on the position and the velocity but also on the initial velocity radial gradient. Indeed, accounting for the finite cone effect requires an estimate of the initial velocity gradient as stated by \eqref{eq:fd_factor} in Chapter \ref{chap:wind}. The results described below proved very robust provided we select an initial velocity, at the stellar surface, of the order of the sound speed ; the first radial step can be chosen to be a fraction of the stellar radius and the corresponding velocity can be derived from a fiducial $\beta$-law velocity profile (see \eqref{eq:beta_wind}) with realistic parameters. Deviations of those values within an order of magnitude yield results dispersed within a range much smaller than the characteristic dependences we want to study. 


\subsection{Numerical scheme \& degrees of freedom}
\label{sec:num_scheme}

The ballistic assumption enables us to solve a simplified equation for the motion, given by equation \eqref{eq:adim_CAK} in Chapter \ref{chap:wind}, while still capturing a representative dynamics of the wind. This expression is valid using the stellar radius $R$ and the modified escape speed $v_{\text{esc,m}}$ (divided by $\sqrt{2}$) as units of length and velocity. At this point, the shape of the velocity profile depends only on the $\alpha$ force multiplier. The dimensionless shape parameter $\aleph$ introduced in section \ref{sec:motion_CAK} is not a degree of freedom and is fixed by the requirement to get a unique solution and the value of $\alpha$. It encapsulates the $Q$ (and $\alpha$) force multiplier, the Eddington parameter $\Gamma$, the stellar luminosity and the mass loss rate. The latter is thus entirely determined by the other aforementioned parameters.

The inclusion of the finite cone effect does not fundamentally alter the previous comments provided $\aleph$ is replaced with its product with the initial value of the finite cone correction factor $D$ : 
\begin{equation}
\aleph\times D\left( r=R, v=v(r=R) ; \alpha \right) \sim \aleph / \left( 1+\alpha \right)
\end{equation} 
so as the new $\aleph$ is $\left( 1+\alpha \right)$ times larger than its expression deduced from \eqref{eq:critical_X} and \eqref{eq:X_crit_al}. This workaround guarantees that the first integration steps are proceeded in the same way as in the case of the point-source problem whose analytical solution is known and requires a precise value to not be ill-defined. Nevertheless, since $D$ is always above its initial value for $r>R$, it means that the acceleration is larger than for a point-source : the terminal speeds are expected to be higher than the ones given by \eqref{eq:vel_inf_CAK} in Chapter \ref{chap:wind}. The effect of this workaround on the mass loss rates and terminal speeds have been discussed in section \ref{sec:fd}. The equations detailed in section \ref{sec:dynamics_CAK} lead\footnote{With the present assumption of a sonic surface identified to the stellar surface} to a modified mass loss rate $\dot{M}'^{\alpha}\propto D\left( r=R, v=v(r=R) ; \alpha \right)$ to account for the finite cone correction, consistent with the mass outflow being halved compared to the plain \cak model, for the values of $\alpha$ we consider. Without this workaround, the numerical launching of the wind is doomed to fail since the critical conditions are not matched at the sonic surface (\ie, with our model, at the stellar surface).

To solve the unidimensional ballistic equation of motion, we applied a fourth order Runge-Kutta scheme on a test-mass starting anywhere on the homogeneous surface of the star. The time steps are chosen to be smaller by a factor of 100 than any time-homogeneous combination of the displacement, the velocity and the acceleration so as to not alter the outcome of the numerical integration. It also proves accurate according to the limited drift of the mechanical energy of each test-mass from its initial value. 

As mentioned above, the only parameter which constrains the shape of the solution is the $\alpha$ force multiplier, even once the finite cone angle is taken into account since the correction parameter depends only on $\alpha$ and on the solution of the equation of motion which itself is set by $\alpha$. In this sense, we call $\alpha$ a shape parameter, as opposed to the scale parameters which only serve to set the scale of the solution (\ie the stellar radius and the modified escape speed divided by $\sqrt{2}$). 


\subsection{Results}
\label{sec:res_1D}

The results of the numerical integration with and without the finite cone angle factor \eqref{eq:fd_factor} (noticed respectively \textsc{fd} and \textsc{cak}) are displayed in the upper panel of Figure\,\ref{fig:vel_prof} (solid lines). As expected, the velocity profile is higher when the finite cone angle is taken into account, what appears more clearly when looking at the terminal speeds computed at a few 10 stellar radii (bottom panel of Figure\,\ref{fig:vel_prof}). The analytical profile \eqref{eq:vel_prof} given for a point-source matches the one we compute numerically within a few percents but not the profile obtained with the finite cone correction. For the latter, we figure out terminal speeds and exponents which enable to fit the numerically computed data and obtain approximately :
\begin{equation}
\label{eq:vr_FD}
v(r)=v_{\infty}\left(1-\frac{R_*}{r}\right)^{0.7} \quad \text{with} \quad v_{\infty}\sim 2.5 v_{\text{esc,m}} \frac{\alpha}{1-\alpha}
\end{equation}
The exponent 0.7 (instead of 0.5 for the analytical solution with a point-source) makes the velocity profile somewhat smoother which is consistent with observations \citep{Villata1992} and sophisticated semi-analytical computation of modified \textsc{cak}-wind profiles \citep{Pauldrach1986,Villata1992,Muller2008,Araya2014a}, though on the lower end of the range. On the other hand, the expression of the terminal speed is in agreement with state-of-the-art hydrodynamical simulations \citep{Muller2008,Noebauer2015} and analytical studies \citep[see][Table 1]{Kudritzki1989a}, though only for temperatures above the bi-stability threshold\footnote{Below, the factor 2.5 drops and so does the intensity of the wind.} at 21,000K \citep{Vink1999}. Since we are interested in donor stars of earlier or similar spectral type as early B stars, we expect to remain in this regime and can fully rely on this terminal speed. Notice that this law leads to generally higher terminal speeds than the one suggested by \cite{Friend1986} but to similar values than the observed ones for class I luminosity stars \citep[see table 3 of ][]{Groenewegen1989}, if one considers $\alpha\sim 0.55$.

\begin{figure}[!h]
\centering
   \begin{subfigure}[b]{0.7\textwidth}
   \includegraphics[width=1\linewidth]{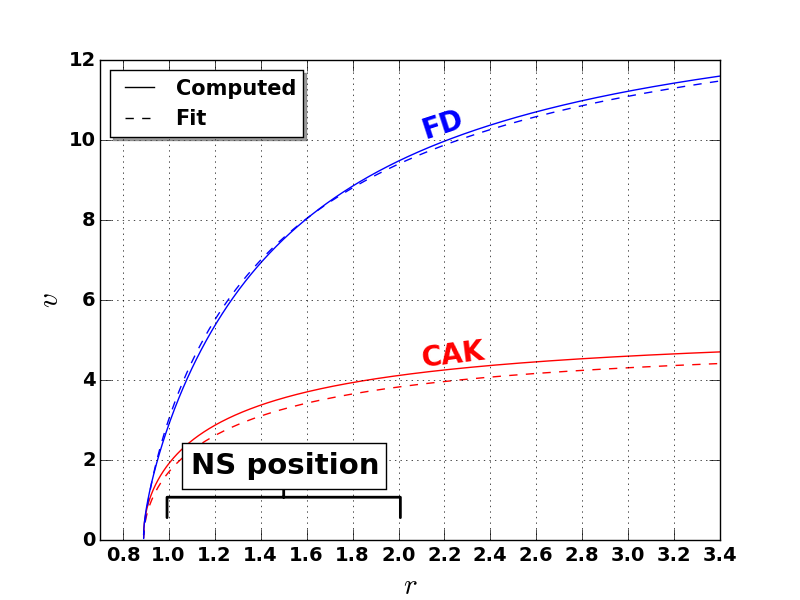}
   \label{fig:sfig1} 
\end{subfigure}
\hspace*{-1cm}
\begin{subfigure}[b]{0.7\textwidth}
   \includegraphics[width=1\linewidth]{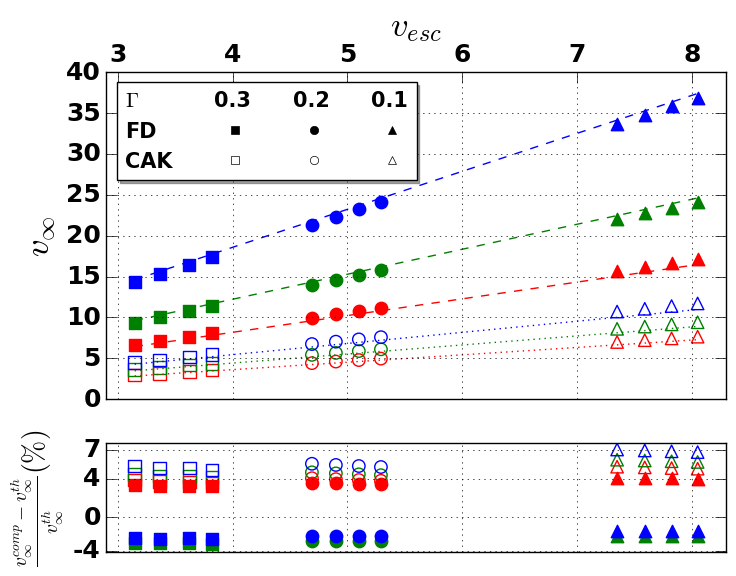}
   \label{fig:sfig2}
\end{subfigure}
\caption[Two numerical solutions]{Summary of the preliminary check-up of the integrator in the case of the wind of an isolated star. On the top, the velocity profile (in the units system given in the text) in the simple case described in section \ref{sec:vel_prof_CAK} in red and in the case where the finite cone effect has been included (blue). The red and blue dashed lines are the corresponding approximated profiles given respectively by \eqref{eq:vel_prof} and \eqref{eq:vr_FD}. At the bottom is gathered the information concerning the terminal speeds (measured in our simulations beyond $r=30$) as a function of the effective escape velocity. Red, green and blue correspond respectively to $\alpha=0.45$, $0.55$ and $0.65$. $\Gamma$, $q$ and $f$ have been tuned so as to explore a wide range of escape velocities, still given in the scaling system described in \ref{sec:num_scheme}. The relative differences between the computed and the expected values is represented below the main plot.}
\label{fig:vel_prof}
\end{figure}


\section{Wind motion in a Roche potential : setting the stage}
\label{sec:set_stage_sgxb}


\subsection{Context \& normalization}
\label{sec:context_sgxb}

\sgx feature low-eccentricity orbits compared to the other classes of \hmxb, which justifies from now on to work with circular orbits\footnote{The introduction of eccentric orbits would not require major modifications in the code but would break up the steady-state assumptions ; each effective separation as the neutron star orbits the supergiant star - at a non constant angular speed - would have to be considered and treated as if the wind unfolded quickly compared to the time between two successive positions on the orbit. It would add a time modulation, at the orbital period, to the possible time variable behaviour within the vicinity of the compact object otherwise fed with a steady-state though inhomogeneous inflow.} (see also section \ref{sec:circ_orb}). So as to reduce the dimensionality of the space of shape parameters and efficiently explore it, we rely on the following normalization :
\begin{enumerate}
\item the stellar Roche lobe radius $R_{\textsc{R,}1}$ as the length scale
\item the mass of the compact object, $M_2$, as the mass scale
\item $GM_2/R_{\textsc{R,}1}^2$ as the acceleration scale
\end{enumerate}  
Unless explicitely stated in the rest of the manuscript, this system of scaling is the one we use. The velocity scale for instance is deduced by the simplest combination of the length and the acceleration scale (\ie the square root of their product, without any additional factor). 

We did not consider the departure from the spherical shape of the stellar surface, more important for bodies close from Roche lobe overflow but still too low to fundamentally alter the results (see Figure\,\ref{fig:aspect_ratios}). We also discard the asynchronous rotation of the star : it is supposed to have reached synchronization with the orbital period (see comments on synchronization in \ref{sec:sync_tides}).

\begin{figure}[!h]
\begin{center}
\def\svgwidth{350pt} 
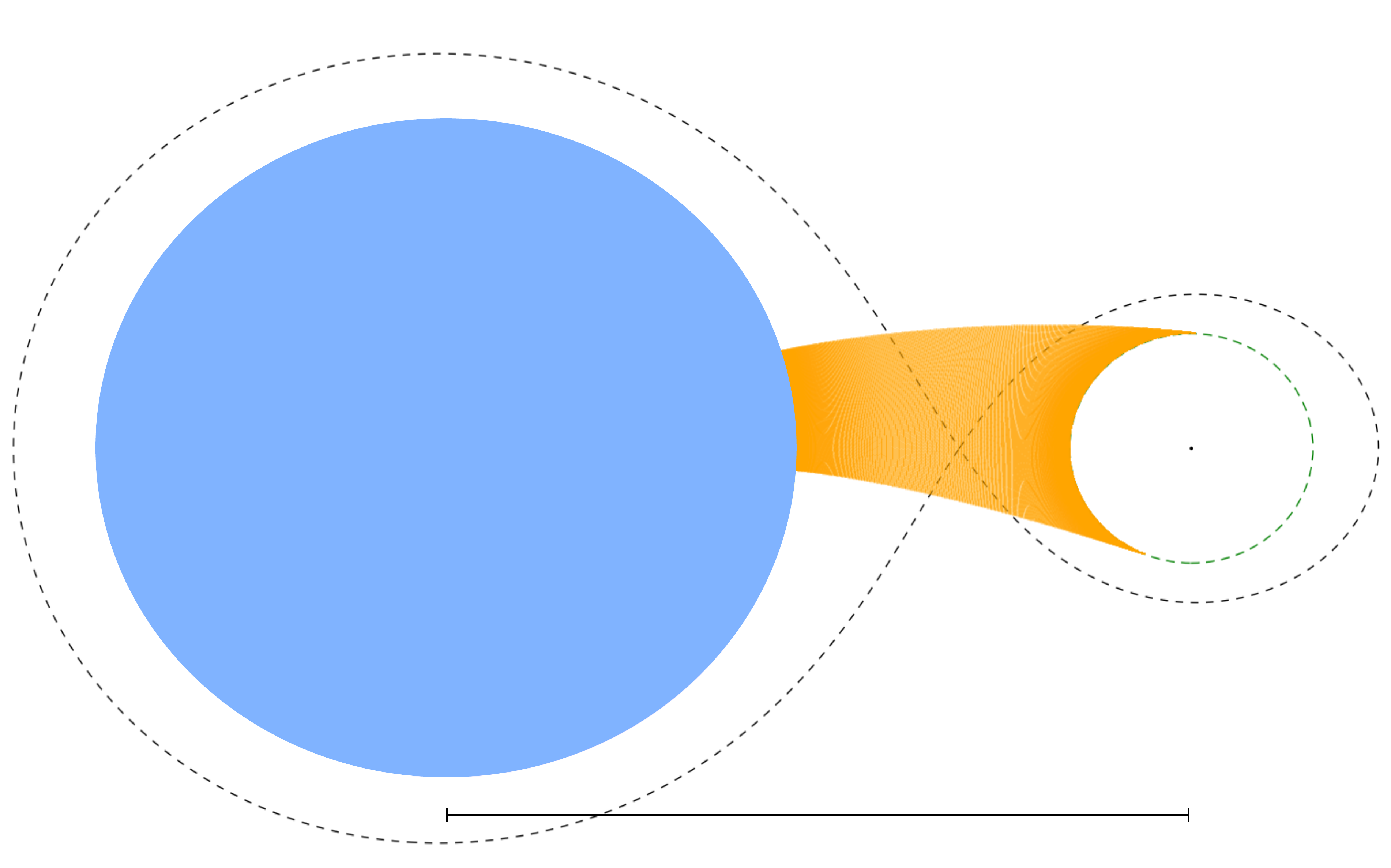 
\caption{Orbital slice of the \sgx configuration we consider with, in blue, the supergiant stellar companion and on the right, the compact object. The eight-shaped black dotted line is the Roche surface passing through the first Lagrangian point and the orange lines are the computed streamlines which enter the vicinity of the compact object. The latter, in dashed green, is defined by its radius $R_{\text{out}}$ as specified in the section \ref{sec:acc_rad_sgxb}. $a$ is the orbital separation. Notice that "CM", which stands for the center of mass, would be well inside the star in the configuration of a \sgx.}
\label{fig:sketch}
\end{center}
\end{figure} 


\subsection{Scale invariant expression of the equation of motion}
\label{sec:scale_inv_exp}

We now set ourselves in the frame co-rotating at the orbital period\footnote{And so do all the velocities we refer to in this chapter.}, with the center of mass (\textsc{cm}) as the origin and use subscript 1 (resp. 2) to refer to the star (resp. to the compact object), as described in Figure\,\ref{fig:sketch}. As reminded in Chapter \ref{chap:roche}, the modified gravitational force of the star, the gravitational force of the compact object and the centrifugal force acting on a test-mass can then be written as the Roche potential, and Kepler's third law can be used to show that its shape depends only on the mass ratio (and on the Eddington factor if we account for the continuous radiative pressure). One must also consider the Coriolis force per mass unit which cannot be written as a potential. At this point, $q$ and $\Gamma$ are the only shape parameters of the problem. The radius of the star does not play a role since in Roche's formalism, both objects are point-like\footnote{We will see in a moment that it enters the equation essentially as an initial condition on the position.}. Numerically, it means that $q$ and $\Gamma$ are the only meaningful parameters for the study of the motion of a test-mass in a Roche potential accounting for radiative pressure, since the scaling back to physical units can be realized afterwhile. 


We can now gather those terms with the absorption lines acceleration. The dimensionless expressions of the positions of the test-mass relatively to the star and the compact object write, with the notations of Figure\,\ref{fig:sketch} and $\mathcal{E}$ the Eggleton function introduced with \eqref{eq:egg} : 
\begin{equation}
\left\{
\begin{array}{ll} 
\mathbf{\tilde{r}_1}=\mathbf{\tilde{r}}-\mathbf{\tilde{R}_1}=\mathbf{\tilde{r}}+\frac{1/\mathcal{E}(q)}{q+1}\mathbf{\hat{x}}\\[10pt]
\mathbf{\tilde{r}_2}=\mathbf{\tilde{r}}-\mathbf{\tilde{R}_2}=\mathbf{\tilde{r}}-\frac{q/\mathcal{E}(q)}{q+1}\mathbf{\hat{x}}
\end{array} 
\right.
\end{equation}
Omitting the tildes and using the radial velocity of the test-mass relatively to the star, $v_1=\mathbf{v}\cdot\left(\mathbf{r_1}/r_1\right)$, we then get the following dimensionless ballistic equation of motion of a radiatively-driven wind in a \sgx Roche potential :
\begin{equation}
\label{eq:main}
\frac{\d \mathbf{v}}{\d t} = - \left[ \frac{q(1-\Gamma )}{r_1^3} 	- \aleph \cdot D\left(r_1,\frac{\d v_1}{\d r_1};\alpha,f\right) \frac{v_1^{\alpha}}{r_1^{3-2\alpha}} \left( \frac{\d v_1}{\d r_1}\right)^{\alpha}  \right] \mathbf{r_1}  - \frac{1}{r_2^3}  \mathbf{r_2} + (1+q)\mathcal{E}^2 \mathbf{r_{\bot}} -2\mathcal{E}^{3/2}\sqrt{1+q}\cdot\mathbf{\hat{z}}\wedge\mathbf{\tilde{v}}
\end{equation}
with $f=R_1/R_{\text{R,}1}$ the filling factor\footnote{Where $R_1$ is the stellar radius, not to be confused with the norm of $\mathbf{R_1}$.}, $\mathbf{\hat{z}}$ the unit vector normal to the orbital plane and oriented in the direction of the orbital rotation vector $\boldsymbol{\Omega}$ and $\mathbf{r_{\bot}}$ the distance of the test-mass to the z axis. Notice the interesting dependence of the finite cone effect correction factor $D$ on the $\alpha$ force multiplier (like in the previous section \ref{sec:num_launching}) but also, now that we no longer normalize the lengths with the stellar radius, on the filling factor. This equation illustrates that the filling factor $f$, the Eddington factor $\Gamma$, the mass ratio $q=M_1/M_2$ and the $\alpha$ force multiplier are the cornerstone parameters of the \sgx toy-model we investigate at large scale in this chapter. The computational sampling of this four-dimensional parameter space is the following :
\begin{enumerate}
\item $\alpha$ is either 0.45, 0.55 or 0.65, in agreement with the range of values computed by theoretical models for supergiant OB stars\footnote{The two force multipliers, $\alpha$ and $Q$, were calculated by \cite{Shimada1994} for OB-Supergiants with 520,000 atomic lines and they depend mostly on the effective stellar temperature $T$ and the surface gravity $\log (g)$ (the decimal logarithm of its value in \textsc{cgs} units is usually considered in the literature).}. 
\item $\Gamma$, the stellar luminosity relatively to the Eddington limit (\aka the Eddington parameter). For OB-Supergiants, we expect it to be below 30\%, as opposed to Wolf-Rayet stars, Luminous Blue Variables or hypergiant stars (like the donor in GX 301-2) which can go beyond. We take 10, 20 and 30\% as the three possible values.
\item $q$, the mass ratio. We ran simulations for 12 integer values of $q$ ranging between 7 and 18.
\item $f$, the filling factor. We picked up 10 values non regularly spaced between 50\% and 99\%.
\end{enumerate} 


\subsection{The modified accretion radius}
\label{sec:acc_rad_sgxb}

As visible in the upper panel of Figure\,\ref{fig:vel_prof}, the neutron star typically lies within the acceleration zone of the stellar wind. This feature makes the case for a specificity of \sgx compared to symbiotic binaries where the wind, driven by a different mechanism whose properties are less precisely known than for winds of hot stars, has nonetheless reached its terminal speed once it gets in the gravitational vicinity of the accretor. To evaluate the volume where the gravitational influence of the neutron star is large enough to significantly deflect the flow and break up the ballistic assumption, we can rely on a modified accretion radius, computed along the lines of the one described in the simplified framework of the \bhl flow (see Chapter \ref{chap:acc_pt-mass}). The bottom line of the derivation of this radius is to wonder which are the streamlines which undergo at some point a change in potential energy comparable to the kinetic energy of the bulk motion at infinity, all this solely due to the action of the accretor. For \sgx, the "infinity" is hazardous to define due to the fact that the orbital separation and the acceleration radius are of the same order. Furthermore, the flow is not planar since it comes from a spherically diluting source whose curvature radius is not negligible and is sheared by the joint action of the Coriolis, the centrifugal\footnote{Indeed, the latter being centered at the center of mass of the system and not at the stellar center, it does not preserve the purely radial velocity of the flow with respect to the star.} and, to a lesser extent, the gravitational force of the compact object (see \ref{sec:wind_formation}). 

In a pragmatic attempt to overcome those difficulties, we compute a modified accretion radius using the following process. We first ask what would be the motion of a streamline starting from the point on the stellar surface the closest from the neutron star, if all the forces but the gravitational attraction of the compact companion were present. It provides information about the kinetic energy transferred to the wind once the streamline reaches the orbital separation, independently from the action of the accretor ; the corresponding velocity is written $v_{\bullet}$. We rely on this kinetic energy to evaluate the critical pseudo impact parameter that we call the modified accretion radius (and still write $\zeta_{\textsc{hl}}$) : 
\begin{equation}
\label{eq:mod_acc_rad}
\zeta_{\textsc{hl}}=\frac{2GM_2}{v^2_{\bullet}}=\frac{2}{\tilde{v}^2_{\bullet}}
\end{equation}
where the last equality is the evaluation of the modified accretion radius using the dimensionless velocity at the orbital separation (in the units system previously specified). Variations around the process described above to compute the modified accretion radius have convinced us that this quantity leads to consistent and robust evaluations of the accretion vicinity. 

Given the numerical simulations performed in the reduced problem of \bhl flow in Chapter \ref{chap:num_sim_BHL}, we can now define an extended accretion radius, $R_{\text{out}}$, which corresponds to the volume where a proper hydrodynamical treatment is required. We evaluated this radius to approximately 8 times the accretion radius in \ref{sec:dyn} to guarantee that the shock has weakened enough in the wake of the accretor to be back in a supersonic weakly perturbed regime. The extended accretion sphere of radius $R_{\text{out}}$ defines the zone around the accretor where the ballistic framework described here gives way to a hydrodynamical simulation which takes as outer boundary conditions the streamlines at the orbital scale. We will therefore consider this volume as a black box within which the present study can not make any definitive conclusion.


\subsection{Numerical implementation}
\label{sec:num_imp_sgxb}

\subsubsection{Sampling}

\begin{figure}[!h]
\begin{center}%
\def\svgwidth{450pt} 
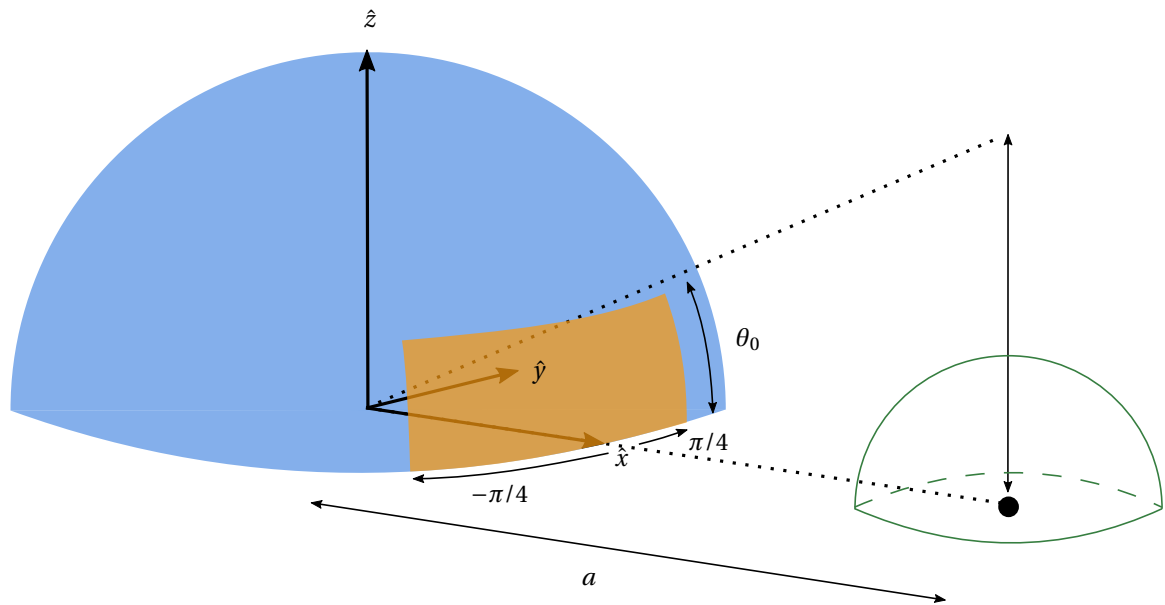 
\caption{Sketch of the area of the stellar surface susceptible to produce a wind accreted by the compact object (orange patch). Its longitudinal extension is of $\pi/2$ centered on the direction of the compact object (black dot in the lower right part of the figure) and its latitudinal extension above the orbital plane, $\theta_0$, computed following \eqref{eq:theta0}. The extended accretion sphere has been represented in dark green around the accretor.}
\label{fig:meshing}
\end{center}
\end{figure}

To appreciate the orbital structure of the steady-state wind from the hot evolved supergiant companion, we first sample the upper half\footnote{The problem admits an up/bottom symmetry with respect to the orbital plane, which enables us to compute the streamlines only in the upper half and to copy and paste the results for the lower part once we want to post-process a subset of streamlines.} of the stellar surface using a standard spherical meshing. Notice that it is not a homogeneous tessellation of the sphere in the sense that neighbouring points delimit patches of varying surface, a point to account for when we want to make any statement on a "fraction" of the streamlines verifying a certain property. The typical initial mesh first admitted a few thousands longitudinal (\ie East-West) departure points on the surface and four times less\footnote{To have an angular aspect ratio of approximatively one at the equator.} in the latitudinal direction. However, given the large velocities of the wind and the slight bending in the parameter space we consider, the points which give birth to streamlines which get close at some point of the compact object are always part of a subset of the stellar surface (the orange region in Figure\,\ref{fig:meshing}) with a $\phi$ longitudinal coordinate between $-\pi/4$ and $\pi/4$ and a $\theta$ latitudinal coordinate between the equator and a $\theta_0$ such as :
\begin{equation}
\label{eq:theta0}
\tan \left( \theta_0 \right) = \frac{2R_{\text{out}}}{a} 
\end{equation}
In particular, the many points well above the orbital plane are not likely to feed directly the compact object and are then of fewer interest in our study. We limit ourselves to streamlines launched from this region ; a safety check assures that the streamlines which intersect the extended accretion sphere start always well within this angular region and not on its border.

\subsubsection{Integration \& selection}

As explained in \ref{sec:num_scheme} in the one-dimensional case of an isolated star, we integrate the trajectories using a fourth order Runge-Kutta. This time though, we stop the integration once one of the following conditions is met :
\begin{enumerate}
\item the test-mass crashes back onto the star. With the prescription described in \ref{sec:num_scheme} to account for the finite cone factor and the fact that the mass ratio is large, this case does not present on the line joining the two centers.
\item the time integration takes an excessively long amount of time.
\item the test-mass is farther from the stellar center than the sum of the orbital separation and the extended accretion radius (\ie the extended accretion sphere has been "missed").
\item the streamline intersects the extended accretion sphere.
\end{enumerate}
The first case is a sanity check to control whether our assumption that the wind can be treated as if the star was isolated, at least at launch, holds. The second one monitors a possible case of "lost streamline" which would be trapped within the potential (unexpected given the large terminal speed compared to the other velocities at stake in the model). The third and fourth cases are more realistic and stand for, respectively, a test-mass which freely passes by the neutron star without being involved in the accretion process and a test-mass which is susceptible to cross a region where hydrodynamical effects are no longer negligible. The latter streamlines are the ones selected for further investigation.

\subsubsection{Refinement}

Ballistic simulations are affordable enough to let us raise the number of streamlines we work with. To do so, we refine the meshing of the departure points on the stellar surface when they yield selected streamlines after integration. We refine in both angular direction, proceed to a second integration for all those new points and reiterate the refinement step until we get a similar number of selected streamlines (a few 10,000) whatever the values of the 4 degrees of freedom. Some of the streamlines closest from the orbital plane and finally selected have been represented in orange in Figure\,\ref{fig:sketch} in a fiducial case - a similar sketch for each configuration $(q,f,\Gamma,\alpha)$ is automatically produced to give the user an idea of the bending of the streamlines, the relative size of the extended accretion sphere, etc.

\subsubsection{Density}

After several refinements, we are now left with arrival points on the virtual extended accretion sphere centered on the accretor\footnote{To make sure that the last iteration brings the test-mass very close from this sphere - not too far within or outside -, we shorten dramatically the timestep as the test-mass approaches it.}. Each of those points has velocity components but we also want to be able to evaluate the shear of the inflow : we must quantify the deformation of the arrival map compared to the departure one. Due to the steady-state assumption, the conservation of mass guarantees that the streamlines do not intersect each other. However, the cross-section of the flux tube which surround each of them evolves, something we can monitor by using the four neighbouring streamlines. The ratio of the surface they delimit on the extended accretion sphere by the one of the surface they delimit on the star yields the evolution of the mass density (times the magnitude of the velocity) of the streamline from an extremity to another. Once we will set the scale parameters and compute the quantities (such as the photospheric mass density and sound speed) in physical units, we will be able to relate each of the arrival point to a certain local density. The Appendix \ref{sec:surf_sph} details how to compute the surface elements on the sphere.

An angular resampling of those primitive quantities (mass density and velocity) on pre-defined meshes provides us with physically-motivated outer boundary conditions for upcoming hydrodynamical simulations on the final fate of this inflow. To each cell of the mesh are associated values of density and velocity components computed as an average over the values associated to all the arrival points, weighted with a Gaussian kernel centered on the cell and of width approximately the size of the cell (in terms of arclength since we are still working on a sphere).\\

\subsubsection{Computational cost}

The whole computation time per simulation on 1 \textsc{cpu}, including the post-processing described in the next paragraph, is of the order of a few hours such as the dimensionless parameter space considered in section \ref{sec:scale_inv_exp} (3$\times$3$\times$12$\times$10$=$1,080 configurations) can be explored within a week with 36 \textsc{cpu}s.
 
\subsubsection{Comment on the computation of the outputs}

All this output values obtained by averaging or integrating combinations of density and velocity values on the extended accretion sphere have been computed in two different ways to guarantee their robustness :
\begin{enumerate}
\item directly with the values of density, velocity and flux tube cross-section of each streamline.
\item using the values resampled on a spherical mesh on the extended accretion sphere with two different resolutions (16$\times$64 and 64$\times$256 usual spherical coordinates).
\end{enumerate}
The computational pipeline has been corrected and improved until the outputs obtained following those three approaches prove close from each other. The final output values given further are obtained by simply arithmetically averaging the three values.


\subsubsection{Visualisation of the outputs}
\label{sec:visu}

Once computed, the plethora of possible outputs requires a visualisation tool to be analyzed. The numerous data sets produced by those simulations are prone to be obstacles to the understanding of the dependencies at stake if one does not care about the question of their visualisation. In particular, we need to be able to explore not only the four-dimensional dimensionless space parameters but also the role of the three scale parameters. Their choice and the effective set up of this tool using the \href{https://github.com/adamhajari/spyre}{Spyre library for Python} developed by A. Hajari (2015) is detailed in Appendix \ref{sec:spyre}. 


\section{Structure and properties of the flow at the orbital scale}
\label{sec:structure_wind}


\subsection{Structure of the accretion flow}
\label{sec:struct_acc_flow}
We now try to quantify what we mean by "wind" accretion. The usual smoking gun to qualify an accretion of wind accretion is to observe a star which does not fill its Roche lobe with an accreting companion. However, it does not tell us much about the actual structure of the flow. Indeed, a wind which would not be fast enough compared to the orbital velocity once it reaches the critical Roche surface would result in a highly collimated flow, mimicking the stream one usually associates to \textsc{rlof} \citep{Nagae2005}. \textsc{rlof} tells us something about the filling factor while wind accretion refers to the structure of the flow, not necessarily incompatible with \textsc{rlof}. Thus, we choose to term "wind" or "stream" an accretion flow depending on the relative size of the extended accretion sphere, where the shock is expected to develop, with respect to the size of the Roche lobe of the compact object $R_{\textsc{r,}2}$. When the wind is slow, the former becomes of the order or even larger than the latter and the whole orbital scale requires a proper hydrodynamical treatment. 

\begin{figure}[!h]
\begin{center}
\includegraphics[height=8cm, width=10cm]{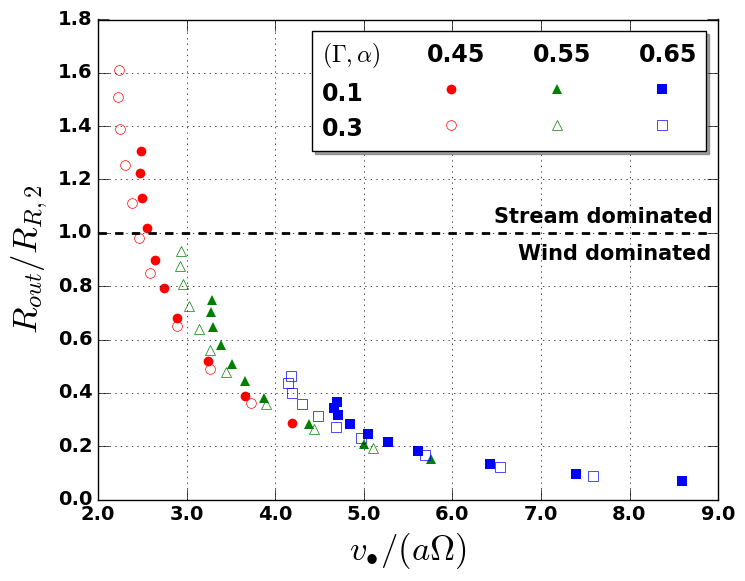}	
\caption{Ratio of the extended accretion radius by the size of the compact object Roche lobe as a function of the ratio of the velocity at the orbital separation by the orbital speed. For each of the six curves, from upper left to lower right, we used values of $(q,f)$ along the line going from $(8,0.99)$ to $(17,0.5)$ in Figure\,\ref{fig:threesome} so as to probe the largest range of $R_{\text{out}}/R_{\text{R,}2}$.}
\label{fig:RLOF_VS_wind}
\end{center}
\end{figure}

To evaluate how slow the wind must be to be in such a configuration, we plotted in Figure\,\ref{fig:RLOF_VS_wind} the evolution of the ratio of the aforementioned length scales compared to the ratio of $v_{\bullet}$ by the orbital speed (where $v_{\bullet}$ has been defined in \ref{sec:acc_rad_sgxb}). Once again, none of these quantities was forced ; instead, they were computed from parameters linked to the launching of the wind and the structure of the potential. A first conclusion is that, even in configurations where the filling factor is close to one, wind accretion is expected for efficient wind acceleration (\ie $\alpha > 0.55$), for any Eddington factor below 30\%. It is a typical situation where \textsc{rlof} cohabits with wind accretion. On the other hand, even for a filling factor as low as 90\%, one can get a stream dominated flow for $\alpha=0.45$, intermediate mass ratios and Eddington factors of 30\%. We observe that lower values of $\Gamma$, \ie of the dimensionless luminosity, favor a wind dominated flow, which can sound paradoxical but is actually consistent with the velocity of the wind scaling as the modified escape speed in \eqref{eq:vr_FD} - see sections 3.2.3 and 8.9.2 of \cite{Lamers1999} for a physical interpretation.

An important conclusion to draw from this Figure\,\ref{fig:RLOF_VS_wind} is also that, as far as the structure of the accretion flow is concerned, the ratio $v_{\bullet}/a\Omega$ is an excellent tracer to evaluate how windy the flow looks, much more reliable than the filling factor which only plays a secondary role here. In the case of radiatively-driven winds, we also advocate in favour of this ratio rather than the ratio of the mean normal velocity of the wind on the critical Roche surface by the orbital speed, $V_R/a\Omega$, as suggested in \cite{Nagae2005} and \cite{Jahanara2005}. Indeed, first, since the accretor lies within the wind acceleration radius, it is more relevant to consider the velocity of the wind at the orbital separation than at the critical Roche surface which must not be considered as seriously in \sgx as in systems hosting low $\Gamma$ stars (symbiotic binaries, \textsc{lmxb}...) : in Figure\,\ref{fig:sketch}, the critical Roche surface is represented for $\Gamma=0$ but in \sgx, the lowering of the effective stellar gravity by the radiative pressure from the continuum opacity is no longer negligible. Second, it is not equivalent to model the wind as described in Chapter \ref{chap:wind} and to consider that the radiative pressure exactly balances the stellar gravitational force at any radius (ie $\Gamma=1$ and $g_{\textsc{cak}}=0$), as used in \cite{Theuns1996d} and \cite{Jahanara2005}. It is legitimate in symbiotic binaries where the mechanism responsible for the driving of the wind is still poorly known but not in \sgx. As a consequence, the conservative threshold on $v_{\bullet}/a\Omega$ we found, of the order of 2 to 3, should prove to be more reliable in \sgx than $V_R/a\Omega\sim 0.5$ derived in \cite{Nagae2005}. 

\subsection{Wind capture and scaled mass accretion rate}
\label{sec:mass_acc_rate}

We introduce several quantities to evaluate the fraction of matter which is accreted by the compact object. The first one, $\beta ^+$\footnote{The "$+$" superscript refers to the fact that not all the matter entering the extended accretion sphere is expected to be accreted - but all the finally accreted streamlines belong to this sample ; the fraction of stellar wind entering the extended accretion sphere is an upper limit on the fraction of stellar wind being accreted, $\beta$.}, is a numerical measure of the fraction of the wind entering the extended accretion sphere (dashed green circle in Figure\,\ref{fig:sketch} and solid green sphere in Figure\,\ref{fig:meshing}) from which we can guess an effectively accreted fraction, $\beta$, using the results of Chapter \ref{chap:num_sim_BHL}. Then, we will confront this quantity and its dependencies on the shape parameters to a theoretical prediction made by neglecting the non planarity of the flow, $\beta_{\textsc{hl}}$, and to a potential indirect tracer $\mathcal{L}$ of the fraction of the wind accreted introduced by \cite{Abate2013} for symbiotic binaries.


\subsubsection{Direct numerical measures}

First, we save $\beta ^+$, the fraction of stellar wind entering the extended accretion sphere. A first underestimate is obtained by comparing the number $n$ of streamlines which were selected (in the sense defined in section \ref{sec:num_imp_sgxb}) to the total number $N$ of streamlines within the launching patchwork considered\footnote{The orange area in Figure\,\ref{fig:meshing}.} and then by extending it to account for the whole sphere :
\begin{equation}
\label{eq:betaplus_first}
\beta ^+ = \frac{n}{N} \times \frac{\frac{\pi}{2}\times \sin\theta_0}{2\pi}
\end{equation}
where the numerator of the second factor of the \rhs is the solid angle covered by the launching patchwork. $\beta ^+$ is an underestimate because it does not account for the fact that the classical meshing of the sphere we rely on is not homogeneous. It is less dense at the equator, from where come most of the streamlines selected, than at the poles because of the\footnote{Where $\theta$ is to be understood as the colatitude, not like $\theta_0$ which is a latitude.} $r\sin\theta$ arclength which separates the points along the longitudinal direction. Another way to evaluate $\beta ^+$ is to compute the surface delimited on the stellar sphere (in blue in Figures\,\ref{fig:meshing} and \ref{fig:sketch}) by the streamlines selected. To do so, we compute the mean arc\footnote{It is an angle, the ratio of the arclength by the radius of the sphere on which it is measured (here, the stellar radius $R_1$). To measure it between two points on the sphere, one just needs to use the dot product between the position vectors which locate those two points ; the arc appears in the cosinus.} $\left\langle l \right\rangle$ between selected streamlines and compute :
\begin{equation}
\beta ^+ = \frac{n \times \left\langle l \right\rangle^2}{2\pi}
\end{equation} 
In practice, because $\theta_0$ remains small, \eqref{eq:betaplus_first} is only a slight underestimate of the order of a few percent compared to the second estimate of $\beta ^+$. Due to the matching between those two independent estimates, we consider their arithmetically averaged value as representative of the actual fraction of the wind entering the extended accretion sphere.  

Concerning the actual fraction of the wind which will take part in the feeding of the compact object and in the subsequent emission of X-rays, we can not make any definitive measure without three dimensional simulations of the evolution of the flow within the extended accretion sphere. However, an indirect measure of $\beta$ can be obtained using the results of \cite{ElMellah2015} described in Chapter \ref{chap:num_sim_BHL}. Indeed, if we write $\dot{M}^+_{\textsc{ec15}}$ the mass inflow which was arbitrarily set as an outer boundary condition in \cite{ElMellah2015} and $\dot{M}_{\text{acc}}$ the mass accretion rate which was then measured for Mach numbers at infinity larger than 4, the fraction of the wind entering the sphere of radius $8\zeta_{\textsc{hl}}$ which was actually accreted is given by :
\begin{equation}
\frac{\dot{M}_{\text{acc}}}{\dot{M}^+_{\textsc{ec15}}}=\frac{0.77\dot{M}_{\textsc{hl}}}{\rho_{\infty}v_{\infty}\pi\left(8\zeta_{\textsc{hl}}\right)^2}=\frac{0.77}{8^2}\sim 1\% \sim \frac{\beta}{\beta^+}
\end{equation}
where $\rho_{\infty}$ and $v_{\infty}$ are the mass density and the velocity at infinity\footnote{Approximation acceptable since the flow is almost not disturbed from planarity when it enters the simulation space at $8\zeta_{\textsc{hl}}$ - see outer boundary conditions \eqref{eq:position} to \eqref{eq:dens_BK}.}. Remarkably enough, we observed in Chapter \ref{chap:num_sim_BHL} that this ratio no longer depends on any parameter and remains fairly constant when the Mach number is above $\sim$4, which is easily verified by the highly supersonic winds in \sgx.

\subsubsection{Theoretical estimates of the accreted wind fraction}

For the theoretical expectations, we have two predictions we can confront to. First, we can use the computed dimensionless velocity of the wind at the distance of the orbital separation, $\tilde{v}_{\bullet}$, to make predictions on the fraction of the wind being accreted by the compact object according to a \textsc{bhl} approach \citep{Boffin1988}. Corrected with an additional factor $0.77$, it has been shown to be accurate within a few percent for Mach numbers at infinity larger than a few \citep{Foglizzo1996,ElMellah2015}. If we assume the wind to be isotropic, we have, using the conservation of mass :
\begin{equation}
R_1^2 \rho _1 v_1 \sim a^2 \rho _{\bullet} v_{\bullet}
\end{equation}
with $\bullet$ refers to the values at the orbital separation $a$ and the subscript 1, to the values at the stellar surface. Then, the fraction of the wind accreted is given by \eqref{eq:betaBHL_SgXB} that we rewrite with the 0.77 factor :
\begin{equation}
\label{eq:betaHL}
\beta_{\textsc{hl}} \hat{=} \frac{0.77\dot{M}_{\textsc{hl}}}{\dot{M}_1} = \frac{0.77\cdot \pi\zeta_{\textsc{hl}}^2 \rho_{\bullet}v_{\bullet}}{4\pi R_1^2 \rho _1 v_1} \sim \frac{0.77}{4}\left( \frac{\zeta_{\textsc{hl}}}{a} \right)^2 \sim 0.77\times\frac{\mathcal{E}^2(q)}{\tilde{v}^4_{\bullet}}
\end{equation}
where $\dot{M}_1$ and $\dot{M}_{\textsc{hl}}$ are respectively the stellar mass outflow and the \textsc{bhl} mass accretion rate. The last equality uses \eqref{eq:mod_acc_rad} and \eqref{eq:egg}. Apart from the factor 0.77, this expression is strictly the same as the equation (7) by \cite{Abate2013} for a wind speed large compared to the orbital speed. A decisive point we emphasized in \eqref{eq:betaHL} above is the independence of $\beta_{\textsc{hl}}$ on the scaling : whatever the period, the orbital separation or the mass of the compact object, the fraction of the wind accreted according to Bondi, Hoyle \& Lyttleton's sketch is the same if the 4 shape parameters remain unchanged. This comment will be of prime importance in the interpretation of the results presented in section \ref{sec:phys_mass_acc_rate}.

\begin{figure}
\floatbox[{\capbeside\thisfloatsetup{capbesideposition={left,top},capbesidewidth=8cm}}]{figure}[\FBwidth]
{\caption{Ratio between the mass accreted by the secondary star and the mass lost by the primary (the \textsc{agb}) star as a function of the ratio between the wind acceleration radius and the radius of the Roche lobe of the primary. The lines correspond to theoretically-motivated predictions for different mass ratios while the markers are the results of hydrodynamical simulations by \cite{Mohamed}. The maximum fraction of the wind accreted has been forced to 50\%. In the case of \sgx, the wind accelerates very quickly but because the star can be close from filling its Roche lobe (while in symbiotic binaries, the \textsc{agb} star is usually well inside its Roche lobe), it can happen that the wind reaches the escape speed only outside the stellar Roche lobe ; the ratio $x$ can then be around the unity value also, where the amplification compared to the \textsc{bhl} approach is significant, albeit for different physical reasons than in a symbiotic binary. Figure from \cite{Abate2013}.}\label{fig:abate_mohamed}}
{\includegraphics[width=8cm]{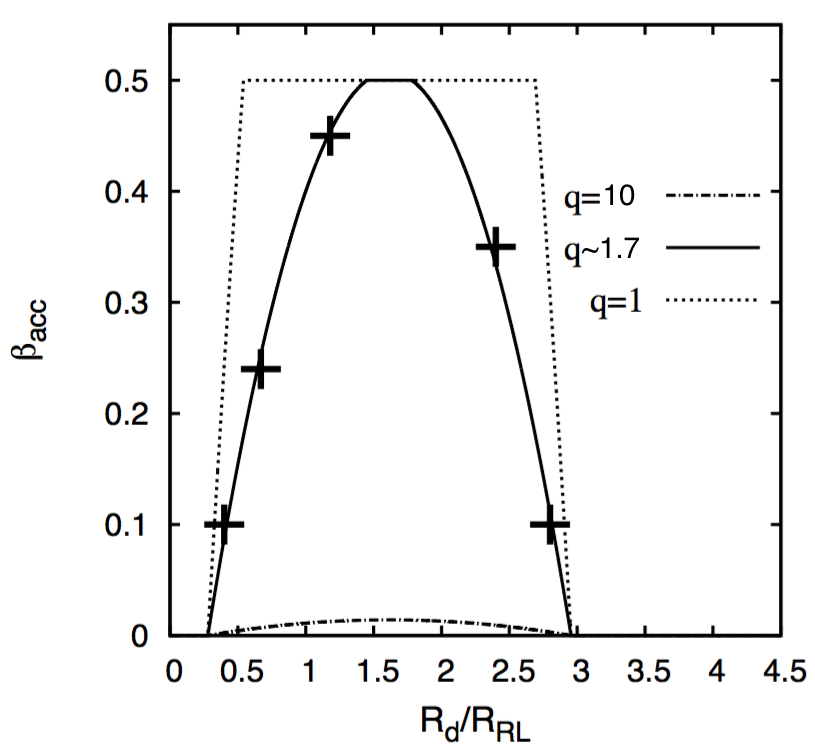}}
\end{figure}

In the context of the wind-\rlof model developed by \cite{Mohamed} for symbiotic stars, it has been suggested that the fraction of the Asymptotic Giant Branch (\textsc{agb}) star outflow accreted by the stellar companion could be higher than the \textsc{bhl} prescription above. In \cite{Abate2013}, an empiric (non monotonic) function of the mass ratio $q$ and of the ratio $x$ of the wind acceleration radius, $R_d$, by the Roche lobe radius of the donor star, $R_{\textsc{r,}1}$ \citep{Abate2013}, represented in Figure\,\ref{fig:abate_mohamed}, is introduced to fit the numerical experiments from \cite{Mohamed}. The wind acceleration radius is defined as the distance to the star where the wind reaches the local escape velocity. For \sgx where the mass ratio is high, we can interpret the latter by comparing the specific kinetic energy of a test-mass to the stellar potential only\footnote{Not to the whole Roche potential.}. In symbiotic binaries, the acceleration radius is believed to scale with the stellar radius \citep[and a ratio of temperatures not relevant in the present discussion, see equation (1) in][]{Abate2013}, which means that their ratio $x$ depends only on the filling factor among the parameters we consider. In particular, defined as such, it does not depend on the mass ratio. To account for the specificities of wind acceleration for hot massive stars and compute $R_d$, we can summon the velocity profile \eqref{eq:vr_FD} and include the dependence on the $\alpha$ force multiplier :
\begin{equation}
\label{eq:L07}
x\hat{=}\frac{R_d}{R_{\textsc{r,}1}}=f \times y\left(\alpha\right) 
\end{equation} 
where $y\left(\alpha\right)$ is the solution of the equation obtained by using \eqref{eq:vr_FD} to spot the distance from the star at which the wind reaches the local escape speed relatively to the modified gravitational potential of the star :
\begin{equation}
y\left(1-\frac{1}{y}\right)^{1.4}=\left(\frac{1-\alpha}{2.5\alpha}\right)^2
\end{equation}
For symbiotic binaries, other expressions of this ratio have been identified using the specific launching mechanism of \textsc{agb} stars leading to a different velocity profile \citep{Hofner2007}. 

Following \cite{Abate2013}, let us finally define the empirical relation represented in Figure\,\ref{fig:abate_mohamed}, between a theoretical value of the fraction of wind accreted, $\beta_{\textsc{sb}}$, the mass ratio $q$ and $x=R_d/R_{\textsc{r,}1}$ (which is a function of the filling factor $f$ and of the $\alpha$ force multiplier) :
\begin{equation}
\label{eq:beta_abate}
\beta_{\textsc{sb}}=\text{max}\left[ \frac{25}{9q^2}\cdot\left( c_1x^2 + c_2x + c_3 \right) ; \beta_{\textsc{hl}} \right]
\end{equation}
where $c_1$, $c_2$ and $c_3$ are three fitting coefficients who \cite{Abate2013} evaluated, using simulation results from \cite{Mohamed} performed with $q\sim 1.7$.
We choose the maximal value between the empirical fit and $\beta_{\textsc{hl}}$ to avoid the polynomial expression to yield non physical values. However, these coefficients can not be expected to fit well the case of \sgx since an essential similarity with symbiotic binaries breaks up here : the kinetic energy of the wind approaching the accretor, compared to the gravitational potential of the latter, is always much larger in \sgx than in symbiotic binaries and will lead to different values of those three fitting coefficients. This statement is directly supported by the dependency of $\beta_{\textsc{hl}}$ on the dimensionless velocity of the wind at the orbital separation in \eqref{eq:betaHL}. The question of the amplification of the mass accretion rate and of the dependencies with the mass ratio, the filling factor and the $\alpha$ force multiplier though subsist and are addressed below.

\subsubsection{Assessment of the theoretical tracers}

\begin{figure}
\centering

\begin{subfigure}[t]{.49\textwidth}
\centering
\hspace*{-1cm}
\includegraphics[height=9cm, width=9cm]{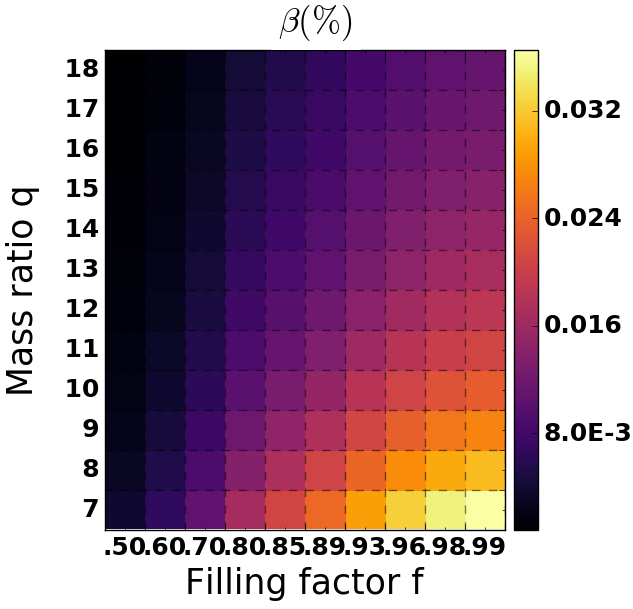}
\end{subfigure}
\begin{subfigure}[t]{.49\textwidth}
\centering
\hspace*{1cm}
\includegraphics[height=9cm, width=7cm]{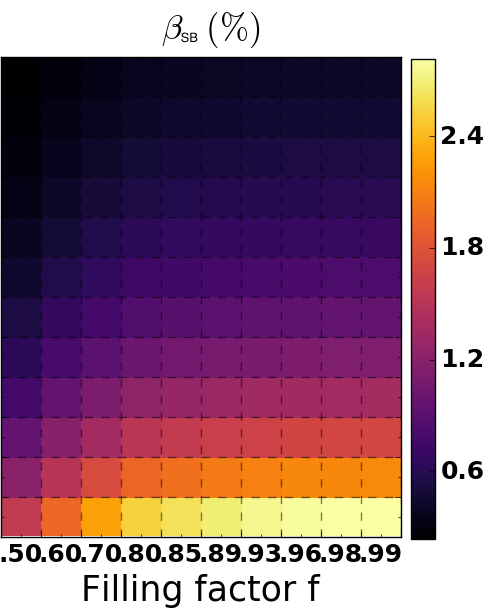}
\end{subfigure}

\medskip

\begin{subfigure}[t]{.49\textwidth}
\centering
\vspace{0pt}
\hspace*{-1.5cm}
\includegraphics[height=10cm, width=10cm]{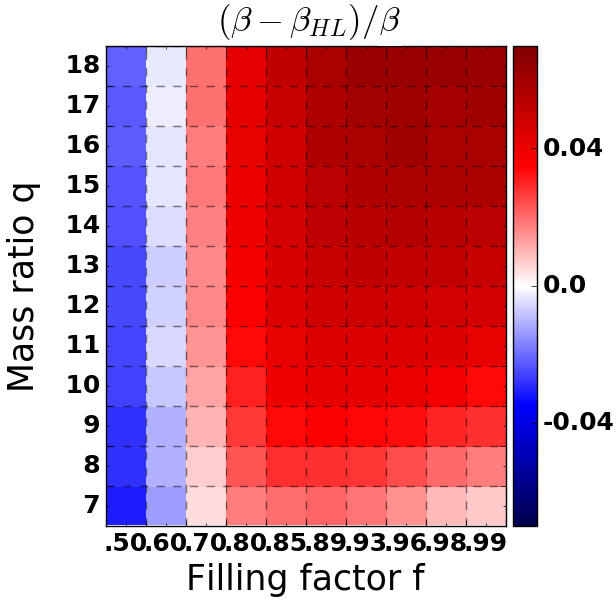}
\end{subfigure}
\begin{minipage}[t]{.49\textwidth}
\caption{(\textit{upper panels}) Measured and predicted fractions of wind accreted according to \eqref{eq:beta_abate} obtained for $\alpha=0.45$ and $\Gamma=0.2$, as a function of the filling factor and the mass ratio. The lower panel displays the relative discrepancy between the measured fraction of wind accreted and the one predicted by the \bhl prescription \eqref{eq:betaHL} used with the computed velocity at the orbital separation $\tilde{v}_{\bullet}$.}
\label{fig:threesome}
\end{minipage}

\end{figure}

The relevance of the wind-\rlof model for symbiotic binaries is a suggestive hint that wind accretion can prove more efficient than what the \textsc{bhl} prescription would suggest. Let us evaluate the relevance of the two previously introduced theoretical predictions, $\beta_{\textsc{sb}}$ and $\beta_{\textsc{hl}}$ by confronting them with the measured $\beta$. Figure\,\ref{fig:threesome} represents in the upper part the measured fraction of wind accreted from our simulations and the wind-\rlof predicted values for symbiotic binaries, both for\footnote{The conclusions drawn in the following paragraphs remain qualitatively valid for $\alpha=0.55$ and 0.65.} $\alpha=0.45$. As expected, $\beta_{\textsc{sb}}$ is several orders of magnitude larger due to wind speeds much lower in symbiotic binaries (a few 10\kms) than in \sgx (a few 100\kms at least). Let however discuss the dependencies of those fractions of wind accreted on the mass ratio and on the filling factor. 

For $\beta$, it is no surprise that, at a given mass ratio, as the star fills a lower fraction of its Roche lobe, the relative distance between the stellar surface and the compact object rises. Since the terminal velocity does not depend on the position of the sonic point and since the stellar gravity is largely overrun by the absorption lines acceleration close to the surface, we can expect the velocity at the compact object position with respect to the unchanged orbital velocity to be larger and thus, a lower extended accretion radius $R_{out}$ compared to the unchanged orbital separation. In other words, the angular size of the extended accretion sphere as seen from the star drops and so does the fraction of the wind being accreted. However, if one considers, at a given filling factor, a rise in the mass ratio, let us say at a given mass of the compact object, things get trickier. Indeed, on one hand the terminal speed is proportional to the escape velocity which rises and on the other hand, relative distance between the stellar surface and the compact object drops, giving the wind less room to accelerate. Thus, it is not possible to conclude a priori about the evolution of $\beta$ with $q$ and numerical computation was indeed required to obtain this tendency : for any filling factor, rising the mass ratio from 7 to 18 leads to a drop in the fraction of wind accreted by a factor 3. This didactic yet insightful reasoning illustrates how misleading qualitative approaches can be in systems where variables are as entangled with each other as \sgx. 

Concerning the dependency of the fraction of wind being accreted by the compact object with respect to $\alpha$, it will be discussed in greater details in the discussion devoted to the predicted X-ray luminosities (section \ref{sec:phys_mass_acc_rate}) but a first observation we make it that it follows the dependency of $R_{\text{out}}/R_{\text{R,}2}$ in Figure\,\ref{fig:RLOF_VS_wind} : as $\alpha$ rises, the wind acceleration gets more efficient and the terminal speed is higher, making the speed at the orbital separation larger and the computed accretion radius smaller. Thus, both $\beta$ and $\beta_{\textsc{hl}}$ drop by approximatively an order of magnitude as $\alpha$ rises from 0.45 to 0.65. Since a rise in $\alpha$, all other things being equal, also means a lower stellar mass outflow according to \eqref{eq:Mdot_MCAK}, we expect much lower X-ray luminosities for $\alpha=0.65$ than for $\alpha=0.45$, statement to be confirmed by Figure\,\ref{fig:array_LXHL} afterwards. Although the fraction of wind being accreted depends on $\alpha$, it is worth noticing that it does not depend on the $Q$ force multiplier : whatever its value, the velocity profile remains unchanged. Eventually, for the influence of $\Gamma$, a rise from 10\% to 30\% only results in a rise of a factor less than 2 for $\beta^+$ and $\beta_{\textsc{hl}}$, but it will play a more important role once its influence on the mass outflow is taken into account.


$\beta_{\textsc{sb}}$ shows little dependency on the filling factor compared to the measured $\beta$. Indeed, for $\alpha=0.45$, $x$ varies from 0.8 to 1.4 for a filling factor going from 50 to 99\%, which corresponds to a rise by a factor of approximately two in Figure\,\ref{fig:abate_mohamed}, too low compared to the rise by a factor of 4 derived from the upper left panel in Figure\,\ref{fig:threesome}. This discrepancy is likely due to the fact that in symbiotic binaries, the specific kinetic energy of the wind is of the order of the Roche potential amplitude which leads to a non isotropic dilution of the wind around the star \citep[see Figure 1 in][]{Mohamed2011} : the wind has time to be beamed towards the accretor for smaller filling factors, which partly compensates the larger distance between the stellar surface and the vicinity of the accretor. This departure from isotropy of the wind is much less pronounced in \sgx, at least outside of the extended accretion sphere. In the formula of \cite{Abate2013}, \eqref{eq:beta_abate}, the dependence on the mass ratio is accounted for empirically, focusing on the fact that, for a given set of trajectories, the accretion radius scales as the mass of the accretor and thus, the fraction of wind accreted rises with the square of the mass of the accretor. This evolution is an overestimate since it neglects the fact that the cross-section of the accretion cylinder is a disk which intercepts the flow of a spherically diluting wind, which expains why $\beta_{\textsc{sb}}$ overestimates the dependence on the mass ratio.

The bottom part of Figure\,\ref{fig:threesome} indicates that the enhancement of the fraction of wind being accreted with respect to the \bhl prescription is much lower in \sgx than in symbiotic binaries. Provided the wind speed at the orbital separation (and not the terminal speed) is used to compute $\beta_{\textsc{hl}}$, we retrieve the measured value of $\beta$ within 6\% for the mass ratios and the filling factors we consider. We then rely on $\beta_{\textsc{hl}}$ from now on to estimate the fraction of wind being accreted.

\subsection{Physical mass accretion rate and X-ray luminosity}
\label{sec:phys_mass_acc_rate}

\subsubsection{Scale parameters}

As a preliminary paragraph to this section, we first discuss the numerical values of the scale parameters. They have been chosen so as to suit observers' needs but also their measuring precision capacities : 
\begin{enumerate}
\item the orbital period $P$, which is precisely known when measured. According to their positions in the Corbet diagram (see section \ref{sec:SgxBeXB}), the orbital periods of \sgx range between 3 and 20 days. 
\item the mass of the compact object $M_2$, lying in a narrow range of values (1 to 2\msun approximately) and of prime importance to constrain the equation-of-state of \ns\footnote{Given the mass ratios we consider, the present investigation applies mostly to accreting \ns. However, since their intrinsic properties do not play a role (except a gravitational one), replacing them by \bh accretors is possible.}.
\item the dimensionless $Q$ force multiplier introduced by \cite{Gayley1995}, which ranges from 500 to 2,000 and amounts to 900 with little dispersion for OB Supergiant stars \citep{Shimada1994}. It serves to compute the amplitude of the mass loss rate.
\end{enumerate}
So as to compute thermodynamical quantities (\eg the mass loss rate, the density, etc), we need to set, in addition to the mass of the compact object and to the orbital period of the system, an additional energy related normalization variable. We chose the Eddington luminosity which makes $\Gamma$ a direct measure of the stellar luminosity and can be directly computed from the value of $q$ and the mass scale parameter.
 
\subsubsection{Mass accretion rate and X-ray luminosity}
 
\begin{figure}
\centering
\hspace*{-1cm}
\includegraphics[height=12cm, width=17cm]{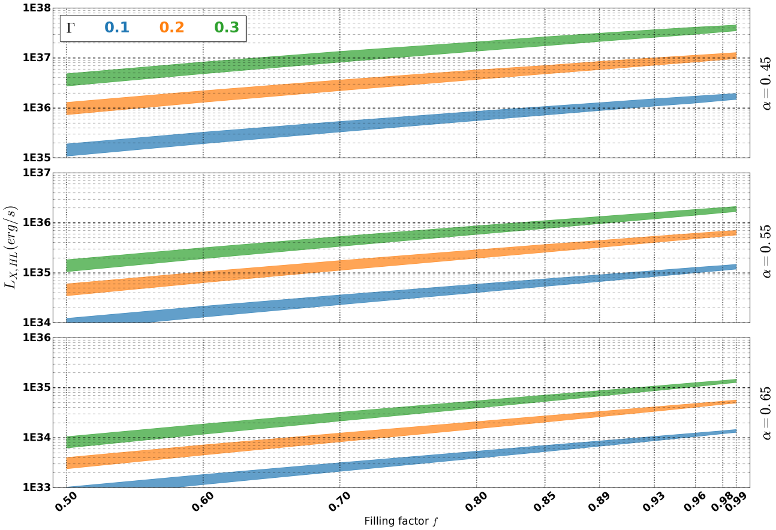}
\caption{Representation of the permanent X-ray luminosity described in section \ref{sec:phys_mass_acc_rate} as a function of the filling factor $f$ (x-axis), for different $\alpha$-force multipliers (different panels) and Eddington factors $\Gamma$ (different colors). The weaker dependence on the mass ratio, which ranges from 7 to 18, is represented by the thickness of each line (the lower limit being for larger mass ratios). Those results are computed for an intermediate value of the $Q$-force multiplier of 900. We also set a fiducial value of $M_2=1.8M_{\odot}$ for the mass of the neutron star and $L_X \propto M_2$. Changing the orbital period while keeping those quantities invariant does not change those luminosity profiles.}
\label{fig:array_LXHL}
\end{figure}

Interestingly enough, Figure\,\ref{fig:threesome} shows that, for a fixed mass of the compact object, the fraction of wind accreted decreases when the mass of the Supergiant star rises. But notice that this statement, along with all the ones made in the previous section, concerns only the fraction of wind which is accreted, not the actual physical mass accretion rate ; since a larger stellar mass also means a larger absolute mass loss rate according to \eqref{eq:Mdot_MCAK}, the evolution of the X-ray luminosity with $q$ could still be different. To compute the physical mass accretion rate, we rely on the expression \eqref{eq:Mdot_MCAK} of the mass loss rate, divided by a factor 2 to take into account the influence of the finite cone angle effect (see section \ref{sec:fd}). On the other hand, we use \eqref{eq:betaHL} for the fraction $\beta_{\textsc{hl}}$ of wind accreted by the compact object. The product of the two yields the physical mass accretion rate and the relation \eqref{eq:Lacc} provides the corresponding luminosity, essentially radiated as X-rays given the compactness of the \ns and the conclusions drawn in the introductory Chapter \ref{chap:acc_comp_obj}.

In Figure\,\ref{fig:array_LXHL} has been represented the evolution of the steady-state X-ray luminosity as a function of the 4 shape parameters. We set a fiducial value of $M_2=1.8M_{\odot}$ for the mass of the accreting \ns and according to\footnote{Provided the stellar mass - encompassed in the Eddington luminosity - is written as the product of the mass ratio by the mass of the accretor.} \eqref{eq:Mdot_MCAK}, the X-ray luminosity scales as the mass of the \ns. Similarly, we set the $Q$ force multiplier to $1,000$ and $L_X \propto Q^{\frac{1-\alpha}{\alpha}}$. Given the restricted range of expected values for $Q$ and $M_2$ and the weak dependency of $L_X$ on them, we chose to highlight the dependence of the X-ray luminosity on the dimensionless parameters. 

The stronger dependence of $L_X$ is on the $\alpha$ force multiplier, with a luminosity divided by 20 each time alpha rises from 0.45 to 0.55 and then 0.65. Each decreasing increment on $\Gamma$ entails a division of the X-ray luminosity by a factor of approximately 6 ; a lower effective gravity makes the wind terminal speed - which scales as the effective escape speed at the stellar surface - decrease, which leads to enhanced accretion of matter. We also notice that larger values of the $\alpha$ force multiplier lower the relative dynamical range of X-ray luminosities : fast winds leave less room to the influence of the other parameters. The luminosity rises by about an order of magnitude from a filling factor of 50\% to a configuration where the star is close to fill its Roche lobe ($f>95$\%). The weakest dependence is on the mass ratio. It turns out that, at a fixed mass of the accretor, larger stellar masses lead to lower X-ray luminosities : in the balance aforementioned, the fact that a larger stellar mass implies a larger wind speed and thus a smaller accretion cross-section prevails over the larger mass loss rate. However, the balance between those two effects weakens significantly the dependence of the X-ray luminosity on the mass ratio, weaker than the one on the filling factor $f$, the Eddington parameter $\Gamma$ and the $\alpha$ force multiplier. Given the low luminosity levels observed for $\alpha=0.65$ and even for $\alpha=0.55$, we can affirm that most of the \sgx we observe verify $\alpha\in\left[0.45;0.55\right]$. If it matches theoretical predictions from stellar atmosphere models for early-type B stars\footnote{For solar metallicity levels, \citep{Shimada1994} gives $\alpha\in\left[0.47;0.52\right]$ for a stellar effective temperature between 20 and 30,000K.} \citep{Shimada1994}, it is a bit below the values expected for O stars ($\alpha\sim 0.6$), which might suggest that the Supergiant companions in \sgx are anomalously inefficient at accelerating their winds given their spectral properties, maybe because of an anomalously low metallicity\footnote{The observational bias introduced by our limited sensibility does anyway artificially enhance the fraction of marginally low metallicity observed donors for radiatively-driven winds.}.

This Figure also enables us to pinpoint the degeneracies between the parameters. Indeed, we notice that a luminous star with a wind opaque enough\footnote{With respect to the metal absorption lines.} to be efficiently accelerated (\ie $\Gamma=0.3$ and $\alpha=0.55$) yields very similar X-ray luminosities, whatever the mass ratio or the filling factor, as a lower luminosity star with a less efficient radiatively driven acceleration (\ie $\Gamma=0.1$ and $\alpha=0.45$). 

A more spectacular property of the mass accretion rate emphasized by this parametrization (with both shape and scale parameters) is the independence, all other parameters being equal, of the X-ray luminosity on the orbital period, in agreement with Figure\,\ref{fig:wind_accretion}. The key point is that, if the orbital period rises, $\beta_{\textsc{hl}}$ is not altered since the accretion radius rises in the same proportions as the orbital separation. Indeed, at a fixed filling factor, it induces a larger stellar radius and alters the modified escape speed at the stellar surface in the same proportions as the reference velocity used as a velocity scale $\sqrt{\left(GM_2\right)/R_{R,1}}$. Because the velocity of the wind simply scales with the modified escape speed at the stellar surface, the dimensionless velocity at a distance of an orbital separation, $\tilde{v}_{\bullet}$, remains the same and so does the fraction of the wind being accreted according to \eqref{eq:betaHL},. On a more straightforward hand, changing the orbital period does not modify the stellar mass outflow\footnote{At least for large mass ratios for which we supposed that the mass of the \ns was too low compared to the stellar mass to alter the launching of the wind, essentially determined by the conditions at the sonic (here, stellar) surface.}. In the end, we are left with strictly the same X-ray luminosity. As a consequence, we claim that any dependency of the observed X-ray luminosities on the orbital periods must be attributed to a departure from the permanent \textsc{bhl}'s purely wind accretion sketch, typically a \textsc{rlof} mechanism, or to an underneath correlation not taken into account, for instance between the Eddington factor $\Gamma$ and the stellar radius, expected to be smaller for shorter period systems not undergoing \textsc{rlof}. In no case the orbital period could be considered as the main culprit for the X-ray luminosity, merely as a correlated quantity whose causal relation to the permanent luminosity must be traced back. The approach of section \ref{sec:X-ray_lum_wind_acc} to account for the observed range of X-ray luminosities is, according to this comment, unsatisfying. It provides an effective empirical way to portray the evolution of the X-ray luminosity with the period but only because stellar evolutionary arguments are likely to make the orbital separation (and so the orbital period) negatively correlated with the filling factor. A proper explanation should rather rely on the latter parameter, more directly involved in the building up of the X-ray emission.


\subsection{Shearing of the accretion flow}

\begin{figure}[!b]
\centering

\begin{subfigure}[t]{.49\textwidth}
\centering
\hspace*{-1cm}
\includegraphics[height=9cm, width=8cm]{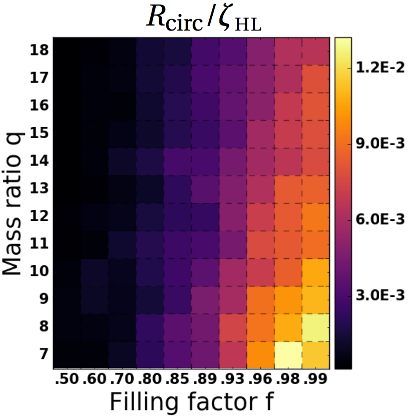}
\end{subfigure}
\begin{subfigure}[t]{.49\textwidth}
\centering
\hspace*{0cm}
\includegraphics[height=9cm, width=8cm]{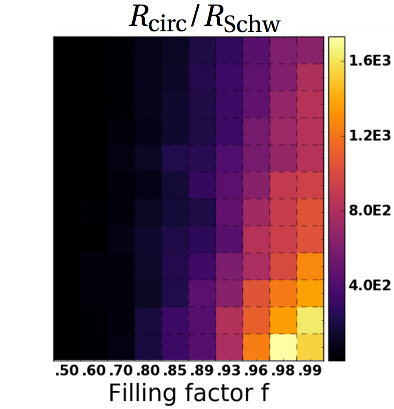}
\end{subfigure}

%
\begin{minipage}[t]{.49\textwidth}
\caption{(\textit{Upper left panel}) Colormap of the relative size of the circularisation radius of the flow entering the extended accretion sphere compared to the accretion radius for $\alpha=0.55$ and $\Gamma=0.2$. (\textit{Upper right panel}) Colormap of the relative size of the circularisation radius of the flow entering the extended accretion sphere compared to the Schwarzschild radius of the accretor for $\alpha=0.55$, $\Gamma=0.2$, $P=9$days and $M_2=2$\msun.}
\label{fig:threesome_2}
\end{minipage}

\end{figure}

The question of the angular momentum accretion rate and of its dependencies with respect to the 4 shape and 3 scale parameters remains. Do the configurations favourable to large mass accretion rates correlate positively with the likelihood to form a disk around the accretor in wind dominated mass transfers? 

To possibly form a disk-like structure around the accretor, the flow needs to gain enough angular momentum as it approaches the compact object. Similarly to the quantities $\beta ^+$ and $\beta$ in the previous section, we introduce numerical tracers of the likelihood of an inflow to form a disk. We also measure quantities both based on a streamline approach and on a mesh approach and check the matching between the two. A first dimensionless quantity of interest is the ratio of the inflow of angular momentum by the amount needed to place the matter on a circular orbit at the accretion radius. The main advantage of this quantity is that it depends only on the shape parameters and in particular, it does not depend on the mass of the accretor. It turns out to evolve in a similar fashion as $\beta$ whose dependencies have been detailed in the previous section. At a given mass ratio, the flow presents higher absolute values of angular momentum at the surface of the extended accretion sphere for lower ratios of the velocity of the wind at the orbital separation by the orbital speed $a\Omega$ (see also Figure\,\ref{fig:RLOF_VS_wind}). Since the angular momenta at stake in our simulations, at a scale where the Coriolis force marked the trajectories, are all negative, it means that the Coriolis force plays a more important role in systems where the wind speed is lower, a result explained by the fact that the force has more time to bend the trajectories, feature which overcompensates for the lower intensity of the force in this regime, whatever $\alpha$ and $\Gamma$. Within the extended accretion sphere, the angular momentum of the flow keeps evolving\footnote{Since the Roche potential is, strictly speaking, not isotropic.} which makes any statement about the tendency of the flow to form co or counter rotating discs around the accretor hazardous. However, for the flows where $R_{out} \ll R_{\textsc{R,}2}$, the ones strongly wind dominated, we can affirm that the extended accretion sphere is deeply enough embedded in the gravitational potential of the compact object to make the angular momentum of the flow approximately constant. It turns out that those pure wind accretion configurations (typically the blue squares in Figure\,\ref{fig:RLOF_VS_wind}) feature the lowest levels of angular momentum inflows. Those configurations are associated, in the order of decreasing influence, to high $\alpha$, low $f$, high $\Gamma$ and high $q$.

We also compute a dimensionless circularization radius given the amount of angular momentum (and of mass) entering the extended accretion sphere per second. For the moment, the absence of scale prevents us from comparing the circularisation radius $R_{\text{circ}}$ to the Schwarzschild radius $R_{\text{Schw}}$ of the neutron star, as the latter is set by the mass of the compact object\footnote{It is also because it summons the speed of light which can not be computed as long as the system units is not fixed.}, a degree of freedom not encompassed by the 4 shape parameters. The circularisation radius can be measured relatively to the accretion radius. Such a comparison sheds light on the way a disk-like structure could possibly form, provided it does. The ratios $\zeta_{\textsc{hl}}/R_{\text{R,}2}$ and $R_{\text{circ}}/R_{\text{R,}2}$ follow the same trend as $\beta$. But more interestingly, the ratio $R_{\text{circ}}/\zeta_{\textsc{hl}}$ also follows this trend, indicating that the bending of the shocked structure is stronger for low $q$, high $f$, low $\alpha$ and high $\Gamma$. In the upper left panel of Figure\,\ref{fig:threesome_2} has been represented this ratio as a function of $q$ and $f$ for $\alpha=0.55$ and $\Gamma=0.20$. Yet, the circularisation radius still remains two orders of magnitude smaller than the accretion radius\footnote{For very unefficient wind acceleration with $\alpha=0.45$, the circularisation radius is still smaller than the accretion radius but more by one order of magnitude only.}, characteristic of the size of the shocked region. We can then conclude that, apart in case of significant gain of angular momentum, a winding up of the whole shocked tail is not to be expected. If a disk forms, it grows within the shocked region \citep[as witnessed in numerical simulations by][]{Blondin2013a} where substantial instabilities are believed to take place \citep{Foglizzo2000}. Whether those instabilities drive variations of the front shock position large enough to bring the Coriolis bending in action remains to be investigated. 

The previous comments would turn out to be of little interest if the inflow crashes on the \ns surface before it can reach its circularisation radius. Indeed, a major difference with symbiotic binaries is that in \sgx, the accretor is small enough to let plenty of room for the shock to develop and possibly, within it, to a disk to form. Once a scale is set (see preliminary paragraph of the previous section on the scale parameters), the ratio $R_{\text{circ}}/R_{\text{Schw}}$ can be represented, along with $R_{\text{circ}}/\zeta_{\textsc{hl}}$ (discussed in the previous paragraph), for intermediate values of the $\alpha$ and $\Gamma$ parameters (see upper panels in Figure\,\ref{fig:threesome_2}). It shows that the circularization radius lies approximately two orders of magnitude below the accretion radius and two to three orders of magnitude above the \ns surface. Given the sketch in the bottom part of Figure\,\ref{fig:threesome_2}, we can affirm that the accretion radius is approximately one hundredth of the orbital separation for the fiducial set of shape parameters considered\footnote{Which has been chosen deliberately to lie on the lower edge of the flows susceptible to form a disk. Higher filling factors and lower mass ratios make the circularization radius still small enough to fit within the shock and large enough to confortably escape the magnetosphere influence.} ; according to the upper panels, the circularization radius is approximately 500 times smaller than this accretion radius and 200 times larger than the Schwarzschild radius of the \ns (so approximately 100 times larger than the \ns itself for a compactness parameter of $\Xi\sim 25\%$). Consequently, we retrieve the ratio 100$\times$500$\times$100$=$5$\cdot$10$^6$ between the orbital separation and the characteristic size of the \ns in a representative \sgx. This intermediate position of the circularisation radius between the shock and the \ns surface tends to immunize a putative disk both against disruptive instabilities at the shock level \citep{Manousakis2015c} and against premature truncation by the magnetosphere (see order of magnitude \eqref{eq:ns_magnetosphere} in section \ref{sec:obj}), provided the angular momentum does not significantly evolve within the extended accretion sphere. More importantly, we notice that for $\alpha$ at $0.55$ or below and for moderately high values of the filling factor, the likelihood to form a disk follows the same trend as the one drawn by the X-ray luminosity : the most luminous systems are also more likely to feature a disk-like structure around the accretor.


\section{Wind accretion on stage}
\label{sec:dim_out}

We now go on by assessing the self-consistency of our toy-model ; how far from the observations can be the plethora of parameters we have access to from a reduced number of key quantities? 


\subsection{A roadmap to self-consistent system parameters}

\begin{figure}
\begin{center}
\includegraphics[height=20cm, width=16cm]{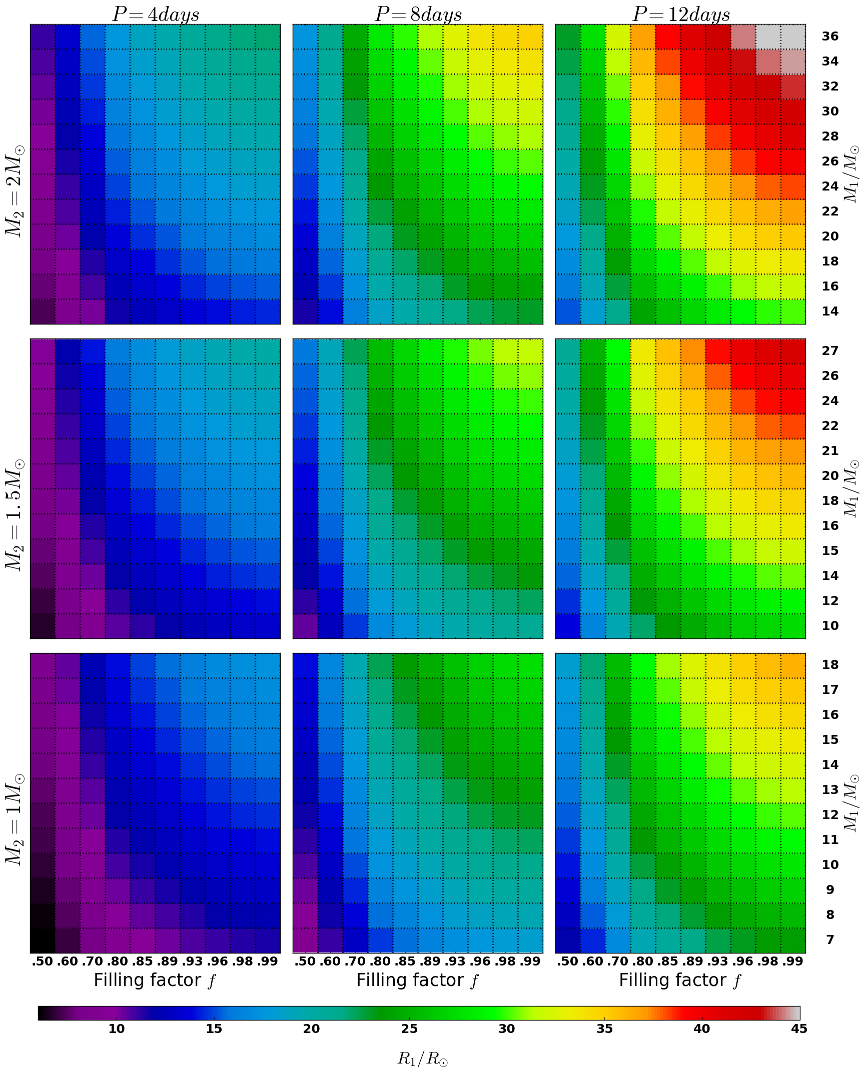}
\caption{All the possible values spanned by the stellar radius $R_1$ as a function of the only parameters which influence it. Indeed, the stellar radius is computed from the filling factor $f$ and the size of the stellar Roche lobe $R_{\text{R,}1}$ with the formula \eqref{eq:egg}. Hence the dimensionless $R_1$ depends on the filling factor only but so as to compare it to solar radii, it requires the knowledge of the length scale $R_{\text{R,}1}$, set by the orbital period $P$ (from left to right column), the mass of the compact object $M_2$ (from upper to lower line), the mass ratio (on the right side, already converted into stellar masses using $M_2$) and the resort to Kepler's third law.}
\label{fig:array_R1}
\end{center}
\end{figure} 

The theoretical four shape parameters our toy-model is based on are theoretical tools which encapsulate all the scale-free dynamics of the problem\footnote{Whether the stellar volume is of the order of the solar one or a bucket does not change the slightest things to the dimensionless results - at least those which do not make use of external constants such as the speed of light.}. However, they do not correspond to the observables we have access to. In this section, we aim at providing a minimal though comprehensive track through the maze of physical quantities in \sgx which fit our model. If the applet we designed (see appendix \ref{sec:spyre}) is the handy carriage to travel through, we now give a map for a quick journey to a consistent overview of a \sgx.

We first introduce the relations we will rely on to deduce the broad set of parameters available in this toy-model. We start with the following set of 6 parameters we call prior parameters, selected for their accessibility to observations and/or their necessity to deduce the other parameters :
\begin{enumerate}
\item the orbital period $P$
\item the mass of the compact object $M_2$
\item the stellar temperature $T$
\item the decimal logarithm of the stellar surface gravity in \textsc{cgs} units $\log (g)$
\item the $\alpha$ force multiplier
\item the $Q$ force multiplier
\item either the terminal speed of the wind $v_{\infty}$ or the average X-ray luminosity of the system $\langle L_X \rangle$
\end{enumerate}
Let us illustrate the procedure we follow starting from the above set of prior parameters. 

Spectroscopic inquiries of the pressure-broadened Balmer and He II lines coupled to stellar atmospheres codes set constraints on the decimal logarithm of the effective surface gravity and on the stellar temperature $T_1$ \citep[see \eg ][]{Clark2002a} : to know the actual surface gravity, $\log(g)$, thus requires to correct for radiative pressure \citep{Sander2015}. One can however rely on the calibrated data for the same spectral type of stars \citep[see \eg][]{Searle2008,Puls2008}. $\log(g)$ features few dependences ($f$ and $P$, plus a weak dependence on $M_2$ and $q$) such as it narrows the range of filling factors possible :
\begin{equation}
\label{eq:filling}
f \sim \underbrace{\frac{0.32q^{1/2}}{\mathcal{E}(q)\cdot(1+q)^{1/3}}}_\text{78\% $\rightarrow$ 82\%} \left( \frac{M_2}{2M_{\odot}} \right)^{\frac{1}{6}} \left( \frac{9\text{ d}}{P} \right)^{\frac{2}{3}} 10^{\frac{3-\log (g)}{2}} 
\end{equation}
with $g$ in \textsc{cgs} units and where the bracket evaluates the factor it embraces as $q$ goes from 7 to 18. The dependence on the mass of the compact object is also weak, accounting for a few percent only for a neutron star. Since the orbital period is precisely known (when measured), we can affirm that $\log(g)$ is a direct measure of the filling factor. Its precise knowledge (\ie with a precision of at most a few percent), along with the orbital period, is a precious and sufficient asset to enclose the value of the filling factor in a $\sim$10 to 20\% precision range. In parallel, we deduce the Eddington factor from the prior of $T$ and $\log(g)$ :
\begin{equation}
\label{eq:Gamma_roadmap}
\Gamma \sim 0.23 \left( \frac{T}{25\text{kK}} \right)^4 10^{3-\log (g)}
\end{equation}
A first sanity check can be realized here checking whether correcting the effective surface gravity with this value of $\Gamma$ does lead to a value of $\log\left(g\right)$ consistent with the one above, from stellar spectral type considerations. The values we get using the two equations above hold whatever the $\alpha$ and $Q$ force multipliers but will participate to set the mass-loss rate and thus, the average X-ray luminosity.

Concerning the two force multipliers listed as prior parameters, they were calculated by \cite{Shimada1994} for OB-Supergiants with 520,000 atomic lines and they depend mostly on $T$ and $\log (g)$. For Solar metallicity and effective temperatures between 20 and 30kK, typical of the early-type B stars we consider in the upcoming section\footnote{Except if explicitly stated.}, $Q\sim900$ and $\alpha\sim0.5$, with little departure from those values. 

Here, a junction shows up : either we make use of a measured velocity of the wind at infinity\footnote{Which does not depend on the $Q$ force multiplier.}, which is not always available nor precise, or we rely on the X-ray luminosity of the system, which requires a precise estimate of the distance to the \sgx\footnote{Generally available thanks to the spectral energy distribution fits accomplished by \cite{Coleiro2013} and \cite{Coleiro2013a}.} and more dramatically, to assume a trivial relation like \eqref{eq:Lacc} between the mass accretion rate and the X-ray luminosity (\eg without reprocessing of the X-rays). The first option leads, once the wind speed at infinity corrected for the finite cone effect \eqref{eq:vr_FD} is accounted for, to the following relation :
\begin{equation}
\frac{q}{\mathcal{E}(q)(1+q)^{1/3}} \sim 0.023\frac{f}{1-\Gamma}\left(\frac{\alpha}{1-\alpha}\right)^2\left[\frac{v_{\infty}}{(GM_2/P)^{1/3}}\right]^2
\end{equation}
The mass ratio is then obtained after a numerical inversion with a Newton-Raphson root finder for instance. This operation however leads to a poor level of precision on the value of $q$ due to a low variation of the \lhs in the range of mass ratios considered\footnote{Approximately from 6 to 11 for $q\in\left[7;18\right]$.}. It is thus preferable to make the most of the whole coupling between phenomena our model relies on and to compare the result obtained above to the one obtained using the X-ray luminosity. The non-trivial coupling of the latter with the other parameters makes a manual monitoring of the $q$ values compatible with the persistent X-ray luminosity observed compulsory but feasible with the \texttt{WASO} interface described in appendix \ref{sec:spyre}. The weak dependence of $L_X$ on $q$ blamed in \ref{sec:phys_mass_acc_rate} does not make this operation very precise neither. The matching between the two values of the mass ratio above is a good hint in favour of a reliable value of the mass ratio, what we will show in the next section with eclipsing \sgx (where the mass ratio is observationally constrained).

In any case, once the mass ratio is determined, we are left with all the shape and scale parameters, opening the doors to all the remaining variables and in particular the stellar radius (see Figure\,\ref{fig:array_R1}), mass, luminosity and mass outflow, the wind mass density, the dimensions of the accretion sphere around the neutron star and the circularisation radius compared to the size of the compact object, a clue which indicates the likelihood of the formation of a disk. One can finally summon additional independent measures (\eg of the stellar radius, of the spectroscopically derived stellar mass) or theoretical expectations (\eg the wind momentum - luminosity relationship of section \ref{sec:WLR}) to discuss the reliability of the initially considered mass of the compact object, the only truly arbitrary prior.


\subsection{Classical persistent \sgx}

\subsubsection{Selection procedure}

We now discuss the case of three persistent \sgx likely being archetypes of this observational category : Vela X-1, XTE J1855-026 and IGR J18027-2016 (\aka SAX 1802.7-2017). For short, we nickname the second one XTE and the third one SAX. We ended up with these three classical \sgx using the following requirements, suitable to fit our toy-model, applied to the sample of $\sim$20 available \sgx :
\begin{enumerate}
\item an orbital period has been measured.
\item the location of the system in the Corbet diagram \citep[][Figure 5]{Walter15} is consistent with \sgx (orbital period below 20 days and spin period of the pulsar above 100 seconds).
\item to avoid any discrepancy on the force multipliers due to metallicity-related issues, we focused on \sgx within the Milky Way \citep{Mokiem2007a}.
\item to avoid the net shift in the fraction of the wind being accreted induced by an eccentric orbit \citep[see equation (6) in][]{Boffin1988}, we selected systems where the eccentricity is below 0.1 \citep[which invalidates for example 4U 1907+097 according to][]{Nespoli2008}.
\item we rejected potential intermediate \textsc{sfxt} such as the eclipsing IGR J16418-4532 and IGR J16479-4514.
\item we rejected the systems where \textsc{rlof} was suspected or confirmed (Cen X-3).
\item we did not retain the systems hosting a black hole candidate (Cyg X-1 or even 4U 1700-377) due to the mass ratios and the assumption of mildly perturbed spherical wind we considered.
\item we selected systems where the star was not excessively evolved (as in GX301-2).
\item we discarded systems featuring a peculiarly slow wind (OAO 1657-415).
\item the neutron star is eclipsed by its stellar companion so as to be able to retrieve additional information from radial velocity measurements.
\end{enumerate}
Using eclipsing systems where the best-fits ephemerides of the light curves set constraints on the mass ratio is a way to compensate for the low precision of the mass ratio obtained from the measures of the terminal wind speed and the average X-ray luminosity. Nevertheless, the former is flawed by the unknown inclination of the system, an issue our approach is not influenced by : in the incoming Table \ref{tab:paramsAll}, the values between parenthesis are the ones if the system is seen edge-on while the other one is the limit-case of \rlof. 

Interestingly enough, a look a posteriori to the Corbet diagram in Figure\,\ref{fig:corbet_diag} indicates that those three systems (identified with an additional blue circle around the marker) lie in a close region of the $P$-$P_{\text{spin}}$ space ; it might support the idea that the properties of the accreting neutron star (age, magnetic field, etc) are similar and indirectly favor a wind dominated mass transfer.

\subsubsection{Observed parameters}

\begin{table*}
\begin{threeparttable}
 \caption{Parameters of the 4 \sgx considered in decreasing order of orbital period. The second set of stellar parameters are deduced from methods not directly based on the stellar spectral type (radial velocity measures, eclipses...). The pairs of numbers for the wind parameters are not the extremal values found in the literature but correspond to the slow and fast regimes we will consider to check self-consistency. The parameters in bold are among the lever quantities we use to deduce the other ones. The X-ray luminosities reported are for an energy range between 3 and 100 keV.}
 \label{tab:paramsAll}
 \begin{tabular}{l|ccc}
  \hline
          & Vela X-1 & XTE J1855-026 & IGR J18027-2016\\
  \hline
   Stellar parameters... & & &\\
   SpT & B0.5Iae & B0Iaep & B1Ib\\ 
   $\mathbf{T/10^3}$\textbf{K} & 25 & 28 & 22\\
   \textit{... from SpT} & & &\\
   $\mathbf{\textbf{log}(g)}$ & 2.9 & 3.0 & 2.7\\
   $R/R_{\odot}$ & 34 & 27 & 35\\
   $\log(L/L_{\odot})$& 5.6 & 5.6 & 5.4\\
   $M_{1\text{,evol}}/M_{\odot}$& 33 & 25 & 22\\
   \textit{... from observations} & & &\\
   $\mathbf{\textbf{log}(g)}$ & 2.9 & 3.1 & 3.1(3.2)\\
   $R/R_{\odot}$ & 32(27) & 22 & 20(17)\\
   $\log(L/L_{\odot})$& 5.6(5.4) & 5.4 & 4.9(4.8)\\
   $M_1/M_{\odot}$& 28(23) & 21 & 19(18)\\   
   Orbital parameters& & &\\
   $\mathbf{P/}$\textbf{day}& 8.96 & 6.07 & 4.47 \\
   $q=M_1/M_2$& 11-13 & 12-16 & 12-16\\
   $M_2$& 1.9(2.3) & 1.4 & 1.3(1.5) \\
   Wind parameters& & &\\
   $v_{\infty}/$\,km$\cdot$\,s$^{-1}$& 700 | 1,700 & 1,600 & 1,100\\
   $\dot{M}/10^{-6}M_{\odot}\cdot$\,yr$^{-1}$& 0.6 | 2 & 1.9 & 1\\
   X-ray luminosity& & &\\
   $D/kpc$& 1.9-2.2 & 9.8-11.8 & 12.4\\
   $F_{14\rightarrow 195}/10^{-11}$\,erg$\cdot$\,s$^{-1}\cdot$\,cm$^{-2}$& 390 & 20 & 7\\
   $F_{17\rightarrow 60}/10^{-11}$\,erg$\cdot$\,s$^{-1}\cdot$\,cm$^{-2}$& 215 & 10.3 & 4.2\\	
   $\langle L_X \rangle/10^{36}$\,erg$\cdot$\,s$^{-1}$& 2.5-3.7 & 3.4-5.1 & 1.9-2.3\\
   & & & \\
   Spin/seconds& 238 & 361 & 140 \\
  \hline
 \end{tabular}
 \begin{tablenotes}
 \small
 \item 1 : \cite{Falanga2015}, 2 : \cite{Coley2015}, 3 : \cite{Clark2014}, 4 : \cite{Quaintrell2003}, 5 : \cite{Coleiro2013}, 6 : \cite{Liu2006}, 7 : \cite{Ducci2009}, 9 : \cite{Prat2008}, 10 : \cite{Hannikainen2007}, 11 : \cite{Walter15}, 12 : \cite{Lutovinov2013}, 13 : \cite{Prinja2010}, 14 : \cite{Searle2008}, 15 : \cite{Mason2011}
 \end{tablenotes}
\end{threeparttable}	
\end{table*}

We searched the literature to gather the parameters displayed in Table \ref{tab:paramsAll}.

The spectral types of Vela X-1 and XTE have been found in \cite{Coleiro2013} who also provide the corresponding effective temperature. For SAX, we used the conclusions of \cite{Torrejon2010}. Following their approach, we read the corresponding stellar parameters in \cite{Searle2008}, Table 5. However, the value obtained from this modeling approach for the surface gravity of SAX leads with \eqref{eq:filling} to a star which largely overflows its Roche lobe, for masses of the compact object larger than even 0.5\msun. We thus discard it and pinpoint an important discrepancy between the surface gravity deduced from the stellar spectral type and the one compatible with the accreting regime of the system. Concerning the observed parameters, we deduced the actual surface gravity from the measured radius and mass.

For the observed stellar parameters and mass of the compact object in Vela X-1, we relied on the radial velocity measurements made by \cite{Quaintrell2003}. The values between parenthesis are the ones if the system is seen edge-on while the other one is the limit case of Roche lobe overflow. The values measured by \cite{Falanga2015} essentially agree, and so do the previous values given by \cite{VanKerkwijk1995}. For the terminal speed of the wind, we report both the low value obtained by \cite{Gimenez-Garcia2016} and \cite{VanLoon2001} and the higher terminal speed deduced by \cite{Watanabe2006}. Yet, we notice that \cite{VanLoon2001} and \cite{Gimenez-Garcia2016} consider stellar masses and/or radii marginally compatible with the ones derived by \cite{VanKerkwijk1995}, \cite{Quaintrell2003} and \cite{Falanga2015}. The mass-loss rates are from \cite{Gimenez-Garcia2016} (lower value) and \cite{Watanabe2006} (upper value). The low value of the mass-loss rate and the high value of the terminal speed are consistent with the theoretical predictions for isolated stars of similar spectral type \cite{Searle2008} albeit not necessarily relevant. 

For the observed stellar parameters and mass of the compact object in XTE, we used the values derived by \cite{Falanga2015}, more precise than the ones derived by \cite{Coley2015}. They based their approach on Monte Carlo best-fit ephemerides of the light curves. However, they tend to determine low inclinations which turn out to be, for XTE and SAX for instance, below the critical ones to avoid \textsc{rlof} derived by \cite{Coley2015}. Observational biases apart, the filling factor does not physically have to be near unity as previously defended in section \ref{sec:struct_acc_flow}. For the parameters of the wind, since they have not been directly measured, to the best of our knowledge, we used the values listed by \cite{Searle2008} for isolated stars of similar spectral type.

For the observed stellar parameters and mass of the compact object in SAX, we used the values derived by \cite{Coley2015} using radial velocities. As for \cite{Quaintrell2003}, the values between parenthesis are the ones if the system is seen edge-on while the other one is the limit-case of Roche lobe overflow. The stellar parameters are consistent with those determined by \cite{Mason2011}. \cite{Falanga2015} obtained slightly larger masses for the star and the compact object. For the parameters of the wind, we used the values listed by \cite{Searle2008} for isolated stars of similar spectral type.

We proceeded in the following way to determine estimates of the average X-ray luminosity. We started from the fluxes : we considered both the ones between 14 and 195keV, $F_{14\rightarrow 195}$ observed by the BAT instrument on Swift (\href{http://swift.gsfc.nasa.gov/results/bs70mon/}{http://swift.gsfc.nasa.gov/results/bs70mon/}), and the ones given by \cite{Walter15} in the hard X-rays, between 17 and 60keV, $F_{17\rightarrow 60}$. We used the estimates of the distances from \cite{Walter15,Coleiro2013,Kaper1998} for Vela, from \cite{Coleiro2013} for XTE and from \cite{Walter15} for SAX. We neglected the absorption of those sources in these wavebands, an assumption legitimate for Galactic sources as explained in appendix \ref{sec:abs_in_nut} (a prescription to account for absorption is also given there). To extrapolate to broadband X-ray fluxes, we notice that \cite{Filippova2005} claims a broadband (3 to 100keV) X-ray flux for XTE approximately 3 times larger than $F_{17\rightarrow 60}$ and 1.5 times larger than $F_{14\rightarrow 195}$ ; assuming similar photon energy distribution for the three sources within this energy band, we then deduce the corresponding X-ray luminosities. Those X-ray luminosities are merely estimates which must not be used but as guidelines. The uncertainty we get for XTE and SAX is obtained by using both aforementioned waveband for the flux and by varying the distance to Vela X-1 from 1.8 to 2.2kpc. Notice that for Vela X-1, a shorter distance has been derived by \cite{Chevalier1998} and would lead to a higher X-ray luminosity. Also, the values we get for its average X-ray flux between 3 and 100keV are slightly below the usual X-ray luminosity found in the literature, 4$\cdot10^{36}$\,erg$\cdot$\,s$^{-1}$, possibly because of the higher absorption at the lower limit of the energy range we consider, not corrected here ; those luminosities are thus lower limits. We expect the use of those X-ray luminosities combined with the measures of the terminal speed to yield mass ratios compatible with the ones measured by light curves fitting given in Table \ref{tab:paramsAll}. 

\subsubsection{Self-consistent sets of parameters}

We start to explore the possible set of self-consistent parameters for Vela X-1 from Table \ref{tab:fill_and_gam}. Given the high value of $\Gamma$, we start by setting it to 30\%. The lower value of the speed at infinity, $700$\kms, requires at most $\alpha=0.45$ and $M2=1$\msun. The associated X-ray luminosity would be larger than 10$^{37}$erg$\cdot$s$^{-1}$, way above the permanent level observed for Vela X-1. The stellar mean density would also be suspiciously low with a 10\msun star having a 17\rsun radius. Given the observational constraints, we discard this possibility and support the upper measured value of the wind speed at infinity of 1,700\kms. With $\alpha=0.5$ (and $Q=900$), typical of early type B stars with effective temperatures between 20 and 30,000K, we get a more realistic overview. However, even for an accretor mass as high as 2\msun though (where the filling factor has to be larger than 83\%), we hardly reach $v_{\infty}=$1,300\kms, and for very high stellar masses (above 35\msun), hardly compatible with the spectral type. The associated X-ray luminosities would also be excessively large by a factor of 2. Rising $\alpha$ to 0.55 makes the X-ray luminosity drop to a value several times too low and an Eddington factor larger than 30\% is unlikely given the stellar spectral type. On the contrary, keeping $\alpha=0.50$ but lowering $\Gamma$ to 20\%, the low end of the range we deduced from the effective temperature and the surface gravity, brings us to $v_{\infty}\sim$1,600\kms for $q>11$ (and still $f>83\%$). To overlap this area with the available X-ray luminosity and the estimated stellar mass outflow, we must privilege a large value of the filling factor ($\sim 95\%$) and a moderately low value of the mass ratio ($q\gtrsim 12$). Coupled to the observational results derived from the eclipses, it means that Vela X-1 is almost seen edge-on and that the lowest values of stellar masses and radii must be privileged. With a circularization radius one hundred times smaller than the accretion radius, the flow can cicrularize once it reaches a radius of approximately one thousand times the Schwarzschild radius of the accretor. Depending on the extension of the magnetosphere, it might be enough to see a disk-like structure maintained in a short range of radii. The orbital structure of the flow is mostly wind-dominated.

\begin{table*}
\begin{threeparttable}
 \caption{Preliminary ranges of filling factors and Eddington parameters possible for each of the three systems considered, using the equations \eqref{eq:filling} and \eqref{eq:Gamma_roadmap}. The subscript of each of the filling factor refers to the mass of the accretor, in \msun. It does not affect the value of the Eddington parameter $\Gamma$. The uncertainties represent on one hand the uncertainty on $\log\left(g\right)$, set to 0.05 around the centered value given in Table \ref{tab:paramsAll}, and on the other hand the variation of the mass ratio from 7 to 18. The lower (resp. upper) edge of each range corresponds to $q=7$ (resp. $q=18$) and a maximum (resp. minimum) value of $\log\left(g\right)$. Each time the filling factor obtained was above 100\%, we set a dash to warn the reader that a \rlof star is unlikely.}
 \label{tab:fill_and_gam}
 \begin{tabular}{c|ccc}
  \hline
        System  & Vela X-1 & XTE J1855-026 &  IGR J18027-2016 \\
  \hline
   $f_1$ & 74 $\rightarrow$ 88\% & 80 $\rightarrow$ 96\% & 88 $\rightarrow$ -\% \\
   $f_{1.5}$ & 79 $\rightarrow$ 94\% & 86 $\rightarrow$ -\% & 94 $\rightarrow$ -\% \\
   $f_2$ & 83 $\rightarrow$ 98\% & 90 $\rightarrow$ -\% & 99 $\rightarrow$ -\%\\
   $\Gamma$ & 22 $\rightarrow$ 38\% & 25 $\rightarrow$ 40\% & 7 $\rightarrow$ 13\% \\
  \hline
 \end{tabular}
\end{threeparttable}	
\end{table*}

\begin{table*}
\begin{threeparttable}
 \caption{Sets of self-consistent fundamental parameters associated to each system. No uncertainty is specified because the systematics dominate and are discussed in the text in more details. Those values serve to compare the three systems to each other, not to specify precisely the parameters of each system. They indicate a trend, not a state.}
 \label{tab:do_it_all}
 \begin{tabular}{c|ccc}
  \hline
        System  & Vela X-1 & XTE J1855-026 &  IGR J18027-2016\\
  \hline
   $q$ & 12 & 14 & 11 \\
   $f$ & 95\% & 89\%  & 98\% \\
   $\alpha$ & 0.50 & 0.50 & 0.45\\
   $\Gamma$ & $\gtrsim$20\% & 30\% & 10\% \\
   $P$ (days) & 8.96 & 6.07 & 4.47 \\
   $M_2$ (\msun) & 2 & $\lesssim$2 & 1.5 \\
   $Q$ & 900 & 900 & 1,100 \\  
  \hline
 \end{tabular}
\end{threeparttable}	
\end{table*}

The situation is very similar for XTE which displays a fast wind : a higher velocity means a larger stellar mass\footnote{Remember that the terminal speed scales with the escape speed.} at a similar filling factor and mass ratio, which implies a drop in the X-ray luminosity both because the wind is faster and because the mass ratio is larger. The larger mass outflow is not enough to compensate for the smaller solid angle delimited by the extended accretion sphere. We must thus either consider heavy accretors and set $M_2$ to 2\msun with the canonical $\alpha$ at 0.5 or intermediate mass accretors with $M_2=1.5$\msun but a larger efficiency of the wind launching with $\alpha=0.55$. Notice that the value of $Q$ does not play any role in the value of $v_{\infty}$ and that $\Gamma$ is somewhat larger than the one derived for Vela X-1 and can less easily be considered as low as 20\%. If the accretor is heavy, the condition on $v_{\infty}$ enforces a filling factor in a narrow range between 85 and 90\% and a mass ratio above 15 ; if $M_2=1.5$\msun, we are left with all the upper right quarter of the $\left( f,q \right)$ space. In the latter case, the X-ray luminosity is hardly compatible with the observed one, merely reaching 1.5$\cdot 10^{36}$erg$\cdot$s$^{-1}$ for $q=13$ and $f=99\%$, a factor of 2 to 3 below the estimated value. However, for a heavier accretor, the X-ray luminosity found in the region compatible with the measured values of $v_{\infty}$ lies between 4 and 6$\cdot10^{36}$erg$\cdot$s$^{-1}$, in agreement with the observations. We thus support larger stellar masses and radii than the ones displayed in Table \ref{tab:paramsAll} on the basis of this analysis, with a stellar mass loss rate above 4$\cdot$10$^{-6}$\msun$\cdot$yr$^{-1}$. The circularization radius is one hundredth of the accretion radius and one thousand times larger than the Schwarzschild radius of the accretor ; in spite of the large velocity of the flow, it sounds like this configuration still leave room to the possibility to form a disk, contrary to Vela X-1 which however has a very similar mass ratio and filling factor.

\begin{figure}
\centering

\begin{subfigure}[t]{1\textwidth}
\centering
\includegraphics[height=6.5cm, width=10cm]{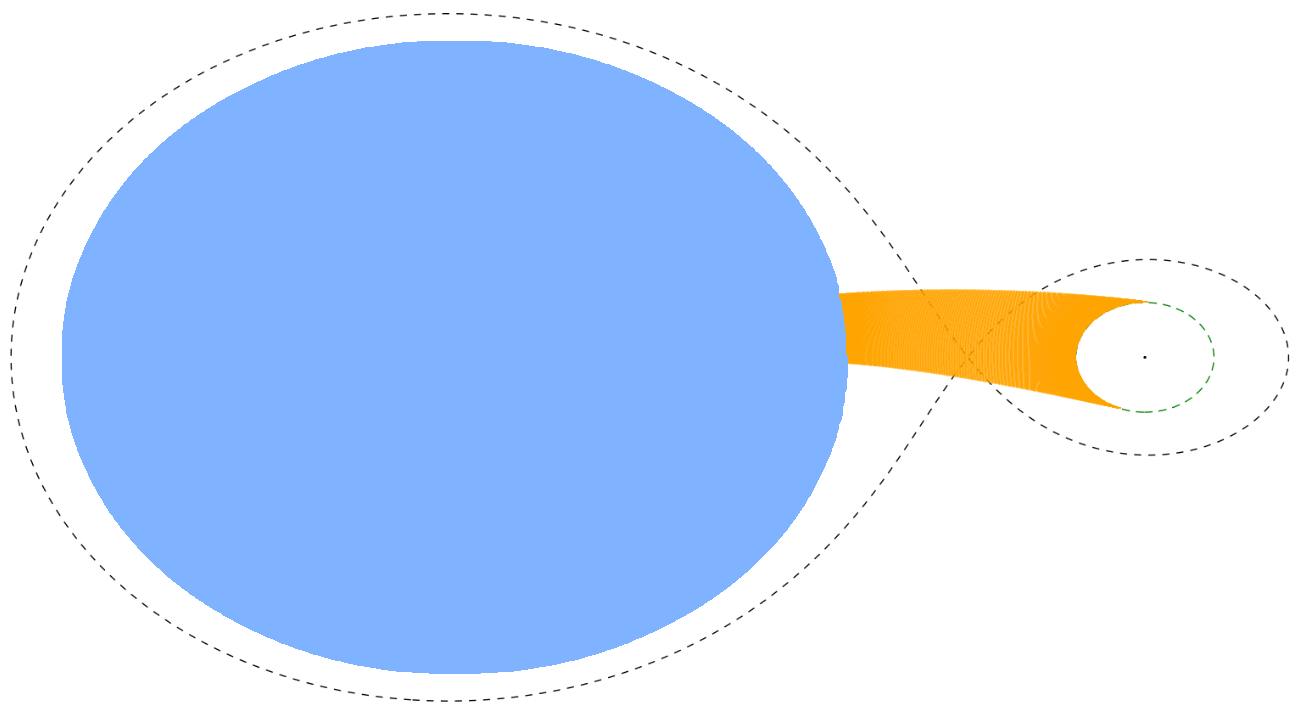}
\end{subfigure}
\begin{subfigure}[t]{1\textwidth}
\centering
\includegraphics[height=6.5cm, width=10cm]{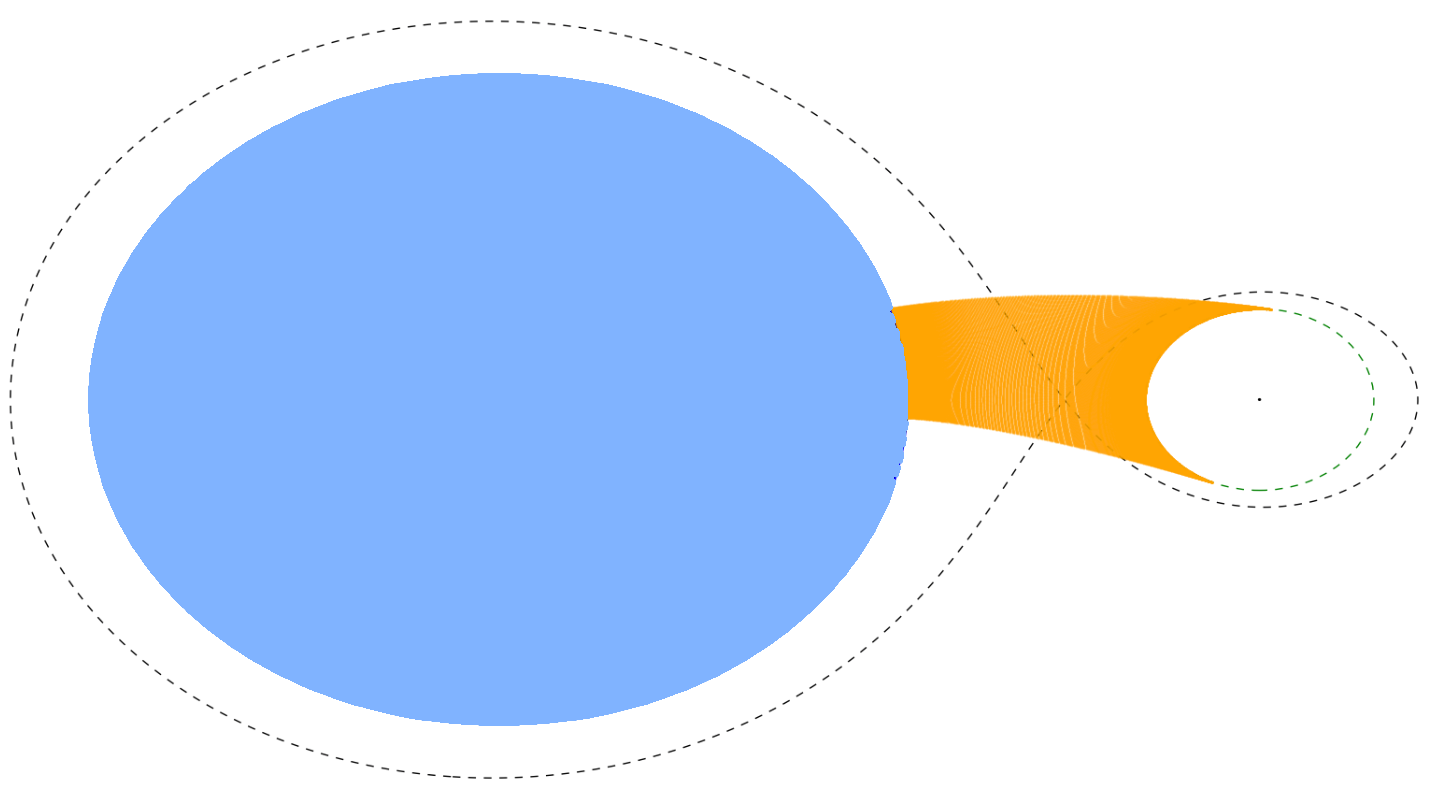}
\end{subfigure}
\begin{subfigure}[t]{1\textwidth}
\centering
\includegraphics[height=6.5cm, width=10cm]{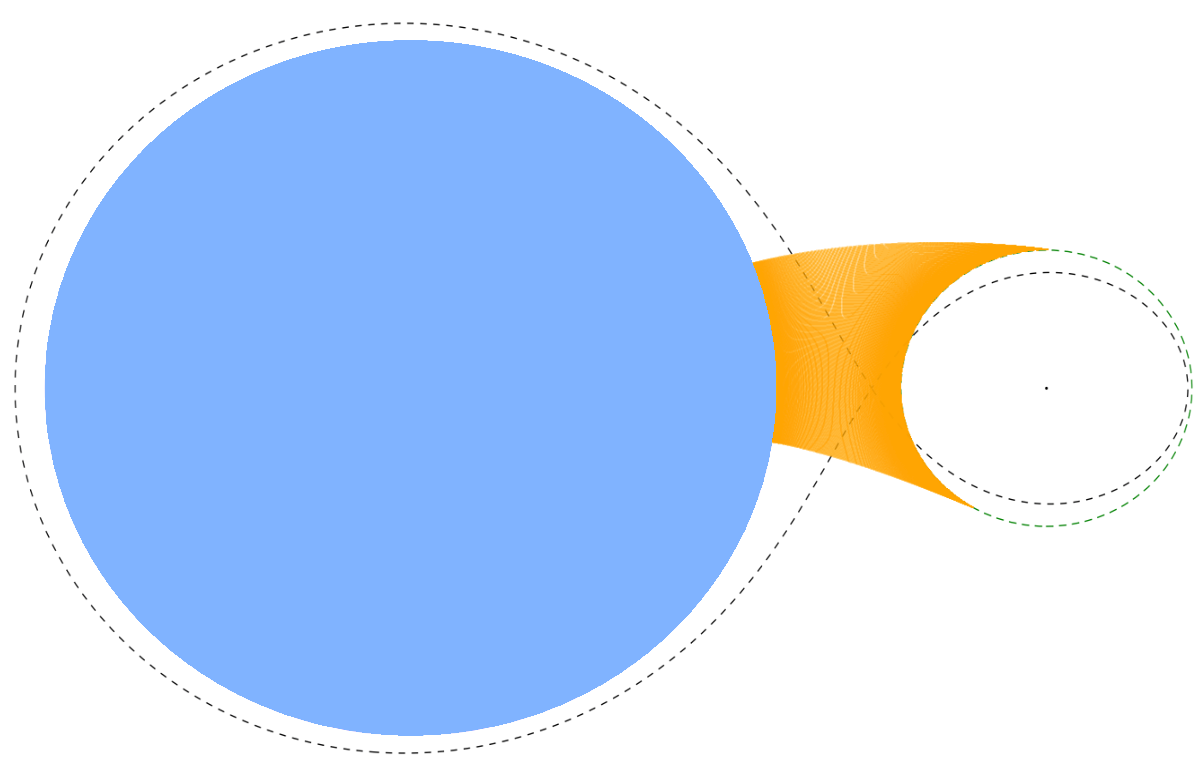}
\end{subfigure}
\begin{minipage}[t]{1\textwidth}
\caption{Schematic scale-free representations of the structure of the flow in three different \sgx. From top to bottom, Vela X-1, XTE J1855-026 \& IGR J18027-2016. Those sketches have been derived following the self-consistency roadmap described in \ref{sec:dim_out} to find the most likely set of 4 shape parameters. The bending of the streamlines provides a clue about the capacity of the flow to carry angular momentum into the extended accretion sphere around the accretor (dashed green), the zone where a proper hydrodynamical treatment is required.}
\label{fig:final_sketches}
\end{minipage}

\end{figure}

For SAX, Table \ref{tab:fill_and_gam} indicates that we deal with a very compact system where the lower radius of the less evolved star is not enough to compensate for the lower orbital separation : whatever the mass of the compact object, we deal with a stellar companion presenting a filling factor above 90\%. Given the narrow range of $\Gamma$ values centered on 10\%, we enforce it to 10\%. Since we do not expect the star, way more massive than the accretor, to be able to fill its Roche lobe and transfer matter in a stable way, we discard the heaviest accretor at 2\msun and focus on the intermediate and lightest cases. Relying on the measured value of the velocity at infinity at a 10\% precision level, we find that compatible values of mass ratios lie :
\begin{enumerate}
\item below $q=11$, in the lower right corner of $\left( f,q\right)$ space we consider.
\item anywhere between $q=7$ and $q=18$ for $M_2=1$\msun.
\end{enumerate}
Concerning the X-ray luminosity and the stellar mass outflows, the values are systematically lower than the values listed in Table \ref{tab:paramsAll}, mostly due to the lower value of the Eddington parameter of this later spectral type star. Solving this discrepancy requires an anomalously large $Q$ force multiplier\footnote{Remember that the wind speed is independent of its value.}, with a larger anomaly for a lower mass accretor. Since the star is less evolved, it is possible that the stellar photosphere is denser which could rise the value of $Q$ by $\sim$20\% \citep[see Table 1 of][]{Gayley1995}, enough to solve the discrepancy for an accretor of 1.5\msun, not for 1\msun. We thus privilege the former option over the latter. For $M_2=1.5$\msun and $Q=1,100$, the configuration with an X-ray luminosity compatible with the observations are confined below $q=12$ and above $f=95$\%, in agreement with the previous argument and in particular with the measure of the wind speed at infinity. The confrontation with the observational constraints inferred using the eclipses shows a marginal overlap with our holistic derivation. If the system is seen close to edge-on, the two approaches match and yield a system with the set of self-consistent fundamental parameters listed in Table \ref{tab:do_it_all} from which all the stellar, orbital, wind and accretion parameters can be derived. A sketch of the most likely configuration is drawn on Figure\,\ref{fig:final_sketches} (bottom panel). As visible on Figure\,\ref{fig:ang_mom_for_disk}, with a wind speed at infinity of 1,100\kms, an orbital period of 4.47 days and the stellar parameters we derived, the system lies approximately on the edge according to criterion \eqref{eq:circ_rad_needed} and thus deserved to be investigated with our toy-model. In the most likely configuration whose parameters are listed above, the shearing properties are in favor of the formation of a wind capture disk : the circularization radius is 6 to 10 thousands times larger than the Schwarzschild radius of the compact object and only 10 to 20 times smaller than the accretion radius. It is then likely that a disk forms within the shocked region. The orbital structure of the flow is stream-dominated.

\section{Discussion}

The present work shows that in some persistent \sgx, the differences from a system to another in terms of parameters of the flow at the orbital scale, of the star, of the wind and of the observed accretion can be understood by the present ballistic model. We have made the case for a dichotomy between stream and wind dominated mass transfers, independently from the a direct value of the filling factor. In systems such as IGR J18027-2016 where the extended accretion sphere is of the order of the Roche lobe of the compact object (see bottom sketch in Figure\,\ref{fig:final_sketches}), the accretion might share similar features with the classical \rlof observed in \lmxb such as the presence of a disk-like structure. 

\begin{wrapfigure}{r}{8cm}
\begin{center}
\includegraphics[height=6cm, width=7.5cm]{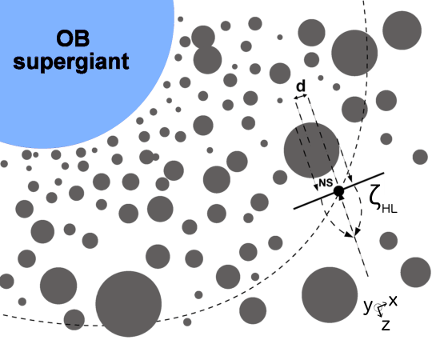}	
\caption{Sketch of the statistical clumpiness model developed by \cite{Ducci2009}. A distribution of clumps are considered in their interaction with the compact object using a cross-section approach (see Figure\,\ref{fig:ducci_cross-section}). The impact parameter is given by $d$ and $\zeta_{\textsc{hl}}$ is the accretion radius. The basis in the lower right corner is given as a reference for the Figure\,\ref{fig:ducci_cross-section}. From \cite{Ducci2009}.}
\label{fig:ducci_clumpy}
\end{center}
\end{wrapfigure}

It has been argued that the time variability of the X-ray luminosity could be due to inhomogeneities in the wind. The time variation of the inflow within the extended accretion sphere can straightforwardly be implemented in the present model via a density modulation of the boundary conditions \ie on the extended accretion sphere. In this way, the present characterization of the streamlines complements the model by \cite{Ducci2009} which does not address the dynamical problem but neatly models the statistical distribution of clumpiness of winds at launching in \sgx, the way the clumps size evolves as they flow away from the star\footnote{This part of the work owes much to the model developed by \cite{Howk2000} for the wind of the main sequence B0.2 star $\tau$ Sco.} and the way they can be accreted through the effective cross-section of the compact object. Within a couple of stellar radii, the clumps are more contrasted with the background wind which makes clumpiness a good candidate for the time variability observed in classical \sgx but not necessarily in \sfxt systems which can feature an accretor orbiting further away from the star. In addition, the density excess in clumps can hardly catch up with the large X-ray dynamics in \sfxt. Either large scale structures or phenomena in the vicinity of the accretor are then required.

\begin{figure}
\begin{center}
\includegraphics[height=4cm, width=15cm]{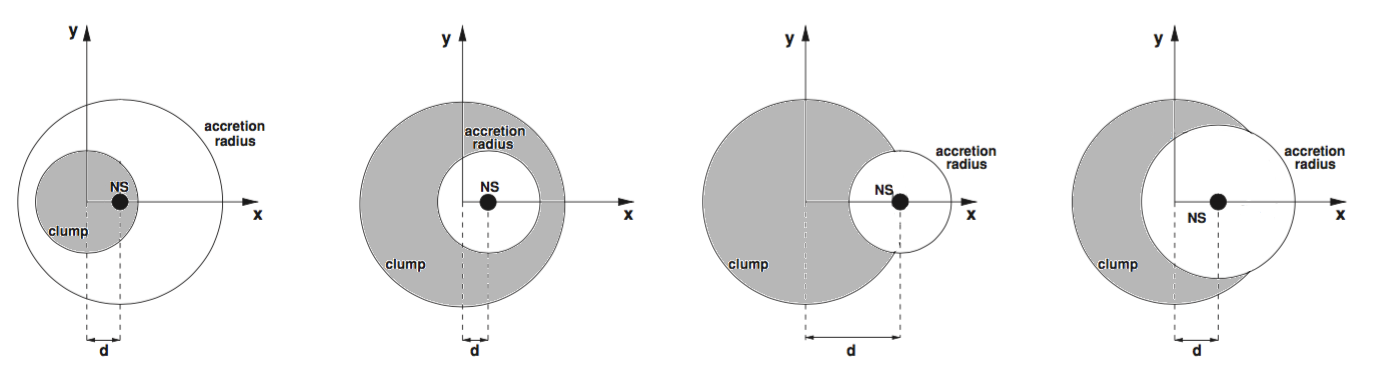}	
\caption{The different cases for the accretion of a spherical clump with an impact parameter $d$ with respect to the accretor. The axis are the ones specified in Figure\,\ref{fig:ducci_clumpy}. The problem reduces to a disk-disk interaction between the geometrical cross-section of the clump (in gray) and the virtual disk of radius the accretion radius (in white), centered on the accretor (the black dot). The smaller of the two is arbitrarily set as a forefront in each case. In the first case, all the clump is accreted while in the three other ones, the overlapping surface must be computed to estimate the fraction of the clump accreted. From \cite{Ducci2009}.}
\label{fig:ducci_cross-section}
\end{center}
\end{figure}

We notice that the X-ray luminosity levels we derive from our pipeline are in agreement with the observations listed in Table \ref{tab:paramsAll}, albeit on the lower side. Since those values are themselves slightly underestimated, and because we consider an efficiency of conversion of mass accretion rate in X-ray luminosity equal to the compactness parameter\footnote{Which is the maximal efficiency of the process ; we did not account for the possibility that part of the gravitational potential energy might not be converted in free radiations.}, the mass accretion rate we derive can be considered as too low. However, we did not account for the ionizing feedback of the accreting source which will, at some point, wiped out the possibility for the wind to be accelerated. Indeed, if the X-ray luminosity reaches levels high enough, a Str\"{o}mgren sphere\footnote{Since it does not take the same energy to fully ionize an hydrogen and a metal, its size and its influence on the wind acceleration will depend on the element considered.} forms around the accretor \citep{Hatchett1977,Ho1987,Stevens1991,Blondin1991,Ducci2010}. Without any excitation level left, a fully ionized metal can no longer absorb any photon and the radiatively-driven acceleration described in Chapter \ref{chap:wind} is quenched. Such a phenomenon must be accounted for once we enter the extended accretion sphere and might lower slightly the value of the specific kinetic energy at the orbital separation, $v_{\bullet}^2$, we used to derive the accretion radius and, by then, the mass accretion rate. Since the mass accretion rate is very sensitive to the value of $v_{\bullet}$, even a small inhibition of the acceleration of the wind could lead to the factor of a few we miss to reach the observed levels. We think the inhibition is not likely to come into play between the extended accretion sphere and the star since otherwise, the quenching of the wind would be so strong that the system would systematically swing between a low and a high luminosity state. The underlying positive feedback follows this principle : the higher the mass accretion rate onto the compact object, the larger the radiative feedback, the larger the Str\"{o}mgren sphere, the slower the wind to the inhibition process described above, the larger the mass accretion rate. The saturation point of this loop is reached once the Str\"{o}mgren sphere is larger than the Roche lobe of the accretor. In this case, the wind is no longer able to penetrate the Roche lobe, the mass accretion rate and, following the same positive feedback, the X-ray luminosity vanishes. We do not address this bi-stability case and rather focus on systems where the Str\"{o}mgren sphere is likely to be confined, most of the time, well within the extended accretion sphere. 

Finally, another scale must be considered once we enter the extended accretion sphere : the magnetosphere. It is also an additional reason to monitor closely the ionization level to evaluate how the flow couples to the magnetic field. If the final fate of the flow, in a disk and along the magnetic field lines, does not essentially alter the total X-ray luminosity, it does modulate the spectrum. Comparing synthetic spectra extracted from \textsc{mhd} simulations within the extended accretion sphere to observe permanent spectra is a possible extension of the present work.




\setlength{\parskip}{0ex} 


\part*{}
\chapter*{Conclusion \& perspectives}
\addcontentsline{toc}{part}{Conclusion \& perspectives}
\adjustmtc

To briefly summarize the work presented in detail in this manuscript, we have characterized the permanent behavior of a hydrodynamical \bhl flow being accreted by a compact object whose typical size is three to five orders-of-magnitude smaller than the scale of the shock formed by the incoming supersonic flow. We then set the conditions to extend this work to a numerical setup representative of a Supergiant X-ray binary namely a realistic acceleration of the radiatively-driven wind and the accounting of the non-inertial forces associated to the orbital motion. \\


\begin{figure}[!b]
\begin{center}
\includegraphics[height=6cm, width=10cm]{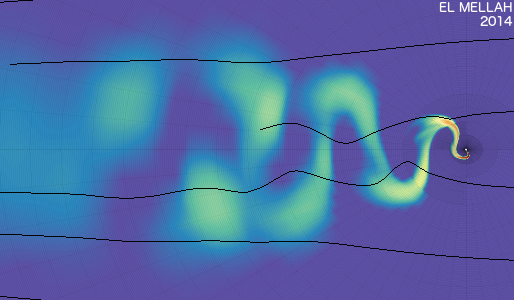}	
\caption{Logarithmic colormap of the density for a two-dimensional simulation with \textsc{amr} of a planar flow onto a point-mass. The flip-flop instability in the wake is clearly visible. In black are represented a few streamlines.}
\label{fig:FIG10}
\end{center}
\end{figure}

Thanks to a customization of the geometry of the mesh and its boundary conditions, an optimization of the load balancing, and an adaptation of the numerical scheme so as to suit the multi-scale needs of the \bhl flow onto a compact object, we were able to span up to 6 orders of magnitude in space, the equivalent of 17 levels of refinement in \textsc{amr}. For the first time, the size of the accretor in these highly consistent simulations matches the relevant physical size of a compact object moving with a realistic relative speed through the wind. Doing so, we brought a conclusive answer to the structure of the flow and the convergence of the mass accretion rate as the inner boundary size is reduced. The flow which relaxed on our 2.5D spherical grid featured a detached bow shock ahead the accretor, with a hollow conic tail at high Mach numbers and no axisymmetric instabilities, provided physically accurate inner boundary conditions are set up. The evolution of the properties of the flow as a function of its Mach number at infinity has been identified and compared to theoretical expectations. We conclude that at Mach numbers larger than 3, the mass accretion rate remains essentially unchanged as the opening angle and the distance of the shock to the accretor reduce. The sonic surface was found anchored into the inner boundary, whatever its size, for supersonic inflows and an adiabatic index of $5/3$ in agreement with a monoatomic gas. This observation matches the topological property derived analytically by \cite{Foglizzo1996} and which lays at the basis of a theoretical evaluation of the mass accretion rate consistent with our measures. For mildly supersonic flows, we show that the fraction of mass accreted without crossing the sonic surface is larger which highlights the need for small inner radii to characterize mildly supersonic flows. Thanks to the unprecedented high dynamical range we have access to, we could also follow the flow down to the vicinity of the accretor.

We now need to carry out a refined study of the axisymmetric stability of the steady-state flow we obtained. Indeed, the stability of a \bhl flow has long been a matter of debate according to the diverging conclusions that numerical groups have drawn since the late 1980's. As first pinpointed by \cite{Foglizzo2000}, a resonant cavity where an advective-acoustic cycle takes place can form between the shock and the sonic surface, the latter being typically 100 to 10,000 times closer from the accretor. The numerical study of this configuration has revealed how important this mechanism can be, with a possible application to the case of core-collapse supernovae. We already interpolated the relaxed state we got using a much finer grid in the central parts and relied on a less diffusive numerical scheme to monitor the non-linear growth of the instability which does appear. This new setup fits the needs for resolving wavelengths of perturbations corresponding to growth rates high enough for an amplification to take place in the cavity on a computationally affordable number of time steps. First results indicate suggestive breathing modes excited by the interplay between entropic disturbances adiabatically advected inwards and outflowing acoustic waves ; the origin of this cycle and its saturation level remain to be numerically investigated. If the instability proves to be amplified enough, it might be able to excite in three dimensions, transverse instabilities responsible for torque reversals of pulsars in wind accreting \hmxb\footnote{The neutron star in the \textsc{sfxt} 4U 1907+09 for instance is believed to undergo spin-up and spin-down phases, possibly due to the accreted material.} \cite{Blondin:2012vf}.\\


Given the robustness of the results detailed in the second part of this manuscript, we decided to relax the axisymmetric assumption and head towards fully three dimensional simulations.
Because we go beyond the ideal \bhl model, the flow is no longer free of net angular momentum and the formation of a disc around the accretor is made possible. To fit the physical conditions found in \sgx, we modeled the launching of the wind following a refined version of the theory of radiatively-driven winds to accurately grasp the acceleration stage within a couple of stellar radii, in a zone where the accretor usually lies. The numerical integrator we designed computes the trajectories of test-particles submitted to radiative accelerations induced by line absorption and scattering by free electrons, in a Roche potential with the Coriolis force. We quantified the fraction of the stellar wind likely to be captured by the compact object using a modified expression of the accretion radius. The toy-model we designed couples the stellar, orbital, accretion and wind properties all together in a simple though comprehensive manner which emphasizes the fours essential flow numbers which shape accretion in \sgx : the mass ratio, the stellar filling factor and Eddington parameter and the $\alpha$ force multiplier of the wind. Thanks to the underneath coupling, this code provides self-consistent sets of parameters which can be visualized through the \texttt{waso} Web interface. We also have access to the amount of angular momentum accreted and thus, to a direct tracer of the likelihood to form a disc-like structure around the accretor. The X-ray luminosity and the likelihood to form a disc rise together and the promising sets of parameters are explicited. 

\begin{figure}[!h]
\floatbox[{\capbeside\thisfloatsetup{capbesideposition={left,top},capbesidewidth=6cm}}]{figure}[\FBwidth]
{\caption{Isodensity surface from a three dimensional simulation of a Roche lobe overflow configuration centered on the accretor. The size of the inner boundary is of the order of a couple of Earth radii \ie one thousand times larger than the accretor and smaller than the orbital separation. The formation of the disc-like structure is clearly visible.}\label{fig:FIG5}}
{\includegraphics[width=10cm]{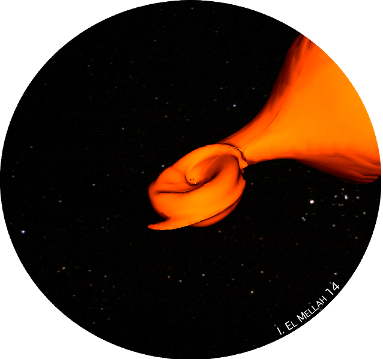}}
\end{figure}

This approach enables us to focus on the configurations which are likely to display a hybrid regime of accretion, in-between the asymptotic \rlof and wind regimes. If the former represented on Figure\,\ref{fig:FIG5} can not be straightforwardly summoned to explain the observations in \sgx, the beaming of the stellar wind in the direction of the accretor can be important enough to qualify the wind mass transfer as stream-dominated, a regime which has been coined by \cite{Mohamed} as wind-\rlof. Our code enables us to evaluate the dimensions of the virtual sphere around the accretor which requires a proper hydrodynamical treatment analogous to the one described in the second part of this manuscript but without the enforced axisymmetry. We are now in a position where we can implement physically-motivated outer boundary conditions (shearing of the inflow, mass accretion rate, velocity field, etc) in a fully three dimensional simulation centered on the accretor. This strategy disentangles the large scale, dominated by the orbital parameters and the properties of the supersonic wind, from the accretion scale, where hydrodynamical simulations come into play.\\


In my PhD research, I have made the most of this junction between a harvest of new wind accreting X-ray binaries and the unprecedented computational power entailed by parallel computing to identify the conditions favorable to the formation of a disc around the accretor. Expected to be different from Shakura \& Sunyaev's $\alpha$-disc model \citep{Shakura1973b} which fits well the disks formed during \rlof, the wind-capture discs may be a fertile ground for new kinds of instabilities involving, for instance, torque reversals. The coupling between the essentially radial incoming wind of the star and the disc-like structure might also display interesting behaviors at the interface between the two structures.\\

\setcounter{chapter}{1}
\setcounter{section}{0}
\renewcommand{\thechapter}{\Alph{chapter}}

\chapter*{Appendixes}
\addcontentsline{toc}{chapter}{Appendixes}
\chaptermark{Appendixes}
\hypersetup{linkcolor=black}
\minitoc
\hypersetup{linkcolor=red}
\setlength{\parskip}{1ex} 

\section{Adiabatic shock jump conditions}
\label{sec:jump}

Rankine-Hugoniot for oblique shocks.
Rq : if we allow the solutions of \eqref{eq:main_Bondi_acc} for a Bondi spherical accretion to form a shock, transonic solutions which do not go through the sonic point can exist.

When one deals with a steady-state shock, the local form of the equations of conservation is no longer valid due to the discontinuity introduced by the shock but an integrated formulation remains meaningful\footnote{In reality, the shock has a non-zero thickness which is set by viscous considerations out of the scope of the present reminder.}. To that purpose, a virtual volume $\mathcal{V}$ surrounded by a closed surface $\Sigma$ in the vicinity of the shock, laterally enclosed by the tube made of infinitely close streamlines, of cross section $S$ and whose extension along the velocity vector is infinitesimal on both sides of the shock\footnote{Except compared to the shock thickness.}, can be defined (dashed line in Figure\,\ref{fig:planar_shock} and \ref{fig:oblique_shock}). The supersonic upstream flow, indexed with 1, and the subsonic downstream flow, indexed with 2, are homogeneous and permanent and we work in the frame where the shock is at rest. We neglect any heating or cooling term in the energy equation.


\subsection{Planar shock}
\label{sec:planar_shock}
We first start to remind the main results concerning the jump conditions at a shock for a planar flow \ie for a flow whose velocity is normal to the shock front (the normal being given by the unit-vector $\mathbf{n}$ and oriented downstream in Figure\,\ref{fig:planar_shock}). The velocities remain in the same direction which justifies a one-dimensional approach along the colinear axis. The integrated conservation of mass, linear momentum and energy in the virtual volume are given by, respectively :
\begin{equation}
\begin{cases}
\oiint_{(\Sigma)}         \rho                    \mathbf{v} \cdot \mathbf{\d S}=0\\
\oiint_{(\Sigma)} \left( \rho v \mathbf{v} + P \mathbf{n} \right) \cdot \mathbf{\d S}=0\\
\oiint_{(\Sigma)} \left( e + P        \right) \mathbf{v} \cdot \mathbf{\d S}=0
\end{cases}
\end{equation}
where $v$ stands for the projection of the velocity vectors along the colinear axis oriented downstream. Notice that the source terms in the two last equations are null since we considered a volume with an infinitely small extension along the streamline on both sides of the shock and the forces responsible for those terms are bounded. If we decompose each of those integrals in a lateral, an upstream and a downstream surface, we have :

\begin{figure}
\begin{center}
\def\svgwidth{250pt} 
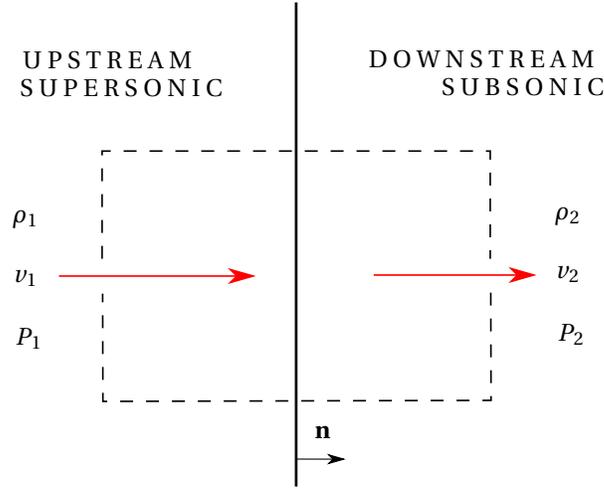
\caption{Configuration of a planar shock separating an upstream region, supersonic, from a downstream region, subsonic. The dashed line stands for the virtual volume $\mathcal{V}$ we consider to integrate the equations of conservation.}
\label{fig:planar_shock}
\end{center}
\end{figure}
\begin{figure}[!h]
\begin{center}
\def\svgwidth{250pt} 
\includegraphics[height=6.5cm, width=10cm]{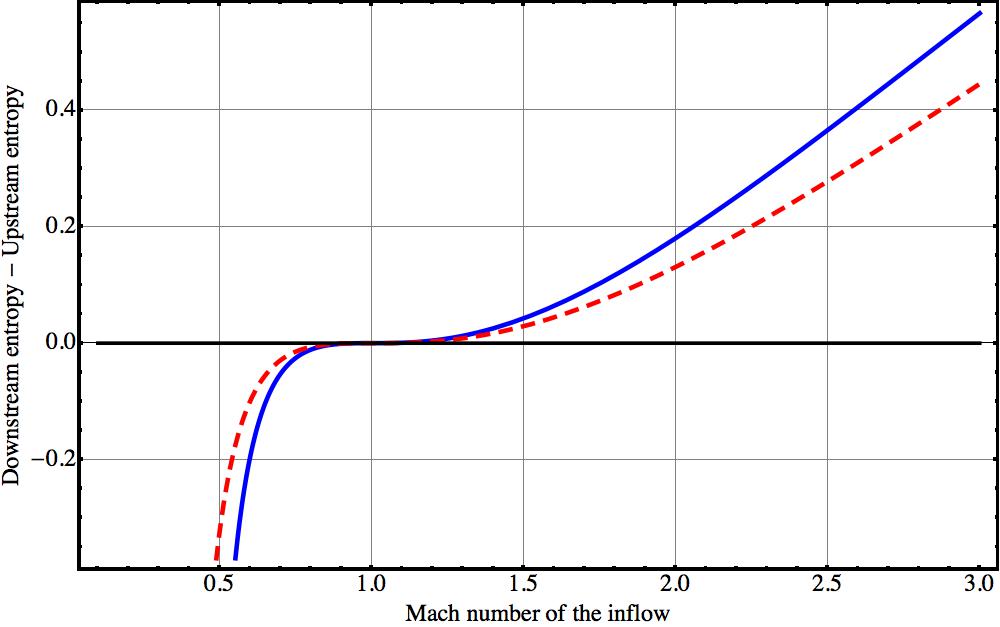}	
\caption{Evolution of the entropy (fiducial unit) as the flow crosses the shock as a function of the Mach number of the upstream flow. It is precisely because of the unphysical decrease of entropy below 1 that a subsonic flow becoming supersonic will in no case do so through a shock discontinuity. The blue solid line is for $\gamma=5/3$, the dashed red for $\gamma=7/5$ and the black horizontal one is the limit for $\gamma\rightarrow 1$.}
\label{fig:entropy_shock}
\end{center}
\end{figure}
\begin{equation}
\begin{cases}
0 - \rho_1 v_1 S + \rho_2 v_2 S = 0\\
0 - ( \rho_1 v_1^2 + P_1 ) S + ( \rho_2 v_2^2 + P_2 ) S = 0\\
0 - ( e_1 + P_1 ) v_1 S + ( e_2 + P_2 ) v_2 S = 0
\end{cases}
\end{equation} 
where the integrals across the lateral surface yield zero since this surface is made of infinitely close streamlines such as $\mathbf{v} \perp \d \mathbf{S_{\text{lat}}}$. Using the expression of the total specific energy reminded in \ref{sec:mass_and_energy_density}, we have the Rankine-Hugoniot conditions which must be verified at the shock : 
\begin{equation}
\begin{cases}
\rho_1 v_1 = \rho_2 v_2\\
\rho_1 v_1 ^2+ P_1 = \rho_2 v_2^2 + P_2\\
\left( \frac{1}{2}\rho_1v_1^2 + \frac{\gamma}{\gamma -1} P_1 \right) v_1 = \left( \frac{1}{2}\rho_2v_2^2 + \frac{\gamma}{\gamma -1} P_2 \right) v_2
\end{cases}
\end{equation} 
We can derive the jump conditions after a bit of algebra on this set of equations :
\begin{equation}
\begin{cases}
\frac{\rho_2}{\rho_1}=\frac{v_1}{v_2}=\frac{\left(\gamma+1\right)\mathcal{M}_1^2}{2+\left(\gamma-1\right)\mathcal{M}_1^2}\\
\frac{P_2}{P_1}=1+\frac{2\gamma\left(\mathcal{M}_1^2-1\right)}{\gamma +1}
\end{cases}
\end{equation}
where $\mathcal{M}_1>1$ is the Mach number of the upstream flow. The ratios between the values of a physical quantity before and after the shock then depend only on the Mach number of the inflow and its adiabatic index $\gamma>1$. Those relations imply, for the Mach number of the flow downstream, the temperatures and the entropies :
\begin{equation}
\label{eq:thermal_jump}
\begin{cases}
\mathcal{M}_2^2=\frac{2+\left(\gamma-1\right)\mathcal{M}_1^2}{2\gamma \mathcal{M}_1^2-\left(\gamma -1\right)}\\
\frac{T_2}{T_1}=\frac{P_2}{\rho_2}\frac{\rho_1}{P_1}\\
S_2-S_1=\ln \left( \frac{P_2}{\rho_2^{\gamma}}\frac{\rho_1^{\gamma}}{P_1} \right)
\end{cases}
\end{equation}
where the latter equation quantifies the production of entropy at the shock. This production, due to the mechanically non reversible transformation of the flow, explains why a subsonic incoming flow becoming supersonic will never produce a shock front. Indeed, imagine a similar configuration but with $\mathcal{M}_1<1$. Then, those equations still hold, at least down to a certain Mach number, but Figure\,\ref{fig:entropy_shock} reveals that entropy would be damped\footnote{It is an actual damping since there is no exchange of entropy with an outside system, the shock being adiabatic.}. As a consequence, a sonic surface is not associated to a shock. 


\subsection{Oblique shock}
\label{sec:oblique_shock}

\begin{figure}
\begin{center}
\def\svgwidth{250pt} 
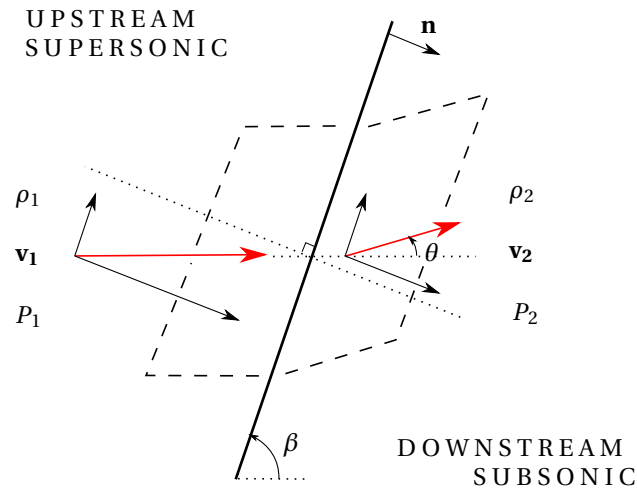
\caption{Configuration of an oblique shock separating an upstream region, supersonic, from a downstream region, subsonic. The velocity vectors, in red, have been decomposed in a normal component, $u$, and a tangential one, $w$. The dashed line stands for the virtual volume $\mathcal{V}$ we consider to integrate the equations of conservation.}
\label{fig:oblique_shock}
\end{center}
\end{figure}
\begin{figure}[!b]
\begin{center}
\def\svgwidth{250pt} 
\includegraphics[height=6cm, width=10cm]{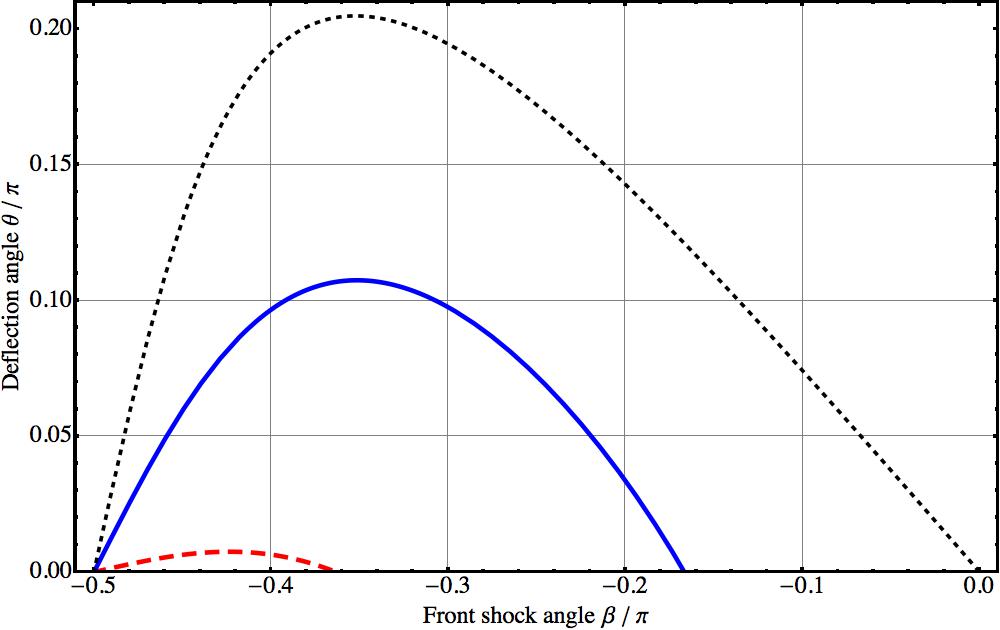}	
\caption{Evolution of the deflection angle $\theta$ as the front shock angle $\beta$ decreases from $\pi/2$ (left) to zero (right). Red dashed, blue solid and black dotted are for a mildly supersonic flow ($M_1=1.1$), a supersonic flow ($M_1=2$) and an infinitely supersonic flow ($M_1=1000$). The relative angle ($>-0.5$) for which the curves reach the x-axis give the front shock angle where the normal Mach number upstream reaches 1 ; it is also the moment where the flow is no longer deflected.}
\label{fig:deflection_angle}
\end{center}
\end{figure}

The more general situation of an oblique shock can be treated similarly provided we decompose the velocity vectors in their normal and tangential components, written respectively $u$ and $w$ in this section. From there, we have strictly the same equations than in \ref{sec:planar_shock} above but with $u$ playing the role of $v$ and the Mach number $\mathcal{M_1}$ being replaced by the normal Mach number $\mathcal{M_{\perp ,1}}$ associated to $u$ so as $\mathcal{M_{\perp ,1}}=\mathcal{M_1} \sin \beta$. Also, the relation above for the Mach number of the flow downstream now gives the normal Mach number $\mathcal{M_{\perp ,2}}=\mathcal{M_{2}}\sin \left( \beta - \theta \right)$. An additional relation is given by the non diagonal term of the tensor in the conservation of linear momentum :
\begin{equation}
\oiint_{(\Sigma)} \rho w \mathbf{v} \cdot \mathbf{\d S} = - \rho_1 w_1 u_1 S + \rho_2 w_2 u_2 S\\
\end{equation}
where the lateral fluxes cancelled out on each side of the shock. Given the relation deduced from the conservation of mass, this expression states that the tangential component of the velocity is left invariant at the crossing of the shock.

We can then deduce the deflection angle $\theta$ of the streamline using $\tan (\beta - \theta) = u_2/w_2$ and $\tan \beta = u_1 / w_1 $ :
\begin{equation}
\label{eq:deflection_angle}
\tan \theta = \frac{2}{\tan \beta} \frac{\mathcal{M}_1^2\sin^2\beta-1}{2+\mathcal{M}_1^2\left[ \gamma+\cos \left( 2\beta \right) \right]}
\end{equation}

Using all those results, let us try to interpret the expected evolution of the shock properties of a detached bow shock as we move from its front to the wake. At the front, the shock angle $\beta$ is $\pi/2$, the shock is planar and the flow downstream is subsonic. As we go further from the front along the shock, $\beta$ decreases which initially makes the deflection angle $\theta$ increase (see curves in Figure\,\ref{fig:deflection_angle}). This deflection angle finally reaches a maximum and soon after, the flow behind the shock is at Mach 1\footnote{To find this $\beta$ and check whether it is indeed beyond the value corresponding to the maximum of $\theta$, one has to numerically solve the implicit equation which descends from the condition $M_2=1$.}. From this point, the shock is no longer able to slow down the inflow to subsonic speeds. Finally, the shock has totally vanished once the deflection angle reaches zero (vertical lines in Figure\,\ref{fig:deflection_angle}). Indeed, with $\mathcal{M_{\perp ,1}}=1$, the relations above indicate an absence of discontinuity between the properties of the flow upstream and downstream.


\section{Absorption in a nutshell}
\label{sec:abs_in_nut}


\subsection{Context and references}

The website of the BAT instrument on the SWIFT satellite provides the information concerning the photon flux they received in photon per time unit per surface unit (\href{http://swift.gsfc.nasa.gov/results/bs70mon/}{http://swift.gsfc.nasa.gov/results/bs70mon/}). Once the spectrum is known, the integrated flux over a specified photon energy range can be derived\footnote{14 to 195keV for BAT.}, in erg per second per centimeter squared. So as to confront the observations of a few \sgx to the predicted time-averaged X-ray luminosities we derived from our simulations, we need to convert those reported X-ray fluxes to absolute luminosities. The distances being known \citep{Coleiro2013}, we can easily account for the spherical dilution and get the pristine luminosity if there was no absorption, the one we call the apparent luminosity\footnote{Not to be confused with the approach used for magnitudes where the apparent magnitudes is the one before a proper scaling for the actual distance.}. To correct for absorption and trace back the absolute luminosity, we relied on the thought experiment described below.


\subsection{Statistical description of the absorption}

\begin{figure}
\begin{center}
\includegraphics[height=5cm, width=10cm]{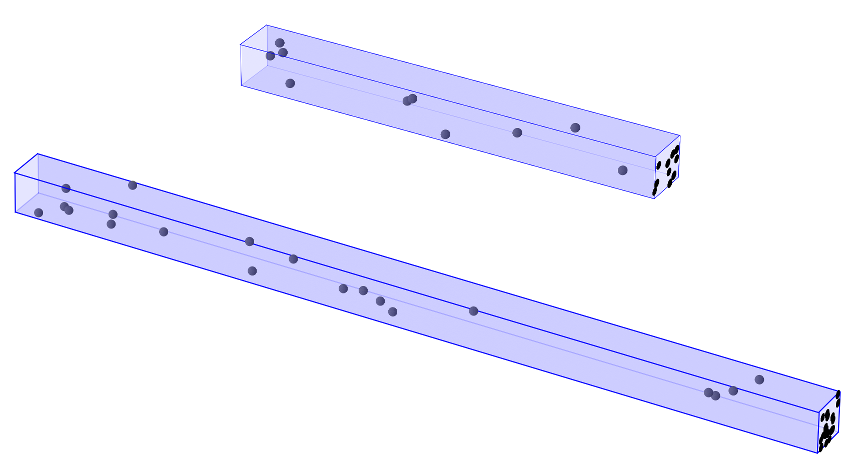}	
\caption{A column of transverse section $S$ and of length the distance $D$ between the observer, at the right end, and the source, at the left end. The black spheres within it represent the absorbers, uniformly distributed, modelled as solid spheres of cross-section $\sigma$. On the observer side is displayed the projected shadow of the absorbers for a planar background luminosity.}
\label{fig:sketch_absorption}
\end{center}
\end{figure}

\begin{figure}
\begin{center}
\includegraphics[height=8cm, width=8cm]{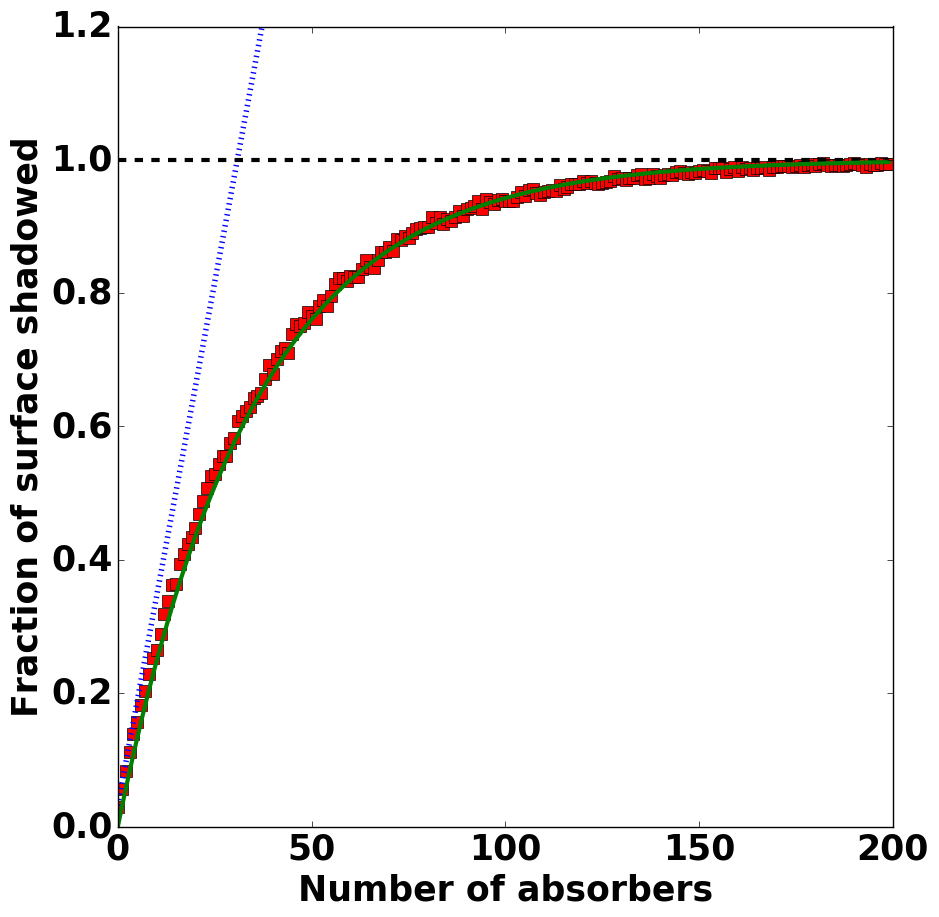}	
\caption{Fraction of the column transverse section shadowed as a function of the number of absorbers (red markers). The linear growth, made up of non-overlapping shadows, is represented in dotted blue and reaches the maximum value of 100\% for $N=N_0=S/\sigma$. The green continuous line is the best fit described in the text.}
\label{fig:comp_abs_log}
\end{center}
\end{figure}

Let us model the absorbers in a given photon energy range by solid spheres of cross-section\footnote{For solid spheres, $\sigma$ is simply the geometrical cross-section.} $\sigma$. We place ourselves in an optically thin framework and thus neglect the diffusion properties of those objects, which will be justified a posteriori in section \ref{sec:mfp}. If the density of absorbers $n$ is assumed to be uniform along the line-of-sight, we can think of their positions as randomly picked up within a column of transverse surface $S$ and of length the distance between the observer and the source, $D$. The ratio between the transverse size and the radius of the absorbing sphere of 10 is considered large enough to neglect the edge effects. In the uniform case, the number of absorbers $N$ evolves linearly with the distance :

\begin{equation}
\label{eq:NtoD}
N=n\times SD
\end{equation}
such as the number of absorbers is a direct measure of the distance to the source. We can now evaluate the fraction of the transverse surface obscured by the absorbers, as a function of the depth of the column. To do so, we simply select randomly the projected position of the absorbers within the transverse section and compute the shadowed surface with a Monte-Carlo method : we pick up at random approximately 1000 test points and check whether they end up in the disk-like shadow of a sphere. It results in the red markers of Figure\,\ref{fig:comp_abs_log}. The best fit one can suggest to represent this curve, in dashed green in Figure\,\ref{fig:comp_abs_log}, is :

\begin{equation}
f(D)=1-\exp ^{-D/D_0}
\end{equation}
where $D$ is linked to the number of absorbers via \eqref{eq:NtoD} and $D_0$ is given by, in agreement with the characteristic number of absorbers $N_0$ mentioned in the caption of Figure\,\ref{fig:comp_abs_log} :

\begin{equation}
D_0=\frac{N_0}{nS}=\frac{1}{n\sigma}
\end{equation}

\subsection{Physical light mean free path}
\label{sec:mfp}
\begin{figure}
\begin{center}
\includegraphics[height=10cm, width=13cm]{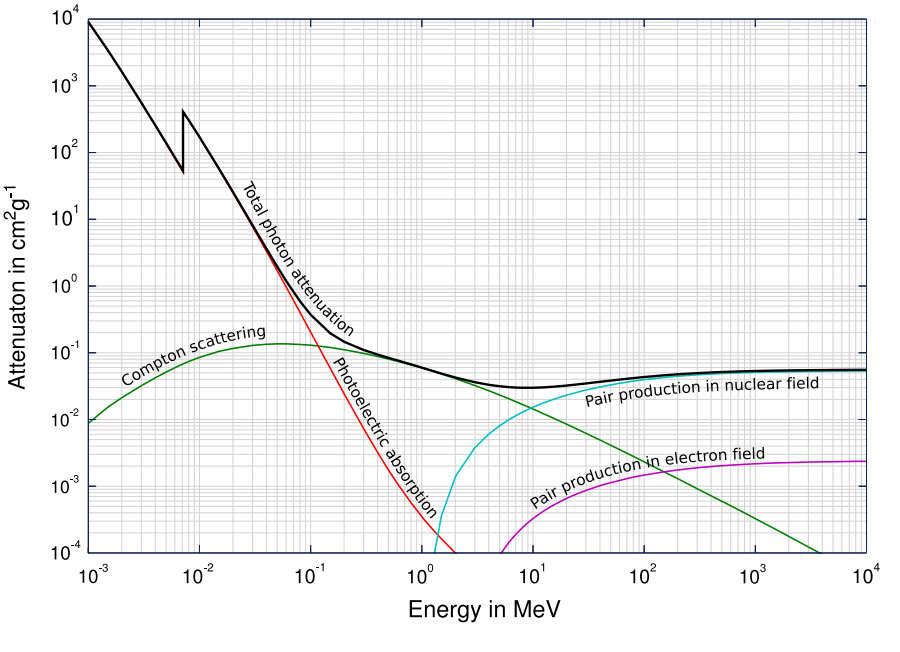}	
\caption{A plot to illustrate the different levels of absorption undergone by photons of different energies. For X-rays, in the lower part of the range considered here, the photoelectric effect is the main responsible for absorption. Credits : \href{http://www.nist.gov/pml/data/xcom/}{http://www.nist.gov/pml/data/xcom/} and Wikimedia Commons}
\label{fig:sigma}
\end{center}
\end{figure}

In the above computation, $D_0$ stands for the optical mean free path. The usual average absorbers density in the Milky Way can be deduced from the Hydrogen column density, typically $10^{22\text{ to }23}$cm$^{-2}$ which correspond, for distances from 1 to 10kpc, to volumic densities\footnote{Except in case of a dense cloud along the line-of-sight, especially in the vicinity of the Galactic center, where the volumic density can jump up to $10^5$cm$^{-3}$. We discard this possibility for the 3 systems we consider.} ranging between $1$ and $10$cm$^{-3}$. At the photon energies probed by BAT above, Compton scattering and the photo-electric effect are the main contributors to the total opacity \citep[][and Figure\,\ref{fig:sigma} below]{Longair2011} ; once the chemistry of the interstellar medium is accounted for, one obtains cross-sections of the order of $10^{-24}$cm$^{-2}$ per particle \citep[see][Figure 1]{Wilms2000}. Then the mean free path for hard X-rays ranges from a few 100 kpc to a few 10 kpc. Given that we consider only \sgx within the Milky Way\footnote{For the systems next door, in the Magellanic clouds, the absorption can be lower due to the sharpness of the Galactic disk and the lower density of absorbers in the intergalactic medium.}, the absorption regime we consider qualifies as optically thin and the toy-model we presented in the previous section holds.


\subsection{Unabsorbed luminosity}

The model described above drives us into identifying the ratio of the apparent by the absolute luminosity to the fraction of the transverse section not being obscured :

\begin{equation}
\frac{L_{\text{app}}}{L_{\mathbb{R}}}=\exp ^{-D/D_0}
\end{equation}
where we assumed that the absolute luminosity is the sum of the absorbed and the unabsorbed one since we do not handle diffusion. Since we are in the optically thin regime, we have :

\begin{equation}
L_{\mathbb{R}}\sim L_{\text{app}}\left( 1+D/D_0 \right)
\end{equation}
To alleviate the sensibility of our model to the optical mean free path $D_0$, we use a well-known system as a pivot, \eg Vela X-1. Adding the subscript V to the quantities corresponding to the latter, we have :

\begin{equation}
L_{\mathbb{R}}\sim L_{\text{app}}\left[ 1+\frac{D}{D_{\text{V}}}\left( \frac{L_{\mathbb{R}\text{,V}}}{L_{\text{app,V}}}-1 \right)\right]
\end{equation}


\subsection{Comments}

The proper treatment of absorption requires to relax the assumption of uniform density of absorbers and to integrate along the line-of-sight accounting for the local density. It provides the optical path length. Another point is the necessity to take into account the wavelength dependence of the cross-section as a function of the photon energy ; from an energy range to another one, the interactions at stake between photons and matter differ (Compton scattering, photoelectric absorption, pair production, etc). Eventually, notice that at lower photon energies ($\sim$ keV), the absorption is much higher and the environment can no longer be treated as optically thin.

\section{Surface on the sphere}
\label{sec:surf_sph}

\begin{figure}
\begin{center}
\def\svgwidth{400pt} 
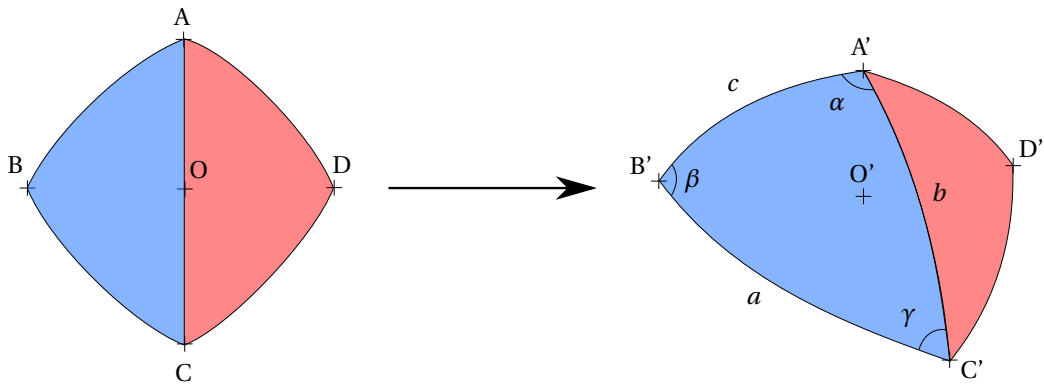
\caption{Detail of a regular spherical mesh of initial points on the stellar surface (5 points centered on a fiducial point O) on the left which becomes, once the streamlines is computed, the mesh on the right, drawn on the extended accretion sphere. The primes stand for the associated points. To compute the evolution of the density corresponding to the point O', we monitor the evolution of the surface formed by the four closest neighbours at departure by looking at its subdivision into two spherical triangles (blue and red areas).}
\label{fig:spherical_triangles}
\end{center}
\end{figure}

To compute the mass density of a streamline with respect to the departure point\footnote{Homogeneous, at the photosphere density of the Supergiant donor star.} as it enters the extended accretion sphere around the accretor, we need an estimate of the surface formed by the closest four points, at departure (\ie on the stellar surface) and at arrival (\ie on the virtual surface of the extended accretion sphere) - see Figure\,\ref{fig:spherical_triangles}. Since the streamlines do not cross each other in the incompressible flow approximation, the neighbouring points at departure and at arrival are related by their associated streamlines. We subdivide the surface delimited by the four neighbours into two triangles on the sphere \aka as spherical triangles, the analogue of the planar triangle on the sphere. 

Those triangles present the interesting property to have a sum of their angles which is larger than $\pi$. If we note $\alpha$, $\beta$ and $\gamma$ the three angles of a spherical triangle (in radians), one can define the spherical excess $S$ given by :
\begin{equation}
S=\alpha +\beta +\gamma -\pi
\end{equation}  
This excess can also be interpreted as the solid angle (in steradians) subtended by the spherical triangle : from now on, we normalize the surfaces by the square of the radius of the sphere and this excess corresponds to the surfaces we are looking for. To find the spherical excess using the information on the angular position of the points, we first compute the arclengths (normalized to the radius) using the fact that the dot products between two (normalized) position vectors on the sphere yields the cosine of the angle between the two (\ie the normalized arclength). The relation between the angles $\alpha$, $\beta$ and $\gamma$ and the semiperimeter $s=\left(a+b+c\right)/2$ (where $a$, $b$ and $c$ are the arclengths) is then given by L'Huilier's theorem :
\begin{equation}
\tan \left(\frac{S}{4}\right) = \sqrt{ \tan \left( \frac{s}{2} \right) \tan \left( \frac{s-a}{2} \right) \tan \left( \frac{s-b}{2} \right) \tan \left( \frac{s-c}{2} \right) }
\end{equation}
This process is the one we used to estimate the distribution of mass densities on the extended accretion sphere for each streamline.

\section{The \texttt{WASO} interface}
\label{sec:spyre}

\href{https://github.com/adamhajari/spyre}{Spyre} is a Python library developed by A. Hajari (2015) to design \href{https://www.youtube.com/watch?v=NPV2hHV6hxY}{customized data visualization interfaces}. It is a canvas which takes as inputs user-defined parameters (“which orbital period for my binary system?”) or options (“should I give the masses or the mass ratios for the y-axis?") and display pre-defined plots (the usual ones you write in Python such as the colormap in the snapshot attached). It is a handy environment to quickly explore a space parameter of too high dimensionality to fully appreciate the intertwined dependences at stake in the model. It is also possible to make a Spyre interface alive with interactive-oriented libraries such as Bokeh.

\begin{figure}
\centering
\includegraphics[width=1\columnwidth]{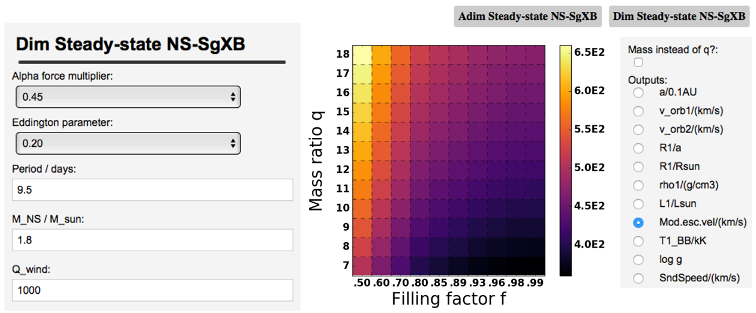}
\caption{A snapshot of the visualization applet to explore the different \sgx configurations, with the effective escape velocity in \,km$\cdot$\,s$^{-1}$ as an output. For the sake of visibility, only orbital and stellar outputs are displayed here.}
\label{fig:spyre}
\end{figure}

Using the Spyre library, we designed a handy web application framework to quickly visualize all the configurations in the scope of the third part of this manuscript, the \texttt{WASO} interface\footnote{The fours letters cover the four physical elements assembled together in our toy-model : the Wind, the Accretion, the Star and the Orbit.} : dimensionless or not, for precomputed sets of shape parameters (the 12 values of $q$, 10 values of $f$, 3 different $\alpha$ and 3 different $\Gamma$) and for user-defined values of the scale parameters ; to bridge the gap between the physical quantities of this model and the observables, we chose to work with the orbital period, the mass of the compact object and the $Q$-force multiplier as scale parameters. The first is precisely known (when measured) for each system, the second is enclosed in a restricted range when we have reasons to think it is a neutron star\footnote{X-ray pulses for instance.} and the latter does not vary much neither and is relatively precisely set by the spectral type of the stellar companion. The interface displays colormaps in the $(q,f)$ space for more than 40 outputs - computed based on the simulations results, the shape parameters and, for the dimensioned outputs, on the user-specified scales - relevant to appreciate the properties of the model we developed and their interplay. Some of them are described in more details in section \ref{sec:structure_wind}. A snapshot of the applet is shown in Figure\,\ref{fig:spyre}. The \texttt{WASO} interface, available on demand, aims at conducting the user towards a better understanding of the trends we observe in \sgx by quickly travelling through a reduced though essential space of parameters. It also entitles the user to evaluate the dependencies of each output with the 7 shape and scale parameters and, where suggested by the figures, to neglect some of those dependencies. The latter then constitutes a minimal system of relations so as to deduce the parameters of the system from a limited amount of data (the stellar temperature, surface gravity and radius, the orbital period, the mass of the compact object and the time-averaged X-ray luminosity of the system for instance).

\newpage

\section{\texttt{MPI-AMRVAC}}
\label{sec:amrvac_doc}

\begin{figure}
\centering
\includegraphics[width=1\columnwidth]{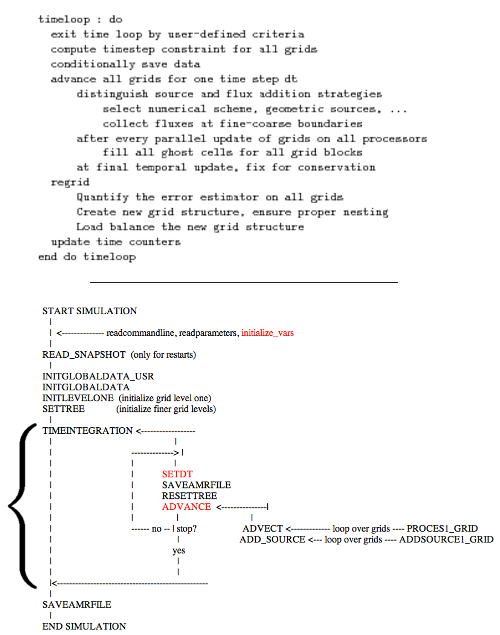}
\caption{An overview of the main nominal steps followed by \vac when it is run with, at the top, the principles of each step of the integration stage (see the brace in the bottom panel) and at the bottom, the corresponding main \vac subroutines. After the initialization steps, the time integration starts and takes generally most of the computing time. Grids are advanced separately and information is communicated between grids (and possibly \cpus) after each step. From the official documentation of \vac.}
\label{fig:UML_diag_VAC}
\end{figure}

\begin{wrapfigure}{l}{0.7\textwidth}
\begin{center}
\includegraphics[height=7cm, width=\textwidth]{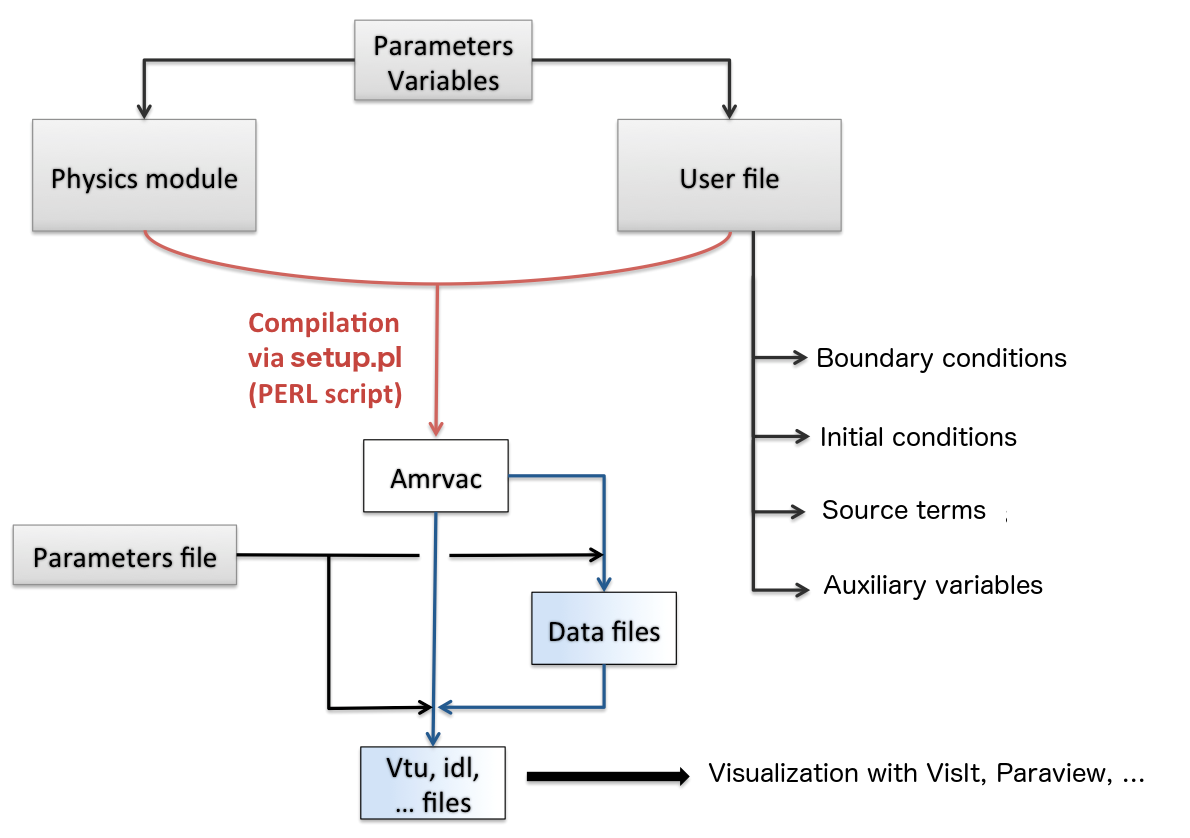}	
\caption{Structural and functional diagram of the \vac code.}
\label{fig:struct_diag}
\end{center}
\end{wrapfigure}

\texttt{MPI-AMRVAC} (hereby \vac) is an \textsc{mpi}-parallelized \textsc{amr} code aims to advance any system of (primarily hyperbolic) partial differential equations by a number of different numerical schemes \citep{Porth:2014wv}. The dimensionality, the geometry of the mesh and the physical equations are handled uniformly so as to make the background code independent of them and to let the user choose and customize them without altering the core subroutines. To do so, the code is written in a Fortan-based language, the \texttt{LASY} syntax \citep{Toth1997a}, and is then converted into proper Fortran files using a \texttt{perl} preprocessor which takes as inputs the dimensionality, the sub-grid size, etc ; it automatically generates Fortran files specifically designed to handle the physical problem the user is interested in. Figure\,\ref{fig:UML_diag_VAC} represents the generic skeleton code, similar to most \textsc{amr} codes. Formally speaking, proper \textsc{uml} diagrams\footnote{\textsc{uml} (for Unified Modeling Language) provides a general approach to software design and is widely used upstream of the code development effort to evaluate the objectives and describe the structure of a code.} would be required to describe the structure and the behaviour of the code (a first sketch is suggested in Figure\,\ref{fig:struct_diag}). More documentation on the code and the different advancing schemes available can be found in the \href{https://homes.esat.kuleuven.be/~keppens/}{official documentation}.


\setlength{\parskip}{0ex} 

%


\bibliographystyle{rusnat}
\addcontentsline{toc}{chapter}{Bibliography}
\bibliography{./PhD_manuscript_no_url}


\clearpage
\thispagestyle{empty}
\newpage

\pagestyle{plain}
\phantom{ab}
\clearpage
\thispagestyle{empty}
\newpage

\clearpage
\thispagestyle{empty}
\begin{Huge}
\begin{center}
\textsc{abstract}
\end{center}
\end{Huge}

X-ray emission associated to accretion onto compact objects displays important levels of photometric and spectroscopic time-variability. When the accretor orbits a Supergiant star, it captures a fraction of the supersonic radiatively-driven wind which forms shocks in its vicinity. The amplitude and stability of this gravitational beaming of the flow conditions the mass accretion rate responsible, in fine, for the X-ray luminosity of those Supergiant X-ray Binaries. The capacity of this low angular momentum inflow to form a disc-like structure susceptible to be the stage of well-known instabilities remains at stake.

Using state-of-the-art numerical setups, we characterized the structure of a Bondi-Hoyle-Lyttleton flow onto a compact object, from the shock down to the vicinity of the accretor, typically five orders of magnitude smaller. The evolution of the mass accretion rate and of the bow shock which forms around the accretor (transverse structure, opening angle, stability, temperature profile…) with the Mach number of the incoming flow is described in detail. The robustness of those simulations based on the High Performance Computing MPI-AMRVAC code is supported by the topology of the inner sonic surface, in agreement with theoretical expectations.

We developed a synthetic model of mass transfer in Supergiant X-ray Binaries which couples the launching of the wind accordingly to the stellar parameters, the orbital evolution of the streamlines in a modified Roche potential and the accretion process. We show that the shape of the permanent flow is entirely determined by the mass ratio, the filling factor, the Eddington factor and the alpha force multiplier. Provided scales such as the orbital period are known, we can trace back the observables to evaluate the mass accretion rates, the accretion mechanism (stream or wind-dominated) and the shearing of the inflow, tracer of its capacity to form a disc around the accretor.


\end{document}

%% file: cells.pdf_tex
\begingroup%
  \makeatletter%
  \providecommand\color[2][]{%
    \errmessage{(Inkscape) Color is used for the text in Inkscape, but the package 'color.sty' is not loaded}%
    \renewcommand\color[2][]{}%
  }%
  \providecommand\transparent[1]{%
    \errmessage{(Inkscape) Transparency is used (non-zero) for the text in Inkscape, but the package 'transparent.sty' is not loaded}%
    \renewcommand\transparent[1]{}%
  }%
  \providecommand\rotatebox[2]{#2}%
  \ifx\svgwidth\undefined%
    \setlength{\unitlength}{595.27559055bp}%
    \ifx\svgscale\undefined%
      \relax%
    \else%
      \setlength{\unitlength}{\unitlength * \real{\svgscale}}%
    \fi%
  \else%
    \setlength{\unitlength}{\svgwidth}%
  \fi%
  \global\let\svgwidth\undefined%
  \global\let\svgscale\undefined%
  \makeatother%
  \begin{picture}(1,0.25238095)%
    \put(0,0){\includegraphics[width=\unitlength,page=1]{cells.pdf}}%
    \put(0.41437313,0.20821242){\color[rgb]{0,0,0}\makebox(0,0)[lb]{\smash{$\text{cell}\#i$}}}%
    \put(0.15326858,0.2289934){\color[rgb]{0,0,0}\makebox(0,0)[lb]{\smash{$\Delta x$}}}%
    \put(0.80529345,0.10920989){\color[rgb]{0,0,0}\makebox(0,0)[lb]{\smash{$t^n$}}}%
    \put(0,0){\includegraphics[width=\unitlength,page=2]{cells.pdf}}%
    \put(0.3393052,0.01827068){\color[rgb]{0,0,0}\makebox(0,0)[lb]{\smash{$x_{i-1/2}$}}}%
    \put(0.47756016,0.01878253){\color[rgb]{0,0,0}\makebox(0,0)[lb]{\smash{$x_{i+1/2}$}}}%
    \put(0.33760398,0.08668372){\color[rgb]{0,0,0}\makebox(0,0)[lb]{\smash{$\Phi_{i-1/2}$}}}%
    \put(0.47143076,0.08777631){\color[rgb]{0,0,0}\makebox(0,0)[lb]{\smash{$\Phi_{i+1/2}$}}}%
  \end{picture}%
\endgroup%

%% file: grid_blocks.pdf_tex
\begingroup%
  \makeatletter%
  \providecommand\color[2][]{%
    \errmessage{(Inkscape) Color is used for the text in Inkscape, but the package 'color.sty' is not loaded}%
    \renewcommand\color[2][]{}%
  }%
  \providecommand\transparent[1]{%
    \errmessage{(Inkscape) Transparency is used (non-zero) for the text in Inkscape, but the package 'transparent.sty' is not loaded}%
    \renewcommand\transparent[1]{}%
  }%
  \providecommand\rotatebox[2]{#2}%
  \ifx\svgwidth\undefined%
    \setlength{\unitlength}{992.12598425bp}%
    \ifx\svgscale\undefined%
      \relax%
    \else%
      \setlength{\unitlength}{\unitlength * \real{\svgscale}}%
    \fi%
  \else%
    \setlength{\unitlength}{\svgwidth}%
  \fi%
  \global\let\svgwidth\undefined%
  \global\let\svgscale\undefined%
  \makeatother%
  \begin{picture}(1,0.57142857)%
    \put(0,0){\includegraphics[width=\unitlength,page=1]{grid_blocks.pdf}}%
    \put(0.11457873,0.0108852){\color[rgb]{0,0,0}\makebox(0,0)[lb]{\smash{$n$}}}%
    \put(0.90037487,0.47068512){\color[rgb]{0,0,0}\makebox(0,0)[lb]{\smash{$\delta$}}}%
    \put(0.00955034,0.26894348){\color[rgb]{0,0,0}\makebox(0,0)[lb]{\smash{$N$}}}%
  \end{picture}%
\endgroup%

%% file: load_balancing.pdf_tex
\begingroup%
  \makeatletter%
  \providecommand\color[2][]{%
    \errmessage{(Inkscape) Color is used for the text in Inkscape, but the package 'color.sty' is not loaded}%
    \renewcommand\color[2][]{}%
  }%
  \providecommand\transparent[1]{%
    \errmessage{(Inkscape) Transparency is used (non-zero) for the text in Inkscape, but the package 'transparent.sty' is not loaded}%
    \renewcommand\transparent[1]{}%
  }%
  \providecommand\rotatebox[2]{#2}%
  \ifx\svgwidth\undefined%
    \setlength{\unitlength}{992.12598425bp}%
    \ifx\svgscale\undefined%
      \relax%
    \else%
      \setlength{\unitlength}{\unitlength * \real{\svgscale}}%
    \fi%
  \else%
    \setlength{\unitlength}{\svgwidth}%
  \fi%
  \global\let\svgwidth\undefined%
  \global\let\svgscale\undefined%
  \makeatother%
  \begin{picture}(1,0.57142857)%
    \put(0,0){\includegraphics[width=\unitlength,page=1]{load_balancing.pdf}}%
    \put(0.09038826,0.0108852){\color[rgb]{0,0,0}\makebox(0,0)[lb]{\smash{$n$}}}%
    \put(0.00955034,0.26894348){\color[rgb]{0,0,0}\makebox(0,0)[lb]{\smash{$N$}}}%
    \put(0,0){\includegraphics[width=\unitlength,page=2]{load_balancing.pdf}}%
  \end{picture}%
\endgroup%

%% file: acc_column.pdf_tex
\begingroup%
  \makeatletter%
  \providecommand\color[2][]{%
    \errmessage{(Inkscape) Color is used for the text in Inkscape, but the package 'color.sty' is not loaded}%
    \renewcommand\color[2][]{}%
  }%
  \providecommand\transparent[1]{%
    \errmessage{(Inkscape) Transparency is used (non-zero) for the text in Inkscape, but the package 'transparent.sty' is not loaded}%
    \renewcommand\transparent[1]{}%
  }%
  \providecommand\rotatebox[2]{#2}%
  \ifx\svgwidth\undefined%
    \setlength{\unitlength}{1162.20472441bp}%
    \ifx\svgscale\undefined%
      \relax%
    \else%
      \setlength{\unitlength}{\unitlength * \real{\svgscale}}%
    \fi%
  \else%
    \setlength{\unitlength}{\svgwidth}%
  \fi%
  \global\let\svgwidth\undefined%
  \global\let\svgscale\undefined%
  \makeatother%
  \begin{picture}(1,0.57804878)%
    \put(0,0){\includegraphics[width=\unitlength,page=1]{acc_column.pdf}}%
    \put(0.11258676,0.55168793){\color[rgb]{0,0,0}\makebox(0,0)[lb]{\smash{$(\rho_{\infty},v_{\infty})$}}}%
    \put(0.3616352,0.28197066){\color[rgb]{0,0,0}\makebox(0,0)[lb]{\smash{$r$}}}%
    \put(0.63836477,0.22641514){\color[rgb]{0,0,0}\makebox(0,0)[lb]{\smash{$s$}}}%
    \put(0.04397689,0.39832286){\color[rgb]{0,0,0}\makebox(0,0)[lb]{\smash{$\zeta$}}}%
    \put(0.04693317,0.4948845){\color[rgb]{0,0,0}\makebox(0,0)[lb]{\smash{d$\zeta$}}}%
    \put(0.46855345,0.12264151){\color[rgb]{0,0,0}\makebox(0,0)[lb]{\smash{d$r$}}}%
    \put(0,0){\includegraphics[width=\unitlength,page=2]{acc_column.pdf}}%
    \put(0.76973836,0.17714733){\color[rgb]{0,0,0}\makebox(0,0)[lb]{\smash{2$\alpha$}}}%
    \put(0,0){\includegraphics[width=\unitlength,page=3]{acc_column.pdf}}%
  \end{picture}%
\endgroup%

%% file: fd.pdf_tex
\begingroup%
  \makeatletter%
  \providecommand\color[2][]{%
    \errmessage{(Inkscape) Color is used for the text in Inkscape, but the package 'color.sty' is not loaded}%
    \renewcommand\color[2][]{}%
  }%
  \providecommand\transparent[1]{%
    \errmessage{(Inkscape) Transparency is used (non-zero) for the text in Inkscape, but the package 'transparent.sty' is not loaded}%
    \renewcommand\transparent[1]{}%
  }%
  \providecommand\rotatebox[2]{#2}%
  \ifx\svgwidth\undefined%
    \setlength{\unitlength}{960.72001953bp}%
    \ifx\svgscale\undefined%
      \relax%
    \else%
      \setlength{\unitlength}{\unitlength * \real{\svgscale}}%
    \fi%
  \else%
    \setlength{\unitlength}{\svgwidth}%
  \fi%
  \global\let\svgwidth\undefined%
  \global\let\svgscale\undefined%
  \makeatother%
  \begin{picture}(1,0.5830627)%
    \put(0.8243817,0.50795226){\color[rgb]{0,0,0}\makebox(0,0)[lb]{\smash{}}}%
    \put(0.79940043,0.58705959){\color[rgb]{0,0,0}\makebox(0,0)[lb]{\smash{}}}%
    \put(0.87434422,0.57456896){\color[rgb]{0,0,0}\makebox(0,0)[lb]{\smash{}}}%
    \put(0,0){\includegraphics[width=\unitlength,page=1]{fd.pdf}}%
    \put(0.74821512,0.4695204){\color[rgb]{0,0,0}\makebox(0,0)[lb]{\smash{$|\mathbf{v}| $}}}%
    \put(0,0){\includegraphics[width=\unitlength,page=2]{fd.pdf}}%
  \end{picture}%
\endgroup%

%% file: sketch_shapiro_perso.pdf_tex
\begingroup%
  \makeatletter%
  \providecommand\color[2][]{%
    \errmessage{(Inkscape) Color is used for the text in Inkscape, but the package 'color.sty' is not loaded}%
    \renewcommand\color[2][]{}%
  }%
  \providecommand\transparent[1]{%
    \errmessage{(Inkscape) Transparency is used (non-zero) for the text in Inkscape, but the package 'transparent.sty' is not loaded}%
    \renewcommand\transparent[1]{}%
  }%
  \providecommand\rotatebox[2]{#2}%
  \ifx\svgwidth\undefined%
    \setlength{\unitlength}{850.39370079bp}%
    \ifx\svgscale\undefined%
      \relax%
    \else%
      \setlength{\unitlength}{\unitlength * \real{\svgscale}}%
    \fi%
  \else%
    \setlength{\unitlength}{\svgwidth}%
  \fi%
  \global\let\svgwidth\undefined%
  \global\let\svgscale\undefined%
  \makeatother%
  \begin{picture}(1,0.7)%
    \put(0,0){\includegraphics[width=\unitlength,page=1]{sketch_shapiro_perso.pdf}}%
    \put(0.91863597,0.5576148){\color[rgb]{0,0,0}\makebox(0,0)[lb]{\smash{$y$}}}%
    \put(0.84121951,0.64383896){\color[rgb]{0,0,0}\makebox(0,0)[lb]{\smash{$z$}}}%
    \put(0,0){\includegraphics[width=\unitlength,page=2]{sketch_shapiro_perso.pdf}}%
    \put(0.30187411,0.61909402){\color[rgb]{0,0,0}\makebox(0,0)[lb]{\smash{$\alpha$}}}%
    \put(0,0){\includegraphics[width=\unitlength,page=3]{sketch_shapiro_perso.pdf}}%
    \put(0.38221643,0.39047076){\color[rgb]{0,0,0}\makebox(0,0)[lb]{\smash{$\alpha$}}}%
    \put(0,0){\includegraphics[width=\unitlength,page=4]{sketch_shapiro_perso.pdf}}%
    \put(0.14803302,0.61096131){\color[rgb]{0,0,0}\rotatebox{-36.9889194}{\makebox(0,0)[lb]{\smash{radial direction}}}}%
    \put(0,0){\includegraphics[width=\unitlength,page=5]{sketch_shapiro_perso.pdf}}%
    \put(0.3419106,0.33867484){\color[rgb]{0,0,0}\makebox(0,0)[lb]{\smash{P(y)}}}%
    \put(0.4684046,0.33869907){\color[rgb]{0,0,0}\makebox(0,0)[lb]{\smash{O}}}%
    \put(0,0){\includegraphics[width=\unitlength,page=6]{sketch_shapiro_perso.pdf}}%
    \put(0.65901387,0.02237454){\color[rgb]{0,0,0}\makebox(0,0)[lb]{\smash{$\zeta_{\textsc{hl}}$}}}%
    \put(0.56545288,0.42092282){\color[rgb]{0,0,0}\makebox(0,0)[lb]{\smash{$\mathbf{v_c}(0)$}}}%
    \put(0.54417873,0.24549075){\color[rgb]{0,0,0}\makebox(0,0)[lb]{\smash{$-\mathbf{v_c}(0)$}}}%
    \put(0.2181301,0.25655227){\color[rgb]{0,0,0}\makebox(0,0)[lb]{\smash{$-\mathbf{v_c}(-\zeta_{\textsc{hl}})$}}}%
    \put(0.89437934,0.23929498){\color[rgb]{0,0,0}\makebox(0,0)[lb]{\smash{$-\mathbf{v_c}(\zeta_{\textsc{hl}})$}}}%
    \put(0.22708555,0.34176033){\color[rgb]{0,0,0}\makebox(0,0)[lb]{\smash{$\sim\mathbf{v_w}(0)$}}}%
    \put(0.87369826,0.35362472){\color[rgb]{0,0,0}\makebox(0,0)[lb]{\smash{$\sim\mathbf{v_w}(0)$}}}%
    \put(0.564948,0.34148461){\color[rgb]{0,0,0}\makebox(0,0)[lb]{\smash{$\sim\mathbf{v_w}(0)$}}}%
    \put(0.40667205,0.21017324){\color[rgb]{0,0,0}\makebox(0,0)[lb]{\smash{$\mathbf{v_{\text{rel}}}(0)$}}}%
    \put(0,0){\includegraphics[width=\unitlength,page=7]{sketch_shapiro_perso.pdf}}%
    \put(0.00735361,0.5872705){\color[rgb]{0,0,0}\rotatebox{-36.9889194}{\makebox(0,0)[lb]{\smash{to the star}}}}%
  \end{picture}%
\endgroup%

%% file: sketch_2.pdf_tex
\begingroup%
  \makeatletter%
  \providecommand\color[2][]{%
    \errmessage{(Inkscape) Color is used for the text in Inkscape, but the package 'color.sty' is not loaded}%
    \renewcommand\color[2][]{}%
  }%
  \providecommand\transparent[1]{%
    \errmessage{(Inkscape) Transparency is used (non-zero) for the text in Inkscape, but the package 'transparent.sty' is not loaded}%
    \renewcommand\transparent[1]{}%
  }%
  \providecommand\rotatebox[2]{#2}%
  \ifx\svgwidth\undefined%
    \setlength{\unitlength}{1095.12001953bp}%
    \ifx\svgscale\undefined%
      \relax%
    \else%
      \setlength{\unitlength}{\unitlength * \real{\svgscale}}%
    \fi%
  \else%
    \setlength{\unitlength}{\svgwidth}%
  \fi%
  \global\let\svgwidth\undefined%
  \global\let\svgscale\undefined%
  \makeatother%
  \begin{picture}(1,0.62400467)%
    \put(0,0){\includegraphics[width=\unitlength,page=1]{sketch_2.pdf}}%
    \put(0.58648398,0.04915428){\color[rgb]{0,0,0}\makebox(0,0)[lb]{\smash{a}}}%
    \put(0.34877365,0.27006023){\color[rgb]{0,0,0}\makebox(0,0)[lb]{\smash{$\mathbf{R_1}$}}}%
    \put(0.79952753,0.26950934){\color[rgb]{0,0,0}\makebox(0,0)[lb]{\smash{$\mathbf{R_2}$}}}%
    \put(0.86400846,0.31533177){\color[rgb]{0,0,0}\rotatebox{-26.53238967}{\makebox(0,0)[lb]{\smash{$R_{\text{out}}$}}}}%
    \put(0.7232084,0.44561326){\color[rgb]{0,0,0}\makebox(0,0)[lb]{\smash{}}}%
    \put(0,0){\includegraphics[width=\unitlength,page=2]{sketch_2.pdf}}%
    \put(0.68953888,0.34865487){\color[rgb]{0,0,0}\makebox(0,0)[lb]{\smash{$\mathbf{r}$}}}%
    \put(0,0){\includegraphics[width=\unitlength,page=3]{sketch_2.pdf}}%
    \put(0.58892443,0.25624999){\color[rgb]{0,0,0}\makebox(0,0)[lb]{\smash{CM}}}%
    \put(0,0){\includegraphics[width=\unitlength,page=4]{sketch_2.pdf}}%
    \put(0.62019884,0.4290541){\color[rgb]{0,0,0}\makebox(0,0)[lb]{\smash{$\mathbf{r_1}$}}}%
    \put(0.78225684,0.41347123){\color[rgb]{0,0,0}\makebox(0,0)[lb]{\smash{$\mathbf{r_2}$}}}%
    \put(0,0){\includegraphics[width=\unitlength,page=5]{sketch_2.pdf}}%
    \put(0.92865702,0.60121619){\color[rgb]{0,0,0}\makebox(0,0)[lb]{\smash{$\mathbf{\hat{y}}$}}}%
    \put(0.701293,0.51501204){\color[rgb]{0,0,0}\makebox(0,0)[lb]{\smash{}}}%
    \put(0.76703921,0.50405434){\color[rgb]{0,0,0}\makebox(0,0)[lb]{\smash{}}}%
    \put(0.97838362,0.5533321){\color[rgb]{0,0,0}\makebox(0,0)[lb]{\smash{$\mathbf{\hat{x}}$}}}%
    \put(0.89621625,0.51376508){\color[rgb]{0,0,0}\makebox(0,0)[lb]{\smash{$\mathbf{\hat{z}}$}}}%
    \put(0,0){\includegraphics[width=\unitlength,page=6]{sketch_2.pdf}}%
    \put(0.73750943,0.58521032){\color[rgb]{0,0,0}\makebox(0,0)[lb]{\smash{$\Omega$}}}%
    \put(0.71986421,0.45979558){\color[rgb]{0,0,0}\makebox(0,0)[lb]{\smash{test-mass}}}%
    \put(0,0){\includegraphics[width=\unitlength,page=7]{sketch_2.pdf}}%
  \end{picture}%
\endgroup%

%% file: meshing.pdf_tex
\begingroup%
  \makeatletter%
  \providecommand\color[2][]{%
    \errmessage{(Inkscape) Color is used for the text in Inkscape, but the package 'color.sty' is not loaded}%
    \renewcommand\color[2][]{}%
  }%
  \providecommand\transparent[1]{%
    \errmessage{(Inkscape) Transparency is used (non-zero) for the text in Inkscape, but the package 'transparent.sty' is not loaded}%
    \renewcommand\transparent[1]{}%
  }%
  \providecommand\rotatebox[2]{#2}%
  \ifx\svgwidth\undefined%
    \setlength{\unitlength}{595.27559055bp}%
    \ifx\svgscale\undefined%
      \relax%
    \else%
      \setlength{\unitlength}{\unitlength * \real{\svgscale}}%
    \fi%
  \else%
    \setlength{\unitlength}{\svgwidth}%
  \fi%
  \global\let\svgwidth\undefined%
  \global\let\svgscale\undefined%
  \makeatother%
  \begin{picture}(1,0.55238095)%
    \put(0,0){\includegraphics[width=\unitlength,page=1]{meshing.pdf}}%
    \put(0.52288888,0.12597715){\color[rgb]{0,0,0}\makebox(0,0)[lb]{\smash{$\hat{x}$}}}%
    \put(0,0){\includegraphics[width=\unitlength,page=2]{meshing.pdf}}%
    \put(0.62587673,0.22671978){\color[rgb]{0,0,0}\makebox(0,0)[lb]{\smash{$\theta_0$}}}%
    \put(0.58578361,0.13942861){\color[rgb]{0,0,0}\makebox(0,0)[lb]{\smash{$\pi/4$}}}%
    \put(0.40274926,0.09528127){\color[rgb]{0,0,0}\makebox(0,0)[lb]{\smash{$-\pi/4$}}}%
    \put(0.45345306,0.19956825){\color[rgb]{0,0,0}\makebox(0,0)[lb]{\smash{$\hat{y}$}}}%
    \put(0.31215633,0.49489012){\color[rgb]{0,0,0}\makebox(0,0)[lb]{\smash{$\hat{z}$}}}%
    \put(0,0){\includegraphics[width=\unitlength,page=3]{meshing.pdf}}%
    \put(0.4950133,0.0256344){\color[rgb]{0,0,0}\makebox(0,0)[lb]{\smash{$a$}}}%
    \put(0,0){\includegraphics[width=\unitlength,page=4]{meshing.pdf}}%
  \end{picture}%
\endgroup%

%% file: planar_shock.pdf_tex
\begingroup%
  \makeatletter%
  \providecommand\color[2][]{%
    \errmessage{(Inkscape) Color is used for the text in Inkscape, but the package 'color.sty' is not loaded}%
    \renewcommand\color[2][]{}%
  }%
  \providecommand\transparent[1]{%
    \errmessage{(Inkscape) Transparency is used (non-zero) for the text in Inkscape, but the package 'transparent.sty' is not loaded}%
    \renewcommand\transparent[1]{}%
  }%
  \providecommand\rotatebox[2]{#2}%
  \ifx\svgwidth\undefined%
    \setlength{\unitlength}{980.78740157bp}%
    \ifx\svgscale\undefined%
      \relax%
    \else%
      \setlength{\unitlength}{\unitlength * \real{\svgscale}}%
    \fi%
  \else%
    \setlength{\unitlength}{\svgwidth}%
  \fi%
  \global\let\svgwidth\undefined%
  \global\let\svgscale\undefined%
  \makeatother%
  \begin{picture}(1,0.7716763)%
    \put(0,0){\includegraphics[width=\unitlength,page=1]{planar_shock.pdf}}%
    \put(0.04090885,0.65242325){\color[rgb]{0,0,0}\makebox(0,0)[lb]{\smash{U P S T R E A M}}}%
    \put(0.56408448,0.65405387){\color[rgb]{0,0,0}\makebox(0,0)[lb]{\smash{D O W N S T R E A M}}}%
    \put(0.07935942,0.6864483){\color[rgb]{0,0,0}\makebox(0,0)[lb]{\smash{}}}%
    \put(0.02436083,0.41832893){\color[rgb]{0,0,0}\makebox(0,0)[lb]{\smash{$\rho_1$}}}%
    \put(0,0){\includegraphics[width=\unitlength,page=2]{planar_shock.pdf}}%
    \put(0.02540783,0.32849722){\color[rgb]{0,0,0}\makebox(0,0)[lb]{\smash{$v_1$}}}%
    \put(0.02983337,0.2338859){\color[rgb]{0,0,0}\makebox(0,0)[lb]{\smash{$P_1$}}}%
    \put(0.8450287,0.42356976){\color[rgb]{0,0,0}\makebox(0,0)[lb]{\smash{$\rho_2$}}}%
    \put(0.84607577,0.3337381){\color[rgb]{0,0,0}\makebox(0,0)[lb]{\smash{$v_2$}}}%
    \put(0.85050123,0.23912677){\color[rgb]{0,0,0}\makebox(0,0)[lb]{\smash{$P_2$}}}%
    \put(0.03666911,0.61008816){\color[rgb]{0,0,0}\makebox(0,0)[lb]{\smash{S U P E R S O N I C}}}%
    \put(0.67462235,0.61195192){\color[rgb]{0,0,0}\makebox(0,0)[lb]{\smash{S U B S O N I C}}}%
    \put(0.48318813,0.09194508){\color[rgb]{0,0,0}\makebox(0,0)[lb]{\smash{$\mathbf{n}$}}}%
    \put(0,0){\includegraphics[width=\unitlength,page=3]{planar_shock.pdf}}%
  \end{picture}%
\endgroup%

%% file: oblique_shock.pdf_tex
\begingroup%
  \makeatletter%
  \providecommand\color[2][]{%
    \errmessage{(Inkscape) Color is used for the text in Inkscape, but the package 'color.sty' is not loaded}%
    \renewcommand\color[2][]{}%
  }%
  \providecommand\transparent[1]{%
    \errmessage{(Inkscape) Transparency is used (non-zero) for the text in Inkscape, but the package 'transparent.sty' is not loaded}%
    \renewcommand\transparent[1]{}%
  }%
  \providecommand\rotatebox[2]{#2}%
  \ifx\svgwidth\undefined%
    \setlength{\unitlength}{980.78740157bp}%
    \ifx\svgscale\undefined%
      \relax%
    \else%
      \setlength{\unitlength}{\unitlength * \real{\svgscale}}%
    \fi%
  \else%
    \setlength{\unitlength}{\svgwidth}%
  \fi%
  \global\let\svgwidth\undefined%
  \global\let\svgscale\undefined%
  \makeatother%
  \begin{picture}(1,0.7716763)%
    \put(0,0){\includegraphics[width=\unitlength,page=1]{oblique_shock.pdf}}%
    \put(0.01818737,0.72210248){\color[rgb]{0,0,0}\makebox(0,0)[lb]{\smash{U P S T R E A M}}}%
    \put(0.58028365,0.07244943){\color[rgb]{0,0,0}\makebox(0,0)[lb]{\smash{D O W N S T R E A M}}}%
    \put(0.07935942,0.6864483){\color[rgb]{0,0,0}\makebox(0,0)[lb]{\smash{}}}%
    \put(0.00152204,0.45748114){\color[rgb]{0,0,0}\makebox(0,0)[lb]{\smash{$\rho_1$}}}%
    \put(0,0){\includegraphics[width=\unitlength,page=2]{oblique_shock.pdf}}%
    \put(0.00256905,0.36764944){\color[rgb]{0,0,0}\makebox(0,0)[lb]{\smash{$\mathbf{v_1}$}}}%
    \put(0.00210055,0.27303811){\color[rgb]{0,0,0}\makebox(0,0)[lb]{\smash{$P_1$}}}%
    \put(0.74877947,0.46272199){\color[rgb]{0,0,0}\makebox(0,0)[lb]{\smash{$\rho_2$}}}%
    \put(0.74982654,0.37289031){\color[rgb]{0,0,0}\makebox(0,0)[lb]{\smash{$\mathbf{v_2}$}}}%
    \put(0.75425204,0.27827899){\color[rgb]{0,0,0}\makebox(0,0)[lb]{\smash{$P_2$}}}%
    \put(0,0){\includegraphics[width=\unitlength,page=3]{oblique_shock.pdf}}%
    \put(0.62258364,0.36763888){\color[rgb]{0,0,0}\makebox(0,0)[lb]{\smash{$\theta$}}}%
    \put(0,0){\includegraphics[width=\unitlength,page=4]{oblique_shock.pdf}}%
    \put(0.4065891,0.07924234){\color[rgb]{0,0,0}\makebox(0,0)[lb]{\smash{$\beta$}}}%
    \put(0,0){\includegraphics[width=\unitlength,page=5]{oblique_shock.pdf}}%
    \put(0.01708865,0.67849015){\color[rgb]{0,0,0}\makebox(0,0)[lb]{\smash{S U P E R S O N I C}}}%
    \put(0.69506018,0.02839861){\color[rgb]{0,0,0}\makebox(0,0)[lb]{\smash{S U B S O N I C}}}%
    \put(0,0){\includegraphics[width=\unitlength,page=6]{oblique_shock.pdf}}%
    \put(0.61571919,0.71222513){\color[rgb]{0,0,0}\makebox(0,0)[lb]{\smash{$\mathbf{n}$}}}%
  \end{picture}%
\endgroup%

%% file: spherical_triangles.pdf_tex
\begingroup%
  \makeatletter%
  \providecommand\color[2][]{%
    \errmessage{(Inkscape) Color is used for the text in Inkscape, but the package 'color.sty' is not loaded}%
    \renewcommand\color[2][]{}%
  }%
  \providecommand\transparent[1]{%
    \errmessage{(Inkscape) Transparency is used (non-zero) for the text in Inkscape, but the package 'transparent.sty' is not loaded}%
    \renewcommand\transparent[1]{}%
  }%
  \providecommand\rotatebox[2]{#2}%
  \ifx\svgwidth\undefined%
    \setlength{\unitlength}{1077.16535433bp}%
    \ifx\svgscale\undefined%
      \relax%
    \else%
      \setlength{\unitlength}{\unitlength * \real{\svgscale}}%
    \fi%
  \else%
    \setlength{\unitlength}{\svgwidth}%
  \fi%
  \global\let\svgwidth\undefined%
  \global\let\svgscale\undefined%
  \makeatother%
  \begin{picture}(1,0.38684211)%
    \put(0,0){\includegraphics[width=\unitlength,page=1]{spherical_triangles.pdf}}%
    \put(0.16317976,0.35521801){\color[rgb]{0,0,0}\makebox(0,0)[lb]{\smash{A}}}%
    \put(0.00517879,0.21435463){\color[rgb]{0,0,0}\makebox(0,0)[lb]{\smash{B}}}%
    \put(0.16380022,0.0180117){\color[rgb]{0,0,0}\makebox(0,0)[lb]{\smash{C}}}%
    \put(0.31256721,0.2144131){\color[rgb]{0,0,0}\makebox(0,0)[lb]{\smash{D}}}%
    \put(0.80359064,0.32678364){\color[rgb]{0,0,0}\makebox(0,0)[lb]{\smash{A'}}}%
    \put(0.593557,0.21337939){\color[rgb]{0,0,0}\makebox(0,0)[lb]{\smash{B'}}}%
    \put(0.90629402,0.02079506){\color[rgb]{0,0,0}\makebox(0,0)[lb]{\smash{C'}}}%
    \put(0.96144129,0.2296872){\color[rgb]{0,0,0}\makebox(0,0)[lb]{\smash{D'}}}%
    \put(0,0){\includegraphics[width=\unitlength,page=2]{spherical_triangles.pdf}}%
    \put(0.78453945,0.28234649){\color[rgb]{0,0,0}\makebox(0,0)[lb]{\smash{}}}%
    \put(0.1771852,0.21007459){\color[rgb]{0,0,0}\makebox(0,0)[lb]{\smash{O}}}%
    \put(0.80146901,0.20604342){\color[rgb]{0,0,0}\makebox(0,0)[lb]{\smash{O'}}}%
    \put(0,0){\includegraphics[width=\unitlength,page=3]{spherical_triangles.pdf}}%
    \put(0.78147957,0.27474121){\color[rgb]{0,0,0}\makebox(0,0)[lb]{\smash{$\alpha$}}}%
    \put(0.64473688,0.20143565){\color[rgb]{0,0,0}\makebox(0,0)[lb]{\smash{$\beta$}}}%
    \put(0.85034449,0.07493583){\color[rgb]{0,0,0}\makebox(0,0)[lb]{\smash{$\gamma$}}}%
    \put(0.68495937,0.29283898){\color[rgb]{0,0,0}\makebox(0,0)[lb]{\smash{$c$}}}%
    \put(0.70325206,0.09044714){\color[rgb]{0,0,0}\makebox(0,0)[lb]{\smash{$a$}}}%
    \put(0.87906507,0.18800812){\color[rgb]{0,0,0}\makebox(0,0)[lb]{\smash{$b$}}}%
  \end{picture}%
\endgroup%